\documentclass[12pt]{book}
\usepackage[utf8]{inputenc}

\usepackage[bookmarks=true,hyperfigures=true]{hyperref}

\usepackage[a4paper,text={155mm,218mm},centering]{geometry}

\usepackage[font=small,labelfont=bf,width=0.85\textwidth]{caption}

\usepackage{amsmath,amssymb}

\usepackage{bbm}
\DeclareSymbolFont{bbold}{U}{bbold}{m}{n}
\DeclareSymbolFontAlphabet{\mathbbold}{bbold} 

\usepackage{color}
\usepackage{graphicx}

\usepackage{array} 
\usepackage{pdflscape}  
\usepackage{nicefrac}		

\usepackage[makeroom,Smaller]{cancel}
\usepackage{ulem} 
\usepackage{verbatim} 
\usepackage{tikz}

\usepackage{yfonts}
\usepackage{mathrsfs}	
\DeclareMathAlphabet{\mathpzc}{OT1}{pzc}{m}{it}

\usepackage{tocbibind}

\allowdisplaybreaks[3]

\makeatletter
\let\old@startsection=\@startsection
\renewcommand{\@startsection}[6]{\old@startsection{#1}{#2}{#3}{#4}{#5}{#6\mathversion{bold}}}
\makeatother

\makeatletter
\newlength{\apb@width}
\newcommand{\autoparbox}[2][c]{\settowidth{\apb@width}{#2}\parbox[#1]{\apb@width}{#2}}

\makeatother

\ifx\href\asklfhas\newcommand{\href}[2]{#2}\fi

\makeatletter
\def\Left#1#2\Right{\begingroup%
   \def\ts@r{\nulldelimiterspace=0pt \mathsurround=0pt}%
   \let\@hat=#1%
   \def\sht@im{#2}%
   \def\@t{{\mathchoice{\def\@fen{\displaystyle}\k@fel}%
          {\def\@fen{\textstyle}\k@fel}%
          {\def\@fen{\scriptstyle}\k@fel}%
          {\def\@fen{\scriptscriptstyle}\k@fel}}}%
   \def\g@rin{\ts@r\left\@hat\vphantom{\sht@im}\right.}%
   \def\k@fel{\setbox0=\hbox{$\@fen\g@rin$}\hbox{%
      $\@fen \kern.3875\wd0 \copy0 \kern-.3875\wd0%
      \llap{\copy0}\kern.3875\wd0$}}%
      \def\pt@h{\mathopen\@t}\pt@h\sht@im%
      \Right}%
\def\Right#1{\let\@hat=#1%
   \def\st@m{\mathclose\@t}%
   \st@m\endgroup}
\makeatother

\newcommand{\lleft}[1]{\Left{#1}\Right.}
\newcommand{\rright}[1]{\Left.\Right{#1}}

\def\dotuline{\bgroup
  \ifdim\ULdepth=\maxdimen  
   \settodepth\ULdepth{(j}\advance\ULdepth.4pt\fi
  \markoverwith{\begingroup
  \advance\ULdepth0.08ex
  \lower\ULdepth\hbox{\kern.15em .\kern.1em}%
  \endgroup}\ULon}

\def\dashuline{\bgroup
  \ifdim\ULdepth=\maxdimen  
   \settodepth\ULdepth{(j}\advance\ULdepth.4pt\fi
  \markoverwith{\kern.15em
  \vtop{\kern\ULdepth \hrule width .3em}%
  \kern.15em}\ULon}

\newcommand{\scr}{\mathscr} 
\newcommand{\pzc}{\mathpzc} 

\newcommand{\hypref}[2]{\ifx\href\asklfhas #2\else\href{#1}{#2}\fi}

\newcommand{\chapref}[1]{Chap.~\ref{#1}}

\newcommand{\secref}[1]{Sec.~\ref{#1}}

\newcommand{\ssecref}[1]{Subsec.~\ref{#1}}

\newcommand{\appref}[1]{App.~\ref{#1}}

\newcommand{\figref}[1]{Fig.~\ref{#1}}

\newcommand{\nn}{\nonumber}

\newcommand{\be}{\begin{equation}}
\newcommand{\ee}{\end{equation}}
\newcommand{\ba}{\begin{aligned}}
\newcommand{\ea}{\end{aligned}}
\newcommand{\bea}{\begin{eqnarray}}
\newcommand{\eea}{\end{eqnarray}}
\newcommand{\bean}{\begin{eqnarray*}}
\newcommand{\eean}{\end{eqnarray*}}

\newcommand{\f}{\frac}
\newcommand{\sfrac}[2]{{\textstyle\frac{#1}{#2}}}		

\renewcommand{\sf}[2]{{\textstyle\frac{#1}{#2}}}		

\newcommand{\qqquad}{{\qquad\qquad\qquad}}

\long\def\symbolfootnote[#1]#2{\begingroup
\def\thefootnote{\fnsymbol{footnote}}\footnote[#1]{#2}\endgroup}

\definecolor{pink}{rgb}{0.7,0,0.7}

\definecolor{grey}{rgb}{0.4,0.4,0.5}
\definecolor{darkgreen}{rgb}{0,0.5,0}
\definecolor{darkred}{rgb}{0.6,0.0,0}
\definecolor{lightbrown}{rgb}{1,0.9,0.8}
\definecolor{brown}{rgb}{0.6,0.3,0.3}
\definecolor{darkblue}{rgb}{0,0,0.8}
\definecolor{darkmagenta}{rgb}{0.5,0,0.5}

\newcommand{\abs}[1]{|#1|}
\newcommand{\Abs}[1]{\left|#1\right|}

\newcommand{\la}{\langle}
\newcommand{\ra}{\rangle}

\newcommand{\vev}[1]{\langle#1\rangle}

\newcommand{\bra}[1]{\langle #1|}	

\newcommand{\ket}[1]{|#1\rangle}

\renewcommand{\a}{\alpha}

\renewcommand{\b} {\beta}

\newcommand{\g}{\gamma}
\newcommand{\G}{\Gamma} 
\newcommand{\GG}{\ensuremath{\text{I}\hspace{-.23em}\Gamma}}

\renewcommand{\d}{\delta}
\newcommand{\D}{\Delta}

\newcommand{\e}{\epsilon}
\newcommand{\eps}{\epsilon}

\newcommand{\z}{\zeta}

\renewcommand{\th}{\theta}
\newcommand{\vt}{\vartheta}
\newcommand{\Th}{\Theta}

\renewcommand{\k}{\kappa}

\renewcommand{\l}{\lambda}

\renewcommand{\L}{\Lambda}

\newcommand{\m}{\mu}

\newcommand{\n}{\nu}

\renewcommand{\r}{\rho}

\newcommand{\s}{\sigma}

\renewcommand{\S}{\Sigma}

\newcommand{\Ups}{\Upsilon}

\newcommand{\om}{\omega}
\newcommand{\Om}{\Omega}

\newcommand{\da}{{\dot\alpha}}
\newcommand{\db}{{\dot\beta}}
\newcommand{\dg}{{\dot\gamma}}
\newcommand{\dr}{{\dot\rho}}

\newcommand{\ta}{{\tilde{a}}}
\newcommand{\tb}{{\tilde{b}}}
\newcommand{\tc}{{\tilde{c}}}

\newcommand{\tV}{{\tilde{V}}}		
\newcommand{\tW}{{\tilde{W}}}

\newcommand{\ts}{\tilde{\sigma}}

\newcommand{\tpm}{{\tilde{\pm}}}

\newcommand{\ul}[1]{\uline{#1}}

\newcommand{\ua}{{\underline{\alpha}}}
\newcommand{\ub}{{\underline{\beta}}}
\newcommand{\ug}{{\underline{\gamma}}}
\newcommand{\ud}{{\underline{\delta}}}

\newcommand{\uA}{{\underline{A}}}
\newcommand{\uB}{{\underline{B}}}
\newcommand{\uC}{{\underline{C}}}
\newcommand{\uD}{{\underline{D}}}

\newcommand{\calC}{\mathcal{C}}

\newcommand{\calE}{\mathcal{E}}

\newcommand{\calF}{\mathcal{F}}	

\newcommand{\cJ}{\mathcal{J}}

\newcommand{\calH}{\mathcal{H}}		
\newcommand{\cH}{\mathcal H}		

\newcommand{\cK}{\mathcal{K}}		
\newcommand{\Kk}{{\mathcal{K}}}		

\newcommand{\calL}{\mathcal{L}}		
\newcommand{\cL}{\mathcal{L}}		

\newcommand{\cM}{\mathcal{M}}		

\newcommand{\calN}{\mathcal{N}}		

\newcommand{\calO}{\mathcal{O}}		

\newcommand{\calP}{\mathcal{P}}		
\newcommand{\cP}{\mathcal{P}}		

\newcommand{\calR}{\mathcal{R}}

\newcommand{\calV}{\mathcal{V}}		
\newcommand{\cV}{\mathcal{V}}		

\newcommand{\cW}{\mathcal{W}}

\newcommand{\X}{\mathcal X}		

\newcommand{\calY}{\mathcal{Y}}
\newcommand{\Y}{\mathcal Y}		

\renewcommand{\v}{\vec}

\newcommand{\vX}{\vec{X}}

\newcommand{\vXr}{\vec{X}_{\rm re}}
\newcommand{\vXi}{\vec{X}_{\rm im}}

\newcommand{\vXz}{\vec{X}_{0}}

\newcommand{\vx}{\vec{x}}
\newcommand{\vxpr}{\vec{x}^{\,}{}'{}}
\newcommand{\vxz}{\vec{x}_{0}}

\newcommand{\cx}{\mathpzc{x}}
\newcommand{\vcx}{\vec{\mathpzc{x}}}
\newcommand{\vcxpr}{\vec{\mathpzc{x}}^{\,}{}'{}}
\newcommand{\vcxz}{\vec{\mathpzc{x}}_{\,0}}

\newcommand{\vY}{\vec{Y}}

\newcommand{\vy}{\vec{y}}
\newcommand{\vypr}{\vec{y}^{\,}{}'{}}
\newcommand{\vyz}{\vec{y}_{0}}

\newcommand{\cy}{\mathpzc{y}}
\newcommand{\vcy}{\vec{\mathpzc{y}}}
\newcommand{\vcypr}{\vec{\mathpzc{y}}^{\,}{}'{}}
\newcommand{\vcyz}{\vec{\mathpzc{y}}_{0}}

\newcommand{\vz}{\vec{z}}

\newcommand{\vP}{\vec{P}}

\newcommand{\vPr}{\vec{P}_{\rm re}}
\newcommand{\vPi}{\vec{P}_{\rm im}}

\newcommand{\vPz}{\vec{P}_{0}}

\newcommand{\vp}{\vec{p}}
\newcommand{\vpx}{\vec{p}_x}
\newcommand{\vpy}{\vec{p}_y}

\newcommand{\vpxz}{\vec{p}_{x,0}}
\newcommand{\vpyz}{\vec{p}_{y,0}}

\newcommand{\vcpx}{\vec{\mathpzc{p}}_\cx}
\newcommand{\vcpy}{\vec{\mathpzc{p}}_\cy}

\newcommand{\vcpxz}{\vec{\mathpzc{p}}_{\cx,0}}
\newcommand{\vcpyz}{\vec{\mathpzc{p}}_{\cy,0}}

\newcommand{\vth}{\vec{\theta}}
\newcommand{\vet}{\vec{\eta}}

\newcommand{\va}{\vec{a}}
\newcommand{\vb}{\vec{b}}
\newcommand{\vc}{\vec{c}}

\newcommand{\vg}{\vec{\gamma}}

\newcommand{\alg}[1]{\mathfrak{#1}}

\newcommand{\su}{\alg{su}}
\newcommand{\sls}{\alg{sl}}
\newcommand{\gls}{\alg{gl}}
\newcommand{\psu}{\alg{psu}}
\newcommand{\un}{\alg{u}}
\newcommand{\so}{\alg{so}}

\newcommand{\fkT}{{\ensuremath{\mathfrak{t}}}}

\newcommand{\SU}{\mathrm{SU}}
\newcommand{\SL}{\mathrm{SL}}
\newcommand{\GL}{\mathrm{GL}}
\newcommand{\PSU}{\mathrm{PSU}}
\newcommand{\Un}{\mathrm{U}}
\newcommand{\SO}{\mathrm{SO}}

\newcommand{\diag}{\mathop{\mathrm{diag}}}
\newcommand{\Span}{\mathop{\mathrm{span}}}

\newcommand{\tr}{\mathop{\mathrm{tr}}}		
\newcommand{\str}{\mathop{\mathrm{str}}}		
\newcommand{\sdet}{\mathop{\mathrm{sdet}}}

\renewcommand{\Re}{\mathop{\mathrm{Re}}}

\newcommand{\sign}{\mathop{\mathrm{sgn}}}

\newcommand{\ord}[1]{\mathcal{O}(#1)}
\newcommand{\Ord}[1]{\mathcal{O}\left(#1\right)}

\newcommand{\dd}{{\mathrm{d}}}		
\newcommand{\p}{\partial}		

\newcommand{\mI}{\mathbbm{1}}		

\newcommand{\mbb}[1]{{\ensuremath{\mathbb{#1}}}}

\newcommand{\Reals}{\ensuremath{\mathbb{R }}}
\newcommand{\Complex}{\ensuremath{\mathbb{C }}}
\newcommand{\Integers}{\ensuremath{\mathbb{Z }}}

\newcommand{\AdSxS}{{\ensuremath{\mathrm{AdS}_5 \times \mathrm{S}^5}}}

\newcommand{\NfourSYM}{{$\mathcal{N}$=4 SYM}}
\newcommand{\gYM}{{\ensuremath{g_\mathrm{YM}}}}

\newcommand\bp{\hbox{\large $\pi$}}		

\newcommand{\cl}{\ensuremath{\mathrm{cl}}}		

\newcommand{\Ms}{M_\alg{s}}		

\newcommand{\sQ}{Q}

\newcommand{\pzcH}{\mathpzc{H}}	

\newcommand{\pzG}{\mathpzc{G}} 
\newcommand{\scG}{\mathscr{G}}
\newcommand{\pzg}{\mathpzc{g}} 
 
\newcommand{\nz}{{0\!\!\!/}}

\newcommand{\AdSxSheader}{{AdS\texorpdfstring{${}_5 \times$}{5x}S\texorpdfstring{${}^5$}{5}}}
\newcommand{\AdSthreexSthreeheader}{{AdS\texorpdfstring{${}_3 \times$}{3x}S\texorpdfstring{${}^3$}{3}}}

\def\ov{\over}

\title{Spectrum and Quantum Symmetries of the \texorpdfstring{$\textrm{AdS}_5 \times \textrm{S}^5$}{AdSfiveS} Superstring}
\author{Martin Heinze}
\date{\today}

\begin{document}
\pagenumbering{gobble}

\begin{titlepage}
 \begin{centering}
	\vspace{5 cm}

	\textsc{\Huge Spectrum and \\ Quantum Symmetries of the\\[.1cm] $\textrm{AdS}_5 \times \textrm{S}^5$ Superstring}\\[1.5cm]

	\textsc{\Large Dissertation}\\
	{\large zur Erlangung des akademischen Grades\\
		\textsc{doctor rerum naturalium}\\
		(Dr. rer. nat.)\\
		im Fach Physik\\[1.2cm]

		eingereicht an der\\ 
		Mathematisch-Naturwissenschaftlichen Fakult{\"a}t\\
		der Humboldt-Universit\"at zu Berlin\\[1cm]

		von \\
		Herrn Dipl.-Phys. Martin Heinze\\
		$~$\\[1cm]

		Pr\"asident der Humboldt-Universit{\"a}t zu Berlin:\\
		Prof. Dr. Jan-Hendrik Olbertz\\
		Dekan der Mathematisch-Naturwissenschaftlichen Fakult{\"a}t:\\
		Prof. Dr. Elmar Kulke}\\[1cm]

\end{centering}
	$\ $\\
	{\large
	\begin{tabular}{l l}
	Gutachter: & 1. Prof. Dr. Jan Plefka\\
	& 2. PD Dr. Thomas Klose\\
	& 3. Prof. Dr. Gleb Arutyunov
	\end{tabular}\\[.1cm]

	\hspace{-.4cm}eingreicht am: 26.09.2014\\[.1cm]
	$\phantom{t}$Tag der m{\"u}ndlichen Pr{\"u}fung: 18.12.2014}

\end{titlepage}


\frontmatter
\pagenumbering{Roman}
\pagestyle{plain}

\setcounter{tocdepth}{1}
\tableofcontents

\chapter{Executive Summary} \label{chap:ExSum}

	The AdS/CFT correspondence \cite{Maldacena:1997re, Witten:1998qj, Gubser:1998bc} has dramatically changed our view on both gauge and string theories, triggering various applications in most of modern theoretical physics. For the initial duality pair, the duality between {\NfourSYM} and the $\AdSxS$ superstring, the underlying quantum integrability has been observed \cite{Minahan:2002ve, Beisert:2003tq}, initiating extensive studies and fascinating insights during the last decade. Assuming the quantum integrability to hold generally allowed to devise powerful methods, which recently culminating in the asserted solution of the spectral problem by the quantum spectral curve \cite{Gromov:2013pga, Gromov:2014caa}.

	Despite all progress due to integrability based methods, our understanding of quantization of the $\AdSxS$ superstring from first principles is still very unsatisfactory. For long semiclassical string solutions perturbative quantization is possible at large coupling and has been a prominent tool \cite{Gubser:2002tv, Frolov:2002av, Frolov:2003qc} since the early days of the AdS/CFT integrability. But for short strings, i.e., finitely excited string states, the perturbative expansion of the Lagrangian density in strong 't Hooft coupling, $\l \gg 1$, formally breaks down, preventing a perturbative quantization. This is the more surprising as in the strong coupling limit the curvature $\AdSxS$ appears to become negligible and one would expect to quantize the string as a perturbation to the well understood string theory in flat space.
	This idea has been proposed more than a decade ago \cite{Gubser:2002tv}, yielding the energy spectrum of a string of level $n$ at leading order, $E = 2\sqrt{n} \l^{1/4} + \ord{\l^0}$, but since then there has been very little progress for short strings.

	The goal of this thesis is to investigate possible routes to take perturbative quantization of the $\AdSxS$ superstring beyond the long-string paradigm, which is, to derive the energy spectrum of short strings to higher accuracy in large 't Hooft coupling. Equally, our aim is to gain better appreciation of the quantum symmetries at play.
	This long-standing problem seems more relevant than ever as currently there are no quantities derived by first principles, which the machinery relying on the surmised quantum integrability could be tested against.
  
	Our study will mostly concentrate on the lowest excited string states, having level $n=1$, which are supposedly dual to the members of the Konishi supermultiplet and for which the energy spectrum, respectively, the scaling dimension has been predicted to order $\ord{\l^{-5/4}}$.

	We start by discussing an asserted derivation of the Konishi anomalous dimension up to $\ord{\l^{-1/4}}$ in the framework of pure spinor superstring theory \cite{Vallilo:2011fj} , where we argue that to follow the work it suffices to consider the bosonic subsector only. Doing so, we find several inconsistencies, most of them due to disregard of the particular scaling of the zero modes already observed in \cite{Passerini:2010xc}.

	The acquired expertise suggests that one still might be able to obtain the energy spectrum to this order  by quantizing semiclassical string solutions whilst keeping tracked of the zero modes. We present the work \cite{Frolov:2013lva}, where working in the bosonic subsector and utilizing static gauge \cite{Jorjadze:2012iy} we constructed a so-called single-mode string solution, a generalization of the pulsating string \cite{deVega:1994yz, Minahan:2002rc}, which shows classical integrability as well as invariance under the isometries at the quantum level. Taking supersymmetric corrections heuristically into account, we indeed manage to recover the Konishi anomalous dimension to $\ord{\l^{-1/4}}$.

	This work naturally asks for implementation of static gauge for the full $\AdSxS$ superstring, where however not even static gauge quantization for the flat-space superstring or the $\AdSxS$ superparticle are understood. Ignoring such subtleties, we proceed in analogy to the setup in uniform light-cone gauge \cite{Frolov:2006cc} and find elegant expressions for the Lagrangian density as well as for the supercharges. We furthermore contemplate on a generalization of the work \cite{Frolov:2013lva} including fermions.

	Not surprisingly, canonical quantization is obstructed by a non-canonical kinetic term for the fermions, which leads us to simplify the problem by looking only at the zero modes, the $\AdSxS$ superparticle. For this, we manage to diagonalize the fermion kinetic term at leading order in fermions. Furthermore, we also propose and test another scheme to derive a transformation to intrinsically canonical variables by utilizing the $\psu(2,2|4)$ symmetry algebra.

	Motivated to acquire better understanding of the $\AdSxS$ superparticle, we conclude by presenting the work \cite{Heinze:2014cga}, where we explored yet another quantization scheme, quantization by use of the isometry group orbits \cite{Alekseev:1988ce}. We consider the bosonic particle and multi-spin solution in ${\rm AdS}_3 \times {\rm S}^3$ and manage to quantize these systems exactly as the isometry generators naturally take a Holstein-Primakoff realization \cite{Holstein:1940zp, Dzhordzhadze:1994np}. Finally, multi-spin solutions corresponding to long- and short-strings are identified.

\chapter{Zusammenfassung} \label{chap:Zus}

	Die Entdeckung der AdS/CFT Hypothese \cite{Maldacena:1997re, Witten:1998qj, Gubser:1998bc} hat unsere Sicht auf sowohl Eich- als auch Stringtheorien grundlegend ver{\"a}ndert, was eine Vielzahl von Anwendungen in vielen Teilbereichen der modernen theoretischen Physik hervorrief. F{\"u}r das Musterbeispiel der Dualit{\"a}t zwischen {\NfourSYM} und dem Typ IIB Superstring in $\AdSxS$ wurde zudem die zugrundeliegende Quantenintegrabilit{\"a}t beobachtet \cite{Minahan:2002ve, Beisert:2003tq}, welche w{\"a}hrend der letzten Jahre zu verschiedenste Applikationen und {\"u}berraschenden Erkenntnissen f{\"u}hrte. Die Annahme, dass die Quantenintegrabilit{\"a}t generell g{\"u}ltig ist, erlaubte die Konstruktion von m{\"a}chtigen Methoden, welche k{\"u}rzlich in der L{\"o}sung des spektralen Problems durch die sogenannte ``quanten-spektrale Kurve'' \cite{Gromov:2013pga, Gromov:2014caa} gipfelten.

	Trotz allen Fortschritts ist unser Verst{\"a}ndnis einer direkten Quantisierung des $\AdSxS$ Superstrings weiterhin mangelhaft. St{\"o}rungstechnische Quantisierung ist f{\"u}r lange, semiklassische Stringl{\"o}sungen m{\"o}glich und wurde schon seit Anfang der AdS/CFT Integrabilit{\"a}t untersucht \cite{Gubser:2002tv, Frolov:2002av, Frolov:2003qc}. F{\"u}r kurze, sprich f{\"u}r endlich angeregte Strings jedoch verliert eine St{\"o}rungsentwicklung der Lagrangedichte in starker 't Hooft-Kopplung, $\l\gg1$, ihre G{\"u}ltigkeit, was eine perturbative Quantisierung verhindert. Dies ist umso {\"u}berraschender, da die Kr{\"u}mmung der $\AdSxS$ Raumzeit bei starker Kopplung vernachl{\"a}ssigbar scheint und man vermuten sollte, dass der $\AdSxS$ String als St{\"o}rung zur wohlbekannten Stringtheorie im flachen Raum gehandhabt werden kann.
	Diese Sichtweise wurde bereits vor einem Jahrzehnt vorgeschlagen \cite{Gubser:2002tv}, wodurch die f{\"u}hrende Ordnung das Energiespektrum eines Strings der Stufe $n$ zu $E = 2\sqrt{n} \l^{1/4}+ \ord{\l^0}$ bestimmt wurde. Seitdem konnte aber kaum Fortschritt f{\"u}r kurze Strings verzeichnet werden.

	Das Ziel dieser Dissertation ist die Untersuchung diverser Ans{\"a}tze zur Ausweitung einer perturbative Quantisierung jenseits des Regimes langer Strings, und somit, zur Bestimmung des Energiespektrums kurzer Strings mit h{\"o}herer Genauigkeit in starker 't Hooft-Kopplung. Gleichzeitig streben wir dadurch an, ein besseres Verst{\"a}ndnis der involvierten Quantensymmetrien zu erlangen. Diese Problemstellung erscheint aktueller den je, da momentan f{\"u}r entsprechende Vorhersagen, welche auf der vermuteten Quantenintegrabilit{\"a}t basieren, jegliche Vergleichswerte fehlen.

	In dieser Arbeit konzentrieren wir uns gr{\"o}ßtenteils auf die niedrigst angeregten Strings, Strings der Stufe $n=1$, welche dual zu den Zust{\"a}nden des Konishi-Super-multiplets sind und f{\"u}r welche die Energie bzw. die konforme Skalendimension bis zur Ordnung $\ord{\l^{-5/4}}$ vorausgesagt wurde.

	Wir beginnen unsere Untersuchung mit einer Diskussion der angeblichen Herleitung der Konishi anomalen Skalendimension zur Ordnung $\ord{\l^{-1/4}}$ im Rahmen der Pure-Spinor-Superstringtheorie \cite{Vallilo:2011fj}. Wir begr{\"u}nden, warum es gen{\"u}gt, die Rechnung im bosonischen Subsektor nachzuvollziehen. Unter dieser Beschr{\"a}nkung finden wir diverse Inkonsistenzen, wobei die meisten durch die Nichtbeachtung des besonderen Skalenverhaltens der Nullmoden \cite{Passerini:2010xc} begr{\"u}ndet sind.

	Die erlangte Erfahrung legt nahe, dass nichtsdestotrotz eine Bestimmung des Spektrums zu dieser Ordnung mittels perturbativer Quantisierung semiklassischer String-l{\"o}sungen m{\"o}glich sein sollte, wobei ein besonderes Augenmerk auf nichtverschwindende Nullmoden gelegt werden muss. Wir pr{\"a}sentieren die Arbeit \cite{Frolov:2013lva}, in der wir den bosonischen String in statischer Eichung \cite{Jorjadze:2012iy} untersuchten. Wir konstruierten eine sogenannte Einzel-Moden Stringl{\"o}sung, eine Verallgemeinerung des pulsierenden Strings \cite{deVega:1994yz, Minahan:2002rc}, welche klassische Integrabilit{\"a}t sowie Invarianz unter Isometrietransformationen aufweist. M{\"o}gliche Korrekturen auf Grund der vernachl{\"a}ssigten Supersymmetrie werden heuristisch ber{\"u}cksichtigt, wodurch wir tats{\"a}chlich die Konishi anomale Skalendimension bis zur Ordnung $\ord{\l^{-1/4}}$ erhalten.

	Dieser Erfolg verlangt nach einer Implementierung der statischen Eichung f{\"u}r den vollst{\"a}ndigen $\AdSxS$ Superstring, wobei jedoch nicht einmal Quantisierung des $\AdSxS$ Superteilchens oder eine Umsetzung der statischen Eichung f{\"u}r den Superstring im flachen Raum verstanden sind. Wir ignorieren diese H{\"u}rden und fahren in Analogie zu der Konstruktion f{\"u}r gleichm{\"a}ßige Lichtkegelquantisierung \cite{Frolov:2006cc} fort, wodurch wir elegante Ausdr{\"u}cke f{\"u}r die Lagrangedichte sowie f{\"u}r die Superladungen erhalten. Des Weiteren sinnen wir {\"u}ber eine m{\"o}gliche supersymmetrische Verallgemeinerung des bosonischen Einzel-Moden Strings.

	Kaum {\"u}berraschend erhalten wir einen nicht-kanonischen kinetischen Term f{\"u}r die Fermionen, was einer kanonische Quantisierung des Systems im Wege steht. Um die Problematik zu vereinfachen beschr{\"a}nken wir uns auf das $\AdSxS$ Superteilchen, f{\"u}r welches der kinetischen Term quadratisch in Fermionen direkt diagonalisiert werden kann. Dar{\"u}ber hinaus erproben wir eine weitere Methode um eine Transformation zu immanent kanonischen Variablen zu erhalten, welche die zugrundeliegende $\psu(2,2|4)$ Algebra ausnutzt.

	Mit der Motivation ein besseres Verst{\"a}ndnis des $\AdSxS$ Superteilchen zu erlangen beschließen wir diese Arbeit mit der Untersuchung einer weiteren Quantisierungs-vorschrift, welche unit{\"a}re irreduzible Darstellungen mittels Bahnen der Isometriegruppe bestimmt \cite{Alekseev:1988ce}. In dem Artikel \cite{Heinze:2014cga} untersuchten wir das bosonische Teilchen und die Multi-Spin String-l{\"o}sung in ${\rm AdS}_3 \times {\rm S}^3$ und erzielen eine exakte Quantisierung der Systeme, da die Isometriegeneratoren auf nat{\"u}rliche Art und Weise durch eine Holstein-Primakoff-Transformation realisiert sind \cite{Holstein:1940zp, Dzhordzhadze:1994np}. Schlie{\ss}lich bestimmen wir Multi-Spin String-l{\"o}sung, welche zu langen und kurzen Strings korrespondieren.

\mainmatter
\pagestyle{headings}

\chapter{Introduction} \label{chap:Intro}

	At the beginning of the last century two major discoveries revolutionized theoretical physics in a yet unsurpassed way: Einstein's theory of {\it special relativity} and the theory of {\it quantum mechanics}.

	Special relativity confronts us with the fact that, contrary to our intuition, time is neither absolute nor detached from the three dimensions of space but instead should be viewed as an additional dimension, together forming four dimensional {\it space-time}. According to {\it general relativity}, the successor of special relativity found shortly after, space-time itself is then warped dynamically due to the energy and momentum of massive objects. In turn, the curvature of space-time determines the geodesics of the object at play, giving an elegant comprehension of the gravitational force. General relativity has been tested extensively and appears to describe nature at large scales, however, with black holes being a crucial exception to this assertion.

	Quantum mechanics, on the other hand, is the fundamental principle characterizing nature at microscopic scales and has become an indispensable tool in most of modern physics. It states that objects are characterized by wave functions rather than classical mass points, which leads to the peculiarity that observations of different properties may depend on the order in which they are performed, they do not commute.

	The concepts of special relativity and quantum mechanics were jointly incorporated in {\it quantum field theory} (QFT), which in the forties led to proper understanding of the electromagnetic force in terms of quantum electrodynamics, i.e., abelian gauge theories. Two decades later, after also the non-abelian setup had been understood in terms of Yang-Mill theories, the weak nuclear force was unified with electromagnetism via the mechanism of electroweak symmetry breaking while the strong nuclear force was described by quantum chromodynamics (QCD). Since then, the merger of these, the so called {\it standard model} of particle physics, has constantly been subject to experimental validation, only to result in agreement within its scope of application. These experimental tests culminated recently in observation of the Higgs boson \cite{Aad:2012tfa, Chatrchyan:2012ufa}, the last of the standard model particles to be confirmed, at the Large Hadron Collider (LHC) near Geneva, Switzerland.

	\newpage

	Yet, despite all compliance with experiments, the standard model has long-standing shortcomings, such as the flavor puzzle, the hierarchy problem, and the absence of dark matter and dark energy. Most unsatisfactory seems though that the standard model does only comprise the quantum descriptions of three fundamental forces, but not gravity. Even without attempting to unify gravity with the electro-weak and strong force, the search for a viable theory of quantized gravity poses what is probably the holy grail of fundamental physics. Popular attacks on the quest of {\it quantum gravity} include the proposals of asymptotic safety as well as loop quantum gravity, for reviews see for example \cite{Niedermaier:2006wt} and \cite{Nicolai:2005mc}.

	However, the most renowned contestant, which attempts to explain gravity as well as the other fundamental forces all in one go, is {\it string theory}. First developed as an alternative formulation for the strong force in the sixties, soon it was realized that the closed string sector always contains a suitable candidate for the graviton, the hypothetical messenger particle of gravity. But it was only after Green and Schwarz showed consistency of a ten-dimensional chiral superstring theory \cite{Green:1984sg}, establishing the possibility to accommodate all known elementary particles, that string theory was taking seriously as a potential {\it ``theory of everything''}. In the enthusiasm of the following years seemingly endless amounts of mathematically beautiful structure were revealed, as M-theory, D-branes, and topological string theory, but an unambiguous derivation of a string theory having the standard model and general relativity as low energy effective theories is still lacking.

	An exciting new field of research was opened up by the {\it AdS/CFT correspondence} \cite{Maldacena:1997re, Witten:1998qj, Gubser:1998bc}. This conjectured duality connects string theory in curved space-times with certain {\it conformal field theories} (CFT), which is why the correspondence is also referred to as gauge-string duality. In particular, it allows for new insights into the strong-coupling behavior of the field theory, a realm which until now has been mostly inaccessible due to our limited understanding of quantization and our reliance on perturbative methods.

	By now the AdS/CFT correspondence has found various applications, ranging from modelling quark-gluon plasma to condensed matter physics. But especially the initial and best understood duality pair, the correspondence between the maximally supersymmetric Yang-Mills theory (\NfourSYM) and the type IIB Green-Schwarz superstring in an $\AdSxS$ background, still holds many unsolved mysteries. Clearly, with studying this rather academic case one hopes to then apply and generalize the acquired comprehension to more realistic systems, a strategy which has been successful in physics at many occasions in the past.

	In \NfourSYM, the field theory side of the duality, calculations can be performed in the usual perturbative manner as long as the coupling is small. Taking additionally the planar limit, the underlying {\it quantum integrability} \cite{Minahan:2002ve, Beisert:2003tq} has been revealed, inducing development of a plethora of theoretical instruments and by this marking a tremendous advance in our understanding of QFTs during the last decade. Generally calculations simplify drastically due to the high amount of symmetry and, altogether, for the duality pair at hand our knowledge of the gauge theory side seems rather broad.


	At strong coupling, methods relying on the conjectured quantum integrability have been paralleled. Furthermore, in this limit the curvature of $\AdSxS$ becomes negligible and the $\AdSxS$ superstring seems to approximate the well understood superstring in flat Minkowski space. Already since the early days of the AdS/CFT correspondence semiclassical string solutions have been investigated \cite{Frolov:2002av, Gubser:2002tv, Berenstein:2002jq}, which in turn served as testing ground for the integrability based predictions. The study of these however relies on diverging quantum charges, such that the the strings become {\it long} or {\it heavy}, whereas for finite charges, i.e., {\it short} or {\it light} strings, the expansion of the string action becomes inconsistent.

	Quantization of the type IIB superstring in $\AdSxS$ is a longstanding problem and it is the goal of this thesis to extend our understanding of the quantum symmetries and perturbative quantization beyond the long string paradigm. With the advance of machinery building on the surmised quantum integrability, in particular with the recent proposition of the quantum spectral curve \cite{Gromov:2013pga, Gromov:2014caa} as a solution to the spectral problem, this seems more relevant than ever, as first principle calculations of the spectrum of short string states are still inaccessible.



	In the remainder of this chapter we are giving a minimal review of the AdS/CFT correspondence with focus on the duality between {\NfourSYM} and the type IIB superstring in $\AdSxS$, by which we introduce the relevant concepts and terminology. Hence, the experienced reader might want to skip most sections up to \secref{sec:In-SpecProb}, where we discuss the spectral problem for short strings in more detail. In chapters \ref{chap:BosString} and \ref{chap:SS} we quickly review the bosonic and superstring and the original results as stated in the executive summary can be found from \chapref{chap:PureSpinor} on.

\section{New Directions in Particle Physics}\label{sec:In-NewDir}
	The standard model of particle physics was formulated in the mid-sixties and, essentially, the theoretical framework underlying it has been settled in the mid-seventies. Since then, also the way observables are computed has hardly changed.
	
	The main tool connecting the conceptually beautiful but notional theory to collider experiments are scattering amplitudes, which when integrated to scattering cross sections give the probabilities of the occuring scattering processes. To calculate these one reads off the Feynman rules from the interactions in the Lagrangian, construct all possible Feynman diagrams for a fixed number of loops and integrates over the internal particles. Eventually, one has to cope with regularization of divergent integrals and renormalization of the theory, but given this simple recipe by now most standard model processes ought to be computed.
	
	In fact, this is far from the case. Especially the strong force causes problems and actually most hadronic QCD processes are only known at leading-order (LO), whereas next-to-leading-order (NLO) and next-to-next-to-leading-order (NNLO) processes have only been computed for small numbers of external particles. Going to higher loop order for increasing number of external legs one is confronted with several problems, namely the number of diagrams and, due to non-abelian gauge interactions, the number of terms in each diagram vastly increase. For instance, even at tree-level for process of two gluons scattering to $n$ gluons the number of diagrams grows factorially in $n$ \cite{Mangano:1990by}.

	All the more surprising is that the final results take remarkably simple forms when expressed in suitable variables. For gluons in four dimensional Minkowski space $\Reals^{1,3}$ one can use the so called spinor helicity formalism\footnote{For a pedagogical presentation of  content discussed in the following the interested reader is referred to the textbooks \cite{Zee:2003mt, Srednicki:2007qs}, the reviews \cite{Mangano:1990by, Dixon:1996wi, Roiban:2010kk, Ellis:2011cr, Elvang:2013cua}, and especially the recent textbook \cite{Henn:2014yza}.}, where the light-like gluon momenta are expressed in terms of two bosonic helicity spinors $\l^\a_i$ and $\tilde{\l}^\da_i$,
	\be \label{eq:In-SpinHel}
		p^\m_i \rightarrow p^{\da \a}_i = p^\m_i \s^{\da \a}_\m = \tilde{\l}^\da_i \l^\a_i~,\qquad\quad \s^{\da\a}_\m = (\mI, \vec{\s} )^{\da \a}~,
	\ee
	$\a,\da = 1,2\,$, which solve the massless Dirac equation. They are fixed only up to a $\Un(1)$ helicity transformation, $\l^\a_i$ having helicity $-\frac{1}{2}$ and $\tilde{\l}^\da_i$ having $\frac{1}{2}$, and one constructs Lorentz-invariant quantities as $ \la i j \ra = \eps_{\a\b} \l^\a_i \l^\b_j$ and $[ i j ] = -\eps_{\da\db} \tilde{\l}^\da_i \tilde{\l}^\db_j$, see \appref{app:Notation} for further conventions.
	Taking all particles as outgoing, the tree-level amplitudes with all positive helicity gluons or only one gluon of negative helicity vanish. The first non-vanishing $n$-gluon amplitude is the one with all but two gluons having positive helicity, the so called maximally helicity violating (MHV) amplitude. Stripping of the color dependence one ends up with the compact expression
	\be
		A^\text{tree}_{g^n}(\ldots,i^-,\ldots,j^-,\ldots) 
			= \frac{\la i j \ra^4}{\la12\ra \la23\ra \ldots \la(n-1)n\ra\la n1\ra}~, \label{eq:In-ParkeTaylor}
	\ee
	which is the renowned Parke Taylor formula \cite{Parke:1986gb}. Indeed, having the MHV amplitude and its parity conjugate, the $\overline{\text{MHV}}$ amplitude, all tree-level non-MHV amplitudes can then be constructed successively using the Britto-Chachazo-Feng-Witten (BCFW) on-shell recursion relation \cite{Britto:2004ap,Britto:2005fq}.

	During the last twenty years there has been major progress in understanding of amplitudes, where apart from various methods to tackle loop integrals, such as integral reduction, integration by parts, and Mellin-Barnes techniques, a promising development is the one of generalized unitarity \cite{Bern:1994cg, Bern:1994zx}. In many cases, as for \eqref{eq:In-ParkeTaylor}, the simplicity of the final results indicates that our understanding of QFTs is still very poor and in particular that, despite the simple manual, evaluation using Feynman diagrams is rather inefficient.

\section{Superconformal Symmetry} \label{sec:In-PSU}
	{\it Supersymmetry} (SUSY), a hypothetical symmetry connecting bosons with fermions, seems to be an agent in the analysis of QFTs. By definition the fermionic SUSY generators $Q_A{}^\a$ and $\tilde{Q}^{\da A}$ fulfill the anticommutation relation\footnote{For convenience we depict the $d=4$ dimensional case only. Note that generally the supercharges form spinors of $d$ dimensional space-time and the number of real supercharges is $\calN \times 2^{\lfloor d/2 \rfloor+1}$ modulo factors of 2 for Majorana and Weyl conditions. For introductions to SUSY see \cite{Wess:1992cp, West:1990tg, Sohnius:1985qm}.}
	\be
		\{Q_A{}^\a, \tilde{Q}^{\da B}\} = \d_A{}^B\,P^{\da\a} 
			= \d_A{}^B\,\s^{\da \a}_\m P^\m ~,
	\ee
	with $\a,\da=1,2$ the $d=4$ Weyl-spinor indices introduced in \eqref{eq:In-SpinHel}, which extends the Poincar{\'e} algebra generated by translations $P_\m$ and Lorentz transformations $J_{\m\n}$,
	\be\ba \label{eq:In-Poincare}
		&[P_\m,P_\n]=0~,\qquad\quad
			[J_{\m\n}, P_\r] = i \big(\eta_{\m\r} P_\n - \eta_{\n\r} P_\m \big)~,\\
		&[J_{\m\n}, J_{\r\s}] = i \big(\eta_{\m\r} J_{\n\s} + \eta_{\n\s} J_{\m\r} - \eta_{\m\s} J_{\n\r} - \eta_{\n\r} J_{\m\s}\big)~,
	\ea\ee
 to the {\it super-Poincar{\'e} algebra}. The indices $A, B =1,\ldots,\calN$ parametrize the amount of SUSY and for $d=4$ dimensions $\calN>1$ induces presence of a $\Un(\calN)$ {\it R-symmetry}, which has generators $R_A{}^B$ acting on the supercharges $Q_C{}^\g$ as
	\be
		[R_A{}^B, Q_C{}^\g] = \d_C{}^B Q_A{}^\g - \frac{1}{4} \d_A{}^B Q_C{}^\g~,
	\ee
	and analogously for $\tilde{Q}^{\dg C}\,$. By this, the case $\calN=4$ is special as $[R_A{}^A, Q_C{}^\g]=0$, i.e., the R-symmetry reduces to $\SU(4)$ instead. 

	SUSY is sought after at the LHC as it might both ease the hierarchy problem as well as yield a suitable dark matter candidate. To this day it has not been observed, indicating that SUSY might be broken at energy scales inaccessible to current collider experiments. SUSY is also a favored ingredient of grand unified theories and superstring theories, which if realized in nature will lie at typical energies close to the Planck scale $m_{\rm Pl} = \sqrt{\hbar c /G} = 1.22 \times 10^{19}\,\text{GeV}$.

	{\it Conformal transformations} are by definition the transformations $x^\m \rightarrow x'^\mu(x)$ leaving the metric invariant up to a scaling factor\footnote{For accounts on CFTs see \cite{DiFrancesco:1997nk}, the original works \cite{Mack:1969rr, Callan:1970ze, Coleman:1970je}, as also the nice presentation in \cite{Wiegandt:2012eu}.}
	\be
		G_{\m\n}(x) \rightarrow G'_{\m\n}(x') = \r(x)G_{\m\n}(x)~.
	\ee
	The cases of $d=1,2$ dimensions are special, as for $d=1$ any transformation is conformal, while for $d=2$ the conformal algebra is infinite dimensional. The latter plays a central role in string theory and will be reviewed from \chapref{chap:BosString} onwards.

	The conformal group of $d>2$ dimensional Minkowski space is $\SO(2,d)$, which is generated by the Poincar{\'e} algebra as well as the dilatation $D$ and the special conformal charges $K_\m$ fulfilling the additional non-vanishing commutator relations
	\be\ba \label{eq:In-ConfAlg}
		\qquad&[D,P_\m] = + i P_\m ~,& 
			&[J_{\m\n}, K_\r] = i \big(\eta_{\m\r} K_\n - \eta_{\n\r} K_\m \big) ~, &\\
		&[D, K_\m] = - i K_\m ~,&
			&[P_\m, K_\m] = 2 i \big(\eta_{\m\n} D - J_{\m\n}\big) ~.&
	\ea\ee

	Combination of SUSY with conformal transformations amounts to the {\it superconformal} group. Naturally, the superconformal algebra consists out of the conformal generators and the supercharges, where $[D,Q_A{}^\a]= +{i\ov2} Q_A{}^\a$ and $[D, \tilde{Q}^{\da A}]= +{i\ov2} \tilde{Q}^{\da A}$. Closure of the algebra then requires additional fermionic  generator, the so-called conformal supercharges,
	\be
		\{ S_\a{}^A, \tilde{S}_{B \da}\} 
			= \d^A{}_B\,\s^\m_{\a\da} K_\m~,\quad
		[D, S_\a{}^A] = -{1\ov2} S_\a{}^A~,\quad
		[D, \tilde{S}_{A \da}] = -{1\ov2} \tilde{S}_{A \da}~.
	\ee
	
	With this, let us review some facts about {\it conformal field theories} (CFT). Classically, field theories are conformal invariant if the action is invariant under conformal transformations. This is the case for field theories without any mass-scale and non-derivative or Yang-Mills interactions \cite{Mack:1969rr}. Especially, for the standard model classical conformal invariance is only broken by the Higgs term.

	At the quantum level however ultraviolet (UV) divergences occur. Regularization of these introduces new mass-scales, which break conformal invariance, except at fixed points of the renormalization group. At these, the renormalization group $\b$-function vanishes, corresponding to non-renormalization of the coupling constants.

	Note that even for a quantum conformal field theory one will encounter divergences. Amplitudes are indeed free of UV divergences but one still faces infrared (IR) divergences. Furthermore, compound gauge invariant operators will be subject to UV divergences, as elaborated in the following.

	For a field $\Phi_j (x)$, with $j$ representing the Lorentz structure, infinitesimal conformal transformations split into the change of coordinates and the change of the field as
	\be
		\Phi'_j (x') - \Phi_j (x) 
			= \d \Phi_j(x) + \d x^\m \p_\m \Phi_j (x) + \ord{\d^2}~.
	\ee
	Here, $\d x^\m = i \eps (J x)^\m$, $J\in\{P_\m, J_{\m\n}, D, K_\m\}$ whereas $\d \Phi(x)_j = i \eps (\mbb{J}\, \Phi(x))_j$ and $\mbb{J} \in  \{\mbb{P}_\m , \mbb{J}_{\m\n}, \mbb{D}, \mbb{K}_\m  \}$ are conformal generators only acting on the field, not on the coordinates. In particular $[\mbb{P}_\m,x^\n] = 0$ and $\mbb{P}_\m$ generates translations of the fields. By
	\be
		\mbb{J}\, \Phi_j(x) = e^{i \mbb{P} x} e^{-i \mbb{P} x} \mbb{J} e^{i \mbb{P} x} \Phi_j(0)
	\ee
	one then deduces $\mbb{J}\, \Phi_j(x)$ from $\mbb{J}\, \Phi_j(0)$ and the fields, hence the conformal representations, are characterized by action of the {\it stability group}, alias the {\it little group},
	\be
		(\mathbb{J}_{\m\n}\phi(0))_j = (\S_{\m\n}\phi(0))_j~,\quad 
		\mathbb{D}\,\phi(0)_j  = i \D \phi(0)_j~,\quad
		(\mathbb{K}_\m \phi(0))_j = (\k_\m \phi(0))_j ~, 
	\ee
	where $\S_{\m\n}$, $\Delta$ and $\k_\m$ again respect the conformal algebra. Let $\Phi_j(x)$ belong to an irreducible representation (irrep) of the Lorentz group. Then, $\D$ commutes with the spin operator $\S_{\m\n}$ and by Schur's lemma is proportional to the identity matrix, which in turn implies that $\k_\m =0$. Therefore, when $\D$ acts on $\Phi_j(0)$ it gives a scalar called the {\it scaling} or {\it conformal dimension}. The name stems from the fact that under finite conformal transformations scalar fields $\phi(x)$ transform as
	\be \label{eq:In-Rescale}
		\phi(x) \rightarrow \phi'(x') = \left|\frac{\p x'}{\p x} \right|^{-\D/d} \phi(x)~,
	\ee
	i.e., for the rescaling $x \rightarrow x' = \l x$ one has $\phi'(x') = \l^{-\D} \phi(x)$. Fields transforming as \eqref{eq:In-Rescale} are called {\it quasi-primary}, see for example \cite{Osborn:1993cr, Erdmenger:1996yc, Jackiw:2011vz} for more details. Correspondingly, a local operator $\calO(x)$ transforming as \eqref{eq:In-Rescale}
	is called {\it conformal primary operator}, which by $\mbb{K}_\m \calO(0) = 0$ are lowest weight states of conformal representations.

	This concept generalizes to {\it superconformal primary operators}, where one demands
	\be
		\mbb{D}\,\calO(0) = i \D\,\calO(0)~,\qquad 
		\mbb{S}_{\a A} \calO(0) = 0~,\qquad
		\tilde{\mbb{S}}^A{}_\da \calO(0) = 0~.
	\ee
	Due to the superconformal algebra, acting with $\mbb{Q}^{A\a}$ and $\tilde{\mbb{Q}}^\da{}_A$ on $\calO(x)$ increases the scaling dimension by ${1\ov2}$ and the corresponding operators are called {\it descendants} of $\calO(x)$.

	It is then an important result for quantum CFTs \cite{Polyakov:1970xd} that only requiring invariance under dilatation and inversion uniquely fixes the two-point correlation function of two scalar (super)conformal primary operators to take the form\footnote{Here, $A$ and $B$ merely label different operators and are not R-symmetry indices.}
	\be \label{eq:In-2pt}
		\la \calO_A(x_1) \calO_B(x_2) \ra = \d_{A B} \frac{C_A}{(x^2_{12})^{\D_A}}~,
	\ee
	for some constant $C_A$. Especially, radiative corrections in the two-point correlators give rise to UV divergences. When renormalized these only invoke quantum corrections to the scaling dimensions, which hence is a function of the coupling constants $g$, and the form of \eqref{eq:In-2pt} persist. The difference between the quantum scaling dimension $\D(g)$ and the classical scaling dimension $\D_0$ is the {\it anomalous (scaling) dimension} $\g(g) = \D(g) - \D_0$, which by the previous argument is equal for all descendants, i.e., for all members of the supermultiplet.  

	Similarly, invariance under dilatation and inversion also fixes the form of three-point correlators of scalar (super)conformal primary operators,
	\be
		\la \calO(x_1) \calO(x_2) \calO(x_3) \ra = 
			\frac{C_{ABC}}{|x_{12}|^{\D_A+\D_B-\D_C} |x_{23}|^{\D_B+\D_C-\D_A} |x_{31}|^{\D_C+\D_A-\D_B}}~,
	\ee
	where $C_{ABC}$ are the so-called {\it structure constants}.

	Higher-point correlators are not fixed completely by conformal symmetry anymore but will instead depend on functions of the {\it conformal cross ratios}.

	Another important tool of CFT is the {\it operator product expansion} (OPE) \cite{Wilson:1969zs}, which states that in the limit $x\rightarrow 0$ the product of two local operators $\calO_A (x) \calO_B(0)$ can be substituted by a linear combination of local operators
	\be
		\calO_A (x) \calO_B (0)\ \stackrel{x\rightarrow0}\longrightarrow\ \sum_C C_{ABC}(x) \calO_C (0)~.
	\ee
	where the operators $\calO_C$ have the same global symmetry quantum numbers as $\calO_A \calO_B$ and by application to three-point correlator one finds $C_{ABC}(x) = C_{ABC} |x|^{\D_C - \D_A - \D_B}$.

	In the same manner, the OPE then allows to determine higher-point correlators successively by their limiting behavior. Hence, it is believed that a CFT is completely determined by the scaling dimensions $\D$ and structure constants $C_{ABC}$, which for this reason are also referred to as the {\it conformal data}.


\section{\texorpdfstring{$\calN = 4$}{Nfour} super Yang-Mills theory} \label{sec:In-NfourSYM}
	$\calN = 4$ super Yang-Mills theory (\NfourSYM) \cite{Gliozzi:1976qd, Brink:1976bc} is the maximally supersymmetric gauge theory in four dimensions. It resembles the standard model as it involves similar particles and interactions and, having gauge group $\SU(N_c)$ with $N_c$ the number of colors, in fact can be viewed as a toy-model for massless QCD.

	Following the conventions in \cite{Henn:2014yza}, the $\SU(N_c)$ gluons $A^a_\m$ are accompanied by their superpartners under the fourfold SUSY, the four spin $1\ov2$ gluinos, $\psi^a_{\a A}$ and $\bar{\psi}^{a \da}_ A$, and the six real scalars $\phi^{a AB} = -\phi^{a BA}$, which all are in the adjoint of the $\SU(N_c)$ color group, $a=1,\ldots, N_c^2-1\,$, while $A,B=1,\ldots,4$ are indices of the $\SU(4)$ R-symmetry. The gluinos and scalars behave under complex conjugation as
	\be
		(\psi^a_{\a A})^\ast = \bar{\psi}^{a A}_\da~,\qquad\quad
		(\phi^{a AB})^\ast = \phi^a_{A B} = {1\ov2} \eps_{ABCD} \phi^{a C D}~.
	\ee
	Neglecting gauge-fixing and ghost terms, the action of {\NfourSYM} reads
	\begin{align}
		S =& {1\ov {\gYM^2}} \int \!\dd^4 x \tr \Big(\!-{1\ov4} F^2_{\m\n} 
			- (D_\m \phi_{AB}) D^\m \phi^{AB} 
			- {1\ov2} \left[\phi_{AB},\phi_{CD}\right]\!\left[\phi^{AB},\phi^{CD}\right]
			\\
		&\qqquad \ \ 
			+ i \bar{\psi}^A_\da \s^{\da\a}_\m D^\m \psi_{\a A}  
			- {i\ov2} \psi^\a_A \left[\phi^{AB},\psi_{\a B} \right]
			- {i\ov2} \bar{\psi}_\da^A \left[\phi_{AB},\bar{\psi}^{\da B} \right]
			\Big)~,\nn
	\end{align}
	where the form is uniquely fixed by SUSY and $D_\m(\cdot) = \p_m (\cdot) -i \gYM [A_\m,(\cdot)]$ and $F_{\m\n}={i\ov \gYM} [D_\m,D_\n]$ are the usual gauge covariant derivative and field strength.

	The only parameter is the gauge coupling $\gYM$, yielding that the theory is classically scale invariant. Moreover, it was found that the renormalization group $\beta$-function vanishes up to tree-loop order \cite{Sohnius:1981sn, Mandelstam:1982cb, Brink:1983pd, Howe:1984sr}, a behavior which is believed to persist to any loop order, see also the discussion in \cite{Sohnius:1985qm}. By this, {\NfourSYM} is a CFT not only classically but even at the quantum level and it is arguably the simplest gauge theory in four dimensions. It is therefore frequently referred to as the {\it ``Hydrogen atom of gauge theories''}, as one hopes to unravel fundamental properties of QFTs for this particular model and then generalize these for more realistic theories.

	More precisely, {\NfourSYM} is invariant under the superconformal algebra $\psu(2,2|4)$, which has been discussed in the previous section and which also will be the subject of \ssecref{sec:SS-PSU}. It has the bosonic subalgebra $\su(2,2)\oplus\su(4) \cong \so(2,4) \oplus \so(6)$ corresponding to the direct sum of the $\Reals^{1,3}$ conformal and $\calN=4$ R-symmetry algebra. 

	Armed with this we can go back to the discussion in \secref{sec:In-NewDir}. It shows useful to assemble the {\NfourSYM} on-shell degrees of freedom into an on-shell superfield,
	\be \label{eq:In-OnShellPhi}
		\Phi(p,\eta) = g_+(p) + \eta^A \tilde{g}_A(p) + {1\ov2!}\eta^A \eta^B S_{AB}(p) + {1\ov3!} \eta^A \eta^B \eta^C \eps_{ABCD} \bar{\tilde{g}}^D(p) + \eta^4 g_-(p)~,
	\ee
	where $\eta^A$ are Grassmann variables, $\eta^4 = {1\ov4!} \eta^A \eta^B \eta^C \eta^D \eps_{ABCD}$, and for example $g_\pm(p)$ are the helicity $\pm {1\ov2}$ gluons. Amplitudes then naturally generalize to super-amplitudes \ $\mathbb{A}_n(\l^\a_i, \tilde{\l}^\da_i, \eta^A_i)$ \cite{Nair:1988bq} and superconformal invariance requires $J\, \mathbb{A}_n(\l^\a_i, \tilde{\l}^\da_i, \eta^A_i) = 0$ for all $J\in\psu(2,2|4)$\footnote{Note that even at tree-level the superconformal invariance is broken due to the holomorphic anomaly when taking the colinear limit \cite{Cachazo:2004by, Cachazo:2004dr, Bargheer:2009qu, Korchemsky:2009hm}.}, where a representation of the generators in on-shell superspace can be found in \cite{Drummond:2008vq, Henn:2014yza}. Imposing invariance under translations $P^{\da\a}$ and supersymmetries $Q^{A \a}$, the super-amplitude takes the form
	\be
		\mathbb{A}_n(\l^\a_i, \tilde{\l}^\da_i, \eta^A_i)
			= \frac{\d^4(p)\d^8(q)}{\la12\ra \la23\ra \ldots \la(n-1)n\ra\la n1\ra} \calP_n (\l^\a_i, \tilde{\l}^\da_i, \eta^A_i) ~, \label{eq:In-SuperAmp}
	\ee
	which generalizes the Parke-Taylor formula \eqref{eq:In-ParkeTaylor}. Due to the $\SU(4)$ R-symmetry, $\calP_n (\l^\a_i, \tilde{\l}^\da_i, \eta^A_i)$ expands in powers of $\eta^4$, with the MHV and $\overline{\text{MHV}}$ amplitudes corresponding to the components in lowest and highest powers, respectively. Similarly one finds the generalization to the super BCFW recursion \cite{ArkaniHamed:2009dn, Brandhuber:2008pf, Bianchi:2008pu}. By this technique analytic formulae for all {\NfourSYM} tree-amplitudes were derived \cite{Drummond:2008cr}, which led to analogues results for massless QCD \cite{Dixon:2010ik}.

\section{The Gauge-String Duality}\label{sec:AdSCFT}
	Until now we only reported on the structure of {\NfourSYM} appearing at tree-level but certainly one actually hopes to retrieve insights on the higher radiative corrections, i.e., the loop-level, or even beyond the scope of perturbation theory. 

	With this in mind, a groundbreaking proposal is the {\it AdS/CFT correspondence} \cite{Maldacena:1997re, Witten:1998qj, Gubser:1998bc}. This conjectured duality, or rather class of dualities, states that superstring theory in $d+1$ dimensional Anti-de Sitter space ${\rm AdS}_{d+1}$ is dual to a CFT living on its $d$ dimensional conformal boundary. Since the CFT in question is often a gauge theory, the correspondence is also referred to as {\it gauge-string duality}. Furthermore, it is also known as {\it holographic duality}, as it connects theories in different dimensions. For reviews of the field see for example \cite{Aharony:1999ti, D'Hoker:2002aw, Plefka:2005bk, Nastase:2007kj, Beisert:2010jr} and the textbook \cite{Becker:2007zj}.

	Anti-de Sitter space with $d+1$ dimensions is defined as a hyperboloid of constant negative curvature in $\Reals^{2,d}$ embedding space,
	\be \label{eq:In-AdSdef}
		\eta_{M N} Z^M Z^N = -Z^2_{0'} - Z^2_0 + Z^2_1 + \ldots + Z^2_{d}= - R^2 ~,
	\ee
	with $Z^M \in \Reals^{2,d-1}$, $M,N=0',0,1,\ldots,d-1\,$, and $R$ being the radius of ${\rm AdS}_{d+1}$. Taking $R^2 \sinh^2(\r)=Z^2_1+\ldots+Z^2_{d}$ and $Z_{0'} + i Z_0 = R \,e^{i\,t}\cosh{\r}$ the metric reads
	\be\label{eq:In-AdSmetric}
		\dd s^2_{\rm AdS} = R^2 \left(-\cosh^2(\r)\dd t^2 + \dd \r^2 + \sinh^2(\r) \dd \Om^2_{d-1} \right)~.
	\ee
	Here, $\dd \Om_{d-1}^2$ is the metric of the $d-1$ dimensional unit sphere and $t$ is the ${\rm AdS}_{d+1}$ time coordinate, which to allow for causality has to be decompactified, $t\in \Reals$. Taking now the radial coordinate $\r$ to infinity one can define a metric by rescaling \eqref{eq:In-AdSmetric} as
	\be
		\dd s_\text{c.b.}^2 = \frac{2 }{R^2 \exp(\r)} \dd s^2_{\rm AdS} \big|_{\r \rightarrow \infty} = -\dd t^2 + \dd \Om^2_{d-1}~,
	\ee
	which is nothing but the metric of $d$ dimensional flat Minkowski space $\Reals^{1,d-1}$ \footnote{For $\r \rightarrow \infty$ the sphere $R^2 \sinh^2{\r}=Z^2_1+\ldots Z^2_{d}$ blows up and indeed becomes asymptotically flat.}. This constitutes the so-called conformal boundary of ${\rm AdS}_{d+1}$.

	The isometry group $\SO(2,d)$ of $\Reals^{2,d}$ leaves ${\rm AdS}_{d+1}$ invariant. However, $\SO(2,d)$ is nothing but the conformal group for $d>2$ dimensional flat space, where in particular the  $\Reals^{2,d}$ isometry generators $J_{M N}$ relate to the conformal generators \eqref{eq:In-ConfAlg} as
	\be\ba \label{eq:In-ISO2d}
		&J_{\m\n} = J_{\m\n}~,\quad J_{0'd} = D~,\quad 
		J_{\m 0'} = {1\ov2} (K_\m + P_\m)~,\quad
		J_{\m d} = {1\ov2} (K_\m - P_\m)~,\\
		&\qquad[J_{M N}, J_{R S}] = i \big(\eta_{M R} J_{N S} + \eta_{N S} J_{M R} - \eta_{M S} J_{N R} - \eta_{N R} J_{M S}\big)~,
	\ea\ee
	This suggests that any field theory on the boundary should be a conformal one. But since we are interested in superstring theory on ${\rm AdS}$, one should expect the boundary theory to be not only conformal but even superconformal.

	In the previous section we just described a theory fulfilling all these requirements for the case of the conformal boundary being $d=4$ dimensional, namely {\NfourSYM}.

	Finally, superstring theory appears to be consistent in ten dimensions and one should actually consider a space-time ${\rm AdS}_{d+1}\times M_{9-d}$, where to not alter the conformal boundary $M_{9-d}$ should be a compact manifold. Since the R-symmetry algebra $\su(4) \cong \su(6)$ of {\NfourSYM} is the same as the isometries of the five dimensional sphere ${\rm S}^5$, the space-time $\AdSxS$, with both having common radius $R$, seems favored.

	Indeed, we heuristically deduced the statement of the AdS/CFT correspondence for the initial and best studied duality pair: $\calN=4$ super Yang-Mills theory with gauge group $\SU(N_c)$ is dual to type IIB Green-Schwarz superstring theory in $\AdSxS$ \cite{Metsaev:1998it}\footnote{For introductions to the $\AdSxS$ superstring see the excellent review \cite{Arutyunov:2009ga} as well as \cite{Tseytlin:2010jv, McLoughlin:2010jw, Magro:2010jx}},
	\be
		\text{$\calN=4$ super Yang-Mills} \qquad \Longleftrightarrow\qquad \text{$\AdSxS$ type IIB superstring}~.\nn\vspace{.2cm}
	\ee

	The correspondence was actually proposed by yet another setup, namely by investigation of a stack of $N_c$ coincident D3-branes, which are $1+3$ dimensional objects. Polchinski showed \cite{Polchinski:1995mt} that D$p$-branes are equivalent to extremal $p$-branes, which are supergravity solutions curving the space-time. Close to the horizon of the D3-branes, the {\it near-horizon} limit, or alternatively in the limit $N_c \rightarrow \infty$, the space-time becomes $\AdSxS$ with the D3-branes lying at the conformal boundary. At the same time, D-branes constitute boundaries on which open strings can end. The low-energy modes of the brane decouple and are exactly described by {\NfourSYM}, which also determines the gauge group to be $\SU(N_c)$. The connection of {\NfourSYM} and closed strings in $\AdSxS$ is then established by slightly relaxing the extremality, resulting in Hawking-radiation. 
	The corresponding process is the one of two open strings on the D3-branes meeting each other to form a closed string, which now 'bubbles' into the bulk of ${\rm AdS}_5$. 

	From all this, one concludes the dictionary of the AdS/CFT correspondence: Three-point functions in {\NfourSYM} are mapped to string three-point functions and, most importantly for this thesis, the conformal dimensions $\D$ of local {\NfourSYM} operators are identifies with the energy spectrum $E$ of the dual string states.
	Furthermore, one identifies the Yang-Mills coupling $\gYM$, the string coupling $g_s$, the Regge slope $\a'$ and the string tension $T_0$ as
	\be \label{eq:In-tHooft}
		\l = N_c\,\gYM^2 = 4\pi N_c\, g_s = \left(\frac{R^2}{\a'}\right)^2 = (2\pi R^2 T_0)^2 ~,
	\ee
	where we introduced the {\it 't Hooft coupling} $\l$ \cite{'tHooft:1973jz}.

	It is then instructive to take the {\it 't Hooft limit},
	\be
		N_c \rightarrow \infty\qquad\qquad \text{for} \qquad\qquad \l = \text{const}~.
	\ee
	The limit is also called {\it planar limit}, as in the field theory non-planar Feynman diagrams are suppressed by factors of $(N_c)^{-2 (k-1)}$, with $k$ the genus of the respective topology. For $\l \ll 1$ the usual perturbative techniques are applicable, rendering {\NfourSYM} the appropriate description of the duality pair at weak 't Hooft coupling.
	
	For the string theory, the 't Hooft limit implies vanishing string coupling, $g_s \rightarrow 0$. By this, higher string field theory corrections are negligible as only the world sheet of lowest genus contributes, which also can be viewed as planarity. Taking then $\l \gg 1$ while keeping $T_0$ constant the radius $R$ blows up and $\AdSxS$ seems to asymptote ten dimensional flat Minkowski space. Since type IIB superstring theory in flat space is well understood one hopes to quantize the $\AdSxS$ superstring perturbatively around the flat background and string theory seems to be the appropriate description at strong 't Hooft coupling. Hence, remarkably, the AdS/CFT correspondence allows insights on the strong coupling behavior of CFTs by calculation of the respective quantities in string theory.

	In this thesis however, the last idea, i.e., that the $\AdSxS$ superstring can be quantized as a perturbation to the flat space superstring, will be subject to critical investigation. Note already here that the physics for a particle in some geometric space hardly change when altering the typical scale. Especially, the dimensionless energy spectrum usually only depends on the respective quantum numbers but not the radius of some manifold. This advices that also the string zero modes, which can be perceived as particle degrees of freedom, should be treated with caution.

	Finally, we should comment that by now, apart from the above, there are various AdS/CFT duality pairs under investigation. The most prominent is probably the one obtained by $N_c$ coincident M2-branes, leading to duality between M-theory on ${\rm AdS}_4 \times S^7$, respectively, type IIA superstrings in ${\rm AdS_4}\times \mathbb{CP}^3$ and $\calN=6$ super Chern-Simons theory with gauge group $SU(N_c) \times \SU(N_c)$, the so-called ABJM theory \cite{Aharony:2008ug}, see also \cite{Klose:2010ki}. For other possible setups, the ones for M5-branes, NS5-branes and other Dp-branes, see once more \cite{Aharony:1999ti}. Furthermore, in analogy to the coset space description of the $\AdSxS$ string, in \cite{Zarembo:2010sg} a classification of potentially interesting coset spaces admitting a $\mbb{Z}_4$ grading was discussed.

	Recently, there has been impressive development for the cases of ${\rm AdS}_3\times {\rm S}^3 \times M_4$ and ${\rm AdS}_2\times {\rm S}^2 \times {\rm T}^6$, as techniques developed for $\AdSxS$ have been applied successfully, see for example the recent works \cite{Borsato:2014exa, Sfondrini:2014via}, respectively, \cite{Hoare:2014kma}, and references therein. Another related interesting direction seems to be the deformations of the $\AdSxS$ superstring build on the homogeneous \cite{Kawaguchi:2014qwa} and the modified \cite{Delduc:2013qra, Arutyunov:2013ega, Delduc:2014kha} classical Yang-Baxter equation.

\section{Integrability} \label{sec:In-Integr}
	As discussed above, for finite $N_c$ the string coupling $g_s$ is non-vanishing and one has to deal with string field theory. However, as long as not even the $\AdSxS$ superstring is quantized and as long as string three-point functions are not understood properly, almost all computations for worldsheets of higher genii are virtually impossible, see also \cite{Kristjansen:2010kg}. For recent notable attacks on string three-point correlators see \cite{Kazama:2012is, Kazama:2013qsa, Bargheer:2013faa, Bajnok:2014sza}.

	On the field theory side, of course, perturbative calculations can be performed for finite $N_c$ as long as the coupling $\gYM$ is small. However, there is another reason why one should be interested in the planar limit, $N_c \rightarrow \infty$, as in this limit {\NfourSYM} appears to be {\it quantum integrable}.

	A theory is called integrable if it is exactly solvable, which for field theories corresponds to existence of an infinite number of conserved charges. If this property persists at the quantum level, the theory is called quantum integrable.

	Since the focus of this thesis lies not on these aspects, let us only mention the most important results for the integrability of the $\AdSxS$ superstring/{\NfourSYM} duality pair, sometimes also referred to as the {\it AdS/CFT integrability}. For reviews see \cite{Plefka:2005bk, Beisert:2010jr}

	In \cite{Berenstein:2002jq} Berenstein, Maldacena, and Nastsase (BMN) considered expansion around point-like strings rotating fast along the great circle of ${\rm S}^5$. For total angular momentum $J\propto \sqrt{\l}$ it was found that one obtains string theory in the pp-wave background. In the seminal work \cite{Beisert:2002ff, Beisert:2003tq} the corresponding {\NfourSYM} operators where investigated. These are long operators only build out of two complex scalars $X$ and $Z$, which constitutes the closed $\SU(2)$ sector, where the state $\tr(Z Z \ldots Z)$ corresponds to the BMN particle. Especially, the importance of the dilatation operator as an effective tool to determine the anomalous dimension was demonstrated. The authors of \cite{Minahan:2002ve} then recognized, that this setup matches with the well known {\it Heisenberg spin chain}, which was solved by Bethe in 1931 \cite{Bethe:1931hc} with an ansatz now referred to as {\it coordinate Bethe ansatz}.

	The generalization to general local {\NfourSYM} operators was established in \cite{Beisert:2003yb, Beisert:2003jj}. For certain limits, one can deduce a set of algebraic equations, known as {\it asymptotic Bethe ansatz} (ABA) \cite{Beisert:2004hm, Beisert:2005tm, Beisert:2005fw}, which allow to deduce all-loop results for the anomalous scaling dimension. This successively led to other sets of algebraic equations, notably the {\it Y-system} and {\it T-system} \cite{Gromov:2009tv} and recently the proposal of the {\it quantum spectral curve} \cite{Gromov:2013pga, Gromov:2014caa} as a solution to the spectral problem for arbitrary states. A related concept is the one of the {\it thermodynamic Bethe ansatz} (TBA), which based on the ideas \cite{Yang:1968rm, Zamolodchikov:1989cf} was applied to $\AdSxS$ in \cite{Bombardelli:2009ns, Arutyunov:2009ur}, giving in particular access to the finite-size string spectrum.

	Semiclassical string states have played an import role, especially  the spinning folded \cite{Gubser:2002tv}, the spinning circular \cite{Frolov:2002av}, and multi-spin solutions \cite{Frolov:2003qc}, which were some of the first instances to show agreement with ABA predictions and led to the proposal of the Beisert-Eden-Staudacher (BES) phase \cite{Eden:2006rx, Beisert:2006ez} and the ``string ABA'' \cite{Arutyunov:2004vx}. In \cite{Callan:2003xr, Callan:2004uv, Arutyunov:2005hd, Frolov:2006cc} the light-cone gauge for the $\AdSxS$ superstring in the near-BMN limit was studied, which led to significant insights on the symmetries of the worldsheet S-matrix \cite{Arutyunov:2006yd, Arutyunov:2006ak, Klose:2006zd}. The study of semiclassical partition functions was initiated by the seminal work \cite{Drukker:2000ep}. Another notable string solution is the so-called giant magnon \cite{Hofman:2006xt} leading to various applications.

	A technique allowing to obtain the string spectrum without having a semiclassical string solution at hand is the {\it spectral curve}, which heavily relies on the classical integrability of the $\AdSxS$ superstring found in \cite{Bena:2003wd}. The idea was proposed and established for $\Reals\times{\rm S}^3$ subspace \cite{Kazakov:2004qf} and then subsequently generalized \cite{SchaferNameki:2004ik, Beisert:2005bm}.

	Yet another facet of the AdS/CFT integrability, connecting to the discussion in \secref{sec:In-NfourSYM}, is the duality between amplitudes and Wilson loops\footnote{There is even the {\it triality} between amplitudes, Wilson loops and correlators, see for example \cite{Eden:2010ce}.}, which schematically can be written as
	\be
		\ln (A_n/A^\text{tree}_n) = \ln \la W(\calC_n) \ra + \text{const.} ~.
	\ee
	Here, the $n$ corners $x_i$ of the light-like polygonal contour $\calC_n$ of the Wilson loop are related to the external momenta $p_i$ of the amplitude as $p_i = x_{i+1} - x_i$.
	
	The duality was first proposed at strong coupling \cite{Alday:2007hr, Alday:2007he}, where planar gluon amplitudes where computed in terms of minimal surfaces in AdS, reassembling the computation of Wilson loops. Surprisingly, the duality persisted at weak coupling for four points \cite{Drummond:2007aua} and $n$ points \cite{Brandhuber:2007yx} at one loop. The two-loop results \cite{Drummond:2007cf, Drummond:2007bm} revealed that from six points onwards the renowned {\it BDS-ansatz} \cite{Bern:2005iz} has to be corrected. 

	Interestingly, the Wilson loops proved to be invariant under a {\it dual superconformal symmetry} \cite{Drummond:2008vq}, which from the amplitude point of view is non-local and in particular differs from the original local superconformal symmetry. The existence of both of these implies an infinite tower of symmetries, which were found to form a Yangian symmetry $Y(\psu(2,2|4))$ in \cite{Drummond:2009fd}. Quiet recently, this has also been observed for Wilson loops \cite{Muller:2013rta}. In the string theory the dual superconformal symmetry seems to originate from T-duality \cite{Beisert:2008iq}. 

	Furthermore, in \cite{Witten:2003nn} it was observed that {\NfourSYM} can be described by string theory in {\it twistor space}\footnote{This string theory does not play the role as in the sense of the AdS/CFT correspondence.}. This led to the {\it Grassmannian} formulation of scattering amplitudes \cite{ArkaniHamed:2009dn, ArkaniHamed:2010kv} and the so-called {\it Amplituhedron} \cite{Arkani-Hamed:2013jha}. The connection between to the Yangian symmetry, Grassmannians and T-duality was investigated \cite{Drummond:2010qh}, which recently led to a natural deformation and by this a natural spectral parameter for {\NfourSYM} scattering amplitudes \cite{Ferro:2013dga}.

	Let us stress once more that many of the developed techniques rely on the conjectured quantum integrability, which seems to hold only in the 't Hooft limit, i.e., for a divergent number of colors, $N_c \rightarrow \infty$.

	A method in principle giving all-order results for finite $N_c$ is the conformal bootstrap program. Proposed already in the seventies \cite{Ferrara:1973vz, Polyakov:1974gs}, the main idea is to constrain the conformal data by symmetries of the scattering $S$-matrix. Recently, initiated by the works \cite{Rattazzi:2008pe, El-Showk2012}, there has been a rekindling interest and similar numerics have been applied to {\NfourSYM} \cite{Beem:2013qxa, Alday:2013opa, Alday:2013bha}, where typically upper bounds on some scaling dimensions are derived, see also \cite{Beem:2013hha}. In this context also analytic understanding of superconformal symmetry \cite{Dolan:2001tt, Dolan:2002zh, Dolan:2003hv, Hogervorst:2013kva} seem to become more relevant again.

\section{The Spectral Problem for Short Strings} \label{sec:In-SpecProb}
	As discussed in the previous section, already since the initial days of the AdS/CFT integrability a major subject of investigation were single trace operators with a diverging number of fields, so-called {\it long} operators. The string states dual to these turn out to be semiclassical string states with some of the quantum charges diverging, $\psu(2,2|4)\ni Q \propto \sqrt{\l}\,$, where some prominent solutions were mentioned in \secref{sec:In-Integr}. For many of these the extension of the classical string becomes large, giving another reason to call them {\it long} string solutions. For instance, in the case of the GKP folded spinning string in ${\rm S}^3 \subset {\rm AdS}_5$ \cite{Gubser:2002tv} for spin $S \propto \sqrt{\l}$ the folded string extends and approaches the boundary of ${\rm S}^3$. 

	The virtue of such semiclassical long string states is that due to the diverging charges the Lagrangian, respectively, the world-sheet Hamiltonian take a well defined expansion and it is customary to work instead with rescaled charges, $\tilde{Q}=Q/\sqrt{\l}$. By this the energy turns out to scale as $E \propto \sqrt{\l}$ as well and the long string states are also referred to as {\it heavy} states.
	
	Throughout the last decade these states, long operators on the field theory side as well as long semiclassical string states, have been investigated thoroughly and one could assert that by now they are rather well understood. In particular, methods relying on the surmised quantum integrability were developed and tested extensively in this regime.

	Another regime is given by Bogomolny–Prasad–Sommerfield (BPS) states, which by definition are preserved under part of the SUSY and hence are also referred to as {\it protected}. On the string theory side of the duality, these correspond to supergravity, viz., superparticle modes\footnote{Also D$p$-branes are $1/2$-BPS objects, see \cite{Aharony:1999ti} but also the seminal work \cite{Lin:2004nb}.}, which are the lightest string states in the sense that for finite charges they have finite energy, $E \propto \l^0$. The inherited symmetry simplifies calculations drastically and the spectrum of BPS states is known, see \cite{Kim:1985ez, Gunaydin:1984fk} and \cite{Metsaev:1999kb, Metsaev:2002vr, Horigane:2009qb}, respectively. Especially, as renowned example one should mention that the vacuum of the BMN string \cite{Berenstein:2002jq}, the particle spinning along the great circle of ${\rm S}^5$, is a $1/2$-BPS state. 

	However, there is yet another regime, the regime of non-protected {\it short} states, for which our understanding is still unsatisfactory. On the field theory side the short operators are single trace operators involving only a finite number of {\NfourSYM} fields. Due to the finite length so-called wrapping corrections have to be taken into account, which obscure the underlying integrability. The prime example of such short operators is the Konishi operator \cite{Konishi:1983hf},
	\be \label{eq:In-KonishiOp}
		\cK = \tr(\phi_I \phi_I)~,
	\ee
	where $\phi_I$ are the six real scalars of {\NfourSYM}. As explained in \secref{sec:In-PSU}, the Konishi anomalous scaling dimension $\g = \Delta - \Delta_0$ is the same for all superconformal descendants, i.e., it is independent of the half-integer classical scaling dimension $\Delta_0$ depending on the particular state of the Konishi supermultiplet.

	The Konishi anomalous dimension $\g$ has been a testing ground for the hidden integrability of the spectral problem. In the weak coupling regime, using perturbation theory $\g$ has subsequently been calculated  up to an impressive 5-loop order \cite{Anselmi:1996dd,
 Bianchi:2000hn, Eden:2000mv, Eden:2000vb, Arutyunov:2000im, Fiamberti:2007rj, Fiamberti:2008sh, Velizhanin:2008jd, Eden:2012fe}. 
	
	Employing methods relying on the conjectured integrability as the thermodynamic Bethe-ansatz for the mirror model \cite{Arutyunov:2009zu, Arutyunov:2009ur, Bombardelli:2009xz, Gromov:2009bc} and the Y-system \cite{Gromov:2009tv} the Konishi anomalous dimension was derived to even higher orders \cite{Bajnok:2009vm, Arutyunov:2010gb, Balog:2010xa, Leurent:2012ab, Bajnok:2012bz},
	with the present record being set at eight \cite{Leurent:2013mr, Gromov:2013pga} or even nine loops \cite{VolinTalk}. So despite complications corresponding to finite length, the spectrum of short operators still seems rather accessible at weak coupling.

	Even more intriguingly, in \cite{Gromov:2013pga, Gromov:2014caa} the authors recently introduced what they call the {\it quantum spectral curve}, an advancement of the integrability based methods which supposedly yields the scaling dimension of {\it arbitrary states}. As the integrability in principle gives results to {\it arbitrary order} in the 't Hooft coupling $\l$ this would constitute a {\it solution of the spectral problem}.  

	Despite all this progress, on the strong coupling side of the duality our knowledge is rather sparse and apart from the Konishi anomalous dimension there is almost no data on the spectrum of other short operators. 
	Again, integrability based methods were used to predict the strong coupling expansion of the Konishi anomalous dimension $\g = E - \Delta_0 $ to the third perturbative order\footnote{As will become apparent in the course of this thesis, expansion in string fields is not the same as expansion in powers of $\l$. By this it is arguable whether one should refer to the order $\l^{-(2n-1)/4}$ in \eqref{eq:In-KoAnDim} as the `$n$-loop correction'.} \cite{Gromov:2011bz, Gromov:2011de, Frolov:2012zv, Gromov:2014bva},
	\begin{equation} \label{eq:In-KoAnDim}
		\g = 2\, \l^{1/4} + 2\, \l^{-1/4} 
			+ \left(\frac{1}{2} - 3\,\zeta_3\right) \l^{-3/4} 
				+ \left(-\frac{1}{2} + 6\,\zeta_3 + \frac{15}{6} \zeta_5\right) \l^{-5/4} + \ord{\l^{-7/4}}~.
	\end{equation}
	
	It is now desirable to rigorously test the conjectured quantum integrability by comparing this result to first principle calculations. However, quantization of the type IIB superstring in {\AdSxS} is poorly understood.

	Assuming no level-crossing 
	of the spectrum, it was argued \cite{Polyakov:2001af} that the string states dual to the Konishi supermultiplet should be the lowest massive string states. Already in the seminal work \cite{Gubser:2002tv} it was noticed that in leading order these coincide with the flat space string states at first level. this observation determines the leading order of the spectrum \eqref{eq:In-KoAnDim} to be
	\be	\label{eq:In-GKPresult}
		E = 2 \l^{1/4} + \ord{\l^0}~,
	\ee
	a behavior which has been a guideline for the integrability based methods and has been reobtained in \cite{Arutyunov:2004vx}. 
	Access to subleading orders is however obscured since for finite $\psu(2,2|4)$ charges the perturbative expansion of the Lagrangian, valid for charges diverging as $\sqrt{\l}$, formally breaks down.

	Such problems have been circumvented in \cite{Beccaria:2011uz, Roiban:2011fe, Beccaria:2012xm, Beccaria:2012kp} by taking results for the scaling dimension of semiclassical string solutions and interpolating these to the finite charge regime, an idea contrived in \cite{Tirziu:2008fk, Roiban:2009aa}. Impressively, taking the finite charge limit of different semiclassical solutions consistently reproduces the first quantum correction to the Konishi anomalous dimension, the order $\l^{-1/4}$. Even the non-zeta values of higher orders have been derived, which have been used as input in \cite{Gromov:2011bz, Gromov:2014bva}. One could object thought that this approach ignores potential order-of-limits ambiguities, which again could be viewed as presuming the integrability.

	A more faithful approach has been the work \cite{Passerini:2010xc}, where the bosonic subsector of the $\AdSxS$ superstring has been investigated in light-cone gauge. A crucial insight was that for finite $\psu(2,2|4)$ quantum charges the string zero modes scale differently in $\l$ than the non-zero modes. In particular, the zero modes turned out to obtain a mass determined by the non-zero mode excitations, in contrast to the free massless zero modes of the flat space string. By this, a concept for a meaningful strong coupling expansion was presented and a partial diagonalization of terms contributing to $\Delta_0$ was achieved. Due to ordering ambiguities the first real quantum corrections to $\g$, order $\l^{-1/4}$, were however out of reach.

	In another work \cite{Vallilo:2011fj} the Konishi anomalous dimension $\g$ was claimed to be calculated to order $\l^{-1/4}$ using pure spinor formalism. However, many questions about the derivation stay unanswered, see \chapref{chap:PureSpinor}.

	The goal of this thesis is to elaborate on these previous works and by this acquire insights on the spectrum and quantum symmetries of the $\AdSxS$ superstring from first principles. In a broader sense, the concern of this thesis is to contrive new means for perturbative quantization of the $\AdSxS$ superstring, where of course one hopes to apply the gained understanding to related systems. 

	This longstanding problem seems to be more relevant than ever, as with the advance of machinery building on the surmised quantum integrability, ironically, the need for comparative first principle calculation seem to fall into oblivion.

\chapter{Bosonic String Theory} \label{chap:BosString} 
	In this chapter we will review bosonic string theory, where mostly we will follow the presentation in \cite{Green:1987sp}. Other account on the matter include the textbooks \cite{Lust:1989tj, Polchinski:1998rq, Zwiebach:2004tj, Becker:2007zj} as well as the enjoyable lecture notes \cite{Tong:2012, Arutyunov:2008}.

	Starting from the point particle, in \secref{sec:BS-Classic} we describe the classical setup of the bosonic string, which for flat space is quantized in the following sections. First, in \secref{sec:BS-OCQ} we review the so-called old covariant quantization to then discuss light-cone gauge in \secref{sec:BS-LCG}.

	In \secref{sec:BS-Static} we go beyond the syllabus of \cite{Green:1987sp} and review quantization of the flat-space bosonic string in static gauge, which has been established only recently \cite{Jorjadze:2012iy}. Especially this last section is of particular importance for the research conducted in the course of this thesis.

\section{The Classical Bosonic String}\label{sec:BS-Classic}

 \subsection{From Particle to String}
	Let us look at some $d$ dimensional space-time manifold with one temporal and $d-1$ spatial directions, having coordinates $X^\m$ and metric $G_{\m\n}(X)$. By definition, at a given instance in time a particle is sitting at a point and throughout time the particle will trace out a smooth path in space-time, which can be parametrized by one variable $\tau$.

	The dynamics of a particle with mass $m$ is then determined by the action principle, by which the invariant length is extremized, $S=-m \int \dd s$, with the coordinate invariant definition for the square of the infinitesimal line element
	\be
		\dd s^2 = -G_{\m\n} \dd X^\m \dd X^\n = - \dd X \cdot \dd X~.
	\ee
	
	Since the path of the particle is given by a smooth map $X^\m(\tau)$ from a one dimensional manifold, the {\it world-line}, to the $d$ dimensional space-time, the action becomes
	\be \label{eq:BS-NGpart}
		S_\text{pp} = -m \int \dd \tau \sqrt{-\dot{X}^2}~.
	\ee
	with $\dot{X}^\m = \p_\tau X^\m = {\p X^\m \ov \p \tau}$ and $\dot{X}^2 = G_{\m\n} \dot{X}^\m \dot{X}^\n$, which has the important property of being invariant under reparametrization on the world-line. By introducing an auxiliary coordinate $e(\tau)$, which is interpreted as the {\it einbein} of the world-line, the action \eqref{eq:BS-NGpart} is brought to the form
	\be \label{eq:BS-Polpart}
		S_\text{pp}' = {1\ov2} \int \dd \tau \left(e^{-1} \dot{X}^2 - e m^2\right)~.
	\ee

	The Euler-Lagrange-equation for $e$,
	\be \label{eq:BS-eEoM}
		\dot{X}^2 + e^2 m^2 = 0~,
	\ee
	corresponds to the mass-shell condition and solving it will give again \eqref{eq:BS-NGpart}. In particular, invariance under $\tau$ reparametrization now corresponds to a gauge freedom in $e(\tau)$, where for example one could choose the gauge $e=1/m^2$.

	The form \eqref{eq:BS-Polpart} has the advantages over \eqref{eq:BS-NGpart} that it is also suitable for the massless case $m=0$ and, especially with the quantum theory in mind, that the square root has been resolved. However, one has to deal with the constraint \eqref{eq:BS-eEoM}.

	From this, it is natural to generalize the point particle case to extended objects, i.e., to look not at a world-line but a higher dimensional base manifold with one temporal direction $\tau$ and several spatial directions $\s^i$, $i=1,2,\ldots,n\ $. We are only concerned with the simplest case of a $1+1$ dimensional base manifold, a {\it world-sheet} with coordinates $\s^0=\tau$ and $\s^1=\s$. Hence, at fixed times the objects under investigation are one dimensional {\it strings}, giving string theory its name.

	In analogy to the particle, the action principle for the string extremizes the area and the generalization of \eqref{eq:BS-NGpart} is the {\it Nambu-Goto action}
	\be \label{eq:BS-NG}
		S_\text{NG} = T_0 \int \dd^2 \s \sqrt{-\det\left(G_{\m\n} \frac{\p X^\m}{\p \s^\a} \frac{\p X^\n}{\p \s^\b}\right)} 
			= T_0 \int \dd^2 \s \sqrt{{\dot{X}}^2 {X'}^2  - (\dot{X}\cdot X')}~,
	\ee
	with the string tension $T_0 = (2\pi \a')^{-1}$ and $\a'$ being the Regge slope.
	
	The generalization of \eqref{eq:BS-Polpart} is the so-called {\it Polyakov action}
	\be \label{eq:BS-Poly}
		S_\text{Pol} = -\frac{T_0}{2} \int \dd^2 \s \sqrt{h} h^{\a\b} G_{\m\n} \p_\a X^\m \p_\b X^\n = -\frac{T_0}{2} \int \dd^2 \s \sqrt{h} h^{\a\b} \p_\a X\cdot \p_\b X~,
	\ee
	where $\p_\a = \frac{\p}{\p \s^\a}$, $\,h^{\a\b}$ is the inverse of the world-sheet metric $h_{\a\b}=h_{\a\b}(\tau,\s)$, we defined $h = |\det(h^{\a\b}(\s))|$ and $\sqrt{h}\,\dd^2 \s$ is the invariant volume form. \eqref{eq:BS-Poly} is invariant under local reparametrization of the world-sheet $\s^\a \rightarrow \xi^\a=\xi^\a(\tau,\s)$, giving
	\begin{align}
		&\d X^\m = \xi^\a \p_\a X^\m~,\qquad \d(\sqrt{h})= \p_\a (\sqrt{h})~, 
			\label{eq:BS-WSrep}\\
		&\d h^{\a\b} = \xi^\g \p_\g h^{\a\b} - h^{\a\g} \p_\g \xi^\b - h^{\b\g} \p_\g \xi^\a = \nabla^\a \xi^\b + \nabla^\b \xi^\a~, \label{eq:BS-WSmetRep}
	\end{align}
	and furthermore, since the world-sheet is two dimensional, under local {\it Weyl scaling}
	\be
		\d h^{\a\b} = \L(\tau,\s) h^{\a\b}~.
	\ee
	In addition, the string will be invariant under the global isometries of the target space. For flat Minkowski space this is the Poincar{\'e} invariance
	\be\label{eq:BS-Poinc}
		\d X^\m = a^\m{}_\n X^\n + b^\m~,\qquad\quad \d h^{\a\b} = 0~.
	\ee
	
 \subsection{Equations of Motion}
	Variation of the action with respect to the $h^{\a\b}$ gives the {\it energy-momentum tensor},
	\be \label{eq:BS-EnMom}
		T_{\a\b} = - {2 \ov T_0} {1 \ov \sqrt{h}} \frac{\d S_\text{Pol}}{\d h^{\a\b}} 
			= \p_\a X \cdot \p_\b X - {1\ov2} h_{\a\b} h^{\g\d} \p_\g X \cdot \p_\d X ~,
	\ee
	which due to Weyl-invariance is traceless, $T^\a{}_\a = h^{\a\b} T_{\a\b}=0$. When the equation of motion (EoM) $T_{\a\b} \propto \d S_\text{Pol}/ \d h^{\a\b} = 0$ is imposed the action \eqref{eq:BS-Poly} reduces to \eqref{eq:BS-NG}.

	Furthermore, when the EoMs for $X^\m$ are imposed, $\d S_\text{Pol}/ \d X^\m = 0$, by \eqref{eq:BS-WSmetRep} and integration by parts the energy-momentum tensor shows to be covariantly conserved
	\be \label{eq:BS-TcovInv}
		\nabla^\a T_{\a\b} = 0~.
	\ee

	The EoMs for $X^\m$ take the form
	\be \label{eq:BS-XEoM}
		\p_\a \left(h^{\a\b} G_{\m\n} \p_\b X^\n\right) = 0~.
	\ee

	A comment on the topology of the world-sheet is in place. For strings to have finite spatial extension, the simplest topologies of the world-sheet are the strip and the cylinder. The strip results in {\it open strings} and the spatial coordinate $\s$ can be chosen to take values in $\s\in[0,\pi]$ while the cylinder corresponds to {\it closed strings} and one can choose $\s \in [0,2\pi]$, where one identifies $\s=\s+2\pi$. More complex topologies correspond to interacting strings, a hardly understood matter covered by {\it string field theory}, which we will not touch upon. The calculations for closed and open strings parallel each other so that here, for brevity, we will only cover the open string. 
	
	For the world-sheet being a strip the variation of the action will additionally give the boundary condition for the open string\footnote{As usual, $\calL$ denotes the Lagrange density, $S=\int \dd^2 \s \calL$.}
	\be \label{eq:BS-Bound1}
		\int \dd \tau \frac{\d \calL_\text{Pol}}{\d X'^\m}\Big|^{\s=\pi}_{\s=0} = 0~.
	\ee

	The two reparametrizations of the world-sheet can be used to bring the world-sheet metric into {\it conformal gauge} $h_{\a\b} = e^{2 \phi} \eta_{\a\b}$ and by additional Weyl scaling we can further restrict to the {\it covariant gauge}
	\be \label{eq:BS-CovarGauge}
		h_{\a\b} = \eta_{\a\b} = \begin{pmatrix}-1 & 0 \\ 0 & 1 \end{pmatrix}~.
	\ee
	Thus, the boundary condition \eqref{eq:BS-Bound1} is fulfilled by
	\be \label{eq:BS-Bound2}
		X'^\m = 0 ~,\qquad\quad \text{for $\s=0$ and $\s=\pi$}~.
	\ee

	Let us furthermore constrain our interest to flat Minkowski space as the target manifold, $G_{\m\n}=\eta_{\m\n}=\diag(-1,1,\ldots,1)$, and introducing world-sheet light-cone variables $\s^\pm=\tau\pm\s$, hence $\p_\pm = {1\ov2}(\p_\tau \pm \p_\s)$. For these the gauge \eqref{eq:BS-CovarGauge} reads
	\be \label{eq:BS-CovarGaugeLC}
		\eta_{+-}=\eta_{-+}=-{1\ov2}~,\qquad \eta_{++} = \eta_{--} = 0
	\ee
	and the EoMs \eqref{eq:BS-XEoM} becomes the two dimensional wave equation
	\be \label{eq:BS-WaveEqnX}
		\Box X^\m = \left(\p^2_\s - \p^2_\tau\right)X^\m = -4 \p_+ \p_- X^\m = 0~.
	\ee
	Taking into account \eqref{eq:BS-Bound2}, this has the general solution
	\begin{align} \label{eq:BS-XmodeExp}
		&X^\m(\tau,\s) = \phi^\m(\s^+) + \phi^\m(\s^-) = x^\m + l^2 p^\m \tau + i l \sum_{n\neq0}{1\ov n} \a^\m_n e^{-i n \tau} \cos(n\s)~,
	\end{align}
	corresponding to standing waves. Here,  we have $\a^\m_{-n}=(\a^\m_{n})^\dag$ by reality of $X^\m (\tau,\s)$ and $l = \sqrt{2 \a'} = 1/\sqrt{\pi T_0}$ is the fundamental string length. Note already that the zero modes $x^\m$ and $p^\m$ parameterizing the center-of-mass analog the particle degrees of freedom, while the non-zero modes $\a^\m_{n\neq0}$ capture the stringy dynamics.

	The equal world-sheet time Poisson brackets for $X^\m$ and $\dot X^\m$ are
	\be \label{eq:BS-XPoisson} 
		\{X^\m(\s),X^\m(\tilde{\s})\}=\{\dot{X}^\m(\s),\dot{X}^\m(\tilde{\s})\}=0~,\ \quad \{\dot{X}^\m(\s),X^\m(\tilde{\s})\}= \frac{1}{T} \d(\s-\tilde{\s})~,
	\ee
	which imply the non-vanishing Poisson brackets
	\be \label{eq:BS-XPBmode}
		\{p^\m,x^\n\}=\eta^{\m\n}~,\qquad\quad \{\a^\m_m,\a^\n_n\}=i m \d_{m+n} \eta^{\m\n}~.
	\ee

	In world-sheet light-cone variables, the equations for the energy-momentum tensor, \eqref{eq:BS-EnMom} and \eqref{eq:BS-TcovInv}, take the simple form (no sum over $\pm$)
	\be
		T_{\pm \mp} = 0~,\qquad T_{\pm\pm} = \p_\pm X \cdot \p_\pm X ~,\qquad 
		\p_\pm T_{\mp \mp} = 0~,
	\ee
	where the first equation corresponds to $T^\a{}_\a=0$ and the last equation implies the existence of an infinite set of conserved charges, $Q_{f,\pm} = \int \dd \s f(\s^\pm) T_{\pm\pm}$ with $\dot{Q}_{f,\pm} = 0$ for any function $f(\s^\pm)$. For the open string however, due to the boundary condition \eqref{eq:BS-Bound2}, $T_{--}$ is the continuation of $T_{++}$ to $\s\in[-\pi,\pi]$ and there is only one set of charges. In particular, it seems convenient to take the Fourier modes of $T_{\pm\pm}$,
	\be \label{eq:BS-Tmodes}
		L_m = T_0 \sum_\pm \int^\pi_0 \dd \s\,e^{\pm i m\s} T_{\pm\pm}
			= \frac{T}{4} \int^\pi_{-\pi} \dd \s (\dot{X}+X')^2 
			= {1\ov2} \sum^\infty_{n=-\infty} \a_{m-n}\cdot\a_n~,
	\ee
	where $\a^\m_0 = l p^\m$ and we evaluated at $\tau=0$. These are the generators of two dimensional conformal transformations and, for later reason, are referred to as {\it Virasoro modes}. Classically, they fulfill the Witt algebra
	\be \label{eq:BS-WittAlg}
		\{L_m, L_n\}= i (m-n) L_{m+n}~.
	\ee
	The EoM $T_{\a\b} \propto \d S_\text{Pol}/ \d h^{\a\b} = 0$ is then equivalent to vanishing of all all modes, $L_m = 0$.

	The conserved charges discussed above actually correspond to a residual gauge freedom left over after the choice \eqref{eq:BS-CovarGauge}, $h_{\a\b} = \eta_{\a\b}$. For this, note that any combination of reparametrization and Weyl scaling by $\L$ for which
	\be \label{eq:BS-ResGauge}
		\p^\a \xi^\b + \p^\b \xi^\a = \L \eta^{\a\b} 
	\ee
	leaves \eqref{eq:BS-CovarGauge} invariant. For the light-cone combinations $\xi^\pm = (\xi^0 \pm \xi^1)$ this implies that $\xi^+(\s^+)$ and $\xi^-(\s^-)$ are arbitrary functions. This residual gauge freedom will play a crucial role in \secref{sec:BS-LCG}.

\subsection{Isometries}
	After the survey of the EoMs led to description of the local symmetries in terms of the charges $L_m$, we are left with the global symmetries, the isometries, which in our case the Poincar{\'e} transformations \eqref{eq:BS-Poinc}. From Noether's theorem one finds the momentum and angular momentum currents
	\be \label{eq:BS-Current}
		P^\a_\m = - \frac{\p \calL}{\p (\p_\a X^\m)} = -T_0 \p^\a X_\m~,	\qquad\quad
		J^{\m\n}_\a = X^\m P^\n_\a - X^\n P^\m_\a~,
	\ee
	which are conserved, $\p_\a P^\a_\m = \p^\a J^{\m\n}_\a = 0$. For the open string the boundary condition \eqref{eq:BS-Bound2} implies that indeed there is no  (angular) momentum flowing out at the ends of the string. The total conserved momentum and angular momentum are obtained by integrating \eqref{eq:BS-Current} over $\s$,
	\begin{align} \label{eq:BS-IsoClass}
		&\mathsf{P}^\m = \int^\pi_0 \dd\s P^\m_\tau = p^\m~,\qquad
			J^{\m\n} = \int^\pi_0 \dd\s J^{\m\n}_\tau = l^{\m\n} + E^{\m\n} \\
		\text{with}\ \quad\qquad& l^{\m\n} = x^\m p^\n -x^\n p^\m~,\qquad
			E^{\m\n} = -i \sum_{n=1}^\infty \frac{1}{n} 
				\left(\a^\m_{-n} \a^\n_n - \a^\n_{-n} \a^\m_n\right)~,\qqquad \nn
	\end{align}
	where for convenience we evaluated at $\tau=0$. In particular, in flat space the center-of-mass momentum is conserved.

	With this notation, it is instructive to switch from space-time back to world-sheet concepts. Notice that $P_\m = P^\tau_\m$ is the momentum density conjugate to $X^\m$. Hence the world-sheet Hamiltonian reads
	\be \label{eq:BS-WSham}
		H = \int_0^\pi \dd \s (\dot{X}\cdot P - \calL) 
			= \frac{T_0}{2} \int_0^\pi \dd \s (\dot{X}+X')^2 = L_0 
			= {1\ov2} \sum^\infty_{n=-\infty} \a_{-n}\cdot\a_n~,
	\ee
	The constraint $H=L_0=0$ then translates to the mass-shell condition $p_\m p^\m = - M^2$, with the mass squared defined as
	\be \label{eq:BS-MassCL}
		M^2 = {1\ov \a'} \sum^\infty_{n \geq 0} \a_{-n}\cdot\a_n~.
	\ee

	For the closed string there are two distinct sets of oscillator variables, $\a^\m_n$ and $\tilde{\a}^\m_\n$. These are also referred to as left- and right-moving modes, as they correspond to the mode expansion in $\s^+$ and $\s^-$, repsectively. Accordingly, there are also two sets of conformal charges, $L_m$ and $\tilde{L}_m$. The Hamiltonian is $H = L_0 +\tilde{L}_0$ whereas the combination $V = L_0 - \tilde{L}_0$ generates rigid rotations in $\s$ and gives the {\it level matching constraint} in the quantum theory.

\section{Old Covariant Quantization}\label{sec:BS-OCQ}
	There are different ways to quantize the bosonic string theory, which when used correctly all are equivalent. In this section we review the so-called {\it old covariant} quantization. Instead of fixing the residual gauge freedom left after assuming \eqref{eq:BS-CovarGauge}, this uses the Virasoro constraints to single out the physical Fock space. This still involves so-called null states which require special treatment.

	Of course, by now this method is superseded by modern covariant quantization, which utilizes the Becchi-Rouet-Stora-Tyutin (BRST) cohomology. However, BRST quantization plays a minor role for the content of this thesis and, will not be reviewed\footnote{For more information see e.g. chapter 3 of \cite{Green:1987sp}.}.

\subsection{Mode Expansion and Commutation Relations} \label{subsec:BS-MEaCR}
	Let us quantize canonically by promoting classical functions to operators and Poisson brackets to commutators, $\{\cdot,\cdot\} \rightarrow -i [\cdot, \cdot]\,$. Therefore, \eqref{eq:BS-XPBmode} becomes
	\be \label{eq:BS-CCR}
		[x^\m, p^\n] = i \eta^{\m\n}~,\qquad\quad [\a^\m_m,\a^\n_{-n}] = [\a^\m_m,(\a^\n_{n})^\dag] = m \d_{m-n} \eta^{\m\n}~,
	\ee
	and $\a^\m_n$ are interpreted as harmonic oscillator raising and lowering operators. More precisely, they are related to properly normalized ladder operators $a^\m_n$ and ${a^{\m}_n}^\dag$ as
	\be \label{eq:BS-HarmOsc}
		\a^\m_n = \sqrt{n}\,a^\m_n~,\qquad\quad \a^\m_{-n} = \sqrt{n}\, {a^\m_n}^\dag \qquad\quad\text{for $n>0$}~.
	\ee
	Furthermore, as apparent from the classical world-sheet Hamiltonian \eqref{eq:BS-WSham}, for flat space the center-of-mass is free and any state will be an eigenstate of $\a^\m_0 = l p^\m$. Hence, the ground state is $\ket{0;p^\m}$, which is characterized by having zero excitations of non-zero modes and center-of-mass momentum $p^\m$.

	There is now a fundamental problem with the Fock space, which is that it contains infinitely many negative norm states, for example ${a^0_n}^\dag \ket{0}$ has norm $\bra{0} a^0_n {a^0_n}^\dag \ket{0} = -1$. This points out that we did not impose all constraints. Doing so will cancel these negative norm\footnote{These are usually called ghosts, which however should not be confused with BRST ghosts.}.

	The constraints we are looking for are vanishing of the energy-momentum tensor, $T_{\a\b}=0$, respectively, of the Virasoro modes, $L_m=0$. At the quantum level, these become operator equations on the states where however it is only consistent to impose half of them, i.e., physical states are singled out by requiring
	\be \label{eq:BS-Virasoro2}
		L_m \ket{\phi}=0 \qquad\qquad \text{for $m>0$}~.
	\ee
	By $L_{-m} = L_m^\dag$ then the matrix elements of $L_{m\neq0}$ for arbitrary physical states $\ket{\phi_{1,2}}$ vanish, $\bra{\phi_1} L_m \ket{\phi_2} = 0$, which is as close one gets to the classical constraint $L_m=0$.

	The zero mode $L_0$ now plays a somewhat special role.
	The expression of the modes $L_m$ in terms of $\a^\m_\n$ is given in \eqref{eq:BS-Tmodes} and, since $[\a^\m_{m-n},\a^\n_n]=0$ except for $m=0$, only for $L_0$ there arises an ordering ambiguity. Hence, let us define $L_0$ to be normal ordered
	\be
		L_0 = {1\ov2} \a^2_0 + \sum^\infty_{n=1} \a_{-n} \cdot \a_n
	\ee
	and analog to \eqref{eq:BS-Virasoro2} for physical states $\ket{\phi}$ we demand
	\be \label{eq:BS-Virasoro3}
		L_0 - a \ket{\phi} = 0 ~,
	\ee
	where the $c$-number $a$ parametrizes the yet to be determined ordering ambiguity. By this, quantization of \eqref{eq:BS-MassCL} gives the mass squared operator
	\be \label{eq:BS-MassQ}
		M^2 = \frac{1}{\a'} \bigg(-a + \sum^\infty_{n =1} \a_{-n}\cdot\a_n \bigg)~.
	\ee
	
	The constraints \eqref{eq:BS-Virasoro2} and \eqref{eq:BS-Virasoro3} are the {\it Virasoro constraints} and states respecting them are called {\it Virasoro primaries}, where $a$ is the {\it Virasoro weight} of the respective primary. The constraints \eqref{eq:BS-Virasoro2} and \eqref{eq:BS-Virasoro3} are also referred to as {\it physical state constraints} as in the initial literature, in particular also in \cite{Green:1987sp}, Virasoro primaries were called {\it physical states}. This notion is however somewhat outdated as the Fock space is still plagued by so-called {\it null states}.

	But before plunging into this, let us look at the quantum version of the isometries \eqref{eq:BS-IsoClass}. Obviously, the total momenta $\mathsf{P}^\m = p^\m$ do not inherit any ordering ambiguity and it turns out, that in the covariant treatment also the total angular momenta are just the normal ordered version of \eqref{eq:BS-IsoClass},
	\be \label{eq:BS-LorentzQ}
		J^{\m\n} = x^\m p^\n - x^\n p^\m - i \sum^\infty_{n=1} {1\ov n} \left(\a^\m_{-n} \a^\n_{n} - \a^\n_{-n} \a^\m_{n}\right)~.
	\ee
	as these fulfill the Poincar{\'e} algebra at the quantum level, see \eqref{eq:In-Poincare},
	\begin{align}
		&[p^\m,p^\n]=0~,\qquad\quad [J^{\m\n},p^\r] = i \big( \eta^{\m\r} p^\n - \eta^{\n\r} p^{\m} \big)~,\\[.2em]
		&[J^{\m\n}, J^{\r\s}] = i \big(\eta^{\m\r} J^{\n\s} + \eta^{\n\s} J^{\m\r} - \eta^{\m\s} J^{\n\r} - \eta^{\n\r} J^{\m\s}\big)~.
	\end{align}
	Since $[L_n,J^{\m\n}]=0$, the physical state conditions \eqref{eq:BS-Virasoro2} and \eqref{eq:BS-Virasoro3} are Lorentz invariant and the physical states form Lorentz multiplets.

 \subsection{Virasoro Algebra and Physical States}
	Let us now turn to the algebra of the Virasoro modes $L_m$. For $m\neq-n$ it is easy to see that the algebra is still the Witt algebra \eqref{eq:BS-WittAlg}. For $m=-n$ however the two sums coming from \eqref{eq:BS-Tmodes} each suffer from normal-ordering ambiguities. These can only contribute as an additional number and the quantum version of \eqref{eq:BS-WittAlg} becomes
	\be \label{eq:BS-VirAlg1}
		[L_m, L_n]= (m-n) L_{m+n} + A(m) \d_{m+n}~.
	\ee
	which is the {\it Virasoro algebra}. 

	Apparently, for \eqref{eq:BS-VirAlg1} we have $A(-m)=-A(m)$ and $A(0)=0$ and from the Jacobi identity for $L_{1}$, $L_{n}$, and $L_{n+1}$ one finds the recursive relation
	\be
		A(n+1) = \frac{(n+2)A(n) - (2n+1) A(1)}{(n-1)}~.
	\ee
	Hence, $A(1)$ and $A(2)$ suffice to determine all $A(n)$ and the general solution reads $A(n)=c_3 m^3 + c_1 m\,$.
	The value of $c_1$ could be shifted by redefinition of $L_0$, by which the constant $c_1$ with $a$ are related to each other. The simplest way to determine the constants $c_1$ and $c_3$ is to evaluate the expectation value of the commutator $[L_m,L_{-m}]$ with respect to the ground state $\ket{0;0}$ with $p^\m=0$. In particular, one finds
	\be
		\bra{0;0}[L_1, L_{-1}]\ket{0;0} =0~,\qquad\quad
		\bra{0;0}[L_2, L_{-2}]\ket{0;0} 
		= \frac{d}{2}~,
	\ee
	which suffices to fix $A(m)$ as
	\be
		A(m)=\frac{d}{12} (m^3-m)~.
	\ee
	Note in particular that $L_1$, $L_0$, and $L_{-1}$ form a closed subalgebra isomorphic to $\SU(1,1)$.

	After providing the complete form of the Virasoro algebra \eqref{eq:BS-VirAlg1}, we are left with the Virasoro weight $a$ and the space-time dimension $d$. For certain values of these there will be negative norm states in the Hilbert space, while for other values there will not. These two regions of parameter space are separated by {\it critical} values of $a$ and $d$ for which there are physical states of zero norm, i.e., for which the physical Hilbert space is positive semi-definite.

	Let us look at the ground state $\ket{0;k^\m}$ with momentum $k^\m$, where the mass-shell condition $L_0 = a$ implies $\a' k^2 = a$. States at the first excited level read $\z\cdot\a_{-1} \ket{0;k^\m}$ and have norm $\z^2$, where $\z^\m(k)$ is a polarization vector. For these the mass-shell condition is $\a' k^2 = a-1$ and imposing \eqref{eq:BS-Virasoro2} for $L_1$ gives $\z\cdot k =0$, leaving $d-1$ allowed polarization components. Taking $k^\m$ to lie in the (0,1)-plane, polarizations in the $d-2$ orthogonal spatial directions obviously give positive norm states. For $a>1$ the states are tachionic, $k^2>0$, and $k^\m$ can be chosen to have no time component, while the last $\z^\m$ lies in the time direction and gives a negative norm state. 
	Similarly, for $a>0$ the momentum $k^\m$ can be chosen to lie in the time direction and the last $\z^\m$ is spatial, giving a positive norm state. Finally, for $a=1$ we have $k^2=0$, $\z^\m$ is proportional to $k^\m$, $\z\cdot k=0$, and we end up with a zero norm state, $\z^2=0$. Therefore, the first condition for absence of ghosts is
	\be \label{eq:BS-aConstr}
		a\leq 1~.
	\ee
	The critical weight $a=1$ results in massless vector particles and a tachionic ground state. The zero norm states appearing at the first excited level are only the first of an infinite series of such states.

	To investigate these let us extend the terminology introduced beneath \eqref{eq:BS-MassQ}. A state $\ket{\psi}$ is called {\it Virasoro descendant} if it can be written as finite linear combinations of products of negative Virasoro modes acting on primaries.\footnote{The definition of Virasoro descendants corresponds to the notion of {\it spurious states}, which instead have orthogonality to physical states \eqref{eq:BS-DescPrime} as their defining property.}
	
	By definition, Virasoro descendants always take the form
	\be \label{eq:BS-Desc1}
		\ket{\psi} = \sum_{i>0} L_{-n_i} \ket{\chi_i}~,\qquad\qquad n_i>0~,
	\ee
	from which it is obvious that descendants $\ket{\psi}$ are orthogonal to all primaries $\ket{\phi}$,
	\be \label{eq:BS-DescPrime}
		\bra{\phi}\psi\ra = \sum_{i>0} \bra{\phi} L_{-n_i} \ket{\chi_i} = \sum_{i>0} (\bra{\chi_i} L_{n_i} \ket{\phi})^\dag = 0 ~.
	\ee
	Furthermore, since $L_{-n}$ for $n>3$ can be expressed as commutators of $L_{-1}$ and $L_{-2}$, \eqref{eq:BS-Desc1} can be brought to the form
	\be \label{eq:BS-Spuri3}
		\ket{\psi} = L_{-1} \ket{\chi_1} + L_{-2} \ket{\chi_2}~.
	\ee
	
	It is now important to note that there are also states which are both Virasoro primaries and Virasoro descendants at the same time. It immediately follows that these have zero norm, which is why they are called ${\it null states}$.

	In particular, lets look at descendants of the form
	\be
		\ket{\psi} = L_{-1} \ket{\phi}~,
	\ee
	for $\ket{\phi}$ a primary. For $\ket{\psi}$ to be physical we need
	\be
		(L_0 - a)\ket{\psi} = L_{-1}(L_0 - a + 1) \ket{\phi} =0~,\qquad
		L_1 \ket{\psi} = L_1 L_{-1} \ket{\phi} = 2 L_0 \ket{\phi} = 0~,
	\ee
	which holds for the critical value $a=1$. The state described above \eqref{eq:BS-aConstr} is only the simplest case with $\ket{\phi} = \ket{0;k^\m}$.

	We can try to find even more null physical states. For this, look now at the state
	\be \label{eq:BS-Desc2}
		\ket{\psi} = (L_{-2} + \g L^2_{-1}) \ket{\phi}~,
	\ee
	with again $\ket{\phi}$ a primary. This state is automatically annihilated by $L_m$ for $m>3$. Taking $a=1$, for it to be physical we need to impose $L_1 \ket{\psi} = L_2 \ket{\psi} = (L_0-1) \ket{\psi} = 0$ and one finds $3-2\g=0$ and $d=4(2+3\g)$, i.e., $\g =3/2$ and
	\be
		d=26\qquad\quad\text{(for $a=1$)}~.
	\ee
	This is the {\it critical dimension} of bosonic string theory\footnote{Generally, it can be shown that the bosonic string is free of ghosts for $a=1$ and $d=26$, the critical values, or $a\leq 1$ and $d\leq 25$.}. 

	The importance of null states now lies in their property that, by definition, they are orthogonal to both the Virasoro primaries and descendants. Adding a null state to a primary state will not alter any inner product with other primaries. Hence, altering primaries by null states will not change any physical expectation values and one concludes that null states correspond to gauge degrees of freedom, which in the scheme of covariant quantization have not been fixed. Therefore, {\it physical states} are defined as the equivalence class of Virasoro primaries of weight $a=1$ modulo the null states. Denoting the physical Hilbert space $\pzcH_{\text{phys}}$, the last statement takes the form
	\be
		\pzcH_{\text{phys}} = \frac{\pzcH_{\text{primary}}}{\pzcH_{\text{null}}}~.
	\ee

	For the closed string the computation works in a similar manner, with the main difference that one has two sets of Virasoro charges, $L_m$ and $\tilde{L}_m$, and oscillator modes, $\a^\m_n$ and $\tilde{\a}^\m_n$. Especially, one finds the same critical values, $a=1$ and $d=26$.

\section{Light-Cone Gauge Quantization}\label{sec:BS-LCG}
	In the previous section, we explored a covariant way to quantize the string, which was not manifestly ghost-free. In this section we exploit the residual gauge freedom to instead find a quantization scheme, light-cone quantization, which is ghost-free but not manifestly covariant. Hence, now we will recover the critical values $a=1$ and $d=26$ by requiring covariance.
	
	Historically, light-cone gauge quantization has been proposed before the old covariant quantization reviewed in the previous section.

 \subsection{Light-Cone Gauge}
	In \secref{sec:BS-Classic} we fixed covariant gauge \eqref{eq:BS-CovarGauge}, $h_{\a\b}=\eta_{\a\b}$, and found that for the open string the coordinates had the mode expansion \eqref{eq:BS-XmodeExp}, where additionally one has to satisfy the Virasoro conditions $T_{++}=T_{--}=0$. Additionally, we noted that there is still a residual gauge freedom \eqref{eq:BS-ResGauge}, which here will be exploited.

	For this we start by introducing space-time light-cone variables
	\be \label{eq:BS-STLCcoord}
		X^\pm = \frac{1}{\sqrt{2}} \left(X^0 \pm X^{d-1} \right)~.
	\ee
	It is important to note that even though these look similar to the worlds-sheet light-cone variables $\s^{\pm}=\tau\pm\s$ introduced in \eqref{eq:BS-CovarGaugeLC} in space-time there are $d-1$ different spatial directions and already singling out \eqref{eq:BS-STLCcoord} is manifestly noncovariant.

	In terms of these the flat Minkowski space metric has the non-vanishing components $\eta_{ij}=1$, with $i,j=1,\ldots,d-2$ parameterizing the spatial components orthogonal to the space-time light-cone directions \eqref{eq:BS-STLCcoord}, and $\eta_{+-}=\eta_{-+}=-1$, in particular $X^+=-X_-$ and $X^-=-X_+\,$.

	The residual gauge freedom \eqref{eq:BS-ResGauge} now allows us to choose any new world-sheet light-cone variables $\s^\pm \rightarrow \tilde{\s}^\pm (\s^\pm)$, by which
	\begin{align}
		&\tilde{\tau} = \frac{1}{2} \left(\tilde{\s}^+(\s^+) + \tilde{\s}^-(\s^-)\right)~,\\
		&\tilde{\s} = \frac{1}{2} \left(\tilde{\s}^+(\s^+) - \tilde{\s}^-(\s^-)\right)~.
	\end{align}
	By the first equation, $\tilde\tau$ fulfills the massless wave equation $\p_+ \p_- \tilde{\tau} = 0$, as did the $X^\m$ \eqref{eq:BS-WaveEqnX}. Light-cone gauge now exploits the residual gauge freedom by the choice $\tilde{\tau} = X^+/p^+ + \text{const.}\,$, which also fixes $\tilde{\s}$ up to a rigid translations. Equivalently, the light-cone gauge is usually stated as setting
	\be \label{eq:BS-LightConeG}
		X^+(\tau,\s) = x^+ + p^+ \tau~,
	\ee
	which classically corresponds to setting all oscillator coefficients $\a^+_n$ for $n\neq0$ to zero.

	With this, the Virasoro constraints $(\dot{X} \pm X')^2 = 0$ become
	\be
		(\dot{X}^- \pm {X'}^-) = \frac{1}{2 p^+} (\dot{X}^i \pm {X'}^i)^2~,
	\ee
	i.e., $X^-$ is given by terms quadratic the transverse direction $X^i$. Plugging in the mode expansion \eqref{eq:BS-XmodeExp} this implies
	\be \label{eq:BS-alphaMinus}
		\a^-_n = \frac{1}{p^+}\bigg({1\ov2}\sum^{d-2}_{i=1} \sum^\infty_{m=-\infty} :a^i_{n-m} a^i_m: -a \d_n \bigg)~,
	\ee
	where for convenience we chose $\a'=1/2$, $:\cdot:$ denotes normal ordering and, as in \eqref{eq:BS-Virasoro3}, for $\a^-_0$ we introduced the normal ordering constant $a$. In fact, since $p^+$ is fixed, $\a^-_0 = p^-$ is identified as the mass-shell condition
	\be
		M^2 = (2 p^+ p^- - p^i p^i) = 2(N-a)~,\qquad\quad 
		N = \sum^\infty_{n=1} \a^i_{-n} \a^i_n
	\ee
	where in contrast to \eqref{eq:BS-MassQ} the {\it level operator} $N$ now only counts excitations of the transverse oscillators.

	Note once more that fixing light cone gauge solved the Virasoro constraints $L_m=0$ exactly. Now, the combinations $p^+ a^-_n$ play a role analogous to the $L_m$, as suggested by \eqref{eq:BS-alphaMinus}. Indeed, these turn out to satisfy the Virasoro algebra\footnote{From the beginning, is is not clear that the $a$'s in \eqref{eq:BS-Virasoro3} and \eqref{eq:BS-alphaMinus} have anything in common. The fact that $p^+ \a^-_n$ fulfill \eqref{eq:BS-alphMinVirAlg} and that we will find the critical value $a=1$ establish the connection.},
	\be \label{eq:BS-alphMinVirAlg}
		[p^+ \a^-_m, p^+ \a^-_n] = (m-n) p^+ \a^-_{m+n} + \left(\frac{d-2}{12}(m^3-m) + 2 a m\right) \d_{m+n}~.
	\ee

 \subsection{Lorentz Algebra and Critical Dimension} \label{subsec:BS-LCG-LA}
	Next, one should wonder how the critical values of $a=1$ and $d=26$ might arise in the light-cone gauge setting. Since we solved the Virasoro constraints exactly, there are no subsidiary constraints to be imposed at the quantum level. Also, string states are obtained by acting with only the spatial ladder operators $\a^i_{-n}$ on the vacuum and the Hilbert space is manifestly ghost-free.

	On the other hand, we noted previously that the choice of light-cone variables \eqref{eq:BS-STLCcoord} obscures covariance. Hence, our best guess is that the critical values of $a$ and $d$ will result from tracking Lorentz invariance.

	It is customary to start with a heuristical argument. The first excited states $\a^i_{-1}\ket{0;p^\m}$ form a $(d-2)$ component vector, a representation of the $\SO(d-2)$ transforming the transverse direction $X^i$. If massive, under Lorentz transformations the state will acquire a longitudinal polarization, which is just the statement that massive vectors transform under $SO(d-1)$ while massless vectors transform under $\SO(d-2)$. This however stands in contrast to the physical states only having transverse excitations only. Hence, we require the states $\a^i_{-1}\ket{0;p^\m}$ to be massless, yielding $a=1$.

	To find the critical dimension, one can try to calculate the anomaly in \eqref{eq:BS-alphaMinus} directly. Assuming the summation to be symmetric, one has
	\be \label{eq:BS-ZetaReg}
		{1\ov2}\sum^{d-2}_{i=1} \sum^\infty_{n=-\infty} a^i_{-n} a^i_n
		={1\ov2}\sum^{d-2}_{i=1} \sum^\infty_{n=-\infty} :a^i_{-n} a^i_n:
		+ \frac{d-2}{2}  \sum^\infty_{n=1} n~.
	\ee
	Obviously, the sum $\sum^\infty_{n\geq1} n$ causing the anomaly diverges.  However, for $\Re(s)>1$ the more general sum $\sum^\infty_{n\geq1} n^{-s}$ converges to the Riemann zeta function $\z(s)$ and we can argue that with $\z(-1)=-1/12$ the second term in \eqref{eq:BS-ZetaReg} should be substituted as
	\be \label{eq:BS-ZetaReg2}
		\frac{d-2}{2} \sum^\infty_{n=1} n 
		\quad\stackrel{\text{$\z$-reg.}}\longrightarrow\quad 
		\frac{d-2}{2} \z(-1) = -\frac{d-2}{24}~,
	\ee
  a scheme known as {\it zeta function regularization}. We already saw that the normal ordering constant in \eqref{eq:BS-alphaMinus} should be $a=1$ and comparison with \eqref{eq:BS-ZetaReg2} provides the critical dimension $d=26$.

	The presented heuristic argument is however unsatisfactory. But the derivation of the critical values for $a$ and $d$ can be put on firm ground by investigating the Lorentz algebra. Especially, the rotations $J^{+-}$ and $J^{i-}$ given by \eqref{eq:BS-LorentzQ} will mix $X^+$ with the other coordinates and one has to perform a {\it compensating gauge transformation} to restore the light-cone gauge \eqref{eq:BS-LightConeG}. 

	In the classical field theory, an infinitesimal Lorentz transformation by $a^\m{}_n$ and a reparametrization $\xi^\a(\tau,\s)$ give rise to
	\be \label{eq:BS-LCGlorTrafo}
		\d X^\m (\tau,\s) = a^\m{}_\n X^\n (\tau,\s) + \xi^\a(\tau,\s) \p_a X^\m (\tau,\s)~,
	\ee
	where additionally $\xi^\a$ should respect covariant gauge $h_{\a\b} = \eta_{\a\b}$, i.e., it should be of the type \eqref{eq:BS-ResGauge}. On the other hand, to respect light-cone gauge \eqref{eq:BS-LightConeG} we require
	\be
		\d X^+ = a^+{}_\n (x^\n + p^\n \tau) = a^+{}_\n x^\n(\tau)~.
	\ee
	Combining these two, one finds for the compensating reparametrization
	\be
		\xi^0 = \frac{a^+{}_\n}{p^+}(x^\n(\tau)-X^\n(\tau,\s))~,
	\ee
	where $\xi^1$ is obtained by integrating $\p_\pm \xi^\mp = 0$ \eqref{eq:BS-ResGauge}. Taking this compensation into account the terms involving $a^+{}_i$ in \eqref{eq:BS-LCGlorTrafo} transform the transverse directions non-linearly. In the quantum level, such non-linearities can then give rise to anomalies, which in turn will give us insight on $a$ and $d$. In particular, we should check whether the $J^{\m\n}$ really fulfill the Lorentz algebra \eqref{eq:In-Poincare}.

	In light-cone gauge, $J^{ij}$ have just the same form quadratic in ladder operators as \eqref{eq:BS-LorentzQ} and furthermore $J^{+\m}=-J^{\m+}$ have no $\s$-dependence as we have $\a^+_n = 0$. Due to \eqref{eq:BS-alphaMinus}, the generators $J^{-i}$ are cubic in oscillators, so that the commutator $[J^{-i},J^{-j}]$ can contain terms quadratic and quartic in oscillators\footnote{$[J^{-i},J^{-j}]$ can not contain a normal ordering constant only, as it has to transform non-trivially.}. However, to respect Lorentz invariance, this commutator has to vanish identically, $[J^{-i},J^{-j}]=0$.

	The terms quartic in oscillators are the same as in the classical Poisson bracket and indeed vanish. Hence, one is left with a potential anomaly of the form
	\be
		[J^{-i},J^{-j}] = -\frac{1}{(p^+)^2} \sum^\infty_{m=1} \D_m 
			(\a^i_{-m} \a^j_m - \a^j_{-m} \a^i_m)
	\ee
	with $\D_m$ being a number. A rather tedious calculation\footnote{For more details see for instance \cite{Green:1987sp} or \cite{Arutyunov:2008}.} reveals that
	\be
		\D_m = m \left(\frac{26-d}{12}\right) + {1\ov m} \left(\frac{d-26}{12}+2(1-a)\right)~,
	\ee
	which vanishes only for the critical value, $a=1$ and $d=26$.

	Let us conclude by mentioning another impressive fact. We argued previously that the level one states $\a^i_{-1}\ket{0;p^\m}$ ought to be massless to transform under $\SO(d-2)$. Higher level states are also constructed in terms of the transverse oscillators, i.e., as $\SO(d-2)$ multiplets. But since they are massive they should transform under $\SO(d-1)$ instead. Indeed, it is a highly non-trivial check of Lorentz invariance that {\it only} for $a=1$ the $\SO(d-2)$ multiplets of fixed level form $\SO(d-1)$ multiplets.

	The simplest example arises at level two. Taking $d=26$, the states $\a^i_{-2}\ket{0;p^\m}$ and $\a^i_{-1}\a^j_{-1}\ket{0;p^\m}$ transform as $\bf 24$, respectively, ${\bf 299}_s + {\bf 1}$ of $\SO(24)$.  But together under Lorentz transformations they form the ${\bf 324}_s$ of $\SO(25)$. 
	
	The ground state $\ket{0;p^\m}$, since it is tachionic, is the singlet of $\SO(1,24)$ instead.

\section{Static Gauge Quantization}\label{sec:BS-Static}
	Another way to quantize the bosonic string in Minkowski space is to assume static gauge, where one identifies the world-sheet and the space-time temporal coordinates $X^0 \propto \tau$. As this gauge is most natural for the particle, it seems that it should have been generalized to the string already in the early days of string theory. However, it was only very recently \cite{Jorjadze:2012iy} that a fully consistent quantization of the bosonic string in static gauge has been achieved.

	As pointed out by the authors, the delay of more than forty years might be due to the fact that static gauge corresponds to taking the least optimal route: It is neither manifestly covariant, as the method presented in \secref{sec:BS-OCQ}, nor does it allow to solve all constraints, as in light-cone quantization \secref{sec:BS-LCG}. Hence, one is left with both, imposing constraints at the quantum level as well as demonstrating the Poincar{\'{e}} symmetry.

 \subsection{Classical Setup -- Hamiltonian Reduction}
	Setting $\a'=1/2$, with the canonical moment $P_\m = \pi\,\p\calL/\p \dot{X}^\m$ already defined above \eqref{eq:BS-WSham} the Polyakov action can be brought into first order form
	\be \label{eq:BS-PolyFOF}
		S = \int \dd \tau \int^\pi_0 \frac{\dd \s}{\pi} \left(P_\m \dot{X}^\m 
			- \xi_1 \left(P_\m P^\m + X'_\m X'{}^\m \right) 
			- \xi_2 \left(P_\m X'{}^\m \right)\right)~,
	\ee
	with the Lagrange multipliers given by $\xi_1 = \sqrt{h}/2 h^{\tau\tau}$ and $\xi_2 = h^{\tau\s}/ h^{\tau\tau}$. Solving the EoM for $P_\m$ and substituting the solution into \eqref{eq:BS-PolyFOF} will lead back to \eqref{eq:BS-Poly}, whereas the equations for $\xi_{1,2}$ give the Virasoro constraints
	\be \label{eq:BS-StatVirasoro}
		P_\m P^\m + X'_\m X'^\m = 0~,\qquad\qquad P_\m X'{}^\m = 0~.
	\ee
	Static gauge corresponds to the gauge choice
	\be \label{eq:BS-StaticGauge}
		X^0 + P_0 \tau = 0~,\qquad\qquad P_0'= 0~,
	\ee
	which by $\dot{X}^0 = P_0$ in turn determines the world-sheet metric. For flat Minkowski space the world-sheet metric turns out to be in covariant gauge \eqref{eq:BS-CovarGauge}, $h_{\a\b}=\eta_{\a\b}$, but for more general space-times this is not the case. 
	
	Defining $\SO(d-1)$ vectors as $\vX = (X^1,\ldots,X^{d-1})$ and $\vP = (P^1,\ldots,P^{d-1})$ and denoting their scalar products as $\vX^2=\vX\cdot\vX=\sum^{d-1}_{i=1} (X^i)^2$, etc., with \eqref{eq:BS-StaticGauge} the action reduces to
	\be \label{eq:BS-StatAct}
		S = \int \frac{\dd \s^2}{\pi} \left(\vP\cdot\vX - \frac{1}{2} P_0^2 \right)~,
	\ee
	where the total derivative $\p_\tau (P_0^2\,\tau)$ has been neglected. Hence, $\frac{1}{2}P_0^2$ is nothing but the world-sheet Hamiltonian for the spatial variables. By use \eqref{eq:BS-StatVirasoro} and \eqref{eq:BS-StaticGauge} one has
	\be \label{eq:BS-StatHamil}
		H = \frac{1}{2} P^2_0 = \frac{1}{2}\int^\pi_0 \frac{\dd \s}{\pi} \left(\vP^2 + \vX'^2\right)~,
	\ee
	describing $d-1$ massless free fields, which additionally have to fulfill the constraints
	\be
		\left(\vP^2 + \vX'^2 \right)'=0~,\qquad\qquad \vP\cdot\vX' = 0~.
	\ee
	The mode-expansion of $X^i$ was given in \eqref{eq:BS-XmodeExp} and the Poisson brackets and commutation relations of the modes were already specified in \eqref{eq:BS-XPBmode} and \eqref{eq:BS-CCR}, respectively. The generators of conformal transformations $L_m$ of the system \eqref{eq:BS-StatAct} and \eqref{eq:BS-StatHamil} take again the form \eqref{eq:BS-Tmodes}, where however only spatial oscillators contribute,
	\be
		L_{m} = \frac{1}{2} \sum^\infty_{n=-\infty} \va_{m-n}\cdot \va_n~,
	\ee
	with $\vec\a_{n\neq0} = \sqrt{|n|} \va_{n\neq0}$ \eqref{eq:BS-HarmOsc} and $\va_0 = \vec{p}\,$, where $L_0$ is just the Hamiltonian  \eqref{eq:BS-StatHamil}. In the quantum theory, these give the physical state constraints \eqref{eq:BS-Virasoro2}, $L_m \ket{\phi} = 0$ for $m>0$. 

	The general form of the isometries was given in \eqref{eq:BS-IsoClass}. Imposing static gauge and evaluating at $\tau=0$, the total momenta and the rotations $J^{k l}$ take the usual form. However, for $\tau=0$ one has $P_0 = p_0$ and $X^0=0$, by which the boosts $J^{0 k}$ degenerate. Hence, it is found that to respect the Lorentz algebra the boosts have to be deformed by the constraints $L_{n\neq0}\,$. Altogether at $\tau=0$ the isometries read
	\begin{align}
		&\mathsf{P}^k = p^k~,&
			&J^{kl} = p^k x^l - p^l x^k + i \sum_{n\neq 0} \frac{a^k_{-n} a^l_n}{n}~,&\\
		&\mathsf{P}^0 = p^0 = \sqrt{2 L_0}~,&
			&J^{0k} = p^0 x^k + \frac{i}{p^0} \sum_{n\neq 0} \frac{a^k_{n}}{n} L_{-n}~.&
	\end{align}
	The invariance of the constraint surface $L_{n\neq 0}=0$ is then ensured by
	\be \label{eq:BS-JLpb}
		\{J^{0 k},L_m\} = 
			\frac{m}{p^0} \sum_{n \neq m,0} \frac{a^k_{n+m}}{n+m} L_{-n}
			- \left(\frac{i(m x^k + p^k)}{p^0} + \frac{m}{(p^0)^3} \sum_{n\neq 0} \frac{a^k_{n}}{n} L_{-n} \right) L_m~.
	\ee
	However, the right hand side contains both positive and negative Virasoro modes and, when quantized, \eqref{eq:BS-JLpb} can not provide Lorentz invariance of the physical Hilbert space defined by \eqref{eq:BS-Virasoro2}. Therefore, the boosts require further deformation in terms of higher powers in constraints. 

	Taking the ansatz
	\be \label{eq:BS-StatCJclass}
		\cJ^{0k} = J^{0k} + \frac{i}{p^0} \sum_{j\geq2} \left(
			\sum_{n_1,\ldots,n_j} f_j(p^0) \frac{a^k_n}{n} L_{-n_1} \ldots L_{-n_j} \right)~,
	\ee
	with $\sum^j_{i=1} n_i = n$ with $n\neq0$ and $n_i\neq0$, one can solve $p^0$ dependent coefficients $f_j(p^0)$ recursively in the $j$, the order in constraints, giving the recursion relation
	\be
		(j+1)(p^0)^2 f_{j+1} + (2j - 1) f_j = 0~.
	\ee
	With $f_1 = 1$ one finds
	\be
		f_j = (-1)^{j-1} \frac{\calC_{j-1}}{e^{j-1}}~,
	\ee
	where $e = 2 (p^0)^2$ and $\calC_{j} = \frac{(2j)!}{j!(j+1)!}$ are the Catalan numbers.

\subsection{Quantization}
	Following the steps in \ssecref{subsec:BS-MEaCR}, the unconstrained Hilbert space is generated by acting with the creation operators $a^k_{-n}$ onto the momentum dependent ground state $\ket{0;\vp}$ fulfilling
	\be
		a^k_0 \ket{0;\vp} = p^k \ket{0;\vp}~,\quad\qquad a^k_n \ket{0;\vp} = 0~,\quad\qquad\text{for $n>0$}~.
	\ee
	Note that we still need to impose the Virasoro constraints, corresponding to the fact that, in contrast to light-cone gauge \secref{sec:BS-LCG}, we were not able to solve all constraints. However, in contrast to covariant quantization \secref{sec:BS-OCQ}, the states of the unconstrained Hilbert space only have positive norm, corresponding to the fact that the gauge has been fixed. Therefore, there are no null states to be modded out, i.e., the states left after imposing the Virasoro constraints are the physical states.

	The Virasoro modes $L_m$ have no ordering ambiguity except for $n=0$ and we define
	\be
		L_0 = {1\ov2} \vp^{\,2} + \sum_{n>0} \va_{-n}\cdot\va_n = {1\ov2} \vp^{\,2} + N~,
	\ee
	with $N$ being the number operator. The Virasoro algebra then reads, c.f. \eqref{eq:BS-VirAlg1},
	\be
		[L_m,L_n] = (m-n)L_{m+n} + \frac{d-1}{12}(m^3-m)~.
	\ee
	Since $L_m$ for $m>3$ can be expressed in terms of $L_1$ and $L_2$ it suffices to impose the physical constraints
	\be
		L_1 \ket{\phi}=0~,	\qquad\qquad L_2 \ket{\phi} =0~.
	\ee
	after imposing these, one finds that at first and second level there are $d-2$ and $\frac{1}{2}(d-2)(d+1)$ physical states in agreement with the other quantization schemes.

	Next one looks at the Poincar{\'{e}} generators. In analogy to \eqref{eq:BS-Virasoro3}, one parametrizes the normal ordering ambiguity of the energy squared operator \eqref{eq:BS-StatHamil} by $a$, $(\p^0)^2 = 2(L_0 - a)$, hence
	\be
		p^0 = \sqrt{\vp^2 + 2(N-a)}~,
	\ee
	which is diagonal on level $N$ states $\ket{N;\vp}$.

	More problematic are the boosts. The classical expression \eqref{eq:BS-StatCJclass} generalizes to operators which act on level $N$ states $\ket{N;\vp}$ as
	\be \label{eq:BS-StatCJquant}
		\cJ^{0k} \ket{N;\vp} = \left(:p^0 x^k: + \frac{i}{p^0} \sum^N_{n=1}
			\sum_{(n_1,\ldots,n_j)} f^{(n_1,\ldots,n_j)}(p^0) L_{-n_1} \ldots L_{-n_j} \frac{a^k_n}{n}  \right) \ket{N;\vp}~,
	\ee
	where $:p^0 x^k:=\frac{1}{2}(x^k p^0 + p^0 x^k) = x^k p^0 - \frac{i p^k}{2 p^0}\,$ and $(n_1,\ldots,n_j)$ are the ordered partitions of $n$, i.e., $\sum^j_{i=1} n_i =n$ and $n\geq n_j \geq \ldots n_1 >0$. The coefficients $f^{(n_1,\ldots,n_j)}$ are the quantum counterpart of the $f_j$ in \eqref{eq:BS-StatCJclass}.

	For the physical Hilbert space to be Lorentz invariant, these have to fulfill the Lorentz algebra and we require
	\be \label{eq:BS-StatCJCJquant}
		[\cJ^{0 k}, \cJ^{0 l}] \ket{N;\vp} = i J^{k l} \ket{N;\vp}~.
	\ee
	Since the operators $\cJ^{0 k}$ preserve the level $N$, this is equivalent obtaining all matrix elements $\bra{N}[\cJ^{0 k}, \cJ^{0 l}] \ket{N;\vp} = i \bra{N}J^{k l} \ket{N;\vp}\,$, with $\bra{N}$ being any other level $N$ state. Note that the calculation of the matrix elements simplifies due to
	\be
		\bra{N}\cJ^{0 k} \cJ^{0 l} \ket{N;\vp} = \bra{N} :p^0 x^k: \cJ^{0 l}~, \ket{N;\vp}
	\ee
	which immediately follows from the form of the boosts \eqref{eq:BS-StatCJquant}. This then suffices to fix the coefficients $f^{(n_1,\ldots,n_j)}$ recursively level by level.

	The vacuum is obviously Lorentz invariant. At the first excited level one has
	\be
		\cJ^{0 k} \ket{1;\vp} = \left(x^k p^0 - \frac{i p^k}{2 p^0} + \frac{i f^{(1)}}{p^0} L_{-1} a^k_{-1} \right) \ket{1;\vp}~,
	\ee
	where $p^0 = \sqrt{\vp^{\,2} +2(1-a)}$. By $[L_1,x^k]=-i a^k_{1}$, requiring $L_1 \cJ^{0 k} \ket{1;\vp} = 0$ gives
	\be
		f^{(1)} = \frac{\vp^{\,2} + 2(1-a)}{\vp^{\,2}}~.
	\ee
	On the other hand, requiring \eqref{eq:BS-StatCJCJquant} one obtains
	\be
		\bra{1} p^0 x^k \frac{i}{p^0} L_{-1} a^l_1 \ket{1;\vp} = - \bra{1} a^k_{-1} a^l_1 \ket{1;\vp}~,
	\ee
	which yields $f^{(1)}=1$ and hence the critical value $a=1$. The corresponding mass operator $M^2 = 2(N-1)$ then reproduces the bosonic string spectrum.

	At second level \eqref{eq:BS-StatCJquant} reads
	\be
		\cJ^{0 k} \ket{2;\vp} = \left(x^k p^0 - \frac{i p^k}{2 p^0} 
			+ \frac{i}{p^0} L_{-1} a^k_{-1}
			+ \frac{i f^{(2)}}{p^0} L_{-2} a^k_{-2}
			+ \frac{i f^{(1,1)}}{p^0} L_{-1}L_{-1} a^k_{-2}\right) \ket{2;\vp}~,\nn
	\ee
	and imposing $L_1 \cJ^{0k} \ket{2;\vp} = 0$ and  $L_2 \cJ^{0k} \ket{2;\vp} = 0$ yields
	\be \label{eq:BS-Stat2nd}
		3 f^{(2)} + 2 (\vp^{\,2}+1)f^{(1,1)} = 2~,\qquad
		(4 \vp^{\,2} + d-1) f^{(2)} + 6 \vp^{\,2} f^{(1,1)} = 4(\vp^{\,2}+5)~,
	\ee
	Imposing \eqref{eq:BS-StatCJCJquant} then gives the additional equation $f^{(2)} - f^{(1,1)} = 1$, which together with \eqref{eq:BS-Stat2nd} imply
	\be
		f^{(1,1)} = \frac{1}{e+1}~,\qquad\qquad f^{(2)} = \frac{e}{e+1}~,
			\qquad\qquad \text{with}\quad e= 2(p^0)^2~,
	\ee
	and determine the critical dimension to be $d=26$.

	The authors of \cite{Jorjadze:2012iy} were not able to derive the general expression for the coefficients $f^{(n_1,\ldots,n_j)}$ but using the recursive algorithm sketched above their values were explicitly determined up to level 8. Furthermore, the connection to the covariant quantization has been elaborated.

\chapter{The {\AdSxSheader} Superstring} \label{chap:SS}
	In the last chapter we explored the bosonic string in flat Minkowski space. Especially, using different quantization schemes we derived that the bosonic string is critical in $d=26$ dimensions.

	Naturally, one tries to supersymmetrize the theory. In \secref{sec:SS-FlatSpace}, we start out by reviewing some facts about superstring theory in flat space. We will elaborate on the notions of world-sheet and space-time SUSY, leading to the Ramond-Neveu-Schwarz (RNS), respectively, the Green-Schwarz (GS) formulations of the superstring. From this, eventually, the critical dimension for the superstring proves to be $d=10$. 

	Superstring theory has supergravity (SUGRA) as its low energy effective theory and $d=10$ dimensional Minkowski space is a maximally supersymmetric solution of type IIB SUGRA. But there is another maximally supersymmetric solution, namely $\AdSxS$. So even without knowledge of the AdS/CFT conjecture, one is led to investigation of the corresponding string theory. This background is supported by a self-dual Ramond-Ramond five-form flux, which deters one from using the RNS formulation for the superstring. In particular, the Ramond-Ramond vertex operator is known to be non-local in terms of world-sheet fields and it is unclear how to couple it to the string world-sheet.

	Therefore, one instead uses the Green-Schwarz formalism which has the advantage of manifestly realizing the space-time SUSY. In practice, constructing the GS superstring action for arbitrary superstring solutions is difficult as one has to determine the full structure of the type IIB superfields starting from the bosonic solution, a problem which has not been solved yet in all generality.

	Fortunately, for the $\AdSxS$ background there is the alternative approach of formulating it as a WZNW-type non-linear sigma on a coset superspace. For flat space this approach has been applied \cite{Henneaux:1984mh} with the coset space being the Poincar{\'{e}} group modded by its stabilizer, the Lorentz group $\SO(1,9)$. For the $\AdSxS$ superstring the appropriate coset shows to be
	\be
		\frac{\PSU(2,2|4)}{\SO(1,4)\times\SO(5)}~,
	\ee
	where the corresponding action was first found in \cite{Metsaev:1998it}.
	
	Indeed, for the rest of this chapter we fully concentrate on the coset model construction of the $\AdSxS$ superstring, where we elaborate only on the classical theory. Our presentation closely follows the excellent review \cite{Arutyunov:2009ga}, see also \cite{Tseytlin:2010jv, McLoughlin:2010jw, Magro:2010jx}.

	In \secref{sec:SS-PSU} we review the superconformal algebra $\psu(2,2|4)$. This complements the description in \secref{sec:In-PSU}, as the construction in terms of supermatrices takes quiet a different guise.

	The coset formulation of the Green-Schwarz supertstring in $\AdSxS$ is then formulated in \secref{sec:SS-GS}, while in \secref{sec:SS-PS} we review the pure spinor formulation of the $\AdSxS$ superstring.

	In this chapter we neither review the integrability of the classical theory \cite{Bena:2003wd} nor do we discuss concrete coset parametrizations. For these topics as well as discussion of the superstring scattering $S$-matrix and quantization using uniform light-cone gauge we once more refer the reader to the review \cite{Arutyunov:2009ga}.

\section{Superstring Theory in Flat Space} \label{sec:SS-FlatSpace}
	There are essentially two different ways to supersymmetrize the flat space bosonic string discussed in \chapref{chap:BosString}. One can either supersymmetrize the space-time coordinates, giving the Green-Schwarz (GS) superstring, or one can supersummetrize the world-sheet coordinates, leading to the Ramond-Neveu-Schwarz (RNS) formulation of superstring theory. 
	
	However, the RNS superstring is inconsistent unless the spectrum is truncated in a very specific manner first proposed by Gliozzi, Scherk and Olive (GSO). After this so-called GSO projection is imposed, the truncated RNS string appears not only to be supersymmetric on the world-sheet but also in space-time, suggesting that indeed the RNS and GS formulations of superstring theory are equivalent.

	In this section we quote the most important results for these two formulations of the superstring in flat space, where again our presentation mostly follows the textbook \cite{Green:1987sp}. Further introductions on the flat space superstring include \cite{Lust:1989tj, Polchinski:1998rr, Becker:2007zj}.

 \subsection{World-Sheet Supersymmetry} \label{subsec:SS-FS-RNS}
	To supersymmetrize the world-sheet one has to extend it by fermionic degrees of freedom, i.e., by Grassmann superspace coordinates $\th^A$\footnote{Here, $A$ is counting the number of fermionic components, not the number of spinors.}. After choosing the gauge \eqref{eq:BS-CovarGauge} the world-sheet is (locally) two dimensional Minkowski space, so the added fermions should form spinors of the respective Clifford algebra
	\be \label{eq:SS-2drhos}
		\{\r^\a,\r^\b\} = -2\eta^{\a\b}~,\qquad\text{with e.g.}\quad
		\r^0 =\begin{pmatrix} 0 & -i\\ i & 0 \end{pmatrix}~,\quad
		\r^1 =\begin{pmatrix} 0 & i\\ i & 0 \end{pmatrix}~.
	\ee
	Hence, $\th^A$ form two-component spinors, $A=1,2$. As the Majorana condition can be imposed, these are real and the Dirac and Majorana conjugate of any $d=2$ spinor $\psi$ are given by
	\be
		\bar\psi = \psi^\dag \r^0 = \psi^T \r^0~.
	\ee
	By this, the bosonic space-time coordinates $X^\m(\s)$ generalize to superfields
	\be
		Y^\m(\s,\th) = X^\m (\s) + \bar{\th} \psi^\m(\s) + \frac{1}{2} \bar{\th}\th B^\m (\s)~,
	\ee
	where we omitted contracted spinor indices. As usual, the expansion of $Y^\m(\s,\th)$ stopped after a finite number of terms due to the anticommutation properties of the fermions. Note furthermore that all components $X^\m(\s)$, $\psi^\m(\s)$ and $B^\m(\s)$ transform as $\SO(1,d-1)$ space-time vectors, whereas on the world-sheet $X^\m(\s)$ and $B^\m(\s)$ are scalar while $\psi^\m(\s)$ is a fermionic spinor. 

	The superfield is manifestly invariant under the world-sheet SUSY generated by
	\be
		Q_A = \frac{\p}{\p \bar{\th}^A} + i (\r^\a \th)_A \p_\a~,
	\ee
	which fulfill the defining property that the commutator of two SUSY transformations generates world-sheet translations
	\be
		[\bar{\e}_1 Q, \bar{\e}_2 Q] = 2 (i \bar{\e}_1 \r^\a \e_2) \p_\a~.
	\ee
	
	Under a SUSY transformation $\d Y^\m = [\bar{\e}Q, Y^\m]$ the components transform as
	\be
		\d X^\m = \bar{\e} \psi^\m~,\qquad 
			\d\psi^\m = -i \r^\a \e \p_\a X^\m + B^\m \e~,\qquad
			\d B^\m = -i \bar{\e} \r^\a \p_\a \psi^\m~,
	\ee
	and SUSY is realized {\it off-shell}, that is without imposing any EoM.
	
	The corresponding superspace covariant derivative is given by
	\be
		D_A = \frac{\p}{\p \bar{\th}^A} - i (\r^\a \th)_A \p_\a~,
	\ee
	which indeed fulfills $\{D_A,Q_B\}=0$.

	With all this, the supersymmetric generalization of the bosonic string \eqref{eq:BS-Poly} is
	\begin{align}
		&\qquad S'_\text{RNS} = \frac{i T}{4} \int \dd^2 \s \dd^2 \th \bar{D} Y^\m D Y_\m~,\\
		\text{where}\qqquad
		&D Y^\m = \psi^\m + \th B^\m - i \r^\a \th \p_\a X^\m + {i\ov2}\bar{\th}\th \r^\a \p_\a \psi^\m~,\\
		&\bar{D} Y^\m = \bar{\psi}^\m + B^\m \bar{\th} + i \p_\a X^\m \bar{\th} \r^\a - {i\ov2}\bar{\th}\th \p_\a \bar{\psi}^\m \r^\a~.\qquad
	\end{align}
	The $\th$ integration can be performed using the Berezin integration rules
	\be
		\int \dd^2 \th (a + \th^A b_A + \th^1 \th^2 c) = c~,
	\ee
	hence $\int \dd^2 \th \, \bar{\th} \th = -2i\,$, giving
	\be
		S_\text{RNS} = \frac{-T}{2} \int \dd^2 \s
			\left( \p_\a X^\m \p^\a X_\m - i \bar{\psi} \r^\a \p_\a \psi_\m 
			- B^\m B_\m \right)~.
	\ee
	Immediately, from this we see that the field $B^\m$ is auxiliary and in particular the field equations imply $B^\m=0$. For the coordinates $X^\m$ one again obtains the wave equation \eqref{eq:BS-XEoM} and the equal $\tau$ Poisson structure \eqref{eq:BS-XPoisson}. For the world-sheet fermions $\psi^\m$ the EoM is the two dimensional Dirac equation $\r^\a \p_\a \psi^\m = 0$ and denoting $\psi^\m = (\psi^\m_-, \psi^\m_+)$ for the choice \eqref{eq:SS-2drhos} one has
	\be \label{eq:SS-2dDiracLC}
		(\p_\s + \p_\tau) \psi^\m_- = 0~,\qquad\qquad
			(\p_\s - \p_\tau) \psi^\m_+ = 0~,
	\ee
	hence, $\psi^\m_\pm$ depends only on $\s_\pm$. The $\tau$ Poisson structure of the fermions is
	\be \label{eq:SS-2dpsiPoisson}
		\{\psi^\m_\pm(\s), \psi^\n_\pm(\ts)\} = \frac{1}{T} \eta^{\m\n} \d(\s-\ts)~,\qquad
		\{\psi^\m_\pm(\s), \psi^\n_\mp(\ts)\} = 0~.
	\ee
	But this implies that the temporal fermions $\psi^0_\pm (\s)$ will give rise to negative norm states. For the bosonic string in \secref{sec:BS-OCQ} we got rid of negative norm states corresponding to the temporal coordinates $X^0(\s)$ by acknowledging the conformal symmetry, i.e., by imposing the Virasoro constraints obtained from the energy momentum tensor $T_{\a\b}$. This however exploits the Virasoro constraints and cancel the negative norm states corresponding to $\psi^0_A(\s)$ we are hoping for a new underlying symmetry.

	Indeed, by Noether's procedure one finds the conserved supercurrent corresponding to local SUSY transformations
	\be
		J_\a = \frac{1}{2} \r^\b \r_\a \psi^\m \p_\b X_\m~,\qquad\qquad
			\p^\a J_\a=0~.
	\ee
	which due to $\r^\a \r^\b \r_\a = 0$ additionally satisfies $\r^\a J_\a = 0$. Together with \eqref{eq:SS-2dDiracLC} this yields that the light-cone combinations $J_\pm = J_\tau \pm J_\s$ have only one non-vanishing spinor-component, which is therefore simply denoted as $J_\pm$ and which turn out to be
	\be
		J_\pm = \psi^\m_\pm \p_+ X_\m~,\qquad\qquad \p_\mp J_\pm = 0
	\ee
	
	Now, also the energy-momentum tensor involves additional terms coming from the fermions $\psi^\m$ and the light-cone combinations become
	\be
		T_{\pm\pm} = \p_\pm X^\m \p_\pm X^\n 
			+ {i\ov2} \bar{\psi}^\m_\pm \p_\pm \psi_{\pm\m}~, \qquad 
		\p_\mp T_{\pm\pm} = 0~, \qquad T_{\pm \mp} = 0~,
	\ee
	where again $T_{\pm \mp} = 0$ corresponds to $T_{\a\b}$ being traceless. The currents are then connected through their Poisson algebra
	\be
		\{J_\pm(\s), J_\pm(\ts)\} = \frac{1}{T} \d(\s-\ts) T_{\pm\pm}(\s)~,
			\qquad\quad \{J_\pm(\s), J_\mp(\ts)\} = 0~.
	\ee
	This suggests that one hardly can set $T_{\pm\pm}$ to zero without also doing so for $J_\pm$, which leads us to the {\it super-Virasoro constraints},
	\be
		T_{\pm\pm} = J_\pm = 0~.
	\ee

	With this, one should again investigate boundary conditions. For the open string variation of the Lagrangian gives a surface term which vanishes for $\psi_+ \d \psi_+ - \psi_- \d \psi_-$ vanishing at the ends of the string. This is satisfied for $\psi_+ = \pm \psi_-$ at each end. Without loss of generality at $\s=0$ one takes
	\be
		\psi^\m_+ (0,\tau) = \psi^\m_- (0,\tau)~,
	\ee
	and $\psi^\m_- (\s,\tau)$ can be viewed as a continuation of $\psi^\m_+ (\s,\tau)$ to $\s\in[-\pi,\pi]$. At $\s=\pi$ there are now two distinct choices, the {\it Ramond} (R) boundary condition,
	\be
		\text{(R)}\qquad\ \ \psi^\m_+ (\pi,\tau) = \psi^\m_- (\pi,\tau)\qquad\ \Rightarrow\qquad
			\psi^\m_\pm (\s,\tau) = {1\ov\sqrt{2}} 
				\sum_{n\in\Integers} d^\m_\n e^{-i n (\tau \pm \s)}~,\ \
	\ee
	or the {\it Neveu-Schwarz} (NS) boundary condition,
	\be 
		\text{(NS)}\qquad \psi^\m_+ (\pi,\tau) = -\psi^\m_- (\pi,\tau)\qquad\Rightarrow\qquad
			\psi^\m_\pm (\s,\tau) = {1\ov\sqrt{2}} 
				\sum_{r\in\Integers+{1\ov2}} b^\m_\n e^{-i r (\tau \pm \s)}~,
	\ee
	and accordingly one refers to the {\it Ramond} and the {\it Neveu-Schwarz sector}.

	For the closed string $\psi_+$ and $\psi_-$ are not connected through the boundary condition but instead they separately have to be periodic or anti-periodic. Thus, for each of them one has to choose between Ramond and Neveu-Schwarz boundary conditions, resulting in R-R, NS-R, R-NS and NS-NS sectors.

	Hence, altogether, the string theory with world-sheet SUSY is also referred to as the {\it RNS-superstring}. It can be quantized both by assuming light-cone gauge, as in \secref{sec:BS-LCG}, or covariantly, i.e., either in the old covariant scheme described in \secref{sec:BS-OCQ} or in the more modern language of BRST-quantization. The quantum theories of the different sectors then turn out to differ substantially, which can be seen as follows.

	By \eqref{eq:SS-2dpsiPoisson} when quantized the fermionic modes fulfill the anti-commutation relations
	\be \label{eq:SS-2dphiCCR}
		\{d^\m_m, d^\n_n\} = \eta^{\m\n} \d_{m+n}~,\qquad\qquad
			\{b^\m_r, b^\n_s\} = \eta^{\m\n} \d_{r+s}~,
	\ee
	and negative and positive modes play the role of fermionic harmonic oscillator raising and lowering operators, respectively. However, in the Ramond sector we also have zero modes $d^\m_0$. These respect the Clifford algebra $\{i\sqrt{2} d^\m_0 , i\sqrt{2} d^\n_0\} = -2 \eta^{\m\n}$ of $d$ dimensional Minkowski space, i.e., of the space-time, and any state has to form an irreducible representation of the Clifford algebra, which are spinors. Thus, the R sector leads to fermionic strings whereas the NS sector, having simple a momentum dependent singlet $\ket{0;\vp}$ as vacuum, leads to bosonic strings.

	We conclude by sketching how old covariant quantization works in the case of the Ramond sector of the open string.
	
	The Fourier modes $L_m$ of $T_{++}$, c.f. \eqref{eq:BS-Tmodes}, and $F_m$ of $J_+$ read
	\begin{align}
		L_m &= \frac{1}{2} \sum^\infty_{n=-\infty} :\a_{-n}\cdot \a_{m+n}: + \Big(n + {1\ov2}m\Big) :d_{-n}\cdot d_{m+n}:~,\\
		F_m &= \frac{1}{2} \sum^\infty_{n=-\infty} \a_{-n}\cdot d_{m+n}~,
	\end{align}
	where normal-ordering only effects $L_0$. By \eqref{eq:BS-CCR} and \eqref{eq:SS-2dphiCCR} they have the algebra
	\begin{align}
		&\qquad\qquad [L_m, L_n] = (m-n) L_{m+n} + A(m) \d_{m+n}~,\\
		&[L_m, F_n] = \Big(\frac{1}{2}m - n \Big) F_{m+n}~,\qquad
			\{F_m, F_n\} = 2 L_{m+n} + B(m) \d_{m+n}~,
	\end{align}
	with anomalies $A(m)=\frac{d}{8}m^3$ and $B(m)=\frac{d}{2}m^2$. Analogous to \eqref{eq:BS-Virasoro2} and \eqref{eq:BS-Virasoro3} one now constrains the Fock space by imposing the super-Virasoro constraints
	\be
		(F_0 - \m)\ket{\phi} = 0~,\qquad\text{and}\qquad 
			F_m\ket{\phi} = L_m \ket{\phi} = 0 \qquad \text{for $m>0$}~,
	\ee
	where $(L_0-\m^2)\ket{\phi}= 0$ due to $F^2_0 = L_0$. By similar steps as in \secref{sec:BS-OCQ} one obtains the critical weight $\m=0$ and the critical dimension of the superstring $d=10$.

	Even though in the NS sector one has a weight $a$ playing a role different from $\m$, also here one finds the critical dimension $d=10$. This finding is confirmed for the closed string and can be reproduced using light-cone gauge quantization. Static gauge quantization of the RNS string is under current investigateion \cite{privateGeorge2014} and appears to work analogously to the bosonic case \secref{sec:BS-Static}.

	Here, we will not expound upon the Gliozzi-Scherk-Olive (GSO) projection\footnote{For further details see e.g. section 4.3.3 of \cite{Green:1987sp}.}. Let us only mention that it truncates the space of physical states by inconsistencies resulting from $\psi^\m$ being world-sheet fermions but a space-time vector. The remaining Hilbert space then seems to suggest that the theory is not only supersymmetric on the world-sheet but also in space-time.

 \subsection{Space-Time Supersymmetry}
	In the previous subsection we discussed flat space string theory with manifest world-sheet SUSY, the RNS superstring, giving the critical dimension $d=10$. We now present flat space string theory with manifest space-time SUSY. For this, it is instructive to first go back to the particle \eqref{eq:BS-Polpart}.

	The particle in $d$ dimensional Minkowski space is invariant under global space-time Poincar{\'{e}} symmetry, which we would like to extend to global {\it super}-Poincar{\'{e}} symmetry. The space-time SUSY will transform the coordinates $X^\m(\tau)$ to fermions. Hence, we have to introduce Grassmann valued space-time spinors $\th^{A a}(\tau)$, with $A=1,2,\ldots,\calN$ counting the amount of SUSY. For general $d$, spinors have $2^{\lfloor d/2\rfloor}$ complex components, which can be subject to the Majorana and Weyl condition. Indeed, in $d=10$, the critical dimension for the RNS string, one can choose the Dirac gamma matrices $\G^\m$ to be in Majorana-Weyl basis\footnote{For an explicit Majorana-Weyl basis of the gamma matrices see e.g. \cite{Green:1987sp}, pp. 220.} and imposing the respective conditions leaves only $a=1,2,\ldots,16$ real components per spinor $\th^A$.

	Global SUSY transformations by the infinitesimal spinor $\e^A$ then take the form
	\be \label{eq:SS-FSGSsusy}
		\d \th^A = \e^A~,\qquad \d \bar{\th}^A = \bar{\e}^A~,\qquad
		\d X^\mu = i \bar{\e}^A \G^\m \th^A~,\qquad \d e = 0~.
	\ee
	Hence, $\dot{X}^\m - i \bar{\th}^A \G^\m \dot{\th}^A$ and $\dot{\th}^A$ are invariant and the simplest supersymmetric generalization of the massless particle \eqref{eq:BS-Polpart} is {\it Brink-Schwarz-Casalbuoni} (BSC) {\it superparticle}
	\be \label{eq:SS-BSC}
		S_\text{BSC} =  \frac{1}{2} \int \dd \tau\,e^{-1} 
			\left(\dot{X}^\m - i \bar{\th}^A \G^\m \dot{\th}^A \right)^2~.
	\ee
	With the definition $P^\m = \dot{X}^\m - i \bar{\th}^A \G^\m \dot{\th}^A = e\,(\p \calL_\text{BSC}/\p \dot X_\m)$ the EoM's read
	\be
		P^2=0~,\qquad \dot{P}^\m = 0~,\qquad 
			\cancel{P}\,\dot{\th}^A = \G\cdot P\,\dot{\th}^A = 0
	\ee
	By $(\G\cdot P)^2 = - P^2 = 0$ half of the eigenvalues of the matrix $\G\cdot P$ vanish. Moreover, $\th^A$ only appears multiplied by $\G\cdot P$, such that half of its components decouple. This is the consequence of a new local fermionic symmetry called $\k$ {\it symmetry}. Indeed, for infinitesimal spinors $\k^{A}(\tau)$, which are local as they depend on $\tau$, the transformation
	\be \label{eq:BS-FSppKappa}
		\d \th^A = i \G\cdot P \k^A~,\qquad 
			\d X^\m = i \bar{\th}^A \G^\m \d \th^A~,\qquad 
			\d e = 4 e \dot{\bar{\th}}^A \k^A
	\ee
	leaves the action \eqref{eq:SS-BSC} invariant. Note that the relative sign between $\d X^\m$ and $\d \th^A$ is opposite compared to the SUSY transformation \eqref{eq:SS-FSGSsusy}. A peculiarity of this symmetry is that the commutator of two $\k$ transformations is again a $\k$ transformation, which holds only on-shell,
	\be \label{eq:SS-FS2kappa}
		[\d_1,\d_2] \th^A = i \G\cdot P \k^A + \text{(EoM's)}~,\qquad\text{for}
			\quad \k^A = 4 \k^A_1 \dot{\bar{\th}}^B \k^B_1 - (1 \leftrightarrow 2)~.
	\ee
	Also, on-shell there are no conserved quantities associated with $\k$.

	The quantization of \eqref{eq:SS-BSC} is obscured by phase-space constraints relating the momenta conjugate to $X^\m$ and $\th^A$. Because of this, for a long time quantization relied on gauge fixing, which manifestly breaks Lorentz invariance. More recently, covariant BRST quantization has been investigated, however leading to an infinite tower of ghost fields, see e.g. \cite{Berkovits:2000fe, Grassi:2000qs} and references therein. 

	With this, we now turn to the string. In analogy to \eqref{eq:SS-BSC} there is an obvious guess for the supersymmetric generalization of \eqref{eq:BS-Poly}, namely
	\be \label{eq:SS-FSGSkin}
		S_1 =  -\frac{T}{2} \int \dd^2 \s \sqrt{h} h^{\a\b} \Pi_\a \cdot \Pi_\b~,
	\ee
	where one defines
	\be
		\Pi^\m_\a = \p_\a X^\m - i \bar{\th}^A \G^\m \p_\a \th^A~.
	\ee
	The action \eqref{eq:SS-FSGSkin} is manifestly invariant under space-time SUSY, the local $\k$-symmetry is however lost. Fortunately, for $\calN\leq2$ one can add the {\it Wess-Zumino} (WZ) {\it term}$\,$\footnote{For $\calN=1$ set either of the $\th^A$ to zero. Here, $\e^{\a\b}$ is the Levi-Civita symbol, see \appref{app:Notation}.},
	\be \label{eq:SS-FSGSwz}
		S_2 = T \k \int \dd^2 \s \e^{\a\b} \Big(-i \p_\a X^\m 
			\left(\bar{\th}^1 \G_\m \p_\b \th^1 - \bar{\th}^2 \G_\m \p_\b \th^2\right) 
			+ \bar{\th}^1 \G^\m \p_\a \th^1 \bar{\th}^2 \G_\m \p_\b
			\Big)~,
	\ee
	with $\k = \pm1$, such that the sum of the {\it kinetic term} \eqref{eq:SS-FSGSkin} and the WZ term \eqref{eq:SS-FSGSwz},
	\be \label{eq:SS-FSGSaction}
		S_\text{GS} = S_1 + S_2~,
	\ee
	is again invariant under local $\k$-transformations, i.e., for which half of the fermions decouple. This is the action of the {\it Green-Schwarz} (GS) {\it superstring} in flat space.

	The WZ term $S_2$ is obviously invariant under global Lorentz symmetry and local reparametrizations of the world-sheet. However, one also has to check that $S_2$ is supersymmetric. The check of SUSY boils down to the requirement that
	\be \label{eq:SS-3SpinorId}
		\G_\m \psi_{[1} \bar{\psi}_2 \G^\m \psi_{3]} = 
			\eps^{i j k} \G_\m \psi_{i}\,\bar{\psi}_j \G^\m \psi_{l} = 0
		\qquad\text{for}\quad (\psi_1,\psi_2,\psi_3) = (\th,\th',\dot{\th})~.
	\ee

	Note in this context that the WZ term $S_2$ is the integral of a two form,
	\be \label{eq:SS-FSwz2form}
		S_2 = \int \Th_2 = \int \dd^2 \s \e^{\a\b} \Th_{\a\b}~.
	\ee
	Therefore, one can formally introduce an additional dimension and consider the exact three-form $\Th_3 = \dd \Th_2$ and by Stokes' theorem one has
	\be
		\int_M \Th_3 = \int_\S \Th_2~,
	\ee
	where the world-sheet $\S$ is the boundary of the three dimensional volume $M$, $\S = \p M$. Concretely, for the three-form one finds
	\be \label{eq:SS-FSwz3form}
		\Th_3 = T \k (\dd \bar{\th}^1 \G_\m \dd \th^1 - \dd \bar{\th}^2 \G_\m \dd \th^2) \Pi^\m~,
	\ee
	which is manifestly invariant under global space-time SUSY. However, now the closure of $\Th_3$, $\dd \Th_3 = 0$, is only given requiring the identity \eqref{eq:SS-3SpinorId}.

	This identity only holds for certain spinors in $d=3,4,6,10$ dimensions, especially for Majorana-Weyl spinors in $d=10$. Therefore, even classically, the GS superstring only exists in these dimensions. In the quantum theory one again finds that $d=10$ plays a special role.

	Let us now outline how $\k$-symmetry works for the GS superstring. Trying to generalize \eqref{eq:BS-FSppKappa}, $\d \th^A = i \G\cdot P \k^A$, now as $\Pi^\m_\a$ has a world-sheet vector index so does the spinor $\k^{A\a}$, where we suppressed the spinor index $a$. The Lorentz group on the world-sheet is abelian and the vector representation is reducible into what is usually called the self-dual and anti-self-dual parts. For this one introduces the projectors
	\be \label{eq:SS-Pproj}
		P^{\a\b}_\pm = \frac{1}{2}\left(\g^{\a\b} \pm \k \e^{\a\b} \right)~,\qquad
		P^{\a\b}_\pm \g_{\b\g} P^{\g\d}_\pm = P^{\a\d}_\pm~,\qquad
		P^{\a\b}_\pm \g_{\b\g} P^{\g\d}_\mp = 0~,
	\ee
	where $\g^{\a\b} = \sqrt{h} h^{\a\b}$ is the Weyl-invariant combination of the world-sheet metric $h^{\a\b}$, $\det \g = -1$.
	The parameters  $\k^{A\a}$ are restricted to be anti-self-dual for $A=1$ and self-dual for $A=2$,
	\be
		\k^{1\a} = \frac{1}{\sqrt{h}} P^{\a\b}_- \k^1_\b~,\qquad\qquad
			\k^{2\a} = \frac{1}{\sqrt{h}} P^{\a\b}_+ \k^2_\b~.
	\ee
	and under infinitesimal $\k$ transformations then fields transform as
	\begin{align}
		&\d \th^A = 2 i \G\cdot\Pi_\a \k^{A\a}~,\qquad
			\d X^\m = i \bar{\th}^A \G^\m \d\th^A~,\qquad
			\d \Pi^\m_\a = 2 i \p_\a \bar{\th}^A \G^\m \d\th^A~,\\[.2em]
		&\qquad\qquad\qquad\d \g^{\a\b} = -16 
			(P^{\a\g}_- \bar{\k}^{1\b} \p_\g \th^1 + P^{\a\g}_+ \bar{\k}^{2\b} \p_\g \th^2)~.
	\end{align}
	Actually, the $\k$-symmetry determines the parameter in the WZ term, $\k=\pm1$ and switching the sign corresponding to exchanging $\th^1$ and $\th^2$. Furthermore, for the GS superstring action $S_\text{GS}$ to be invariant under $\k$-symmetry one again requires the identity \eqref{eq:SS-3SpinorId} to hold, which once more underlines the importance of $d=3,4,6,10$.

	With all this, let us comment on the different possible superstring theories. 

	{\it Type I} superstring theory is based on open superstrings, for which the boundary conditions reduce the space-time SUSY to $\calN=1$. Attaching charges at the end of the string, the {\it Chan-Paton factors}, in the quantum theory one finds that these have to gauge under $\SO(32)$. Since open strings can meet and form closed strings, also unoriented closed strings, with $\th^1$ and $\th^2$ of the same chirality and their left- and right-moving modes symmetrized, have to be added.

	{\it Type II} string theories are closed superstring theories with $\calN=2$. Hence, the two spinors $\th^A$ can either have opposite or the same handedness. The first case, $(1,1)$ SUSY, gives {\it type IIA} string theory, which is left-right symmetric and describes oriented closed strings. The second case, $(2,0)$ SUSY, results in {\it type IIB} string theory, which describes unoriented chiral closed strings.

	Finally there are the $\SO(32)$ and ${\rm E}_8\times {\rm E_8}$ {\it heterotic} string theories, which only utilize one spinor, $\calN=1$, and are supersymmetric only in the left- or in the right-movers. 

	All these different superstring theories were shown to be connected via S- and T-duality as well as certain limits of eleven dimensional M-theory.

	To quantize the GS superstring essentially for the same reasons mentioned beneath \eqref{eq:SS-FS2kappa} one assumes light-cone gauge, namely \eqref{eq:BS-CovarGauge} and \eqref{eq:BS-LightConeG}. Furthermore, the $\k$ gauge symmetry is exploited by setting
	\be
		\G^+ \th^A = 0~,\qquad \text{for} \quad 
			\G^\pm = \frac{1}{\sqrt{2}}\left(\G^0 \pm \G^9\right)~.
	\ee
	Hence, solving again $X^-(\tau,\s)$ in terms of $X^+(\tau)$ and $X^i(\tau,\s)$, both the coordinate vector $X^\m$ as well as the spinors $\th^A$ have only eight non-trivial components and form three different eight dimensional representations of Spin(8). By this one eliminates all ghosts from the spectrum. However, as in \secref{sec:BS-LCG}, the Lorentz invariance is manifestly broken. When trying to restore it, i.e., requiring $[J^{i-},J^{j-}]=0$, once more one arrives at the critical dimension for the superstring $d=10$.

	Rather recently there has been substantial progress in the covariant BRST quantization of string theory with space-time SUSY \cite{Berkovits:2000fe}, for reviews see e.g. \cite{Berkovits:2002zk, Oz:2008zz}. As the Green-Schwarz action has to be extended by several terms involving pure spinor BRST ghost fields, this formalism is referred to as the {\it pure spinor superstring}. In flat Minkowski space it has been shown to produce the same spectrum as the GS superstring \cite{Berkovits:2000nn, Aisaka:2008vw}. But for general backgrounds it is not clear, weather the GS and the pure spinor superstring are equivalent at the quantum level.

	We are not going to review the pure spinor string for the flat background but in \secref{sec:SS-PS} we will mention the most important properties for the case of $\AdSxS$, which is relevant for us. Semiclassical equivalence between the GS and the pure spinor string has recently been investigated in \cite{Aisaka:2012ud, Cagnazzo:2012uq, Tonin:2013uec}, for reviews see also \cite{Mazzucato:2011jt, Magro:2010jx}.

\section{The Superconformal Algebra} \label{sec:SS-PSU}
	To be able to construct the $\AdSxS$ superstring as a non-linear string sigma-model with the coset $\PSU(2,2|4)/\SO(1,4)\times\SO(5)$, in this section we first establish the main properties of the superconformal algebra $\psu(2,2|4)$. This discussion should be viewed as complementary to the one in \secref{sec:In-PSU} as the presentation appears quiet different.

 \subsection{Matrix Realization of \texorpdfstring{$\su(2,2|4)$}{su224}}
	Let us start by considering the superalgebra $\gls(4|4)$ over $\Complex$. It is spanned by $8 \times 8$ matrices $M$, which are written in $4 \times 4$ blocks as
	\be
		M = \begin{pmatrix}
					m & \th \\ \eta & n
		    \end{pmatrix}~.
	\ee
	It is endowed with a $\Integers_2$-grading, under which the matrices $m$ and $n$ are even while $\th$ and $\eta$ are odd and can be thought of as fermionic Grassmann variables. The supertrace on $\gls(4|4)$ is then defined as
	\be
		\str M = \tr m - \tr n = 0~,
	\ee
	while for elements of the corresponding super-Lie algebra $\cM \in \GL(4|4)$ one can take the superdeterminant, alias the Berezian, which is defined as
	\be
		\sdet(\cM) 
			= \det(m - \th\,n^{-1} \eta)\det(n)^{-1}
			= \det(m)\det(n - \eta\,m^{-1} \th)^{-1}~.
	\ee
	The super-Lie group $\SL(4|4)$ is obtained from $\GL(4|4)$ by requiring $\sdet(\cM)=1$. Correspondingly, the superalgebra $\sls(4|4)$ is the subalgebra formed by the supermatrices $M\in\gls(4|4)$, which have vanishing supertrace,  $\str(M)=0$. 

	The superalgebra $\su(2,2|4)$ is the non-compact real form of $\sls(4|4)$. it is the set of fixed points $M^\star = M$ under the Cartan involution $M^\star = - H M^\dag H^{-1}$. Hence, a matrix $M$ from $\su(2,2|4)$ fulfills the reality condition
	\be \label{eq:SS-RealCond}
		M^\dag H + H M = 0~,
	\ee
	where $M^\dag = (M^t)^\ast$ is the adjoint of $M$ and $H$ is the hermitian matrix
	\be \label{eq:SS-HSigMat}
		H = \begin{pmatrix}
		     \S & 0 \\ 0 & \mI_4
		    \end{pmatrix}~,\qquad\qquad\text{with}\qquad		
		\S = \begin{pmatrix}
		     \mI_2 & 0 \\ 0 & -\mI_2
		    \end{pmatrix}~,
	\ee
	and $\mI_n$ denotes the $n \times n$ identity matrix. Note furthermore that for odd elements $\th$ the conjugation acts as $\Complex$-anti-linear-involution,
	\be
		(c \th)^\ast = \bar{c} \th^\ast~,\qquad \th^{\ast \ast} = \th~,\qquad
		(\th_1 \th_2)^\ast = \th_2^\ast \th_1^\ast~,
	\ee
	and by this $(M_1 M_2)^\dag = M_2^\dag M_1^\dag$, which in particular ensures that anti-hermitian supermatrices form a Lie superalgebra.

	In terms of the $4\times4$ blocks, \eqref{eq:SS-RealCond} becomes
	\be \label{eq:SS-RealCond2}
		m^\dag = - \S n \S~,\qquad n^\dag = n~,\qquad \eta^\dag = -\S \th~.
	\ee
	Hence, $m$ and $n$ form the unitary subalgebras $\un(2,2)$ and $\un(4)$, respectively. The algebra $\su(2,2,|4)$ also contains the $\un(1)$-generator $i \mI_8$, which indeed has vanishing supertrace and fulfills \eqref{eq:SS-RealCond}. By this, the bosonic subalgebra of $\su(2,2,|4)$ is
	\be
		\su(2,2) \oplus \su(4) \oplus \un(1)
	\ee

	The superalgebra $\psu(2,2|4)$ is then defined as the {\it quotient algebra} of $\su(2,2|4)$ over the $\un(1)$ factor. Note that the quotient algebra $\psu(2,2|4)$ has no realization in terms of $8\times8$ supermatrices.

	Let us fix a convenient basis for the bosonic subalgebra $\su(2,2) \oplus \su(4)$. In contrast to \cite{Arutyunov:2009ga} we choose the five dimensional gamma matrices as
	\begin{align} \label{eq:SS-5dgammas}
	  &\g_i = \begin{pmatrix} 0 & \s_i\\ \bar\s_i & 0 \end{pmatrix}~,\qquad
		\s_i = (-i \vec \s, \mI)~,\qquad \bar\s_i = (-i \vec \s, \mI)~,\\
		\label{eq:SS-5dgammas2}
		&\qquad\; \g_5 = - \g_1 \g_2 \g_3 \g_4 = \begin{pmatrix} \mI_2 & 0 \\ 0 & \mI_2 \end{pmatrix}=\S~,\qquad \g_0 = i \g_5~,
	\end{align}
	with $i=1,\ldots,4$ and $\vec \s = (\s_1, \s_2, \s_3)$ the usual Pauli matrices, see also \eqref{eq:In-SpinHel}.

	For $i,j=1,\ldots,5$ the gamma matrices fulfill the $\SO(5)$ Clifford algebra
	\be
		\{\g_i , \g_j\} = \g_i \g_j + \g_j \g_i = 2\d_{ij}~,\qquad
			\text{for $i,j=1,\ldots,5$}~.
	\ee
	They are hermitian, $(\g^i)^\dag = \g^i$, such that $i \g^i$ belong to $\su(4)$. The spinor representation of $\so(5)$ is spanned by the generators $n_{ij} = \frac{1}{4} [\g_i , \g_j]$, which satisfy
	\be \label{eq:SS-su5alg}
		[n_{ij}, n_{kl}] = \d_{jk} n_{il} + \d_{il} n_{jk} - \d_{ik} n_{jl} + \d_{jl} n_{ik}~,\qquad n_{ij} = n_{ji}~.
	\ee
	If one adds the generators $n_{i6} = \frac{i}{2}\g_i$ the $n_{ij}=-n_{ji}$ fulfill \eqref{eq:SS-su5alg} where now $i,j=1,\ldots,6$. Hence, they generate an irreducible Weyl spinor representation of $\so(6) \sim \su(4)$, where the other Weyl representation corresponds to the choice $n_{i6} = -\frac{i}{2}\g_i$.

	Analogously, for $i,j=0,\ldots,4$ the $\g_i$ generate the Clifford algebra of $\SO(1,4)$,
	\be
		\{\g_i , \g_j\} = \g_i \g_j + \g_j \g_i = 2\eta_{ij}~,\qquad
			\text{for $i,j=0,\ldots,4$}~,
	\ee
	the hermitian conjugate is $(\g_i)^\dag = \eta_{ij} \g_j$ and  $m_{ij} = \frac{1}{4} [\g_i , \g_j]$ satisfy the $\so(1,4)$ algebra
	\be \label{eq:SS-su14alg}
		[m_{ij}, m_{kl}] = \eta_{jk} m_{il} + \eta_{il} m_{jk} - \eta_{ik} m_{jl} + \eta_{jl} m_{ik}~,\qquad m_{ij} = m_{ji}~.
	\ee
	Adding $m_{i,-1} = m_{i 0'} =\frac{1}{2} \g_i$ the $m_{ij} = m_{ji}$ satisfies \eqref{eq:SS-su14alg} for $i,j=0',0,\ldots,4$, i.e., they generate $\so(2,4)\sim \su(2,2)$.

	Therefore, $\su(2,2)$ and $\su(4)$ are regarded as real vector spaces spanned as
	\begin{align} \label{eq:SS-su22basis}
		\su(2,2) &\sim \Span{}_\Reals \big(\sfrac{1}{2}\g_i, \sfrac{1}{4}[\g_i,\g_j] \big)~,\qquad\text{for $i,j=0,\ldots,4$}~,\\
		\su(4) &\sim \Span{}_\Reals \big(\sfrac{i}{2}\g_i, \sfrac{1}{4}[\g_i,\g_j] \big)~,\qquad\text{for $i,j=1,\ldots,5$}~.
	\end{align}
	Together with the center element $i\mI$ this provides an explicit basis of the bosonic subalgebra of $\su(2,2|4)$.

	Note that for $i,j=1,\ldots,4$ the generators $\frac{1}{2} \g_i$ and $\sfrac{1}{4}[\g_i,\g_5]$ are block off-diagonal, i.e., they take the form
	\be
		\begin{pmatrix} 0 & \bullet \\ \bullet & 0 \end{pmatrix} \in \su(2,2)~.
	\ee
	The $\sfrac{1}{4}[\g_i,\g_j]$ are block diagonal and form a $\so(4) = \su(2)\oplus\su(2)$, where one $\su(2)$ sits in the upper and the other one in the lower block. Finally, $\frac{1}{2}\g_0$ is diagonal and its centralizer in $\su(2,2)$ is the maximal compact subalgebra $\su(2)\oplus\su(2)\oplus\un(1) \subset \su(2,2)$. The generator is also referred to as the ``conformal Hamiltonian''.

	For an identification of the generators \eqref{eq:SS-su22basis} with the conformal algebra see \cite{Arutyunov:2009ga}, where due to the change in convention in our case we have $\frac{1}{2}\g_4 = -i D$. This connection can also be established by comparison of \eqref{eq:In-ISO2d} with \eqref{eq:SS-su14alg}.

	Finally we define the following matrix
	\be \label{eq:SS-Kmat}
		K = \g_1\g_3 = \begin{pmatrix}
		                  0 & -1 & 0 & 0 \\
											1 & 0 & 0 & 0 \\
											0 & 0 & 0 & -1 \\
											0 & 0 & 1 & 0
		                 \end{pmatrix}~,\qquad K^2 = - \mI
	\ee
	which will play a central role in the following discussion. One can check that the gamma matrices satisfy the relations
	\be \label{eq:SS-gammaK}
		(\g_i)^t = K \g_i K^{-1}~,\qquad \text{for $i=0,\ldots,5$}~.
	\ee
	
	One can also define the charge conjugation matrix $C=-\g_2 \g_4$, which commutes with $K$ and satisfies the relations
	\be
		C^2 =  - \mI~,\quad C \g_5 C^{-1} = (\g_5)^t~,\quad
			C \g_i C^{-1} = -(\g_i)^t~,\quad \text{for $i=1,\ldots,4$}~.
	\ee

 \subsection{\texorpdfstring{$\Integers_4$}{Z4}-grading}
	The outer automorphism group $\mathrm{Out}(\sls(4,4))$ contains a finite subgroup. For this consider the continuous group $\{ \d_\r, \r \in \Complex^*\}$ acting on elements $M$ as
	\be
		\d_\r(M) = \begin{pmatrix}
		            m & \r \th \\ \sfrac{1}{\r} \eta & n
		           \end{pmatrix}~,
	\ee
	which leaves the bosons invariant while acting as a dilatation on the fermions. This transformation is generated by the so-called hypercharge $\Ups$,
	\be \label{eq:SS-UpsMat}
		\Ups = \begin{pmatrix} \mI_4 & 0 \\ 0 & -\mI_4 \end{pmatrix}~,\qquad\qquad
		e^{\frac{1}{2}\Ups \log\r} = \begin{pmatrix} \r^{1/2} \mI_4 & 0 \\ 0 & \r^{-1/2} \mI_4 \end{pmatrix}~,
	\ee
	and can be written as $\d_\r(M) = e^{\frac{1}{2}\Ups \log\r} M e^{-\frac{1}{2}\Ups \log\r}$. Note that the hypercharge itself is not an element of $\sls(4|4)$ as it has non-vanishing supertrace. However, $e^{\frac{1}{2}\Ups \log\r}$ has the superdeterminant $\r^4$ and for $\r^4 = 1$ the automorphism $\d_\r$ is in fact inner. By this, the continuous family of outer automorphisms of $\sls(4|4)$ is the factor group $\d_r / \{\d_\r: \r^4 = 1\}$ and it can be restricted to $\su(2,2|4)$ by furthermore requiring $|\r|=1$.

	The finite subgroup of $\mathrm{Out}(\sls(4,4))$ coincides with the Klein four-group $\Integers_2 \times \Integers_2$, which is generated by the two transformations
	\be
		M = \begin{pmatrix} m & \th \\ \eta & n \end{pmatrix}
			\rightarrow \begin{pmatrix} n & \eta \\ \th & m \end{pmatrix}~,
		\qquad\qquad M \rightarrow - M^{st}~,
	\ee
	where the supertransposed $M^{st}$ is defined as
	\be
		M^{st} = \begin{pmatrix} m^t & -\eta^t \\ \th^t & n^t \end{pmatrix}~.
	\ee
	Note that $M \rightarrow - M^{st}$ is an automorphism of order four. However,
	\be
		-(-M^{st})^{st} = \begin{pmatrix} m & -\th \\ -\eta & n \end{pmatrix} 
			= \d_{-1} (M)
	\ee
	is an inner automorphism, and in the group of outer automorphisms instead it is of order two.

	The forth order automorphism $M \rightarrow - M^{st}$ endows $\sls(4,4)$ with th structure of a $\Integers_4$-graded Lie algebra. Instead, we will consider an equivalent forth order automorphism, which is related by a similarity transformation. Namely, we choose the automorphism
	\be \label{eq:SS-OmAut}
		M \rightarrow \Om(M) = - \cK M^{st} \cK^{-1}~,\qquad\text{with}\qquad
		\cK = \begin{pmatrix} K & 0 \\ 0 & K \end{pmatrix}~,
	\ee
	and the $4\times4$ matrix $K$ given in \eqref{eq:SS-Kmat}. This has the important property that
	\be \label{eq:SS-OmAutComm}
		\Om(M_1 M_2) = - \Om(M_2) \Om(M_1)~,\qquad 
		\Om\big([M_1, M_2]\big) = \big[\Om(M_1),\Om(M_2)\big]~.
	\ee

	With the new notation $\scr{G}=\su(2,2|4)$ we define
	\be
		\scr{G}^{(k)} = \left\{M\in\scr{G}: \Om(M)=i^k M \right\}~.
	\ee
	By this the vector space $\scr{G}$ can be split up into the direct sum of graded spaces
	\be
		\scr{G} = \scr{G}^{(0)} \oplus \scr{G}^{(1)} \oplus \scr{G}^{(2)} \oplus \scr{G}^{(3)}
	\ee
	and by \eqref{eq:SS-OmAutComm} we have $[\scr{G}^{(k)},\scr{G}^{(l)}]\subset \scr{G}^{(k+l)}$ modulo $\Integers_4$. For any matrix $M \in \scr{G}$ its projections $M^{(k)} \in \scr{G}^{(k)}$ is given by
	\be \label{eq:SS-MgradComp}
		M^{(k)} = \frac{1}{4} \left(M + i^{3k} \Om(M) + i^{2k} \Om^2(M) + i^{k} \Om^3(M) \right)
	\ee
	and it is easy to see that $M^{(0)}$ and $M^{(2)}$ are even while $M^{(1)}$ and $M^{(3)}$ are odd.

	Note that generally $(M^{st})^\dag \neq (M^\dag)^{st}$ and in particular one finds
	\begin{align}
		&\Om(M)^\dag = \Om(M^\dag)~,& &\text{for $M$ even}&~,\\
		&\Om(M)^\dag = -\Om(M^\dag)~,& &\text{for $M$ odd}&~,
	\end{align}
	which by use of \eqref{eq:SS-UpsMat} and \eqref{eq:SS-RealCond} can be reexpressed in the single formula
	\be
		\Om(M)^\dag = \Ups \Om(M^\dag) \Ups^{-1} 
			= - (\Ups H) \Om(M) (\Ups H)^{-1}~,
	\ee
	assuming that $M\in\su(2,2|4)$. By this it can be shown that the grading defined through \eqref{eq:SS-OmAutComm} can be restricted to $\su(2,2|4)$, i.e., that any $M \in \su(2,2|4)$ can be decomposed according \eqref{eq:SS-MgradComp} and that the components $M^{(k)}$ are again elements of $\su(2,2|4)$.

	According to our discussion, under the action of $\Om$ the bosonic subalgebra $\su(2,2)\oplus\su(4)\oplus\un(1)$ is decomposed into the two even components $M^{(0)}$ and $M^{(2)}$. Working out these projections explicitly we find
	\begin{align} \label{eq:SS-M0expl}
		M^{(0)} = \frac{1}{2}\begin{pmatrix}
		           m - K m^t K^{-1} & 0 \\ 0 & n - K n^t K^{-1} 
		          \end{pmatrix}~,\\
		\label{eq:SS-M2expl}
		M^{(2)} = \frac{1}{2}\begin{pmatrix}
		           m + K m^t K^{-1} & 0 \\ 0 & n + K n^t K^{-1}
		          \end{pmatrix}~,
	\end{align}
	where similar expressions for $M^{(1)}$ and $M^{(3)}$ can be found in \cite{Arutyunov:2009ga}.

	Now we can make us of the explicit basis \eqref{eq:SS-5dgammas} and by \eqref{eq:SS-gammaK} we have
	\be \label{eq:SS-Kgamma2}
		\g_i = K (\g_i)^t K^{-1},\qquad [\g_i,\g_j] = -K [\g_i,\g_j]^t K^{-1}~,
			\qquad \text{for $i,j=0,\ldots,5$}~.
	\ee
	From this it follows immediately that $\scr{G}^{(0)}$ coincides with the subalgebra $\so(1,4) \oplus \so(5)$ while $\scr{G}^{(2)}$ comprises the remaining bosonic generators, which lie along the directions of the corresponding cosetspace
	\be \label{eq:SS-AdSxScoset}
		\frac{\SU(2,2)\times \SU(4)}{\SO(1,4)\times \SO(5)} = \AdSxS~,
	\ee
	as well as the central element $i \mI$.

	In particular, in terms of $8\times8$ supermatrices the generators $\S_\m$ of $\AdSxS$ are
	\be	 \label{eq:SS-AdSxSgens}
		\S_{\m=0,\ldots,4} = \frac{1}{2}\begin{pmatrix} \g_\m & 0 \\ 0 & 0 \end{pmatrix}~,\quad
		\S_{\m=5,\ldots,9} = \frac{i}{2}\begin{pmatrix} 0 & 0 \\ 0 & \g_\m \end{pmatrix}~,\quad
		\str(\S_\m \S_\n) = \eta_{\m\n}~,
	\ee
	with e.g. $\S_0$ generating the ${\rm AdS}_5$ time. Similarly, the generators of $\so(1,4) \oplus \so(5)$ lying in $\scr{G}^{(0)}$ are,
	\begin{align}
		&\S_{\m \n} = \frac{1}{4}\begin{pmatrix} [\g_\m, \g_\n] & 0 \\ 0 & 0 \end{pmatrix}~,&\qquad&\text{for $\m,\n = 0,\ldots,4$}~,&\\
		&\S_{\m \n} = -\frac{1}{4}\begin{pmatrix} 0 & 0 \\ 0 & [\g_\m, \g_\n] \end{pmatrix}~,& &\text{for $\m,\n = 5,\ldots,9$}~,&
	\end{align}
	and $\S_{\m\n} = 0$ otherwise. For the latter it is also convenient to introduce the matrices
	\be \label{eq:SS-mIas}
		\mI_\alg{a}  = \begin{pmatrix} \mI_4 & 0 \\ 0 & 0 \end{pmatrix}~,\qquad\qquad
		\mI_\alg{s}  = \begin{pmatrix} 0 & 0 \\ 0 & \mI_4 \end{pmatrix}~,
	\ee
	which for bosonic matrices $M=\diag(m,n)$ project out the $\SU(2,2)$ and $\SU(4)$ part, respectively.
	such that the central element and Hypercharge can be written as
	\be
		i \mI = i(\mI_\alg{a} + \mI_\alg{s})~,\qquad	\qquad		\Ups = \mI_\alg{a} - \mI_\alg{s}~,
	\ee

\section{The Green-Schwarz Superstring in {\AdSxSheader}} \label{sec:SS-GS}
	After establishing the superalgebra $\su(2,2|4)$ in terms of supermatrices and introducing the $\Integers_4$-grading in the previous section, in this section we will state the type IIB Green-Schwarz superstring theory action in $\AdSxS$ as a non-linear coset model and comment in particular on the $\k$-symmetry.

 \subsection{Lagrangian} \label{subsec:SS-Lagrangian}
	The supersstring action will be investigated in terms of its Lagrangian density, where the action is defined as
	\be
		S = \int \dd \tau \dd \s \calL~.
	\ee
	Here, the integration over the spatial world-sheet coordinate is constrained, where for the closed string one typically takes $0\leq\s\leq 2\pi$, as in \chapref{chap:BosString}, or $-\pi \leq \s \leq \pi$.
	
	It will furthermore show useful to introduce the effective dimensional string tension
	\be \label{eq:SS-effST}
		\pzg = R^2 T_0 = \frac{R^2} {2\pi \a'} = \frac{\sqrt{\l}}{2\pi}~,
	\ee
	see also \eqref{eq:In-tHooft}.

	Let now $g$ be an element of the supergroup $\SU(2,2|4)$. The $\su(2,2|4)$ valued Maurer-Cartan one-form is then defined as
	\be
		A = - g^{-1} \dd g = A^{(0)} + A^{(1)} + A^{(2)} + A^{(3)}~,
	\ee
	where on the rhs. we annotated the $\Integers_4$-decomposition of $A$, cf. \eqref{eq:SS-MgradComp}. By definition, the one-form $A$ has vanishing curvature $F = \dd A - A \wedge A = 0$ or in components
	\be \label{eq:SS-Aflat}
		\p_\a A_\b - \p_b A_\a - [A_\a, A_\b] = 0~.
	\ee
	With this, we postulate the Lagrangian density of the $\AdSxS$ superstring to be
	\be \label{eq:SS-AdSxSlag}
		\calL_{\rm GS} = -\frac{\pzg}{2} \left[\g^{\a\b} \str\left(A^{(2)}_\a A^{(2)}_\b\right) + \k \e^{\a\b} \str\left(A^{(1)}_\a A^{(3)}_\b\right) \right]~,
	\ee
	which, in analogy to \eqref{eq:SS-FSGSaction}, is the sum of kinetic and Wess-Zumino term. Here, $\e^{\a\b}$ is once more the Levi-Chivita symbol, $\e^{01}=1$, and $\g^{\a\b}=h^{\a\b} \sqrt{h}$ was introduced beneath \eqref{eq:SS-Pproj}. In conformal gauge one has $\g^{\a\b}=\eta^{\a\b}$, see above \eqref{eq:BS-CovarGauge}.

	Requiring the parameter $\k$ to be real\footnote{As for flat space, requiring of $\k$-symmetry will soon give us $\k=\pm1$.}, using $(\th_1 \th_2)^\ast = \th^\ast_2 \th^\ast_1$ as well as the cyclicity of the supertrace, by \eqref{eq:SS-RealCond} the Lagrangian density indeed shows to be real, $\calL^\ast_{\rm GS} =\calL_{\rm GS}\,$.

	In analogy to the discussion for flat space \eqref{eq:SS-FSwz2form}, the WZ term in \eqref{eq:SS-AdSxSlag} results from the $\SO(1,4)\times\SO(5)$-invariant three-form
	\be
		\Th_3 = \str\left(A^{(2)}\wedge A^{(3)}\wedge A^{(3)} - A^{(2)}\wedge A^{(1)}\wedge A^{(1)}\right)~.
	\ee
	Closure of this is easily shown by flatness of $A$ \eqref{eq:SS-Aflat}. However, since the third cohomology group of the superconformal group is trivial the three-form $\Th_3$ appears to be exact,
	\be
		2 \Th_3 = \dd \str\left(A^{(1)}\wedge A^{(3)}\right)~,
	\ee
	by which the WZ term is reduced to the two-dimensional form in \eqref{eq:SS-AdSxSlag}.

	Next, for $h\in \SO(1,4)\times\SO(5)$ consider the transformation
	\be
		g \mapsto g h\qquad\Rightarrow\qquad A \mapsto h^{-1} A h + h^{-1} \dd h~.
	\ee
	It is the easy to see that $A^{(0)}$ undergoes a gauge transformation, while the other $\Integers_4$-graded components transform by the adjoint action
	\be
		A^{(0)} \mapsto A^{(0)} \mapsto h^{-1} A h + h^{-1} \dd h~,\qquad
		A^{(1,2,3)} \mapsto h^{-1} A^{(1,2,3)} h~.
	\ee

	Hence, even though the Lagrangian density \eqref{eq:SS-AdSxSlag} depends on the $\SU(2,2|4)$ group element $g$, under right multiplication of $g$ with a local element $h \in \SO(1,4)\times\SO(5)$ the homogeneous components $A^{(1)}$, $A^{(2)}$ and $A^{(3)}$ undergo similarity transformations leaving the \eqref{eq:SS-AdSxSlag} invariant. Therefore, the Lagrangian density depends on a coset element from $\SU(2,2|4)/\SO(1,4)\times\SO(5)$.

	Recall that the central element $i \mI$ occurs in the projection $A^{(2)}$, such that under right multiplication of $g$ with this generator $A^{(2)}$ undergoes the shift
	\be
		A^{(2)} \mapsto A^{(2)} + c\,\mI~.
	\ee
	But since both the supetrace of $\mI$ and $A^{(2)}$ vanish, this leaves the Lagrangian density invariant. Hence, additionally to $\so(1,4)\times\so(5)$ we have an extra local $\un(1)$ symmetry, which can be used to gauge away the trace of $A^{(2)}$.

	As depleting  $\un(1)$-symmetry corresponds to the step from $\su(2,2|4)$ to $\psu(2,2|4)$, altogether we find that the Lagrangian density \eqref{eq:SS-AdSxSlag} describes a non-linear sigma-model over the coset space
	\be
		\frac{\PSU(2,2|4)}{\SO(1,4)\times\SO(5)}~.
	\ee
	Note once more that when the fermions are set to zero, this coset space reduces to $\AdSxS$ \eqref{eq:SS-AdSxScoset}, which justifies the name ``$\AdSxS$ superstring''. 

	The group of global symmetries of \eqref{eq:SS-AdSxSlag} is the group $\PSU(2,2|4)$ as can bee seen as follows. Let $G\in\PSU(2,2|4)$ such that the action of $G$ on a coset representative $g$ is defined as
	\be
		G:~g \mapsto g'~,
	\ee
	where the new coset representative $g'$ is determined by
	\be \label{eq:SS-globalPSU}
		G g = g' h
	\ee
	with $h$ a ``compensating'' local element from $\SO(1,4)\times\SO(5)$. Hence, because of the local invariance under $\SO(1,4)\times\SO(5)$ the Lagrangian density is also invariant under global $\PSU(2,2|4)$ transformations.

	We now will investigate the EoM's resulting from \eqref{eq:SS-AdSxSlag}. For this, note that for two supermatrices $M_1$ and $M_2$ we have
	\be
		\str(\Om^k(M_1)M_2) = \str(M_1\Om^{4-k}(M_2))~,
	\ee
	by which the variation of the Lagrangian density can be brought into the form
	\be
		\d\calL_{\rm GS} = - \str(\d A_\a \L^\a)~,\qquad
			\L^\a = \pzg \left[\g^{\a\b} A^{(2)}_\b - \sfrac{1}{2}\k\,\e^{\a\b}(A^{(1)}_\b - A^{(3)}_\b)\right]~,
	\ee
	with the coupling $\pzg$ given in \eqref{eq:SS-effST}. Taking into account that
	\be
		\d A_\a = - g^{-1} \d g A_\a - g^{-1} \p_\a(\d g)
	\ee
	and integration by parts one finds
	\be
		\d\calL_{\rm GS} = - \str \left[g^{-1}\d g (\p_\a \L^\a - [A_\a,\L^\a])\right]~.
	\ee
	Hence, regarding $\p_\a \L^\a - [A_\a,\L^\a]$ as an element of $\psu(2,2|4)$, the EoM's become
	\be \label{eq:SS-AdSxSeom}
		\p_\a \L^\a - [A_\a,\L^\a] = 0~.
	\ee
	This equation can be projected on the $\Integers_4$-components. First of all, the $\scG^{(0)}$-component of \eqref{eq:SS-AdSxSeom} identically vanishes. For the $\scG^{(2)}$-component one gets
	\be
		\p_\a (\g^{\a\b} A^{(2)}_\b) - \g^{\a\b} [A^{(0)}_\a, A^{(2)}_\b]
			+ \sfrac{1}{2}\k\,\e^{\a\b} 
				\left([A^{(1)}_\a, A^{(1)}_\b] - [A^{(3)}_\a, A^{(3)}_\b] \right)
			= 0 ~,
	\ee
	whereas for the projections on $\scG^{(1,3)}$ one finds
	\begin{align}
		&\g^{\a\b} [A^{(3)}_\a, A^{(2)}_\b] 
			+ \k\, \e^{\a\b} [A^{(2)}_\a, A^{(3)}_\b] 
			= 2 P^{\a\b}_- [A^{(2)}_\a, A^{(3)}_\b] = 0~,\\
		&\g^{\a\b} [A^{(1)}_\a, A^{(2)}_\b] 
			+ \k\, \e^{\a\b} [A^{(2)}_\a, A^{(1)}_\b] 
			= 2 P^{\a\b}_+ [A^{(2)}_\a, A^{(1)}_\b]= 0~,
	\end{align}
	where we used the projectors $P^{\a\b}_\pm$ defined in \eqref{eq:SS-Pproj}.

	There is a close connection of the EoM's to the conserved charges. In particular, the Noether current corresponding to the global $\PSU(2,2|4)$-symmetry is given by
	\be \label{eq:SS-GSnoether}
		J^\a = g \L^\a g^{-1} =
			\pzg\, g \left[\g^{\a\b} A^{(2)}_\b - \sfrac{1}{2}\k\,\e^{\a\b}(A^{(1)}_\b - A^{(3)}_\b)\right] g^{-1}~,
	\ee
	and conservation, $\p_\a J^\a=0$, follows from \eqref{eq:SS-AdSxSeom}. The corresponding conserved charge $Q$ is given by the integral
	\be \label{eq:SS-PSUcharges}
		Q = \int^\pi_{-\pi} \dd \s J^\tau = \pzg \int^\pi_{-\pi} \dd \s\,
			g \left[\g^{\tau\b} A^{(2)}_\b - \frac{\k}{2}
				(A^{(1)}_\s - A^{(3)}_\s)\right] g^{-1}~.
	\ee
	Note that since $J^\a \in \su(2,2|4)$ only its traceless part is conserved. From this the $\AdSxS$ components are extracted as 
	$Q^\m = \eta^{\m\n} \str(\S_\n Q)$ and in particular for the energy we have
	\be \label{eq:SS-GSenergy}
		E = \str\big(\S_0 Q\big) = 
			\int^\pi_{-\pi} \dd \s \str \big(\S_0 J^\tau \big)~.
	\ee

	Finally, the EoM's for the world-sheet metric give the stress energy tensor
	\be
		\str(A^{(2)}_\a A^{(2)}_\b) -\g_{\a\b} \g^{\g\d} \str(A^{(2)}_\g A^{(2)}_\d) = 0~,
	\ee
	which once again gives the Virasoro constraints reflecting the invariance of the string action under two dimensional reparametrizations.

 \subsection{Kappa-symmetry}
	Recall that the invariance under global $\PSU(2,2|4)$-transformations was realized by multiplication from the left \eqref{eq:SS-globalPSU}. In this respect, the $\k$-symmetry transformations can be seen as {it right local} action of $G = \exp \e$ on the coset representative $g$,
	\be \label{eq:SS-AdSxSkappa}
		g\,G = g' h~,
	\ee
	where $\eps(\tau,\s)$ is a local fermionic element of $\psu(2,2|4)$ and $h$ is again a compensating element in $\SO(1,4) \times \SO(5)$. In contrast to the global $\PSU(2,2|4)$-invariance, the superstring action is not generally invariant under \eqref{eq:SS-AdSxSkappa}. Hence, one has to find appropriate conditions for the parameter $\e$.

	For infinitesimal $\e$ the Maurer-Cartan one-form transforms under \eqref{eq:SS-AdSxSkappa} as
	\be
		\d_\e A = -\dd \e + [A,\e]~,
	\ee
	which, with $\e = \e^{(1)} + \e^{(3)}$, implies for the $\Integers_4$-components
	\begin{align}
		&\d_\e A^{(1)} = -\dd\e^{(1)} + [A^{(0)},\e^{(1)}] + [A^{(2)},\e^{(3)}]~,\nn\\
		&\d_\e A^{(3)} = -\dd\e^{(3)} + [A^{(2)},\e^{(1)}] + [A^{(0)},\e^{(3)}]~,\nn\\
		&\d_\e A^{(2)} = [A^{(1)},\e^{(1)}] + [A^{(3)},\e^{(3)}]~,\nn\\
		&\d_\e A^{(0)} = [A^{(3)},\e^{(1)}] + [A^{(1)},\e^{(3)}]~.\nn
	\end{align}
	The flatness condition \eqref{eq:SS-Aflat} implies
	\begin{align}
		&\e^{\a\b} \p_\a A^{(1)}_\b = \e^{\a\b}[A^{(0)}_\a, A^{(1)}_\b] + \e^{\a\b} [A^{(2)}_\a, A^{(3)}_\a]~,\\
		&\e^{\a\b} \p_\a A^{(3)}_\b = \e^{\a\b}[A^{(0)}_\a, A^{(3)}_\b] + \e^{\a\b} [A^{(2)}_\a, A^{(1)}_\a]~,
	\end{align}
	and together, after integration by parts, one finds for variation of the Lagrangian
	\be
		-\frac{2}{\pzg} \d_\e \calL_{\rm GS} = \d_\e \g^{\a\b} \str\left(A^{(2)}_\a A^{(2)}_\b\right) - 4 \str\left([A^{(1),\a}_+, A^{(2)}_{\a,-}]\e^{(1)} + [A^{(3),\a}_-, A^{(2)}_{\a,+}]\e^{(3)}\right)\,,
	\ee
	where for any world-sheet vector $V^\a$ we introduced the projections
	\be
		V^\a_\pm = P^{\a\b}_\pm V_\b ~.
	\ee
	Note furthermore, that by $P^{\a\b}_\pm A_{\b,\mp}=0$ one has the relation
	\be \label{eq:SS-AtpmAspm}
		A_{\tau,\pm} = -\frac{\g^{\tau\s}\mp \k}{\g^{\tau\tau}} A_{\s,\pm}~.
	\ee

	We now take the following ansatz for the $\k$-symmetry parameters
	\begin{align}
		&\e^{(1)} = A^{(2)}_{\a,-} \k^{(1),\a}_+ + \k^{(1),\a}_+ A^{(2)}_{\a,-}~,\\
		&\e^{(3)} = A^{(2)}_{\a,+} \k^{(3),\a}_- + \k^{(3),\a}_- A^{(2)}_{\a,+}~, 
	\end{align}
	where $\k^{(i),\a}_\pm$ are new independent parameters of the $\k$-symmetry of degree $i=1,3$, respectively. Here, the correct degree of the $\e^{(i)}$ is inherited from the properties of $\Om$ and 
	$\e^{(i)} \in \psu(2,2|4)$ is provided for $\k^{(i)}$ satisfying the reality condition
	\be
		H \k^{(i)} - (\k^{(i)})^\dag H = 0~,\qquad \text{for $i=1,3$}~.
	\ee
	According to the discussion beneath \eqref{eq:SS-Kgamma2}, the component $A^{(2)}$ can be expanded as
	\be
		A^{(2)} = \begin{pmatrix}
		           m^i \g^i & 0 \\ 0 & n^i \g^i
		          \end{pmatrix}~,
	\ee
	where for $i=1,\ldots,5$ the components $n^i$ and $m^0$ are imaginary while $m^i$ for $i=1,\ldots,4$ are real. Using \eqref{eq:SS-AtpmAspm} one finds that
	\be
		A^{(2)}_{\a,\pm} A^{(2)}_{\b,\pm} = \begin{pmatrix}
		           m^i_{\a,\pm} m^i_{\b,\pm} &0\\0& n^i_{\a,\pm} n^i_{\b,\pm}
		          \end{pmatrix}
			= \frac{1}{8} \Ups \str(A^{(2)}_{\a,\pm} A^{(2)}_{\b,\pm})
				+ c_{\a\b} \mI_8~,
	\ee
	with $2 c_{\a\b} =  m^i_{\a,\pm} m^i_{\b,\pm} + n^i_{\a,\pm} n^i_{\b,\pm}$ and the hypercharge $\Ups$ defined in \eqref{eq:SS-UpsMat}. By this, we find for the variation of the Lagrangian density
	\be
		- \frac{2}{\pzg} \d_\e \calL_{\rm GS} =\d_\e \g^{\a\b} \str\left(A^{(2)}_\a A^{(2)}_\b\right) 
			-\frac{1}{2}\sum_\pm\str\left(A^{(2)}_{\a,\pm} A^{(2)}_{\b,\pm}\right)
			 \str\left(\Ups[\k^{(2\pm1),\b}_\mp,A^{(2\pm1),\a}_\mp]\right)\,,\nn
	\ee
	which vanishes provided that the variation of the world-sheet metric is
	\begin{align} \label{eq:SS-AdSxSkappaWS}
		\d_\e \g^{\a\b} &= \frac{1}{4}
			\str\left(\Ups\big([\k^{(1),\a}_+,A^{(1),\b}_+] + 
				[\k^{(3),\a}_-,A^{(3),\b}_-]\big)\right) + (\a \leftrightarrow \b)\\
			&= \frac{1}{2}
			\str\left(\Ups\big([\k^{(1),\a}_+,A^{(1),\b}_+] + 
				[\k^{(3),\a}_-,A^{(3),\b}_-]\big)\right)~,\nn
	\end{align}
	where for the second line we used the identity
	\be \label{eq:SS-PprojId}
		P^{\a\g}_\pm P^{\b\d}_\pm = P^{\b\g}_\pm P^{\a\d}_\pm~.
	\ee

	The derived variation \eqref{eq:SS-AdSxSkappaWS} indeed suffices the identity $\g_{\a\b} \d_\e \g^{\a\b} = 0$ and has real components due to the reality conditions for $A$ and the $\k^{(1,3)}$.

	It is now important to point out that the derived $\k$-symmetry transformations exploited the orthogonality of the projectors $P^{\a\b}_\pm$ \eqref{eq:SS-Pproj}, which in turn relies on the prefactor of the Wess-Zumino term taking the values $\k=\pm1$. Therefore, only for these values the $\k$-symmetry is realized.


\section{The Pure Spinor Superstring in {\AdSxSheader}} \label{sec:SS-PS}
	The flat space pure spinor superstring \cite{Berkovits:2000fe} is manifestly super-Poincar{\'e} invariant and by this it is straightforward to generalize it to a non-linear sigma model describing strings propagating in any generic supergravity background \cite{Berkovits:2001ue}.

	For $\AdSxS$ it is again favorable to use the coset space formulation established in the last two sections. The connection between the GS and the pure spinor superstring actions is given by the following, see also \cite{Berkovits:2002zk, Oz:2008zz, Magro:2010jx, Mazzucato:2011jt}.
	
	Taking the parameter $\k=1$ in \eqref{eq:SS-AdSxSlag} and covariant gauge for the world-sheet metric \eqref{eq:BS-CovarGauge}, $h^{\a\b}=\diag(-1,1)$, by use of the world-sheet light-cone coordinates \eqref{eq:BS-CovarGaugeLC} the $\AdSxS$ GS superstring action becomes
	\begin{align} \label{eq:SS-GSaction}
		S_{\rm GS} &= \pzg \int \dd^2\s \str 
			\left( 2 A^{(2)}_+ A^{(2)}_- 
			- \left(A^{(1)}_+ A^{(3)}_- - A^{(3)}_+ A^{(1)}_-\right) \right)~,
	\end{align}
	where $A^{(i)}_\pm = (g^{-1}\p_\pm g)^{(i)} = \sfrac{1}{2}(A^{(i)}_\tau \pm A^{(i)}_\s)$ and still $\dd^2 \s = \dd \tau \dd \s = \sfrac{1}{2} \dd \s_+ \dd \s_-$. From this one obtains the pure spinor action by adding two terms,
	\begin{align}\label{eq:SS-PSaction}
		S_\text{PS} &= S_\text{GS} + 2 \pzg \int \dd^2\s \str 
			\left(A^{(1)}_+ A^{(3)}_- +  A^{(3)}_+ A^{(1)}_-\right) + S_{\rm gh}~,\\
		\label{eq:SS-GhostAction}
		S_\text{gh} &= 2 \pzg \int \dd^2\s \str \left(w^{(3)} [D_-, l^{(1)}] +\hat{w}^{(1)} [D_+, \hat{l}^{(3)}] - N_l \hat{N}_l \right)~,
	\end{align} 
	where $D_\pm = [\p_\pm + A^{(0)}_\pm,\cdot]$. In the language of general supergravity backgrounds \cite{Berkovits:2001ue} the second term in \eqref{eq:SS-PSaction} results from integrating out the auxiliary fermion momenta. The third term $S_\text{gh}$ is the {\it ghost action} and depends on the supermatrices
	\begin{align}
		&l^{(1)}=l^\a \S_\a~,\qquad \hat{w}^{(1)}=-\GG^{\hat{\a}\a} \hat{w}_{\hat{\a}} \S_\a& &\in~\scr{G}^{(1)}~,&\\
		\qquad&\hat{l}^{(3)}=\hat{l}^{\hat{\a}} \S_{\hat{\a}}~,\qquad w^{(3)}=\GG^{\hat{\a}\a} w_\a \S_{\hat{\a}}& & \in~\scr{G}^{(3)}~,\qquad&
	\end{align}
	where $\GG^{\hat{\a}\a}=(\G_{01234})^{\hat{\a}\a}$ is the five-form flux, $\S_\a$ are the 16 generators of $\scr{G}^{(1)}$ and $\S_{\hat{\a}}$ are the 16 generators of $\scr{G}^{(3)}$, respectively. Here, $l^\a(\s_+)$ and $\hat{l}^{\hat{\a}}(\s_-)$ are the left- and right-moving pure spinors ghost fields and $w_\a(\s_+)$ and $\hat{w}_{\hat{\a}}(\s_-)$ are the respective conjugate momenta, where $\a,\hat{\a}=1,\ldots,16\,$ are spinor indices. Despite describing degrees of freedom in $\scr{G}^{(1)}$ and $\scr{G}^{(3)}$, these are bosonic, which is a typical feature of $\b\g$-ghost systems. Additionally, $l^\a$ and $\hat{l}^{\hat{\a}}$ have to satisfy the $\SO(1,4)\times\SO(5)$ pure spinor constraints,
	\be \label{eq:SS-PureSpin}
		\{l^{(1)},l^{(1)}\} = l^{\a} \G^\m_{\a\b} l^{\b} = 0~,\qquad
			\{\hat{l}^{(3)},\hat{l}^{(3)}\} = \hat{l}^{\hat{\a}} \G^\m_{\hat{\a}\hat{\b}} \hat{l}^{\hat{\b}} = 0~,
	\ee
	giving pure spinor superstring theory its name. 

	Finally, $N_l$ and $\hat{N}_l$ are the $\SO(1,4)\times\SO(5)$ generators of the pure spinor sector
	\be
		N_l = - \{w^{(3)}, l^{(1)}\}~,\qquad 
			\hat{N}_l = - \{\hat{w}^{(1)}, \hat{l}^{(3)}\}~,\qquad
		\in \scr{G}^{(0)}~.
	\ee
	
	In analogy to \eqref{eq:SS-GSnoether} one then finds the $\PSU(2,2|4)$ Noether currents
	\begin{align}
		J_+ = \pzg\,g \left(A^{(2)}_+ + \frac{1}{2} A^{(1)}_+ + \frac{3}{2} A^{(3)}_+ + N\right) g^{-1}~,\\
		J_- = \pzg\,g \left(A^{(2)}_- + \frac{3}{2} A^{(1)}_- + \frac{1}{2} A^{(3)}_- + \hat{N}\right) g^{-1}~,
	\end{align} 
	and especially the energy is again given by \eqref{eq:SS-GSenergy} where now $J^\tau$ is altered,
	\be \label{eq:SS-PSenergy}
		E = \int^\pi_{-\pi} \dd \s \str(\S_0 J^\tau)~, \qquad 
		J^\tau= -\pzg\,g\left(A^{(2)}_\tau + A^{(1)}_\tau + A^{(3)}_\tau + N + \hat{N}\right)g^{-1}\,.
	\ee

	The characteristic property of the pure spinor superstring is the invariance under the BRST transformations. On-shell these take the form
	\be
		\d_B g = (l^{(1)}+\hat{l}^{(3)})~,\quad
		\d_B w^{(3)} = -J^{(3)}_+~,\quad 
		\d_B \hat{w}^{(1)} = -J^{(1)}_-~,\quad
		\d_B l^{(1)} = \d_B \hat{l}^{(3)} = 0~.\nn
	\ee
	The associate BRST charge naturally splits into left- and right-mover components,
	\be \label{eq:SS-PSBRSTcharge}
		{\cal Q}_B = {\cal Q} + \bar{\cal Q}~,\quad	
		{\cal Q} = \int \dd \s^+ \str(l^{(1)} J^{(3)}_+)~,\quad
		\bar{\cal Q} = \int \dd \s^- \str(\hat{l}^{(3)} J^{(1)}_-)~.
	\ee
	This charge is only nilpotent due to the pure spinor condition \eqref{eq:SS-PureSpin}, where nilpotency is of course required in order to have a BRST cohomology.
	
	One should note that the pure spinor action is neither invariant under $\k$-symmetry nor is it accompanied by subsidiary Virasoro constraints one would have to impose. Both these features are replaced by the BRST symmetry. In particular, for the non-vanishing components of the stress energy tensor one finds
	\begin{align} \label{eq:SS-PSstrentens}
		T_{++} = \{{\cal Q}_B,b\} = \{{\cal Q},b\} = \frac{1}{2} \str\left(A^{(2)}_+ A^{(2)}_+ 
			+ 2 A^{(1)}_+ A^{(3)}_+ + w^{(3)}[D_+,l^{(1)}]\right)~,\\
		T_{--} = \{{\cal Q}_B,\hat{b}\} = \{\bar{\cal Q} ,\hat{b}\}
			= \frac{1}{2} \str\left(A^{(2)}_- A^{(2)}_- 
			+ 2 A^{(1)}_- A^{(3)}_- + \hat{w}^{(1)}[D_-,\hat{l}^{(1)}]\right)~,
	\end{align}
	where $b$ and $\hat{b}$ are the composite $b$-ghosts, which are expressable in terms of the currents $A^{(i)}_\pm$, the ghosts $l^{(1)}$ and $\hat{l}^{(3)}$, and their momenta $w^{(3)}$ and $\hat{w}^{(1)}$. The $b$-ghosts have ghost number (-1,0) and (0,-1), respectively. They are not strictly left- and right-moving but their derivations $\p_- b$ and $\p_+ \hat{b}$ are BRST trivial.

\chapter{Konishi Dimension from Pure Spinor Superstring} \label{chap:PureSpinor}
	In this chapter we review the asserted derivation of the Konishi anomalous dimension \eqref{eq:In-KoAnDim} up to order $\l^{-1/4}$ using pure spinor superstring theory \cite{Vallilo:2011fj}, which has been reviewed by one of the authors in \cite{Mazzucato:2011jt}\footnote{I would like to thank Brenno Vallilo for constructive discussion of the analysis in this chapter.}.

	For this, in \secref{sec:PS-PertTheo} we start out by some general remarks on perturbation theory. These might seem trivial at first but will be crucial for the remaining discussion in this chapter as well as the chapters to come.

	In order to be able to comment the calculations in \cite{Vallilo:2011fj} and \cite{Mazzucato:2011jt}, in \secref{sec:PS-VMresults} we review briefly the works, where sections \ref{sec:PS-LagHamDens} to \ref{sec:PS-Scale} are then devoted to recalculations. As the fermionic and ghost fields seem only to ensure cancellation of normal ordering ambiguities, with especially the order $\l^{-1/4}$ in $E$ determined by first order corrections of purely bosonic operators, we restrict ourself to investigation of the bosonic subsector.

	After deriving the Lagrangian and Hamiltonian densities in \secref{sec:PS-LagHamDens} and after diagonalizing the quadratic Hamiltonian in \secref{sec:PS-ModeExDiag}, in sections \ref{sec:PS-Energy} and \ref{sec:PS-Scale} we face several open questions, which causes us to distrust the result of \cite{Vallilo:2011fj}.

	For convenience, let us collect the found discrepancies:
	\begin{itemize}
		\item It is not taken into account that the spatial $\text{AdS}_5$ zero modes $\vX_0$ and $\vP_{X,0}$ scale differently in 't Hooft coupling than the non-zero modes, a fact already established in \cite{Passerini:2010xc}. Doing so reveals that the operator $(\p^\a X\p_aX) X^2$ is not the only operator contributing to the spectrum at $\ord{\l^{-1/4}}$ via first order perturbations. Even more devastating, there are operators contributing to the same order in $E$ via second order perturbation theory, even for states dual to the Konishi multiplet.
		
		\item It is claimed that the operator $(\p^\a X\p_aX) X^2$ does not suffer from ordering ambiguities due to matching with the vanishing one-loop $\b$-function. But for second order perturbations found to contribute at the same order in $\l$, which effectively corresponds to operator sextic in fluctuations, i.e. two-loop contributions, this argument does not apply. Also, it is not clear to the author, whether the original argument still holds, as due to the particular rescaling of the zero modes various one-loop contributions now have different power in $\l$.
		
		\item Exploiting the Virasoro constraint $\tilde{H}=0$, the authors of \cite{Vallilo:2011fj} determine a parameter $E_\cl$, which however is only the energy of the classical point-like string solution. It is not the energy $E$ of the full quantum string and we observe a discrepancy even at $\ord{\l^0}$.

		\item It is ignored that the scaling of other zero modes, in particular the 'longitudinal' fluctuations $T_0$ and $P_{T,0}$, is not even determined\footnote{For instance, in the current setting such subtleties play a role even for the matching between the world-sheet Hamiltonian and the Virasoro constraint at quadratic order in fluctuations, see \eqref{eq:PS-tHamDens0}.}.
	\end{itemize}
	Let us comment on the last point that in calculations of semiclassical partition function {\`a} la Drukker, Gross and Tseytlin \cite{Drukker:2000ep} such longitudinal fluctuations usually cancel with ghosts and hence correspond to gauge degrees of freedom. In the setting of pure spinor string theory, an analogous statement would have to result from the BRST-cohomology, which is to be demonstrated. The fact that $T$ is not cyclic suggests though that the used coordinates, i.e., the used coset parametrization, might not be suitable for this task.

\section{On Perturbation Theory} \label{sec:PS-PertTheo}
	We start our survey on the flat space limit of the $\AdSxS$ superstring with what may seem as a platitude, namely discussion of perturbation theory. This will however prepare us for the discussion of the following chapters.

	Let us look at the Schr{\"o}dinger equation for the $1+1$ dimensional free particle\footnote{Here, we set the mass of the particle to one to prevent later confusion.},
	\be \label{eq:PS-FreePart}
		H_0 \ket{\psi_0} = \frac{P^2}{2}  \ket{\psi_0} = E_0 \ket{\psi_0}~.
	\ee
	This has eigenstates $\ket{P}$ of the momentum operator and the continuous spectrum
	\be \label{eq:PS-FreePartSpec}
		E_0 = \frac{P^2}{2}~,
	\ee
	where now $P$ is not the momentum operator but the eigenvalue corresponding to $\ket{P}$.
	
	What happens now if we add an $X$-dependent perturbation? In particular, we take
	\be \label{eq:PS-QHO}
		H = H_0 + \e^2\,\d H =
			\frac{P^2}{2} + \e^2 \frac{X^2}{2}~,
	\ee
	with $\e\ll1$. Applying thoughtlessly the usual formulas of perturbation theory,
	\begin{align}
		&\quad \ket{\psi} = \ket{\psi^{(0)}} + \e^2\ket{\psi^{(1)}} + \ord{\e^4}~,\qquad
		E = E^{(0)} + \e^2\,E^{(1)} + \ord{\e^4}~,\\[.3em]
		& \bra{\psi^{(1)}}\psi^{(0)}\ra = 0~,\qquad 
			E^{(0)} = \bra{\psi^{(0)}} H_0 \ket{\psi^{(0)}}~,\qquad 
			E^{(1)} = \bra{\psi^{(0)}}\d H \ket{\psi^{(0)}}~,&\nn
	\end{align}
	this suggests that at the quantum level \eqref{eq:PS-FreePartSpec} will only receive a correction at order $\e^2$.

	However, of course we know the Hamiltonian \eqref{eq:PS-QHO}: It is the harmonic oscillator with $\e$ corresponding to the angular frequency. The quantum theory is solved in terms of the ladder operators
	\be \label{eq:PS-QHOladder}
		a = \frac{P - i \e X}{\sqrt{2\e}}~,\quad
		a^\dag = \frac{P + i \e m X}{\sqrt{2\e}}~,\quad
		[a,a^\dag]=1~,\quad
		H = \e \left(a^\dag a +\frac{1}{2}\right)~,
	\ee
	where the normal ordering constant in $H$ is determined by requiring the underlying $\sls(2,\Reals)$ algebra to hold at the quantum level. The eigenstates of $H$ are eigenstates of the number operator $a^\dag a$,
	\be
		a \ket{0} = 0~,\qquad \ket{n} = \frac{(a^\dag)^n}{\sqrt{n!}}\ket{0}~,\qquad a^\dag a \ket{n} = n \ket{n}~,
	\ee
	and we have the discrete spectrum
	\be \label{eq:PS-QHOspec}
		E = \bra{n} H \ket{n} = \e \left(n +\frac{1}{2}\right)~.
	\ee
	
	So even though for $\e \ll 1$ the energy \eqref{eq:PS-QHOspec} asymptotes a continuous spectrum, as in the case of the free particle \eqref{eq:PS-FreePartSpec}, inclusion of the `perturbation' $\e^2 \d H$ in \eqref{eq:PS-QHO} makes a fundamental difference, as perturbation theory fails dramatically. Apparently, this discrepancy stems from the fact that the continuous set of momentum eigenstates as a basis of perturbation theory is incompatible for a system having as its solution a discrete, countable basis of eigenstates.

	Let us furthermore consider the case where beyond the harmonic oscillator \eqref{eq:PS-QHO} there is another term
	\be
		H' = H + \e'\,\d H'(P,X)~,
	\ee
	Note, that the condition $\e'\ll\e\ll1$ does not suffice to ensure that $\d H'$ is a perturbation of $H$. Only the ladder operators \eqref{eq:PS-QHOladder} are $\e$-independent, so we first have to express $\d H'$ through them and see what powers of $\e$ are left over. For example for
	\be
		\d H' = 4 X^4 = \e^{-2}[(a-a^\dag)^4 +\text{(norm.ord.)}]
			= \e^{-2} \d H''
	\ee
	we need $\e''=\e'\e^{-2} \ll \e \ll 1$ instead. Equivalently, we notice that $P$ and $X$ have a scaling behavior in $\e$ defined by \eqref{eq:PS-QHOladder}, which can be stripped off by the natural rescaling
	\be \label{eq:PS-NatResc}
		P \mapsto p = \sqrt{\e}P = \frac{1}{\sqrt{2}} (a + a^\dag)~,\qquad
		X \mapsto x = \frac{1}{\sqrt{\e}}X = \frac{i}{\sqrt{2}} (a - a^\dag)~.
	\ee
	and one needs that $\e''\ll\e$ for $\e'' \d H''(p,x) = \e' \d H'(P,X)$ to be a perturbation.

	On the other hand, for the free particle \eqref{eq:PS-FreePart} $P$ as well as $X$, which is cyclic, do not have any defined scale and can be rescaled {\it ad libitum}. 

	Without anticipating too much of the discussion to come, the reason we ruminate over this trifle is that the role of the particle above will be played by the zero modes of the superstring: In the flat space paradigm the $\AdSxS$ superstring is expanded for $\l\gg1$ and the leading order reassembles the flat space superstring, such that implicitly the zero modes are treated as free and massless. However, the zero modes are {\it not} free and massless. The subleading term of the expansion introduces a mass-squared term for the zero modes of the form $\d H$ in \eqref{eq:PS-QHO}, where the mass is actually determined by the non-zero mode excitations. In the following discussion of this chapter the role of the angular frequency, respectively, mass $\e$ is connected to the energy as $\e^2 \propto E^2/{\l}$ and in the next chapter it will be determined by the non-zero mode flat space Hamiltonian, $\e^2 \propto H_{n\neq0}/\sqrt{\l}$, both cases being closely related. For short strings, one has $E \propto \l^{1/4}$ 
and so one finds $\e^2 \propto \l^{-1/2}$ such that the rescaling \eqref{eq:PS-NatResc} introduces the typical scale $\l^{1/8}$ for the zero modes. 

	More explicitly, in \secref{sec:SM-Decoupling} we are going to find that after a canonical transformation incorporating the rescaling \eqref{eq:PS-NatResc} the static gauge world-sheet Hamiltonian of the bosonic string in $\AdSxS$ takes the form
	\begin{align}
		&\frac{E^2}{2\sqrt{\l}}\ket{\psi} = H \ket{\psi} 
		= \left(H_{n\neq0} + \l^{-1/4} \d H + \ord{\l^{-1/2}}\right) \ket{\psi}~, \\
		&\d H
		= \frac{1}{2}\left(\vp^{\,2}_0 + H_{n\neq0}\, \vx^{\,2}_0\right) + \d H_{n\neq0} ~,
	\end{align}
	with $\vx_0$ and $\vp_0$ being the spatial ${\rm AdS}_5$ zero modes, $H_{n\neq0}$ the non-zero mode part of the static gauge flat space world-sheet Hamiltonian, the non-zero mode part of \eqref{eq:BS-StatHamil}, and $\d H_{n\neq0}$ only depending on non-zero modes.

	Hence, at leading order the `unperturbed Hamiltonian' $H_{n \neq 0}$ determines the non-zero modes to be harmonic oscillators, while the zero mode Hilbert-space is still completely undetermined. Considering the effects of $\l^{-1/4} \d H$ on the eigenstates of $H_{n \neq 0}$ one is dealing with {\it degenerate perturbation theory}, where the degeneracy does not only correspond to non-zero mode states of same level $\bra{\psi^{(0)}}H_{n\neq0} \ket{\psi^{(0)}}=n$ but in particular to the whole zero mode Hilbert space.

	At first order perturbation theory, within each $H_{n\neq0}$ eigenspace, i.e., for $\ket{\phi^{(0)}}$ and $\ket{\psi^{(0)}}$ having the fixed level, one then has to find a basis which diagonalizes the matrix elements $\bra{\phi^{(0)}} \d H \ket{\psi^{(0)}}$. But for the zero modes one can diagonalize $\frac{1}{2} (\vp^{\,2}_0 + H_{n\neq0}\, \vx^{\,2}_0)$ exactly, determining the zero modes to be harmonic oscillators with mass-squared given by the level. For the string states dual to the Konishi state one is interested in level $n=1$ states and we expect a similar picture to appear for the fermionic zero modes when treating the superstring properly.

	Finally, with the consideration at the beginning of this section one should worry that further terms might again alter parts of the zero mode spectrum from discrete to continuous. Note however that the zero modes obtained a mass depending on the non-zero mode excitation and the exact solution for the zero modes might behave somewhat as the massive superparticle in $\AdSxS$, where we will elaborate on this idea more in the later discussion. For massive superparticles in $\AdSxS$ we know that the spectrum is discrete, see \cite{Metsaev:1999kb, Metsaev:2002vr, Horigane:2009qb} but also the initial works on supergravity \cite{Kim:1985ez, Gunaydin:1984fk}, so for non-vanishing level a countable zero mode Hilbert space seems favored.

	To our knowledge, apart from the works \cite{Passerini:2010xc} and \cite{Frolov:2013lva} the presented considerations on perturbation theory, in particular the consequences of the rescaling \eqref{eq:PS-NatResc}, have not been taken into account elsewhere.

\section{Previous Results} \label{sec:PS-VMresults}
	In \cite{Vallilo:2011fj} it was claimed that the Konishi anomalous dimension \eqref{eq:In-KoAnDim} has been derived up to the first quantum correction, up the order $\l^{-1/4}$, using pure spinor superstring theory in $\AdSxS$ (cf. \secref{sec:SS-PS}), where the calculation has been recapitulated in the review \cite{Mazzucato:2011jt} by one of the authors.

	The ansatz of \cite{Vallilo:2011fj} has been to look at quantum fluctuations around a point-particle of energy $E_\cl$ sitting in the center of ${\rm AdS}_5$, which is a classical sting solution. The energy is then subject to the Virasoro constraints and one hopes that these suffice to fix the energy up to order $\l^{-1/4}$, an idea which has also been investigated in \cite{Passerini:2010xc}.

	The ansatz has been facilitated by choosing the background coset representative as
	\be \label{eq:PS-ClassSol}
		\tilde{g} = {\rm exp}\left(\frac{-E_\cl \tau}{\sqrt{\l}} \S_0 \right)~, \qquad \tilde{A}_\tau = -\tilde{g}^{-1} \p_\tau \tilde{g} = \frac{E_\cl}{\sqrt{\l}} \S_0~, \qquad \tilde{A}_\s = 0~,
	\ee
	which manifestly is in $\scr{G}^{(2)}$.
	Here, the constant parameter $E_\cl$ is the energy of the classical solution, as can be seen from \eqref{eq:SS-PSenergy} (recall \eqref{eq:SS-effST}, $2\pi \pzg = \sqrt{\l}$), 
	\be \label{eq:PS-ClassCurr}
		\tilde{J}^\tau = -\pzg\,\tilde{g} \tilde{A}_\tau \tilde{g}^{-1}= -\frac{E_\cl}{2\pi} \S_0~,\qquad E = \int^\pi_{-\pi} \dd \s \str(\S_0 \tilde{J}^\tau) = E_\cl~.
	\ee
	For the classical solution the Virasoro constraint \eqref{eq:SS-PSstrentens} reads 
	\be \label{eq:PS-VMVirClass}
		0  = \int\dd\s\,\pzg \big(T_{++} + T_{--}\big) = \frac{\sqrt{\l}}{2} \str(\tilde{A}_\tau \tilde{A}_\tau + \tilde{A}_\s \tilde{A}_\s) = - \frac{E^2_\cl}{2 \sqrt{\l}} ~,
	\ee
	by which classically the energy is bound to vanish, $E_\cl = 0$. However, inclusion of quantum fluctuations gives rise to further terms on the rhs. and satisfying the Virasoro constraint will imply a non-vanishing value of $E_\cl$. For the additional terms being finite the energy will take values $E_\cl \propto \l^{1/4}$, which is the regime of short strings. 

	Quantum fluctuations are included by taking the coset representative,
	\be \label{eq:PS-VMgrel}
		g = \tilde{g} e^\mbb{X}~,\qquad\qquad
			\mbb{X} = X^\m \S_\m + \th^\a \S_\a + \hat{\th}^{\hat{\a}} \S_{\hat{\a}}~,   
	\ee
	where $\th^\a$ and $\th^{\hat{\a}}$ are the 32 fermions parameterizing $\scr{G}^{(1)}$ and $\scr{G}^{(3)}$. The ten bosonic coordinates $X^\m = (T, \vX, \vY, \Phi)$ parametrize $\scr{G}^{(2)}$, where $T$ is the temporal and $\vX$ are the spatial ${\rm AdS}_5$ directions and $\vY$ and $\Phi$ parameterize ${\rm S}^5$. By this, the current becomes
	\be \label{eq:PS-CurrBgPert}
		A = - g^{-1} \dd g = e^{-\mbb{X}} \tilde{A} e^\mbb{X} - e^{-\mbb{X}} \dd e^\mbb{X} =  A_{\rm b.g.} + A_{\rm pert}~,
	\ee
	with $A_{\rm b.g.}$ the background and $A_{\rm pert}$ the perturbation part of the current.
	
	Expanding the pure spinor action \eqref{eq:SS-PSaction} in fluctuations, in \cite{Mazzucato:2011jt} it is stated that the quadratic terms read\footnote{We corrected the sign for the $\vX$ mass squared term, which appears to be a typo in \cite{Mazzucato:2011jt}.} 
	\begin{align} \label{eq:PS-VMSquadr}
		S_{X^2} = 2\pzg\int \dd^2 \s &\bigg(\eta_{\m\n} \p_+ X^\m \p_- X^\n 
			- \left(\frac{E_\cl}{2\sqrt{\l}}\right)^{\!2} \vX^2
			+ \left(w_\a \p_+ l^\a 
				+ \hat{w}_{\hat{\a}} \p_- \hat{l}^{\hat{\a}}\right) \\
			&+\GG_{\a\hat{\a}} \p_+\th^\a \p_- \hat{\th}^{\hat{\a}}
			- \left(\frac{E_\cl}{4\sqrt{\l}}\right)^{\!2}\!\!\left(\d_{\a\b}\th^\a\p_+ \th^\b
				+ \d_{\hat{\a} \hat{\b}}\hat{\th}^{\hat{\a}}\p_- \hat{\th}^{\hat{\b}} \right)\!\bigg)~,\nn
	\end{align}
	with in particular $\vX^2 = (X^1)^2+\ldots+(X^4)^2$. From this, it is deduced that the four spatial ${\rm AdS}_5$ directions $\vX$ obtain the mass-squared $m^2_X = E^2_\cl / \l$ and that also 16 of the 32 fermions become massive with mass-squared $m^2_\th = E^2_\cl /4\l$. The corresponding equations of motion are stated in \cite{Mazzucato:2011jt} as\footnote{We corrected further apparent typos. Recall in this context that $\p_\pm = \sfrac{1}{2}(\p_\tau \pm \p_\s)$, cf. \eqref{eq:PS-ClassCurr}.}
	\be \label{eq:PS-VMeoms}
		\p_+ \p_- \vX + \frac{m^2_X}{4} \vX = 0~,\quad
		\p_+ \p_- \hat{\th} + \frac{m^2_\th}{4} \GG \p_+ \th = 0~,\quad
		\p_+ \p_- \th - \frac{m^2_\th}{4} \GG \p_- \hat{\th} = 0~,
	\ee
	and the respective mode expansions are given, where in particular
	\begin{align} \label{eq:PS-VMosc}
		&\vX = \vX_0 \cos(m_X\tau) + \frac{\vP_0}{m_X} \sin(m_X\tau)
			+ \sum_{n\neq0} \frac{i}{\om_n} 
			\Big(\varphi_n \v\a_n + \tilde{\varphi}_n \tilde{\v\a}_n\Big)~,\\
		\text{with}\qquad&\varphi_n = e^{-i(\om_n \tau + k_n \s)}~,\quad
			\tilde{\varphi}_n e^{-i(\om_n \tau - k_n \s)}~,\quad 
			\om_{n} = \sign(n) \sqrt{k_n^2 + m^2_X}~,\nn
	\end{align}
	$k_n = 2\pi n$, and $\v\a_n$ and $\tilde{\v\a}_{n}$ being left- and right-movers.

	The authors of \cite{Vallilo:2011fj} now quantize canonically, i.e., they calculate the canonical momenta $P_X = \p \calL/\p \dot{X}$ and $P_\th = \p \calL/\p \dot{\th}$ and impose the equal time commutation relations
	\be
		[P_X(\s), X(\ts)] = -i \d(\s-\ts)~,\qquad
			\{P_\th(\s), \th(\ts)\} = -i \d(\s-\ts)~,
	\ee
	and similarly for $\hat{\th}$. In particular, for the fermionic zero modes this implies
	\be
		\{\th_0, \th_0\} = \{\hat{\th}_0, \hat{\th}_0\} = -\frac{1}{m_\th} \mI~,
			\qquad\qquad \{ \th_0, \hat{\th}_0\} = 0~.
	\ee
	The vacuum $\ket{E_\cl}$ is a scalar and is annihilated by all positive modes including the zero modes of the ghost momentum, $w$ and $\hat{w}$, which ensures that the ghost $\SO(1,4)\times\SO(5)$ generators, $N_l$ and $\hat{N}_l$, annihilate the vacuum.

	The first correction to \eqref{eq:PS-VMVirClass} comes then from the central charge, viz., the conformal anomaly. In a manner similar to the calculation in \ssecref{subsec:BS-LCG-LA} it is found as
	\be\label{eq:PS-VMnormOrd}
		\frac{2 E_\cl}{\sqrt{\l}} + \f{1}{2} \sum^\infty_{n=1} \left(4 \sqrt{n^2+ \frac{E_\cl^2}{\l}} + 6\sqrt{n^2} 
		- 16 \sqrt{n^2} - 16 \sqrt{n^2 + \frac{E_\cl^2}{4\l}} + 22\sqrt{n^2}\right)  \,, 
	\ee
	where the first term originates from the spatial ${\rm AdS_5}$ zero modes, while there is no contribution from the massive fermionic zero modes. The first term in the sum corresponds to the non-zero modes of the respective massive bosons and the second term comes from the remaining six massless bosonic coordinates. The next two terms correspond to the massless and massive fermionic non-zero modes, while the last term comes from the ghosts.

	It is intriguing that the terms in the sum cancel each other not only at leading order but even at order $E_\cl^2/\l$. Hence, up to the order we are interested in one only has to keep the first term, $2 E_\cl/\sqrt{\l}$. The non-vanishing central charge corresponds to the fact that, contrary to the BMN vacuum, the vacuum under consideration is not BPS.
	
	The authors now constrain themselves to a particular string state dual to some member of the Konishi supermultiplet, namely they look at the state
	\be \label{eq:PS-VMstate}
		\ket{V} =  a^+_{-1} \ta^+_{-1} \ket{l\hat{l}}~,
	\ee
	with $a^\pm_{n}$ and $\ta^\pm_{n}$ the left- and right-moving ladder operators in the complex directions $X^\pm = X^1 \pm i X^2$. Here, the state
	\be \label{eq:PS-llState}
		\ket{l\hat{l}}	 = \str(l \hat{l}) \ket{E_\cl}~,
	\ee
	is the simplest state of ghost number (1,1) and by this ensures that the unintegrated vertex operator $\ket{V}$ is physical. The ghost state $\ket{l\hat{l}}	$ corresponds to the radius changing operator dual to the Yang-Mills Lagrangian, as reviewed in \cite{Mazzucato:2011jt}, and it is deduced that it has classical scaling dimension $\D_0 = 4$, such that $\ket{V}$ has $\D_0 = 6$. Furthermore, $\ket{l\hat{l}}$ is a $\SU(2,2)\times\SU(4)$ singlet, so that $\ket{V}$ is a $\SU(4)$ singlet with Lorentz spin two.

	With respect to the quadratic fluctuations \eqref{eq:PS-VMSquadr}, by the left- and right-moving excitations of the first modes of $X^+$ the state  \eqref{eq:PS-VMstate} contributes to \eqref{eq:PS-VMVirClass} with
	\be \label{eq:PS-VMH0contrib}
		2 \sqrt{1 + m^2_X} = 2 \sqrt{1 + \frac{E_\cl^2}{\l}}~,
	\ee
	resulting in
	\be \label{eq:PS-VMgkpResult}
		-\frac{E^2_\cl}{2\sqrt{\l}} + 2 m_X + 2\sqrt{1+m^2_X} + \Ord{\l^{-1/2}} = 0~,
	\ee
	hence $E_\cl = 2 \l^{-1/4} + (\D_0-4) + \ord{\l^{-1/4}}$, with the leading order being the initial result by GKP \eqref{eq:In-GKPresult}.\footnote{In \cite{Vallilo:2011fj}, the offset of $-4$ to the classical scaling dimension has been deduced by consistency. However, generally it not clear whether there should be an offset and which value it should take \cite{privateSergey}.}

	Now it is claimed that to find the first quantum corrections to the energy, the order $\l^{-1/4}$, one has to investigate the perturbative contribution to \eqref{eq:PS-VMVirClass} coming from quartic terms of the form $(\p \mbb{X})^2 \mbb{X}^2$ only.

	To clarify this statement, the authors are treating the zero mode of \eqref{eq:PS-VMVirClass}, which up to subtleties discussed in \secref{sec:PS-Energy} is the world-sheet Hamiltonian, perturbatively,
	\be \label{eq:PS-VMpertth}
		0 = \int \dd \s\,\pzg \big(T_{++}+T_{--}\big) = \tilde H = \tilde H_0 + \d \tilde H~,
	\ee
	where the terms up to quadratic fluctuations are viewed as unperturbed Hamiltonian $\tilde H_0$ while higher terms are the perturbation $\d \tilde H$. The claim is now, that first order perturbation theory, $\bra{V}\d \tilde H \ket{V}$, of the terms of the form $(\p \mbb{X})^2 \mbb{X}^2$ suffices to determine the energy to order $\l^{-1/4}$, in particular that terms of the form $\frac{E}{\sqrt{\l}}(\p \mbb{X}) \mbb{X}^3$ and $\frac{E^2}{\l}\mbb{X}^4$ as well as higher orders in fluctuations do not contribute.

	Next, it is argued that there are no normal order ambiguities of the terms $(\p \mbb{X})^2 \mbb{X}^2$ as these would have to be proportional to the one-loop $\b$-function, which vanishes \cite{Vallilo:2002mh}. Giving very little information on the actual calculation, it is then stated that due to the specific state \eqref{eq:PS-VMstate} only the bosonic operator
	\be \label{eq:PS-VM4corr}
		\f{1}{6}\left(\str [\mbb{P}^{(2)}, \mbb{X}^{(2)}]^2-\str[(\mbb{X}^{(2)})', \mbb{X}^{(2)}]^2\right) 
	\ee
	contributes, with apparently $\mbb{P}^{(2)} = \eta_{\m\n} \dot{X}^\m \S_\n + \ord{X^2}$, which gives the correction $-2/\sqrt{\l}$ in \eqref{eq:PS-VMVirClass}. Hence, altogether the authors find that
	\begin{align} \label{eq:PS-VMresults2}
		&-\frac{E_\cl(E_\cl-4)}{2\sqrt{\l}} + 2\sqrt{1+m_X^2} - \f{2}{\sqrt{\l}} + \Ord{\l^{-3/4}} = 0 \\
		\Leftrightarrow\qquad & E_\cl = 2\l^{1/4} + 2 + \f{2}{\l^{1/4}} + \Ord{\l^{-1/2}} \ ,
	\end{align}
	which concurs with the result for the Konishi anomalous dimension \eqref{eq:In-KoAnDim} discussed in the introduction.

\section{Lagrangian and Hamiltonian densities} \label{sec:PS-LagHamDens}
	By reviewing \cite{Vallilo:2011fj} and \cite{Mazzucato:2011jt} in the previous section one should note that the fermions and ghost fields seem to play only an indirect role as they only assure cancellation of normal ordering. In particular, the order $\l^{-1/4}$ in the energy is claimed to be determined by taking first order perturbations of the bosonic operator \eqref{eq:PS-VM4corr} only. Hence, to comprehend the calculations we choose to focus on the bosonic subsector, the bosonic string in $\AdSxS$, only and argue about normal ordering heuristically.

	As they are bosonic, equations \eqref{eq:PS-ClassSol} and \eqref{eq:PS-ClassCurr} describing the classical solution stays the same and the first change occurs for the supermatrix $\mbb{X}\in\psu(2,2|4)$ in \eqref{eq:PS-VMgrel}, which now only contains bosonic fluctuations,
	\be
		\mbb{X} = X^\m \S_\m = (T \S_0 + \vX\cdot\v\S_X) + (\vY \cdot \v\S_Y + \Phi \S_9) = \mbb{X}_\alg{a} + \mbb{X}_\alg{s}~,
	\ee
	Here, $\mbb{X}_\alg{a}=\mI_\alg{a} \mbb{X}$ and $\mbb{X}_\alg{s}=\mI_\alg{s} \mbb{X}$ are supermatrices containing only the ${\rm AdS}_5$ and the ${\rm S}^5$ degrees of freedom.
	By the properties of the $\scr{G}^{(2)}$ generators \eqref{eq:SS-AdSxSgens} one finds
	\be
		\mbb{X}^2 = f_\alg{a}^2 \mI_\alg{a} + f_\alg{s}^2 \mI_\alg{s} =
			\begin{pmatrix} f_\alg{a}^2 \mI_4 & 0 \\ 0 & -f_\alg{s}^2 \mI_4 \end{pmatrix}\,,\quad 
		f_\alg{a}^2 = \frac{-T^2 + \vX^2}{4}~,\quad
		f_\alg{s}^2 = \frac{\Phi^2 + \vY^2}{4}~,
	\ee
	and the group element \eqref{eq:PS-VMgrel} becomes
	\be
		g = \tilde{g} e^\mbb{X} = \tilde{g}\left(\cosh(f_\alg{a}) \mI_\alg{a} + \frac{\sinh(f_\alg{a})}{f_\alg{a}}\mbb{X}_\alg{a}
		+ \cos(f_\alg{s}) \mI_\alg{s} + \frac{\sin(f_\alg{s})}{f_\alg{s}}\mbb{X}_\alg{s} \right)~.
	\ee
	The general form of the current \eqref{eq:PS-CurrBgPert} however turns out to be quiet cumbersome, especially when projecting out the $\scr{G}^{(2)}$ component.

	Instead, let us utilize the fact that the commutator respects the $\Integers_4$-grading \eqref{eq:SS-OmAutComm}. For this we introduce compact notation for nested commutators, alias the adjoint,
	\be
		\Left[A, B\Right]_{n} = \left[\Left[A, B\Right]_{n-1},B\right]=\text{ad}^{-n}_B A~,\qquad \Left[A, B\Right]_0 = A~.
	\ee
	In particular, in $\Left[A, B\Right]_{n}$ the parameter $n$ counts powers of $B$ and only $\Left[A, B\Right]_1 = [A, B]$ is antisymmetric under $A \leftrightarrow B$.

	With this one finds the Baker-Campbell-Hausdorf like formula for the current \eqref{eq:PS-CurrBgPert}
	\be \label{eq:PS-CurrBgPert2}
		A_\a = e^{-\mbb{X}} \tilde{A}_\a e^{\mbb{X}} - e^{-\mbb{X}}\p_\a e^{\mbb{X}} = \sum^\infty_{n=0} \frac{1}{(n+1)!} \left(
			\Left[\tilde{A}_\a,\mbb{X}\Right]_{n+1} - 
			\Left[\p_\a \mbb{X},\mbb{X}\Right]_{n} \right)\,.
	\ee
	Note that in our case both $\tilde{A}_\a$ and $\mbb{X}$ are in $\scr{G}^{(2)}$ such that by \eqref{eq:SS-OmAutComm} the nested commutators $\lleft[\tilde{A}_\a,\mbb{X}\rright]_n$ and $\Left[\p_\a \mbb{X},\mbb{X}\Right]_n$ are in $\scr{G}^{2(n+1)\!\!\mod 4}$. Hence, projection of the current on $\scr{G}^{(2)}$ is obtained by simply  dropping out all odd $n$ nested commutators,
	\begin{align} \label{eq:PS-A2exp}
		A^{(2)}_\a &= \sum^\infty_{n=0} \left(\frac{1}{(2 n)!} 
			\Left[\tilde{A}_\a,\mbb{X}\Right]_{2n} - 
			\frac{1}{(2n+1)!}\,\Left[\p_\a \mbb{X},\mbb{X}\Right]_{2n} \right)\\
		&= \d_\a^\tau \frac{E_\cl}{\sqrt{\l}}\Big(\S_0
			+ \frac{1}{2}\Left[\S_0,\mbb{X}\Right]_2 + \ldots \Big) - \Big(\p_\a \mbb{X} + \frac{1}{6} \Left[\p_\a \mbb{X},\mbb{X}\Right]_2 +\ldots\Big)~,\nn
	\end{align}
	where we plugged in \eqref{eq:PS-ClassCurr}. It proves convenient to evaluate
	\begin{align} \label{eq:PS-help1}
		\str\left(\dot{\mbb{X}}\lleft[\dot{\mbb{X}},\mbb{X}\rright]_2\right)\
			&= \str\left(4\,\dot{\mbb{X}}^2 \mbb{X}^2
				-2\{\dot{\mbb{X}},\mbb{X}\} \dot{\mbb{X}} \mbb{X}\right)\\
			&= (\dot{\vX}^2-\dot{T}^2) (\vX^2 - T^2) 
				- (\dot{\vX}\cdot\vX-\dot{T} T)^2 + (\text{S$^5$ contrib.})~,\nn
	\end{align}
	and analogous for $\dot{\mbb{X}}\mapsto\mbb{X}'$. Note that the term $T^2 \dot{T}^2$ cancels exactly.
	Setting $\dot{\mbb{X}}\mapsto \S_0$, i.e., $\dot{X}^\m \mapsto (1,0,0,\ldots,0)$, for one or both two of the $\dot{\mbb{X}}$ we get
	\begin{align}
		\label{eq:PS-help2}
		\str\left(\Left[\S_0,\mbb{X}\Right]_2 \dot{\mbb{X}}\right) 
			&= \str\left( \S_0 \lleft[\dot{\mbb{X}},\mbb{X}\rright]_2 \right) 
			=  T \vX\cdot\dot{\vX} -\dot{T}\,\vX^2~,\\
		\label{eq:PS-help3}
		\str\left(\S_0 \Left[\S_0,\mbb{X}\Right]_2 \right) 
			&= - 4 f_\alg{a}^2 - T^2 = -\vX^2~.
	\end{align}
	Here, the first equality in \eqref{eq:PS-help2} is a special case of the identity
	\be \label{eq:PS-NestComId}
		\str\left(\Left[A,C\Right]_{m+k} \Left[B,C\Right]_n \right) = (-1)^k \str\left(\Left[A,C\Right]_m \Left[B,C\Right]_{n+k} \right)~,
	\ee
	which follows immediately from $\str([A,C]B)=-\str(A[B,C])\,$.

	By \eqref{eq:PS-help1} to \eqref{eq:PS-help3}, in covariant gauge, $\g^{\a\b}=\eta^{\a\b}$, the expansion of the bosonic Lagrangian density $\calL = -\frac{\pzg}{2} \g^{\a\b} \str(A^{(2)}_\a A^{(2)}_\b)$ in fluctuations reads
	\begin{align} \label{eq:PS-Lag1}
		\frac{2}{\pzg}\calL 
		=& \str\bigg(
			-\Big(\p^\a \mbb{X} \p_\a \mbb{X} 
				+\frac{1}{3} \p^\a \mbb{X} \Left[\p_\a\mbb{X},\mbb{X}\Right]_2 \Big) 
		+ \frac{E^2_\cl}{\l}\left(\S^2_0 +\S_0 \Left[\S_0,\mbb{X}\Right]_2 \right) \nn\\
		&\quad\qquad- \frac{2 E_\cl}{\sqrt{\l}}\left(\S_0 \dot{\mbb{X}} + \frac{1}{2}\Left[\S_0,\mbb{X}\Right]_2 \dot{\mbb{X}} + \frac{1}{6}\S_0 \lleft[\dot{\mbb{X}},\mbb{X}\rright]_2 \right) + \Ord{X^4}\bigg) \\
		=& -\frac{E^2_\cl}{\l} 
				+\frac{2 E_\cl}{\sqrt{\l}}\,\dot{T} 
				-\left(\eta_{\m\n}\p^\a X^\m \p_\a X^\n
					+ \frac{E^2_\cl}{\l} \vX^2 \right)
					+ \frac{2 E_\cl}{\sqrt{\l}}\,\dot{T}\,\vX^2
			\nn\\
		& -\frac{1}{3} \left((\p^\a\vX\cdot\p_\a\vX)(\vX^2-T^2) 
			- (\p^\a\vX\cdot\vX)(\p_\a\vX\cdot\vX) \right)\\
		&  +\frac{1}{3}\left(\p^\a T \p_\a T \vX^2 - 2 T \p^\a T \vX\cdot\p_\a\vX \right)+\Ord{X^4}~, \nn
	\end{align}
	where we dropped quartic terms not relevant for the later discussion, in particular terms coming from $(E_\cl^2/\l)\S_0\lleft[\S_0,\mbb{X}\rright]_4$ and terms in ${\rm S}^5$ fields coming from $\p^\a\mbb{X}\lleft[\p_a \mbb{X}, \mbb{X}\rright]_2\,$.
	
	Furthermore, for the cubic terms we integrated by parts,
	\be \label{eq:PS-ValliloIbP}
		-\frac{4 E_\cl}{3\sqrt{\l}} \left(T \vX\cdot\dot{\vX} -\dot{T}\,\vX^2\right) = -\frac{2}{3\sqrt{\l}} \left(\p_\tau ( E_\cl T\,\vX) - \dot{E}_\cl T\,\vX - 3\,\dot{T}\,\vX^2\right)
	\ee
	and ignored the total derivative as well as the term involving $\dot{E}_\cl$.\footnote{We thank Brenno Vallilo for suggesting this step. However, as elaborated in \eqref{sec:PS-Energy}, $E_\cl$ is not the string energy $E$ and is generally not conserved. It is even unclear whether $\p_\tau E_\cl \ll E_\cl$ as long as the scaling of $T_0$ is not determined, see also the discussion before \eqref{eq:PS-ZMops}, but we will assume so for now.}

	The quadratic terms concur with \eqref{eq:PS-VMSquadr}. Especially, the EoM for $\vX$ reads
	\be \label{eq:PS-Xeom}
		\ddot{\vX} - \vX'' + \frac{E_\cl^2}{\l} \vX = 0~,
	\ee
	in agreement with \eqref{eq:PS-VMeoms}.

	As in \cite{Vallilo:2011fj} we want to quantized canonically. For this we have to derive the world-sheet Hamiltonian. The canonical momenta $P_\m = \p \calL / \p\dot{X}^\m$ read
	\begin{align}
		\label{eq:PS-PT1}
		&\frac{ P_T}{\pzg} = \frac{ P_0}{\pzg}
			= \frac{E_\cl}{\sqrt{\l}}\left(1 + 
				\vX^2 \right) - \dot{T}
			- \frac{1}{3}\left(\dot{T}\vX^2 - T \vX\cdot\dot{\vX} \right) 
			+ \ord{X^4}~,\\
		&\frac{\vP_X}{\pzg} = \dot{\vX} 
			+\frac{1}{3}\left((\vX^2 - T^2)\dot{\vX}
				- (\vX\cdot\dot{\vX}- T\dot{T}) \vX \right)
			+ \ord{X^4}~,
	\end{align}
	and $ P_{\m>4} = \pzg \dot{X}^\m + \ord{X^3}$. Note that the quartic terms omitted in \eqref{eq:PS-Lag1} do not contribute to $ P_T$ and $\vP_X$. By this the Hamiltonian density becomes
	\begin{align} \label{eq:PS-HamDens1}
		\calH =  P_\m \dot{X}^\m - \calL
		=& \frac{E_\cl  P_T}{\sqrt{\l}} 
			+ \left(\frac{\eta^{\m\n}}{2 \pzg}  P_\m  P_\n + \frac{\pzg}{2}\eta_{\m\n} X'^\m X'^\n \right) + \ord{X^3}
	\end{align}
	We can get rid of the term linear in $ P_T$ by the shift
	\be \label{eq:PS-PTshift}
		 P_T \mapsto  P_T + \frac{\pzg E_\cl}{\sqrt{\l}}~,
	\ee
	which for $T$ unchanged is a canonical transformation. Hence, instead of \eqref{eq:PS-PT1} we have
	\begin{align}
		\label{eq:PS-PT2}
		&\frac{ P_T}{\pzg}
			= - \dot{T} + \frac{E_\cl}{\sqrt{\l}} \vX^2
			- \frac{1}{3}\left(\dot{T}\vX^2 - T \vX\cdot\dot{\vX} \right) + \ord{X^4}~,\\
		\Leftrightarrow\quad& \dot{T} = -\frac{ P_T}{3 \pzg} \left(3 - \vX^2\right) 
			+ \frac{E_\cl}{\sqrt{\l}} \vX^2 
			+ \frac{1}{3 \pzg} T \vX\cdot\vP_X + \ord{X^4}~,\\
		&\dot{\vX} = \frac{\vP_X}{3\pzg}\left(3 - (\vX^2 -T^2)\right) 
			+\frac{\vX}{3 \pzg} (\vP_X \cdot \vX + P_T T) + \ord{X^4}~,
	\end{align}
	which do not containing terms constant in fluctuations. Notice now that
	\begin{align}
		& P_T \dot{T} + \frac{\pzg}{2}\dot{T}^2 = -\frac{ P_T^2}{2\pzg} + \Ord{\frac{E_\cl^2}{\l}X^4},\quad~
		\vP_X\cdot\dot{\vX} - \frac{\pzg}{2}\dot{\vX}^2 = \frac{\vP_X^2}{2\pzg} + \Ord{\frac{E_\cl^2}{\l}X^4},
	\end{align}
	such that from \eqref{eq:PS-Lag1} the Hamiltonian density becomes 
	\begin{align} \label{eq:PS-HamDens2}
		\calH 
			=& \left( P_T + \frac{\pzg E_\cl}{\sqrt{\l}}\right)\dot{T} 
			+ \vP_X\cdot\dot{\vX} + \vP_Y\cdot\dot{\vY} + P_\Phi\,\dot\Phi - \calL
			\\
		=&\,\frac{\pzg E^2_\cl}{2\l} 
			+ \left(\frac{\eta^{\m\n}}{2\pzg} P_\m  P_\n + \frac{\pzg}{2}\eta_{\m\n}X'^\m X'^\n + \frac{\pzg E_\cl^2}{2\l} \vX^2\right)
			+\frac{E_\cl}{\sqrt{\l}}\,T \vX\cdot\vP_X \nn\\
		& -\frac{1}{6\pzg} \left((\vP^2_X - \pzg^2\vX'^2)(\vX^2-T^2) 
			- (\vP_X\cdot\vX)^2 + \pzg^2(\vX'\cdot\vX)^2 \right)\nn\\
		& +\frac{1}{6\pzg}\left(( P_T^2 - \pzg^2 T'^2) \vX^2 + 2\,T \vX \cdot(\vP_X  P_T + \pzg^2 \vX' T') \right)+\Ord{X^4}~, \nn
	\end{align}
	where once more we neglected quartic terms of order $\Ord{(E_\cl^2/\l)X^4}$ or corresponding to ${\rm S}^5$ fields. In particular, due to the shift \eqref{eq:PS-PTshift} not only the term $E_\cl  P_T /\sqrt{\l}$ in \eqref{eq:PS-HamDens1} but even $\pzg E_\cl \dot{T} /\sqrt{\l}$ canceled exactly.
	
	By use of \eqref{eq:PS-NestComId} and \eqref{eq:PS-help1} one can check now that, up to higher corrections, the quartic operators in the last two lines of \eqref{eq:PS-HamDens2} correspond to
	\be
		\frac{\pzg}{6} \eta^{\a\b} \str\left(\p_\a \mbb{X}\lleft[\p_\b \mbb{X},\mbb{X}\rright] \right)
		= \frac{\pzg}{6}\left(\str[\dot{\mbb{X}},\mbb{X}]^2 - \str[\mbb{X}',\mbb{X}]^2\right)~,
	\ee
	which can be traced back to the Lagrangian density \eqref{eq:PS-Lag1}.
	Assuming that $\mbb{P} = P_\m \S_\m\,$, due to $P_T = -\dot T + \ord{X^2}$ this is not quiet the quartic term \eqref{eq:PS-VM4corr}, see \eqref{eq:PS-tHamDens}.

\newpage

\section{Mode Expansion and Diagonalization} \label{sec:PS-ModeExDiag}
	As commented on in \eqref{eq:PS-VMpertth}, we want to quantize the $\AdSxS$ string perturbatively by splitting the Hamiltonian $H$ obtained from \eqref{eq:PS-HamDens2} into an unperturbed Hamiltonian $H_0$, which is diagonalized, as well as a perturbation $\d H$,
	\be
		H = \int^\pi_{-\pi} \dd\s\,\calH = H_0 + \d H ~,
	\ee
	From the density \eqref{eq:PS-HamDens2} a natural choice for the unperturbed Hamiltonian is to consider only the constant and quadratic terms,
	\be \label{eq:PS-UnpertHam}
		H_0 = H_{X^0} + H_{X^2} = \frac{E^2_\cl}{2\sqrt{\l}} + \int^\pi_{-\pi} \dd\s
			\left(\frac{\eta^{\m\n}}{2\pzg} P_\m  P_\n + \frac{\pzg}{2}\eta_{\m\n}X'^\m X^\n + \frac{\pzg E_\cl^2}{2\l} \vX^2\right)~.
	\ee
	As it will be crucial for the later discussion, let us demonstrate rather explicitly how diagonalization of $H_{X^2}$ works. Omitting dependence on $\tau$, for the phase space variables we choose the Fourier mode expansion as
	\begin{align} \label{eq:PS-XPmodeExp}
		X^\m(\s) &= \sum^\infty_{m=-\infty} X^\m_n e^{i n \s} 
			= X^\m_0 + X^\m_\nz(\s)~,\\
		P_\m(\s) &= \eta_{\m\n} \sum^\infty_{n=-\infty} P^\n_n e^{i n \s}
			= \eta_{\m\n}\big(P^\m_0 + P^\m_\nz(\s)\big)~,
	\end{align}
	where $X^\m_\nz(\s)$ and $P^\m_\nz(\s)$ comprise the non-zero modes, $\int \dd \s X^\m_\nz(\s) = \int \dd \s P^\m_\nz(\s) = 0$. Reality of $X^\m(\s)$ and $P_\m(\s)$ requires $(X^\m_n)^\dag = X^\m_{-n}$ and $(P^\m_n)^\dag = P^\m_{-n}$, respectively. Imposing canonical commutator relations takes the form
	\be
		[X^\m(\s),P^\n(\ts)] = i \eta^{\m\n} \d(\s-\ts)\qquad\Leftrightarrow
			\qquad [X^\m_m,P^\n_n] = \frac{i}{2\pi} \eta^{\m\n} \d_{m n}~.
	\ee
	Plugging these into the quadratic Hamiltonian we have
	\be \label{eq:PS-HamSq1}
		H_{X^2} = 2\pi\, \eta_{\m\n} \sum^\infty_{n=-\infty} \left(\frac{1}{2\pzg} P^\m_n P^\n_{-n} + \frac{\pzg}{2} \left(n^2 + m^2_\m\right) X^\m_n X^\n_{-n} \right)~,
	\ee
	where we define the mass parameter $m_\m$ and frequencies $\om^\m_n$ as
	\be
		m_\m =\left\{\begin{matrix}
		            m_X = \frac{E_\cl}{\sqrt{\l}}~ &\text{for~~ $\m=1,\ldots,4$}\\ 0~~~~ &\text{for~~ $\m \neq 1,\ldots,4$}
		           \end{matrix}\right.~,
		\qquad\quad
		\om^\m_n = \sign(n) \sqrt{n^2 + m^2_\m}~.
	\ee
	By this, for the non-zero modes we define the normalized ladder operators
	\begin{align} \label{eq:PS-NZMops}
		&a^\m_{n\neq0} = \sqrt{\frac{\pi}{\pzg \abs{\om^\m_n}}}
			\left(P^\m_n -i\,\pzg\om^\m_n X^\m_n \right)~,& 
		&\ta^\m_{n\neq0} = \sqrt{\frac{\pi}{\pzg \abs{\om^\m_n}}}
			\left(P^\m_{-n} -i\,\pzg\om^\m_n X^\m_{-n} \right)~,&\\
		&P^\m_{n\neq0} = \sqrt{\frac{\pzg \abs{\om^\m_n}}{4\pi}} \left(a^\m_n + \ta^\m_{-n}\right)~,&
		& X^\m_{n\neq0} = \frac{i \sign(n)}{\sqrt{4\pi \pzg\abs{\om^\m_n}}} \left(a^\m_n - \ta^\m_{-n}\right)~,&
	\end{align}
	with $a^{\m\dag}_n = a^\m_{-n}$ and $\ta^{\m\dag}_n = \ta^\m_{-n}$, which indeed fulfill
	\be
		[a^\m_m, a^\n_n]
		=[\ta^\m_m, \ta^\n_n]= \sign(m) \eta^{\m\n} \d_{m+n}~,\qquad\qquad [a^\m_m, \ta^\n_n] = 0~,
	\ee
	
	The zero modes have to be treated separately. The ${\rm AdS}_5$ time $T$ and the ${\rm S}^5$ coordinates $\vY$ and $\Phi$ are cyclic in \eqref{eq:PS-HamSq1} and the corresponding Hilbert spaces are simply given by the eigenstates of the respective momenta.

	We elaborated in \secref{sec:PS-PertTheo} that this picture is faulty once higher order terms are considered. For instance, second order perturbations of the cubic term $\frac{E_\cl}{\sqrt{\l}}\,T \vX\cdot\vP_X$ as well as the quartic term $-\frac{1}{6\pzg} (\vP^2_X - \pzg^2\vX'^2)(\vX^2-T^2)$ in \eqref{eq:PS-HamDens2} appear to give a massterm to $T_0$, with the corresponding powers in $\l$ suggesting that $T_0$ and $P_{T,0}$ are not rescaled in $\l$ at all\footnote{Note that similarly in \eqref{eq:SM-DoOM-RescY} the spatial ${\rm S}^5$ zero modes $\vY_0$ are not rescaled in $\l$.}. However, $T$ corresponds to longitudinal fluctuations\footnote{This statement is most likely true only in leading order as $T$ is not cyclic.}, which in studies of semiclassical partition functions {\` a} la \cite{Drukker:2000ep} decouple, i.e., correspond to gauge degrees of freedom. Let us therefore ignore this inconsistency for now.
 
	The spatial ${\rm AdS}_5$ zero modes are massive with $\pzg m_X = E_\cl/2\pi$ playing the role of the frequency. By this, we define the normalized ladder operators
	\begin{align} \label{eq:PS-ZMops}
		&\va_0 = \sqrt{\frac{\pi}{\pzg m_X}}
			\left(\vP_{X,0} -i\,\pzg m_X \vX_0 \right)~,\\
	\Leftrightarrow\qquad
		&\vP_{X,0} = \sqrt{\frac{\pzg m_X}{4\pi}} \left(\va_0 + \va^\dag_0\right)~,\qquad
		\vX_{0} = \frac{i}{\sqrt{4\pi \pzg m_X}} \left(\va_0 - \va^\dag_0\right)~,
	\end{align}
	Plugging all these into \eqref{eq:PS-HamSq1} we get
	\begin{align} \label{eq:PS-HamSq2}
		&H_{X^2} = H_{X^2,0} + H_{X^2,\nz}\\[.4em]
		\label{eq:PS-HamSq2z}
		&H_{X^2,0} = m_X \left(\va^\dag_0 \cdot \va_0 + 2\right) + \frac{\pi}{\pzg}\left(-(P^0_0)^2 + (P^5_0)^2 + \ldots (P^9_0)^2\right)~,\\
		\label{eq:PS-HamSq2nz}
		&H_{X^2,\nz} =\eta_{\m\n} \sum_{n > 0} \om^\m_n \left(a^{\m\dag}_n a^{\n}_n + \ta^{\m\dag}_n \ta^{\n}_n\right)\,+\, \Ord{m_X^4}~,
	\end{align}
	where we split up the quadratic Hamiltonian into its zero and non-zero mode parts, $H_{X^2,0}$ and $H_{X^2,\nz}$, respectively, and we employed the result of \eqref{eq:PS-VMnormOrd} that the normal ordering contributions from the bosonic non-zero modes should be canceled up to $\ord{m_X^4}$ by the fermionic and ghost non-zero modes while the massive boson normal ordering constant $2 m_X$ persists.

	The Vacuum of $H_0$ has no excitations of the non-zero and massive zero modes and is characterized by $E_\cl$, which should be determined by the Virasoro constraint, as well as the momenta $P^{\m\neq1,\ldots,4}_0$, which we will take to be zero, $P^{\m\neq1,\ldots,4}_0 = 0$. In particular, the state corresponding to \eqref{eq:PS-VMstate} reads
	\be \label{eq:PS-MyState} 
		\ket{\Psi^{(0)}} =  a^+_{-1} \ta^+_{-1} \ket{E_\cl}~,\qquad\qquad 
		\va_0 \ket{\Psi^{(0)}} =P^{\m\neq1,\ldots,4}_0 \ket{\Psi^{(0)}} = 0~,
	\ee
	where $\sqrt{2} a^\pm_n = a^1_n \pm i a^2_n$ and $\sqrt{2} \ta^\pm_n = \ta^1_n \pm i \ta^2_n\,$.

\newpage

\section{Coordinate Transformations and the Energy} \label{sec:PS-Energy}
	Acting with the unperturbed Hamiltonian $H_0 = H_{X^0} + H_{X^2}$ on the state \eqref{eq:PS-MyState} we get
	\be
		H_0 \ket{\Psi^{(0)}} = \left(\frac{E^2_\cl}{2\sqrt{\l}} + 2 \frac{E_\cl}{\sqrt{\l}} + 2 \sqrt{1+\frac{E^2_\cl}{\l}} + \ord{\l^{-1/2}}\right) \ket{\Psi^{(0)}}~.
	\ee
	Note now that the Virasoro constraint is {\it not} $H \ket{\Psi^{(0)}} = 0$. Assuming so, in contrast to \eqref{eq:PS-VMgkpResult} one would find the nonsensical value $E_\cl = \pm 2 i \l^{1/4} - 2 + \ord{\l^{-1/4}}$, which is due to the changed sign for the constant term, cf. \eqref{eq:PS-VMVirClass}.

	Instead the Virasoro constraint is defined as $0=\tilde{H}=\int \dd\s\,\pzg T_{00}$. Calculating the respective density in analogy to \eqref{eq:PS-Lag1} we have
	\begin{align} \label{eq:PS-tHamDens}
		\tilde{\calH} =&\, \pzg T_{00} = 
			\frac{\pzg}{2}\str\left(A^{(2)}_\tau A^{(2)}_\tau + A^{(2)}_\s A^{(2)}_\s\right)\\
		=&\,\tilde{\calH}_0	+ \frac{E_\cl}{6 \sqrt{\l}} \left(4 P_T \vX^2 - 3 T \vP_X\cdot\vX \right) \nn\\
		& -\frac{1}{6\pzg} \left((\vP_X^2 - \pzg^2\vX'^2)(\vX^2-T^2) 
			- (\vP_X\cdot\vX)^2 + \pzg^2(\vX'\cdot\vX)^2 \right)\nn\\
		& +\frac{1}{6\pzg}\left((P_T^2 - \pzg^2 T'^2) \vX^2 - 2\,T \vX \cdot(\vP_X  P_T - \pzg^2 \vX' T') \right)+\Ord{X^4}~,\nn\\
		\label{eq:PS-tHamDens0}
		\tilde{\calH}_0
			=&\, -\frac{\pzg E^2_\cl}{2\l} 
				-\frac{E_\cl}{\sqrt{\l}}{ P_T}
				+ \left(\frac{\eta^{\m\n}}{2\pzg} P_\m  P_\n + \frac{\pzg}{2}\eta_{\m\n}X'^\m X'^\n 
				+ \frac{\pzg E^2_\cl}{2 \l} \vX^2 \right)~,
	\end{align}
	where $\tilde{\calH}_0$ was defined in analogy to \eqref{eq:PS-UnpertHam}. Now the constant term has the right sign. Also, the quadratic terms coincide with the ones in \eqref{eq:PS-UnpertHam}, where the prefactor of the mass-squared term crucially depends on the integration by parts \eqref{eq:PS-ValliloIbP}. The term linear in fluctuations, $- \frac{E_\cl}{\sqrt{\l}} P_T\,$, suggests that $P_{T,0}$ is subleading in $\l$, which though appears to be in contrast to the discussion above \eqref{eq:PS-ZMops}.

	Furthermore, in comparison to the last line of \eqref{eq:PS-HamDens2} there is a flipped sign in the last term. Only due to this sign change the last two lines concur with the ${\rm AdS}_5$ terms coming from
	\be \label{eq:PS-tHam4corr}
		\f{1}{6 \pzg}\left(\str [\mbb{P}, \mbb{X}]^2 + \pzg^2 \str[(\mbb{X})', \mbb{X}]^2\right)~,
	\ee
	with $\pzc{G}^{(2)} \ni \mbb{P} = P_\m \S_\m$, which corresponds to the term stated in \eqref{eq:PS-VM4corr}.

	It is helpful to understand the connection between the Hamiltonian $H$ and the Virasoro constraint $\tilde{H}$. From expansion of \eqref{eq:PS-VMgrel} one notices that the expansion around a classical solution corresponds to a coordinate transformation from old coordinates $\tilde{X}^\m$ to new ones $X^\m$, which in linear order reads
	\be
		\tilde{T} = \frac{-E_\cl \tau}{\sqrt{\l}} + T + \ord{X^2}~,\qquad
		\tilde{X}^{\m>0} = X^\m + \ord{X^2}~.
	\ee
	Hence, the respective phase space variables, old ones $\{\tilde{X}^\m,\tilde{P}_\m\}$ and the new ones $\{X^\m, P_\m\}$ are connected via a canonical transformation and $\tilde{H}$ and $H$ are the corresponding Hamiltonians. Including the shift \eqref{eq:PS-PTshift} the canonical transformation is generated by a type II generating functional $F = \int\frac{\dd\s}{2\pi} \calF$ with density
	\be \label{eq:PS-GenFunc}
		 \calF = \left(P_T + \frac{\pzg E_\cl}{\sqrt{\l}}\right)\left(\frac{E_\cl \tau}{\sqrt{\l}} + \tilde{T} \right)
			+ \tilde{P}_{\m>0} X^{\m>0} + \ord{X^2}~,
	\ee
	which in particular is {\it non-restricted} as it explicitly depends on the world-sheet time. One can now check that up to quadratic corrections we indeed have
	\be
		\tilde \calH + \frac{\p \calF}{\p \tau} = 
		\left(-\frac{\pzg E^2_\cl}{2\l} -\frac{E_\cl}{\sqrt{\l}}{ P_T} \right) 
		+ \left(P_T + \frac{\pzg E_\cl}{\sqrt{\l}}\right)\frac{E_\cl}{\sqrt{\l}} + \ord{X^2} =  \calH ~.
	\ee
	But as the ${\rm AdS}_5$ time $T$ is subject to a non-restrictive canonical transformation, we should start to question whether $E_\cl$, which is the energy for the classical solution \eqref{eq:PS-ClassSol}, really is the energy $E$ of the full quantum string we are looking for.

	In this context we note once more that the Hamiltonian $H$ still contains the ${\rm AdS}_5$ time coordinate $T$ and its conjugate momentum $P_T$. From terms cubic in fluctuations on the zero mode $T_0$ is not cyclic and the conjugate momentum $P_{T,0}$ is not conserved anymore. Hence, there seems to be no direct link between $P_{T,0}$ and the energy either.

	Indeed, we can calculating the energy by \eqref{eq:SS-PSenergy} finding
	\begin{align} \label{eq:PS-Energy}
		&E = -\int^\pi_{-\pi} \dd \s \pzg\,\str(\S_0\, g A^{(2)}_\tau g^{-1}) \\
		=\,&E_\cl + 2\pi P_{T,0} - \frac{1}{6}\int^\pi_{-\pi} \dd \s  \left(\left(\frac{E_\cl}{2\pi}+P_T\right) (3T^2 + \vX^2) + 2 T \vX\cdot\vP_X  \right) + \ord{X^4}~.\nn
	\end{align}
	Hence, the parameter $E_\cl$ is {\it not} the energy of the string, as was assumed in \cite{Vallilo:2011fj, Mazzucato:2011jt}. In particular, for the state at hand \eqref{eq:PS-MyState} the quadratic term lead to the first order correction
	\begin{align} \label{eq:PS-EnCorr}
		-\frac{E_\cl}{12 \pi} \bra{\Psi^{(0)}} \int \dd \s \vX^2 \ket{\Psi^{(0)}} &\approx -\frac{E_\cl}{12 \pi} \bra{\Psi^{(0)}} \frac{2}{\pzg m_X} + \sum_{n \neq 0} \frac{\va_n\cdot\va_{-n}+\v\ta_n\cdot\v\ta_{-n}}{2\,\pzg\, n}\ket{\Psi^{(0)}}\\
		&= -\frac{E_\cl}{6 \sqrt\l}\left(\frac{2 \sqrt{\l}}{E_\cl} + 2\right) = -\frac{1}{3}\left(1 + \frac{E_\cl}{\sqrt{\l}}\right)~,\nn
	\end{align}
	hence $E = E_{\cl}(1-\frac{1}{3\sqrt{\l}}) - \sfrac{1}{3} + \ldots~$, where again we ignored non-zero mode normal ordering contributions\footnote{Of course, here we also ignored subtleties which could arise from proper treatment of the zero modes $T_0$ and $P_{T,0}$, as discussed above \eqref{eq:PS-ZMops}.}.
	Note that once again the GKP result \eqref{eq:In-GKPresult} is not spoiled as \eqref{eq:PS-EnCorr} is subleading. This fact stems from the rescaling \eqref{eq:PS-NZMops} between $\vX$ and the ladder operators $\va_n$, which brings us to our next point.

\newpage

\section{Scaling and Perturbations} \label{sec:PS-Scale}
	We argued in \secref{sec:PS-PertTheo} that for harmonic oscillators the ladder operators determine the scale for the phase space variables, which is striped off by the rescaling \eqref{eq:PS-NatResc}. 
	
	In \secref{sec:PS-ModeExDiag} we found that for the case at hand the non-zero and massive zero modes have to be treated on different grounds. From \eqref{eq:PS-NZMops} we see that, disregarding higher corrections, applying \eqref{eq:PS-NatResc} to the non-zero modes we get the rescaling
	\be
		X^\m_n \mapsto x^\m_n = \l^{1/4}\sqrt{n} X^\m_n~,\qquad\qquad
		P^\m_n \mapsto p^\m_n = \frac{\l^{-1/4}}{\sqrt{n}} P^\m_n~,
	\ee
	while for the massive zero modes equation \eqref{eq:PS-ZMops} implies
	\be
		\vX_0 \mapsto \vx_0 = \l^{1/4} \sqrt{m_X} \vX_0~,\qquad\qquad
		\vP_0 \mapsto \vp_0 = \frac{\l^{-1/4}}{\sqrt{m_X}} P^\m_0~,
	\ee
	whith $m_X \approx E /\sqrt{\l}$.

	For long stings one has $E \propto \sqrt{\l}$ and the non-zero and massive zero modes are both rescaled by the same power in 't Hooft coupling, $\l^{\pm1/4}$ for each field. For short strings however one has $E \propto \l^{1/4}$ and rescaling of the zero modes will introduce prefactors of different power in $\l$ than the ones coming from rescaling of the non-zero modes. Namely, the non-zero modes are still rescaled by $\l^{\pm1/4}$ while the massive zero modes are now rescaled by $\l^{\pm1/8}$. By this it follows that for short strings {\it expansion in fields is not the same as expansion in 't Hooft coupling} $\l$. Similarly, the `$n$-loop' correction, which usually can be traced back to corrections involving a fixed power of fluctuations $\ord{X^N}$, can not be identified with a certain power in 't Hooft coupling. 

	In terms of the splitting into zero and non-zero mode part of the phase space variables \eqref{eq:PS-XPmodeExp} the leading $\l$-dependence is now stripped off by the substitution
	\be \label{eq:PS-Resc}
		\vX(\s) = \l^{-1/8}\vx_0 + \l^{-1/4} \vx_\nz(\s)~,\qquad
		\vP_X(\s) = \l^{1/8}\vp_{x,0} + \l^{1/4}\vp_{x,\nz}(\s)~.
	\ee
	For the massless zero modes the scaling is undetermined and could be chosen freely.

	Employing \eqref{eq:PS-Resc} one easily sees that the non-zero mode part of the quadratic Hamiltonian \eqref{eq:PS-HamSq2nz} lies at order $\l^0$ while the term in \eqref{eq:PS-HamSq2z} corresponding to the massive zero mode comes with a prefactor of $\l^{-1/4}$ and is hence subleading, which is the scenario described in  $\secref{sec:PS-PertTheo}$.

	Recall now that in \cite{Vallilo:2011fj} it was claimed that only first order perturbations of the quartic terms \eqref{eq:PS-VM4corr} did contributing to the first quantum corrections of the Konishi anomalous dimension, see also \eqref{eq:PS-tHam4corr}. To argue about the scaling of this operator let us focus on the term
	\be \label{eq:PS-VM4corr2}
		\frac{1}{6\pzg}\left(\vP^2_X + \pzg^2 \vX'^2 \right)\vX^2~, 
	\ee
	For all phase space variables taking non-zero modes we get
	\be \label{eq:PS-VM4corrOrd}
		\frac{1}{6\pzg}\left(\vP^2_{X,\nz} + \pzg^2 \vX'^2_\nz \right)\vX_\nz^2 ~\propto~
		\frac{1}{\sqrt{\l}}\left(\vp_{x,\nz}^2 + \vx'^2_{\nz} \right)\vx_\nz^2~, 
	\ee
	where we also ignored the relative factor of $(2\pi)^2$. This is in accordance to the statement that \eqref{eq:PS-VM4corr} gives the contribution $-2/\sqrt{\l}$. Potential normal ordering ambiguities would contribute at the same order, which however are said to vanish \cite{Vallilo:2011fj}.

	Note now that as in \cite{Vallilo:2011fj} we neglected quartic terms of the form
	\be
		\frac{\pzg E^2_\cl}{\l}\str(\S_0\lleft[\S_0,\mbb{X}\rright]_4) =
		\frac{\pzg E^2_\cl}{\l}\str\lleft[\S_0,\mbb{X}\rright]_2^{\,2} \propto \frac{E^2_\cl}{\sqrt{\l}} \left(\vX^2 - T^2\right)\vX^2~.
	\ee
	For all $\vX$ taking zero modes due to $E_\cl \propto \l^{1/4}$ the corresponding operator scales as
	\be \label{eq:PS-X04op}
		\frac{E^2_\cl}{\sqrt{\l}} \vX_0^4 \,\propto\, \frac{1}{\sqrt{\l}} \vx^4_0~
	\ee
	and hence contribute at the same order as \eqref{eq:PS-VM4corr}.

	Even more disastrous, apart from these additional contributions via first order perturbation theory there are also operators which potentially contribute via second order perturbation theory.

	For this, let us denote the eigenvalue of some eigenstate $\ket{\Phi^{(0)}}$ of the unperturbed Virasoro constraint $\tilde{H}_0$ as $\calE^{(0)}_\Phi$,
	\be \label{eq:PS-UnpVirConstrEWeq}
		\tilde{H}_0 \ket{\Phi^{(0)}} = \int\dd\s\,\tilde{\calH}_0 \ket{\Phi^{(0)}} = -\frac{E^2_\cl}{\sqrt{\l}} + H_0 \ket{\Phi^{(0)}} = {\calE}^{(0)}_\Phi \ket{\Phi^{(0)}}~,
	\ee
	while eigenstates and eigenvalues of the full Virasoro contraint $\tilde{H}$ are referred to as $\ket{\Phi}$ and $\calE_\Phi$.
	In \eqref{eq:PS-UnpVirConstrEWeq} we assumed that $P_{T,0} \ket{\Phi^{(0)}} = 0$ or at least subleading and $\tilde{\calH}_0$ was defined in \eqref{eq:PS-tHamDens0}. Note that states $\ket{\Phi^{(0)}}$ of the {\it same level} but {\it different number of massive zero mode excitations} than the state \eqref{eq:PS-MyState} the respective eigenvalues of $\tilde{H}_0$ differ only by
	\be \label{eq:PS-calEdiff}
		\calE^{(0)}_\Psi - \calE^{(0)}_\Phi = \bra{\Psi^{(0)}}H_{X^2,0}\ket{\Psi^{(0)}} - \bra{\Phi^{(0)}}H_{X^2,0}\ket{\Phi^{(0)}} \,\propto\, \l^{-1/4} 
	\ee
	Next, we look at the cubic terms \eqref{eq:PS-tHamDens}. In the Virasoro constraint $\tilde{H} = \int\dd\s \tilde{\calH}$ certainly one $\vX$ could be a zero mode while the other two fields could take non-zero modes. These operators then turn out to scale as
	\be \label{eq:PS-3corrOrd}
		\frac{E_\cl}{\sqrt\l}P_{T,\nz}\,\vX_\nz \cdot \vX_0 \,\propto\,
		\l^{-3/8} p_{T,\nz}\,\vx_\nz \cdot \vx_0~,\quad
			\frac{E_\cl}{\sqrt\l}T_\nz\,\vP_{X,\nz} \cdot \vX_0 \,\propto\, \l^{-3/8} t_\nz\,\vp_{x,\nz}\cdot \vx_0 ~,
	\ee
	and corrections due to second order perturbation theory are of order
	\be \label{eq:PS-3corrOrd2}
		\calE^{(2)}_\Psi \supset \sum_{\ket{\Phi^{(0)}}\neq \ket{\Psi^{(0)}}} 
			\frac{\Abs{\bra{\Phi^{(0)}}\l^{-3/8} t_\nz\, \vp_{x,\nz}\cdot \vx_0 \ket{\Psi^{(0)}}}^2}{\calE^{(0)}_\Psi -\calE^{(0)}_\Phi } \propto \frac{1}{\sqrt{\l}}~,
	\ee
	where the leading order contributions comes from the states fulfilling \eqref{eq:PS-calEdiff}. Hence, these contribute again at the same order as the first order contribution \eqref{eq:PS-VM4corrOrd}. We furthermore note that the operators \eqref{eq:PS-3corrOrd} do not necessarily change the number of excitations and could in particular give a contribution even for the lowest excited states, dual to the Konishi supermultiplet.

	By the same argument also the operators \eqref{eq:PS-VM4corr} for one $\vX$ taking zero modes contribute via second order perturbation theory, for example in case of the operator \eqref{eq:PS-VM4corr2} we have
	\be \label{eq:PS-VM4corrOrd2}
		\frac{1}{6\pzg}\left(\vP^2_{X,\nz} - \pzg^2 \vX'^2_\nz \right)\vX_\nz\cdot\vX_0 ~\propto~
		\l^{-3/8}\left(\vp_{x,\nz}^2 + \vx'^2_{\nz} \right)\vx_\nz\cdot\vx_0~, 
	\ee
	which however could not contribute for the dual Konishi states, as it is cubic in non-zero modes. Note that \eqref{eq:PS-VM4corr2} for $\vX^2$ taking zero modes is of order
	\be \label{eq:PS-VM4corrOrd3}
		\frac{1}{6\pzg}\left(\vP^2_{X,\nz} - \pzg^2 \vX'^2_\nz \right)\vX_0^2 ~\propto~
		\l^{-1/4}\left(\vp_{x,\nz}^2 + \vx'^2_{\nz} \right) \vx_0^2~. 
	\ee
	and actually gives another contribution to the zero mode mass-squared term, where the mass is determined by the quadratic non-zero mode Hamiltonian. This is exactly the behavior which we will find in \eqref{sec:SM-Decoupling}.

	We conclude that perturbation theory seems to be hardly under control and that \eqref{eq:PS-VM4corr} is by far not the only operator contributing to the order $\l^{-1/2}$ in the eigenvalues of the Virasoro constraint $\tilde{H}$, which determines the energy $E$ at order $\l^{-1/4}$. As even for the dual Konishi states there are contributions via second order perturbation theory the knowledge of one particular state is not enough anymore and to determine the order $\l^{-1/4}$ in $E$ one needs at least to understand all states of the same level. 

	Additionally, the effective operator involved in the second order perturbations \eqref{eq:PS-3corrOrd2} is sextic in fluctuations and corresponds to a two-loop contribution. The statement that ordering ambiguities are under control, which was deduced from vanishing of the one-loop $\b$-function \cite{Vallilo:2002mh}, does therefore not apply. To the author it is even unclear whether the argument is still valid for one-loop contributions, as due to the particular rescaling of the spatial $\text{AdS}_5$ zero modes the various one-loop contributions come now with different powers in $\l$. For instance, it seems that in this argument the operator \eqref{eq:PS-X04op} has been neglected previously.

	Finally, we recall once more that in most of the presented analysis we ignore the fact that the scaling of other non-zero modes, in particular $T_0$ and $P_{T,0}$, has not even been determined. For the bosonic subsector, this problem is taken into account properly in \chapref{chap:SingleMode}, were we use first order formalism from the beginning, hence omitting the non-restricted canonical transformation in \secref{sec:PS-Energy}. Utilizing static gauge and coordinates in which $T$ is cyclic, we manifestly work with the constant energy $E = P_{T,0}$ and are not confronted with problems due to longitudinal fluctuations, the $T_0$ and $P_{T,0}$ in this section. We elaborate on the scaling behavior of the remaining $\text{S}^5$ zero modes in \secref{sec:SM-Decoupling}.  

\chapter{The Bosonic {\AdSxSheader} String in Static Gauge} \label{chap:SingleMode} 
	From the investigation in the previous chapter we have learned that the zero and non-zero modes have to be handled separately and that, despite subtleties, in first order formalism a consistent perturbative treatment seems to be feasible. Also, we saw that the zero modes contribute at order $\l^{0}$ in the string energy $E$ and one still nourished the hope that the the first honest quantum correction, the order $\l^{-1/4}$ in $E$, is determined by first order perturbations of bosonic operators only, i.e., that fermions only assure cancellation of normal ordering ambiguities and that the second order perturbations found in the previous chapter somehow can be bypassed. Most of these ideas where already formulated in \cite{Passerini:2010xc} and applied to the bosonic $\AdSxS$ string in uniform light-cone gauge.

	In this chapter we are reviewing the work \cite{Frolov:2013lva}, where we followed the previous arguments and also constraining ourselves to the bosonic subsector of the $\AdSxS$ superstring. From the very beginning we work in first order formalism with the aim to quantize canonically. In contrast to \cite{Passerini:2010xc} we are employing static gauge, see \secref{sec:SM-StringDyn}, where consistent quantization for the flat space bosonic string was derived \cite{Jorjadze:2012iy}, see \secref{sec:BS-Static}. The static gauge has the advantage of granting direct access to the energy squared $E^2 \propto H$ instead of the typical light-cone gauge combinations $P_- \propto E-J \propto H$, which causes problems for finite total angular momentum $J$ on ${\rm S}^5$.

	Furthermore, hoping that only first order perturbations determine the order $\l^{-1/4}$ in $E$ and that fermionic modes only cancel normal ordering leads one to project out all but the excited modes and treat normal ordering heuristically. However, again the zero modes ought to be treated with care, as they seem to play a substantially different role as the non-zero modes. The step of cutting all but one non-zero mode is reviewed in \secref{sec:SM-SMSol}. This leads us to what we call a {\rm single-mode string}, which turns out to be a generalization of the pulsating string \cite{deVega:1994yz, Minahan:2002rc} including non-trivial zero modes.
	We find that the system effectively behaves as a massive particle in ${\rm AdS}_5$, with the ${\rm AdS}_5$ Casimir determined by the non-zero mode dynamics as well as the ${\rm S}^5$ Casimir. By this the system shows to be classically integrable and invariant under the isometries $\SO(1,4)\times\SO(5)$, which persists at the quantum level.

	Quantization is then performed in \secref{subsec:SM-QNZM}, where due to the picture of having a massive particle in ${\rm AdS}_5$ we are left with quantization of the ${\rm AdS}_5$ Casimir, i.e., quantization of the non-zero modes. Arguing about normal ordering heuristically, we ultimately find the spectrum of the form
	\be\label{eq:SM-EnSpectrum}
		E_{(N,m,n,J)}=2\,\l^{1/4}\sqrt{m\,n} + N-2 + \frac{10\,n^2-6\,n+4+{\cal M}^2_S}{4\,\l^{1/4}\sqrt{m\,n}}+\ord{\l^{-1/2}} \, ,
	\ee
	where $N$ accounts for the particle-like excitations in ${\rm AdS}_{5}$, and ${\cal M}^2_{\rm S}$ is the contribution due to the motion on the ${\rm S}^5$ which for the bosonic string is equal to the SO(6) Casimir ${\cal M}^2_{\rm S} = J(J+4)$. It is expected that the fermion contribution would modify the $J$ dependence to  ${\cal M}^2_{\rm S} = J^2 +\ord{\l^{-1/4}}$.  
	For the lightest stringy state with $N=0$, $m = n = 1$ and $J=0$ which is dual to the Konishi operator \eqref{eq:In-KonishiOp}, we find agreement with the integrability based results \eqref{eq:In-KoAnDim} up to the order of interest $\l^{-1/4}$.

	Lowest excitations of higher modes, $n=1$ but $m>1$, fail to agree with the integrability based results derived in \cite{Frolov:2012zv}. This is not too surprising as the level-truncation we have effectively applied will not agree with the full quantum fluctuation result as higher intermediate levels will become important. Indeed, in \secref{sec:SM-Decoupling} we return to the full bosonic action and identify operators corresponding to cut-out modes which could invalidate our result. After performing a canonical transformation, which does not affect the single-mode solution, we find that only for the lowest excited states dual to the Konishi supermultiplet there are no further contributions.

	Apart from \secref{sec:SM-Comp} the content of this chapter follows closely presentation in the original work \cite{Frolov:2013lva}.


\section{Closed Bosonic String Dynamics}
 \label{sec:SM-StringDyn}

	First we consider the static gauge approach to string dynamics in a generic static space-time. Analogously to \secref{sec:BS-Static}, we use only a part of the constraints to exclude the time components of the canonical variables.
	This relates the energy square functional to the Hamiltonian of the system and excludes negative norm states on the quantum level. The remaining constraints specify the physical Hilbert space. We then focus on ${\rm AdS}_{N+1}\times{\rm S}^M$, where we explore the dynamical integrals of the isometry group and analyze their behavior at large coupling.

\subsection{String in a Static Space-Time} \label{subsec:SSST} 
	Bosonic string dynamics in a curved space-time with coordinates $X^\m$, $\m=0,1,...,D-1$, and metric tensor $G_{\m\n}(x)$ is described by the Polyakov action \eqref{eq:BS-Poly}.
	In first order formalism, for a closed string, this action  is equivalent to
	\be\label{eq:SM-SSST-Action}
		S=\int \dd \tau \int_0^{2\pi} \frac{\dd \s}{2\pi}\, \Big(P_\m\,\dot X^\m + \frac{\xi_1}{2T}\left(G^{\m\n}\,
			P_\m P_\n + T^2 G_{\m\n}\,X'{\,^\m} X'{\,^\n}\right) 
			+ \xi_2\,\,P_\m X'^\m\Big)~.
	\ee
	Here, $P_\m = 2\pi \d S/\d X^\m$ are the momentum variables conjugate to $X^\m$, $T\equiv2\pi\,T_0$ and $\xi_1=\frac{1}{\sqrt{-h}\,h^{00}}$ and $\xi_2=\frac{h^{01}}{h^{00}}$ play the role of Lagrange multipliers. Their variations provide the Virasoro constraints
	\be\label{eq:SM-SSST-Constr0}
		G^{\m\n}\,P_\m P_\n + T^2 G_{\m\n}\,X'{\,^\m} X'{\,^\n}=0~,\quad\quad P_\m X'{\,^\m}=0~.
	\ee

	Let us consider a static space-time. Its metric tensor can be brought into the form
	\be\label{eq:SM-SSST-GenMetric}
		G_{\m\n}=\left(\begin{array}{cc}
              -\L & 0 \\
              0 & G_{kl}
            \end{array}\right)~,
	\ee
	with $\L>0$ and positive definite $G_{kl}$, $k, l=1,2,...,D-1$, both $X^0$-independent. By Noether's theorem one then has a gauge invariant conserved energy
	\be\label{eq:SM-SSST-Energy}
		E=- \int_0^{2\pi} \frac{\dd \s}{2\pi}\, ~P_0~.
	\ee

	The static or temporal gauge is defined by the gauge fixing conditions
	\be\label{eq:SM-SSST-GaugeFix}
		X^0=-\a\,P_0\,\tau~, \qquad P_0^{\,\prime}(\tau,\s)=0~,
	\ee
	with positive constant $\alpha$. As the space-time coordinates have the dimension of length we take $\a=1/T$. In static space-times with intrinsic length scale $R$, such as ${\rm AdS}_{N+1}$, one may rescale to dimensionless coordinates leading to an effective $\alpha=1/(R^2 T)$.


	In the static gauge the first order action \eqref{eq:SM-SSST-Action} reduces to\footnote{We neglect the boundary term related to the time derivative $-\p_\tau\left(\frac{\a}{2}\,P_0^2\tau\right)$.}
	\be\label{eq:SM-SSST-RedAction}
		S=\int \dd \tau \int_0^{2\pi}\frac{\dd \s}{2\pi}\,\left(P_k\,\dot X^k-\frac{\a P_0^2}{2}\right)~,
	\ee
	where the square of the energy density is fixed by the first Virasoro constraint in \eqref{eq:SM-SSST-Constr0},
	\be\label{eq:SM-SSST-EnSquared}
		P_0^2={\L}\Big( G^{kl}P_k P_l + T^2\,G_{kl}\,X'{\,^k} X'{\,^l}\Big)~.
	\ee
	Taking into account the second constraint in \eqref{eq:SM-SSST-Constr0} and the gauge fixing conditions \eqref{eq:SM-SSST-GaugeFix}, we find that the Hamiltonian system \eqref{eq:SM-SSST-RedAction}-\eqref{eq:SM-SSST-EnSquared} has to be further reduced to the constraint surface
	\be\label{eq:SM-SSST-Constr1}
		\calH'(\s)=0~, \quad\quad 
		\calV(\s)\equiv P_k(\s) X'{\,^k}(\s)=0~,
	\ee
	where ${\cal H}(\s)=\frac{\a P_0^2}{2}$ is the Hamiltonian density in \eqref{eq:SM-SSST-RedAction} and  we call ${\cal{V}}(\s)$ the level matching density. Thus, ${\cal H}(\s)$ has vanishing non-zero modes and on the constraint surface one gets with $H=\int^{2\pi}_0 \frac{d\sigma}{2\pi}\, \mathcal{H}(\sigma)$
	\be\label{eq:SM-SSST-EnSq&Ham}
		E^2=\f{2H}{\a}=\int_0^{2\pi} \frac{\dd \s}{2\pi}\,\,{\L}\Big( G^{kl}P_k P_l+
		T^2\,G_{kl}\,X'{\,^k} X'{\,^l}\Big)~.
	\ee
	In fact on the constraint surface the integrand in the above is constant.

	As reviewed in \secref{sec:BS-Static}, construction of independent canonical variables on the constraint surface \eqref{eq:SM-SSST-Constr1} is not an easy task even for Minkowski space with $G_{kl}=\d_{kl}$ and $\L=1$. 
	To follow the same scheme for a generic static space-time one has to analyze the Poisson bracket structure of the constraints \eqref{eq:SM-SSST-Constr1}.

	The canonical relations $\{P_k(\s), X{^l}(\tilde\s)\} = 2\pi\d_k^l\d(\s-\tilde\s)$\footnote{The appearance of the $2\pi$ in comparison to \eqref{eq:BS-CCR} is explained by the fact that we included a factor of $1/2\pi$ into the measure in \eqref{eq:SM-SSST-Action}, cf. also the altered definition of the string tension $T\equiv 2\pi T_0$ and of the conjugate momenta $P_\m = 2\pi \d S/\d X^\m$.} provide the Poisson brackets
	\bea\nn
		&&\{{\cal V}(\s) ,{\cal V}(\tilde\s)\}=-2\pi\left[{\cal V}\,'(\s)\,\delta(\s-\tilde\s)+
			2{\cal V}(\s)\,\delta'(\s-\tilde\s)\right]~,
		\\ [.4 em] \label{eq:SM-SSST-ConstrAlgebra}
		&&\{{\cal V}(\s),{\cal H}(\tilde\s)\}=-2\pi\left[{\cal H}'(\s)\,\delta(\s-\tilde\s)+
		2{\cal H}(\s)\,\delta'(\s-\tilde\s)\right]~,
		\\ [.4 em] \nn
		&&\{{\cal H}(\s), {\cal H}(\tilde\s)\}=-2\pi \a^{2}[(\L^2(\s){\cal V}(\s))'\,\delta(\s-\tilde\s)+
		2\L^2(\s){\cal V}(\s)\,\delta'(\s-\tilde\s)]~.
		\eea
	Equation \eqref{eq:SM-SSST-Constr1}, thereby, defines the second class constraints in Dirac's classification, as in the flat case, and they are preserved in dynamics
	\be\label{eq:SM-SSST-ConstrDynamics}
		\{H,{\cal V}(\s)\}={\cal H}'(\s)~,\qquad \quad \{H,{\cal H}'(\s)\}=\a^{2}(\L^2(\s){\cal V}(\s))''~.
	\ee

	To quantize this system one has to realize the quantum version of these Poisson bracket relations and use the constraint operators to select the physical Hilbert space as in the flat case. The spectrum of the Hamiltonian on the physical states then will define the energy spectrum of the system on the basis of the relation \eqref{eq:SM-SSST-EnSq&Ham}.

\subsection{String in {\AdSxSheader}} \label{subsec:SM-AdSS} 

	Let us consider an ${\rm AdS}_{N+1}\times {\rm S}^M$ space with common radius $R$. As already stated in \eqref{eq:In-AdSdef}, ${\rm AdS}_{N+1}$ is realized as a $(N+1)$-dimensional hyperbola ${\cal Z}_A {\cal Z}^A=-R^2$ in $\Reals^{2,N}$ with embedding coordinates ${\cal Z}^A,$ $A = 0',0,1,\ldots,N$, and ${\rm S}^M$ as $M$-dimensional sphere ${\cal Y}_I {\cal Y}^I = R^2$ in $\Reals^{M+1}$, with $I=1,\ldots,M+1$.

	We parameterize the embedding coordinates of $\Reals^{2,\,N}$ by dimensionless $X^0,$ $X^a$ as
	\be\label{eq:SM-AdSS-AdSCoords}
	{\cal Z}^{0'}=R \sqrt{1+\v{X}^2}\,\,\sin(X^0)~,\quad~ 
		{\cal Z}^{0}=R\sqrt{1+\v{X}^2}\,\,\cos(X^0)~,\quad~ 
		{\cal Z}^a =R\,X^a~,
	\ee
	where $X^0$ is the ${\rm AdS}_{N+1}$ time coordinate and $ \v{X}^2 \equiv X^b X^b$, $a, b =1,\ldots,N$.

	For the spherical part one can use the coordinates of the stereographic projection
	\be\label{eq:SM-AdSS-SCoords}
		{\cal Y}^{i}= \frac{R\,Y^i}{1+\v{Y}^2/4}~,\qquad\quad {\cal Y}^{M+1} = R\,\frac{1-\v{Y}^2/4}{1+\v{Y}^2/4}~,
	\ee
	with $Y^i=X^{N+i}$ and $\v{Y}^2 \equiv Y^j Y^j$, $i, j = 1,\ldots,M$.

	The induced metric on ${\rm AdS}_{N+1}\times {\rm S}^M$ takes the following block structure
	\begin{align}\label{eq:SM-AdSS-IndMetric}
		&g_{\m\n}=\left(\begin{array}{ccc}
              -\L & 0 & 0\\
              0 & G_{ab}&0\\
              0 &0 &G_{ij}
            \end{array}\right)~,\qquad  \text{where} \\
	\L=R^2(1+ \v{X}^2)~,\qquad  &G_{ab}= R^2\left(\d_{ab}-\frac{X^a\,X^b}{1+\v{X}^2}\right)~,\qquad
	G_{ij}= \frac{R^2\, \d_{ij}}{(1+\v{Y}^2/4)^2}~,
	\end{align}	
	and the corresponding inverse matrices read
	\be\label{eq:SM-AdSS-InvMetric}
		G^{ab}= \frac{1}{R^2}\left(\d_{ab}+ {X^a\,X^b}\right)~,\qquad
			G^{ij}=\frac{1}{R^2} \left(1+\v{Y}^2/4\right)^2\,\d_{ij}~.
	\ee
	
	We denote the momentum variables conjugated to $X^a$ and $Y^i$  by $P_{a}$ and $ P_{Y^i}= P_{N+i}$, respectively. Taking into account the structure of the metric tensors in \eqref{eq:SM-AdSS-IndMetric}, it is convenient to treat $X^a$ and $P_a$ as vectors of $\Reals^N$, $Y^i$ and $P_{Y^i}$ as vectors of $\Reals^M$, and use standard notations of vector algebra for $O(N)$ and $O(M)$ scalars: 
	$\v{P}_X^{\,2}\equiv \v{P}^{\,2}\equiv P_aP_a$, $\v{P}_Y^2\equiv P_{Y^i}P_{Y^i}$, {\it etc.}~as in \eqref{eq:SM-AdSS-AdSCoords} and \eqref{eq:SM-AdSS-SCoords}.

	Applying the static gauge \eqref{eq:SM-SSST-GaugeFix}, with $\a=1/(R^2T)$, to the string in ${\rm AdS} \times {\rm S}$, by \eqref{eq:SM-SSST-EnSq&Ham} we obtain $E^2 = 2 \sqrt{\l} H$, such that
	\be\label{eq:SM-AdSS-EnSq&Ham}
		E^2 
			=\int_0^{2\pi}\frac{\dd \s}{2\pi}\,\,(1+ \v{X}^2)\left[ \v{P}^{\,2} + (\v{P}\cdot \v{X})^2 +
			\l\left( \v{X}'^{\,2}-\frac{( \v{X} \cdot  \v{X}')^2}{1+ \v{X}^2}\right) +
		{\cal M}_{\rm S}^2\right]~,
	\ee
	where the 't Hooft coupling $\l=T^2 R^4$ was defined in \eqref{eq:In-tHooft} and \eqref{eq:SS-effST}, respectively, and ${\cal M}_{\rm S}^2$ denotes the spherical part
	\be\label{eq:SM-AdSS-SHam}
		{\cal M}_{\rm S}^2=(1+\v{Y}^2/4)^2\,\v{P}_Y^{\,2}+\l\,\frac{\v{Y}'^{\,2}}{(1+\v{Y}^2/4)^2}~.
	\ee
	Analyzing the Hamiltonian system defined by \eqref{eq:SM-AdSS-EnSq&Ham}-\eqref{eq:SM-AdSS-SHam} one has to recall that the integrand in \eqref{eq:SM-AdSS-EnSq&Ham} is $\s$-independent and that the level matching density vanishes
	\be\label{eq:SM-AdSS-LevelMatch}
		{\cal V}=\v{P}\cdot\v{X}'+\v{P}_{Y}\cdot\v{Y}'=0~.
	\ee

	The dynamical integrals related to the rotations in $\Reals^{2,N}$ and $\Reals^{M+1}$,
	\be\label{eq:SM-AdSS-IoM}
		J_{A\,B}=\int_0^{2\pi}{\frac{\dd \s}{2\pi}\,}\,{ V}_{AB}^\mu P_\mu~, \qquad
		L_{I\,J}=\int_0^{2\pi}{\frac{\dd \s}{2\pi}\,}\,{ V}_{IJ}^\mu P_\mu~,
	\ee
	generate the isometry group SO$(2,N)\times$SO$(M+1)$ and they are dimensionless, as is the energy.
	The index $\mu$ in \eqref{eq:SM-AdSS-IoM} incorporates the three blocks $\m=(0,\,a,\,N+i)$, $P_0=-E$ and ${V}_{AB}^\m$, ${ V}_{IJ}^\mu$ are the components of the Killing vector fields in ${\rm AdS}_{N+1}\times {\rm S}^M$,
	\bea\label{eq:SM-AdSS-VectFields}
		{ V}_{AB}^0 = G^{00}({\cal Z}_B \p_0 {\cal Z}_A - {\cal Z}_A\p_0 {\cal Z}_B)~, \qquad
		{ V}_{AB}^a = G^{ab}({\cal Z}_B\p_b {\cal Z}_A - {\cal Z}_A \p_b {\cal Z}_B)~,\\ [.4 em] \nn
		{ V}_{IJ}^{i+N}=G^{ij}({\cal Y}_J\p_j {\cal Y}_I -{\cal Y}_I\p_j {\cal Y}_J)~.~~~~~~~~~~~~~~~~~~~~~~~~
	\eea
	Using then \eqref{eq:SM-AdSS-AdSCoords}-\eqref{eq:SM-AdSS-InvMetric} and the notation  $\om \equiv E/\sqrt\l$, we find
	\begin{align}\label{eq:SM-AdSS-SO24Rots}
		&{J}_{0'\,0} = E ~, \qquad \quad \qquad
			\qquad\ \ {J}_{a\,b}=\int_0^{2\pi}{\frac{\dd \s}{2\pi}\,}
			\left( P_{a}\,X^b - P_{b}\,X^a\right)~,\\
		\label{eq:SM-AdSS-SO24Boosts}
		&{J}_{a\,0'} =E \left(\int_0^{2\pi}\frac{\dd \s}{2\pi}\,\frac{X^a}{\sqrt{1+ \v{X}^2}}\right)
		\cos\left(\om\,\tau\right)
		-\left(\int_0^{2\pi}\frac{\dd \s}{2\pi}\,
		\sqrt{1+\v{X}^2}\,\,P_{a}\right)\sin\left(\om\,\tau\right), ~
		\\
		&{J}_{a 0} =-E\left(\int_0^{2\pi}\frac{\dd \s}{2\pi}\,\frac{X^a}{\sqrt{1+ \v{X}^2}}\right) \sin\left(\om\,\tau\right)
		-\left(\int_0^{2\pi}\frac{\dd \s}{2\pi}\,
		\sqrt{1+\v{X}^2}\,\,P_{a}\right)\cos\left(\om\,\tau\right),\\
		\label{eq:SM-AdSS-SO6Rots}		
		&L_{i\,j} = \int_0^{2\pi}\frac{\dd \s}{2\pi} \left( P_{Y^i}\,Y^j - P_{Y^j}\,Y^i\right),\\
		&L_{i\,M+1} = \int_0^{2\pi}\frac{\dd \s}{2\pi} \Big(\big(1-\v{Y}^2/4\big)P_{Y^i}+\frac{1}{2}\,(\v{P}_Y \cdot\v{Y})Y^i\Big).
	\end{align}

	Our aim is to analyze the $\AdSxS$ string dynamics at large coupling $\l \gg 1$. The Hamiltonian of the system defined by \eqref{eq:SM-AdSS-EnSq&Ham} allows the expansion in powers of $1/\sqrt\l$
	\be\label{eq:SM-AdSS-HamExpand}
		H = H^{(0)}+\frac{1}{\sqrt\l}\,H^{(1)}+\f{1}{\l}\,H^{(2)}+\dots ~,
	\ee
	which is easily obtained if one uses the rescaled phase space coordinates
	\be\label{eq:SM-AdSS-Rescale}
		x^k=\l^{1/4}\,X^k~, \qquad \quad p_k=\l^{-1/4} P_k~,
	\ee
	where $X^k=(X^a,\,Y^i)$, $P_k=(P_a,\,P_{Y^i})$, $k=1,2,\dots ,9$. 
	The leading term in \eqref{eq:SM-AdSS-HamExpand} coincides with the static gauge string Hamiltonian in $10$-dimensional Minkowski space
	\be\label{eq:SM-AdSS-HamLead}
		H^{(0)}=\frac{1}{2}\int_0^{2\pi}\frac{\dd \s}{2\pi}\left(p^2(\s)+x'^{\,2}(\s)\right)~.
	\ee
	The rotation generators $J_{a\,b}$ and $L_{i\,j}$ in \eqref{eq:SM-AdSS-SO24Rots}-\eqref{eq:SM-AdSS-SO6Rots} are invariant under the rescaling \eqref{eq:SM-AdSS-Rescale} and, therefore, they are $\l$-independent.
	The generators  $J_{a\,0'}$ are expanded in powers of $1/\sqrt\l$ like the Hamiltonian. The leading terms of their expansion correspond to four boosts in $x^a$ ($a=1,\dots , 4$) directions of the 10-dimensional Minkowski space.
	Other symmetry generators in \eqref{eq:SM-AdSS-SO24Rots}-\eqref{eq:SM-AdSS-SO6Rots} are singular at $\l\rightarrow \infty$, however, after their rescaling by the factor $\l^{-1/4}$, they also become analytic in $1/\sqrt{\l}$. It is easy to check that the corresponding leading terms define the translation generators of 10-dimensional Minkowski space.
	In fact, the zero modes will have to be scaled differently than stated in (\ref{eq:SM-AdSS-Rescale}) as was already noted in the previous chapter and will be discussed in detail in the next section.

	With a supersymmetric extension, the leading order part is quantized without anomalies. This part of the symmetry generators and the commutation relations of the isometry group can be used as a basis for perturbative quantum calculations.


\section{Single-Mode Strings in {\AdSxSheader}} \label{sec:SM-SMSol}

	In this section we introduce a class of $\AdSxS$ string configurations with excited zero modes both in the AdS and the spherical parts and only one excited non-zero mode in the AdS part.
	We call these configurations single-mode strings. In the first part of the section we describe the Hamiltonian reduction to the single-mode ansatz, where we find canonical physical variables of the system and verify the SO(2,4)$\times$SO(6) symmetry of single-mode strings.
	Using a canonical transformation to new variables, which separates the non-zero mode part, we then show the integrability of the single-mode configurations. They turn out to describe pulsating strings.

\subsection{Hamiltonian reduction to the single-mode ansatz} \label{subsec:SM-HRed}

	The Hamiltonian density of the $\AdSxS$ string is expressed in terms of O(4) and O(5) vectors denoted by $(\v P$, $\v X)$ and $(\v{P}_Y$, $\v Y)$, respectively (see \eqref{eq:SM-AdSS-EnSq&Ham}-\eqref{eq:SM-AdSS-SHam}). We introduce the above mentioned single-mode ansatz in the following form
	\bea\label{eq:SM-HRed-OMAnsatz}
		&\v{X} (\tau,\s) = \v{X}_0(\tau) + \v{X}_+(\tau)\, e^{i m \s} + \v{X}_-(\tau)\, e^{-i m \s} &
		\quad \text{and} \qquad     \v Y(\tau,\s) =\v Y(\tau)~,\qquad \\ \nn
		&\v{P} (\tau,\s) = \v{P}_0(\tau) + \v{P}_+(\tau)\, e^{i m \s} + \v{P}_-(\tau)\, e^{-i m \s} &
		\quad \text{and} \qquad	\v P_Y(\tau,\s) =\v P_Y(\tau) \, ,
	\eea
	where the positive integer $m > 0$ denotes the mode number of the single-mode ansatz. Due to $\v Y'(\tau,\s)=0$, the spherical part of the Hamiltonian \eqref{eq:SM-AdSS-EnSq&Ham} is only given by the first term in \eqref{eq:SM-AdSS-SHam}, rendering the Hamiltonian polynomial in phase space variables. In particular, we are left with particle dynamics on ${\rm S}^5$ and \eqref{eq:SM-AdSS-SHam} coincides with the $SO(6)$ Casimir number calculated from \eqref{eq:SM-AdSS-SO6Rots},
	\be\label{eq:SM-HRed-SO6Casimir}
		{\cal M}_{\rm S}^2=(1+\v{Y}^2/4)\v{P}_Y^2 = 
			\frac{1}{2}\,L_{i\,j}L_{i\,j} + L_{i\,M+1}L_{i\,M+1}~.
	\ee

	Since the AdS string dynamics is nonlinear, an excitation of one non-zero mode generally excites other modes and we have to find conditions, which preserve the single-mode ansatz \eqref{eq:SM-HRed-OMAnsatz} in dynamics. In addition the constraints in \eqref{eq:SM-SSST-Constr1} should be satisfied, that is the Hamiltonian density  should be  $\s$-independent and the level matching density \eqref{eq:SM-AdSS-LevelMatch} should vanish. 
	In the Hamilton equations, which follow from \eqref{eq:SM-AdSS-EnSq&Ham}, scalar combinations as $\v{X}^2$, $\v{P}^2$, $\v{P}\cdot\v{X}$, {\it etc.} play the role of coefficients of the vectors $\v{P},\,$ $\,\v{X}$ and their derivatives. One can show that the vanishing of the level matching density and the stability of the single-mode ansatz in dynamics require these scalar combinations to be $\s$-independent.
	By this, the Hamiltonian density also becomes  $\s$-independent. Explicitly, we find the following conditions on the excited modes
	\bea\label{eq:SM-HRed-ComplConstr}
		&\v{X}_\pm\cdot \v{P}_0=\v{P}_\pm\cdot \v{X}_0=0 ~,
			\qquad \v{X}_\pm\cdot \v{X}_0=\v{P}_\pm\cdot \v{P}_0=0~,\\ \nn
		&\v{X}_\pm^2=\v{P}_\pm^2=0~,
			\qquad \v{P}_\pm\cdot \v{X}_\pm=0~,
			\qquad \v{P}_+\cdot \v{X}_- -\v{P}_-\cdot \v{X}_+=0~.
	\eea

	It is clear from these conditions that the zero mode vectors $\v P_0$, $\v X_0$ may be considered as unconstrained. Then, introducing real and imaginary parts of the non-zero modes, $\,\,\v{X}_\pm=\frac{1}{\sqrt 2}(\v{X}_{\rm re}\pm i\v{X}_{\rm im})$ and $\,\v{P}_\pm=\frac{1}{\sqrt 2}(\v{P}_{\rm re}\pm i\v{P}_{\rm im})$, one sees that they  lie in the plane orthogonal to the plane spanned by the zero mode vectors $\v P_0$, $\v X_0$.\footnote{For any solution of the classical equations of motion this zero mode plane turns out to be defined by the boosts and, therefore, has a $\tau$-independent orientation in $\Reals^4$, as is shown below.}
	In addition, the four non-zero mode vectors $\v X_{\rm re}$, $\v X_{\rm im}$, $\v P_{\rm re}$, $\v P_{\rm im}$ satisfy the constraints
	\bea\label{eq:SM-HRed-RealConstr}
		&\v{X}_{\rm re}^2=\v{X}_{\rm im}^2~, \quad ~~~\v{X}_{\rm re}\cdot\v{X}_{\rm im}=0~, \qquad
		&\v{P}_{\rm re}^2=\v{P}_{\rm im}^2~, \quad \v{P}_{\rm re}\cdot\v{P}_{\rm im}=0~,\\
		&\v{X}_{\rm re}\cdot\v{P}_{\rm im}=0~, \quad \v{P}_{\rm re}\cdot\v{X}_{\rm im}=0~, \qquad &\v{P}_{\rm re}\cdot\v{X}_{\rm re}=\v{P}_{\rm im}\cdot\v{X}_{\rm im}~.\nn
	\eea
	Since we are considering ${\rm AdS}_5$, the non-zero mode vectors can be parameterized as
	\be\label{eq:SM-HRed-DefPQ}
		\v{X}_{\rm re}=\frac{Q}{\sqrt 2}\,\v{\bf e}_{\rm re}~, \quad  \v{P}_{\rm re}=\frac{P}{\sqrt 2}\,\v{\bf e}_{\rm re}~,\qquad
		\v{X}_{\rm im}=\pm\frac{Q}{\sqrt 2}\,\v{\bf e}_{\rm im}~, \quad \v{P}_{\rm im}=\pm\frac{P}{\sqrt 2}\,\v{\bf e}_{\rm im}~,
	\ee
	where $\v{\bf e}_{\rm re}$ and $\v{\bf e}_{\rm im}$ are orthonormal vectors with the standard relative orientation. One can always choose  $\v{\bf e}_{\rm re}$ to be in the direction of  $\v{X}_{\rm re}$ which implies $Q \geq 0$. 
	Then, there is no restriction on the sign of $P$,  i.e. $P\in \Reals^1$. Note that the two solutions for $\v{X}_{\rm im}$ and $\v{P}_{\rm im}$ are related to each other by the reflection $\s\mapsto -\s$, which is a symmetry of the string action. 
	In what follows we choose for definiteness the solution with all positive signs. 

	Inserting the parametrization \eqref{eq:SM-HRed-DefPQ} in the ansatz \eqref{eq:SM-HRed-OMAnsatz}, we find
	\be\label{eq:SM-HRed-RedOMAnsatz}
		\v{X} (\tau,\s) = \v{X}_0(\tau) + Q(\tau)\left[\v{\bf e}_{\rm re} \cos m \s + \v{\bf e}_{\rm im}\sin m \s\right]~,
	\ee
	and a similar expression for $\v{P} (\tau,\s)$. The reduction of the AdS part of the initial canonical symplectic form to the single-mode ansatz \eqref{eq:SM-HRed-OMAnsatz} yields
	\be \label{eq:SM-HRed-RedSymplForm1}
		\Omega=\int_0^{2\pi}\frac{\dd \s}{2\pi}\, \dd \v{P}(\s)\wedge \dd \v{X}(\s)= \dd \v{P}_0\wedge \dd \v{X}_0+
		\dd \v{P}_{\rm re}\wedge \dd \v{X}_{\rm re}+\dd \v{P}_{\rm im}\wedge \dd \v{X}_{\rm im}~.
	\ee
	Using then the parametrization \eqref{eq:SM-HRed-DefPQ} and the identities $\,\dd \v{\bf e}_{\rm re} \wedge \dd \v{\bf e}_{\rm re}=\dd \v{\bf e}_{\rm im}\wedge \dd \v{\bf e}_{\rm im}=0$ as well as $\,\,\v{\bf e}_{\rm re}\cdot \dd \v{\bf e}_{\rm re}=\v{\bf e}_{\rm im}\cdot \dd \v{\bf e}_{\rm im}=0$, we find the reduced symplectic form
	\be\label{eq:SM-HRed-RedSymplForm2}
		\Omega=\dd \v{P}_0\wedge \dd \v{X}_0+\dd P\wedge \dd Q~.
	\ee
	It is independent of the orientation of the vector $\v{\bf e}_{\rm re}$ because the non-zero mode parts of $\v{X} (\tau,\s)$ and $\v{P} (\tau,\s)$ are collinear. This independence  may also be  interpreted as the residual gauge symmetry $\s\mapsto \s+f(\tau)$ of the single-mode ansatz as the level matching constraint is satisfied.
	Even including all the other excitations, one can always perform a canonical transformation which for the vectors $\v{X} (\tau,\s)$ and $\v{P} (\tau,\s)$ takes the form of a rotation in the non-zero mode plane, and change the orientation of  $\v{\bf e}_{\rm re}$ arbitrarily.

	Thus, the Hamiltonian reduction of the AdS part leads to a ten-dimensional phase space with the canonical coordinates $(\v{P}_0,\,\v{X}_0,\,P,\,Q)$, where the pair $(P,\,Q)$ lives on the half-plane $Q\geq 0$.

	Calculating now the string energy square and the symmetry generators, where the boosts are treated as 4-vectors $\v J_{0'}\equiv J_{a\,0'},$  $\,\v J_0 \equiv J_{a\,0}$, from \eqref{eq:SM-AdSS-EnSq&Ham} and \eqref{eq:SM-AdSS-SO24Rots}-\eqref{eq:SM-AdSS-SO24Boosts} we obtain (recall $\omega=E/\sqrt{\lambda}$)
	\be\label{eq:SM-HRed-OMEnSquared}
		E^2 
		= \left(1+\v{X}_0^2+Q^2\right)\left(\v{P}^{\,2}_0 + P^2+(\v{P}_0\cdot\v{X_0}+P Q)^2+{\cal M}_{\rm S}^2+{\l}\,m^2 Q^2\right)~,
	\ee
	\be\label{1-mode AdS rotations}
		J_{0'\,0} = E~,  \qquad \qquad \qquad J_{a\,b} = P_{0,a}\, X^b_0 - P_{0,b}\, X^a_0~,
	\ee
	\be\ba\label{eq:SM-HRed-OMAdSBoosts}
		&\v{J}_{0'} = \frac{E\, \v{X}_{0}}{\sqrt{1+\v{X}_0^2+Q^2}} \cos(\om\,\tau)-
		\sqrt{1+\v{X}_0^2+Q^2}\,\,\v{P}_0\sin(\om\,\tau)~,& \\
		&\v{J}_{0} =-\frac{E\, \v{X}_{0}}{\sqrt{1+\v{X}_0^2+Q^2}}\sin(\om\,\tau)-
		\sqrt{1+\v{X}_0^2+Q^2}\,\,\v{P_{0}}\,\cos(\om\,\tau)~.&
	\ea\ee

	From \eqref{eq:SM-HRed-OMEnSquared} and \eqref{eq:SM-HRed-OMAdSBoosts} follow the Poisson brackets
	\be\label{eq:SM-HRed-PB-EnSq-Boosts}
		\{E^2,J_{a\,0'}\} =-2E\,J_{a\,0}~,  \qquad \{E^2,J_{a\,0}\} = 2E\,J_{a\,0'}~,
	\ee
	which are equivalent to a part of the $\mathfrak{so}(2,4)$ commutation relations. The validity of the remaining algebra is straightforward.	

	In contrast to the particle dynamics, the SO$(2,4)$ Casimir
	\be\label{eq:SM-HRed-SO24Casimir}
		C = E^2 + \f{1}{2} J_{a\,b} J_{a\,b} - \v{J}_{0'}^{\,2} - \v{J}_0^{\,2}
	\ee
	resulting from \eqref{eq:SM-HRed-OMEnSquared}-\eqref{eq:SM-HRed-OMAdSBoosts} is not a number, but a function on the phase space,
	\be\label{eq:SM-HRed-OMSO24Casimir}
		C=(P+\tilde D\,Q)^2+(1 + Q^2) ({\cal M}^2_{S}+\l m^2\,Q^2 )~,
	\ee
	where $\tilde D=P\,Q+\v{P}_0\cdot\v{X}_0$ generates dilatations in the ten-dimensional phase space.

	Note that for vanishing zero modes the SO$(2,4)$ Casimir and the energy square \eqref{eq:SM-HRed-OMEnSquared} coincide.

	In quantum theory we will be interested in the spectrum of the corresponding operators corrected by vacuum fluctuations and possible supersymmetric corrections.

\subsection{Integrability of single-mode strings} \label{subsec:SM-IOMS}
	The Hamilton function of the single-mode string defined by \eqref{eq:SM-HRed-OMEnSquared} has the structure of a Hamiltonian of a non relativistic particle. The kinetic part, given by the terms quadratic in momenta, corresponds to a free-particle on the five dimensional unit half-sphere\footnote{More precisely, it is a quarter of the sphere, since $Q\geq 0$}.
	The term ${\cal M}_{\rm S}^2$ in the potential corresponds to the SO(6) Casimir and can be treated as a constant mass-square parameter. The potential contains terms quadratic and quartic in ($\v{X}_0,\, Q)$. Formally taking $m=0$ in  \eqref{eq:SM-HRed-OMEnSquared} one only has the quadratic terms, which corresponds to the Higgs-potential on a sphere discussed in \cite{Higgs:1978yy}. This model corresponds to the ${\rm AdS}_6$ particle with mass ${\cal M}_S$ in the static gauge \cite{Dorn:2005vh, Dorn:2010wt, Jorjadze:2012jk} and it is exactly solvable in terms of trigonometric functions. Below we show that the quartic terms do not destroy the integrability of the system, however the solution has a more complicated structure. To show the integrability, we introduce a canonical map to new coordinates, which allows to separate the non-zero mode variables and express the Casimir \eqref{eq:SM-HRed-OMSO24Casimir} only through them.

	Taking into account the form of \eqref{eq:SM-HRed-OMSO24Casimir}, we introduce a transformation of the phase space coordinates $(\v{P}_0,\,\v{X}_0,\,P,\,Q)\leftrightarrow (\v{p}_0,\,\v{x}_0,\,p,\,q)$ in the following form
	\be\label{eq:SM-IOMS-CanTrafo}
		\v{P}_{0}=\frac{\v{p}_0}{\sqrt{1+q^2}}~,   \quad  \v{X}_0=\v{x}_0 \sqrt{1+q^2}~,
			\quad P=p-\frac{q(\v{p}_0 \cdot\v{x}_0)}{1+q^2}~, \quad  Q=q~.
	\ee
	It is straightforward to check that \eqref{eq:SM-IOMS-CanTrafo} is a canonical transformation
	\be\label{eq:SM-IOMS-CanonSymplForm}
		\dd \v{P}_0\wedge \dd \v{X}_0+\dd P\wedge \dd Q=\dd \v{p}_0\wedge \dd \v{x}_0 + \dd p\wedge \dd q~,
	\ee
	related to the change of coordinates $(\v X_0, Q)\leftrightarrow (\v{x}_0, q)$.

	In the new variables, the SO(2,4) dynamical integrals \eqref{eq:SM-HRed-OMEnSquared}-\eqref{eq:SM-HRed-OMAdSBoosts}
	take the form of a free massive particle in ${\rm AdS}_5$
	\bea\label{eq:SM-IOMS-EnSquared}
		&E^2 = 2 \sqrt{\l}\,H = \left(1+\v{x}_0^{\,2}\right)\left(\v{p}_0^{\,2}+(\v{p}_0\cdot\v{x}_0)^2 +{\cal M}^2\right)~,\\[.4em]
		\label{eq:SM-IOMS-OMAdSRots}
		&J_{0'\,0} = E~,  \qquad \qquad \qquad 
			J_{a\,b} = p_{0,a}\, x^b_0 - p_{0,b}\, x^a_0~,\\[.3em]	
		&\ba\label{eq:SM-IOMS-OMAdSBoosts}
		&\v{J}_{0'} = \frac{E\, \v{x}_0}{\sqrt{1+\v{x}_0^{\,2}}}\cos(\om\,\tau)-
		\sqrt{1+\v{x}_0^{\,2}}\,\,\v{p}_0\sin(\om\,\tau)~,&
		\\
		&\v{J}_{0} =-\frac{E\, \v{x}_0}{\sqrt{1+\v{x}_0^{\,2}}}\sin(\om\,\tau)- \sqrt{1+\v{x}_0^{\,2}}\,\,\v{p}_0\,\cos(\om\,\tau)~,&
	\ea\eea
	and the mass-squared term depends only on the non-zero modes $q$ and $p$,
	\be\label{eq:SM-IOMS-MassSquared}
		{\cal M}^2=(1+q^2)\left[p^2+p^2 q^2+{\cal M}_{\rm S}^2+{\l}\,m^2 q^2\right]~.
	\ee
	This term coincides with the SO(2,4) Casimir calculated from \eqref{eq:SM-IOMS-EnSquared}-\eqref{eq:SM-IOMS-OMAdSBoosts}
	and, therefore, it is $\tau$-independent, like the SO(6) Casimir ${\cal M}_{\rm S}^2$.

	It follows from \eqref{eq:SM-IOMS-OMAdSBoosts} that $\sqrt{1+\v{x}_0^{\,2}}\left[\v{J}_{0'}\cos(\om\,\tau)-\v{J}_{0}\sin(\om\,\tau)\right]=E\, \v{x}_0$.
	Squaring of this relation defines the scalar $\v{x}_0^{\,2}$ in terms of the boosts and $\tau$-dependent trigonometric functions.
	The pair $(\v{x}_0,\,\v{p}_0)$ is then obtained from \eqref{eq:SM-IOMS-OMAdSBoosts} algebraically, as a solution of the linear system, with given $\tau$-dependent coefficients.

	Note that the integrals of motion \eqref{eq:SM-IOMS-OMAdSRots}-\eqref{eq:SM-IOMS-OMAdSBoosts} are related by $E\,J_{a\,b}=J_{a\,0'}\,J_{b\,0}-J_{b\,0'}\,J_{a\,0}$. This, relation together with \eqref{eq:SM-HRed-SO24Casimir}, allows to express $E$ and $J_{ab}$ through the boosts \cite{Dorn:2005ja}.
	Thus, the pair $(\v{x}_0,\,\v{p}_0)$ is parameterized by $(\v J_{0'}$, $\v J_0$) and the Casimir number ${\cal M}^2$ only.

	The dynamics of the $(p,\,q)$ pair is governed by the Hamiltonian
	\be\label{eq:SM-IOMS-Hq}
		H_m=\frac{1+\v{x}_0^{\,2}(\tau)}{2\sqrt\l}\,{\cal M}^2~,
	\ee
	where $\v{x}_0^{\,2}(\tau)$ has been calculated above and ${\cal M}^2$ is given by \eqref{eq:SM-IOMS-MassSquared}. Eliminating the momentum variable $p$ from the EoM for $\dot q$ by solving \eqref{eq:SM-IOMS-MassSquared} for $p$, one finds
	\be \label{eq:SM-IOMS-qdot}
		\pm\sqrt\l\,\,\dot q=(1+\v{x}_0^{\,2}(\tau))(1+q^2)\sqrt{{\cal M}^2-(1+q^2)({\cal M}_{\rm S}^2+{\l}\,m^2 q^2)}~.
	\ee
	This equation is separable in $(q,\,\tau)$ and can be integrated directly, therefore proving the asserted classical integrability of the single-mode system. The $q$-integration however results in an elliptic integral of third kind and the solutions can not be written in terms of elementary functions.

	To find solutions at large $\l$ and vanishing zero modes, we rescale the phase space variables as
	\be\label{eq:SM-IOMS-pqRescale}
		q\mapsto\frac{q}{\l^{1/4}\sqrt{m}}~, \qquad  p\mapsto\l^{1/4}\sqrt{m}\,p~,
	\ee
	which is the usual rescaling for non-zero modes, cf. the discussions in \cite{Passerini:2010xc} and also in \secref{sec:PS-Scale}. By this,  in the leading order we obtain the oscillator Hamiltonian
	\be\label{eq:SM-IOMS-HqExpand}
		H_m=\,\frac{m}{2}\,(p^2+q^2)+{\cal O}(\l^{-1/2}) ~,
	\ee
	which provides time oscillations of the $q$ coordinate. Since this coordinate is non-negative and it describes the length of the vectors $\v{X}_{\rm re}$ and $\v{X}_{\rm im}$ in the plane of non-zero modes, one gets a circular string with an oscillating radius.

	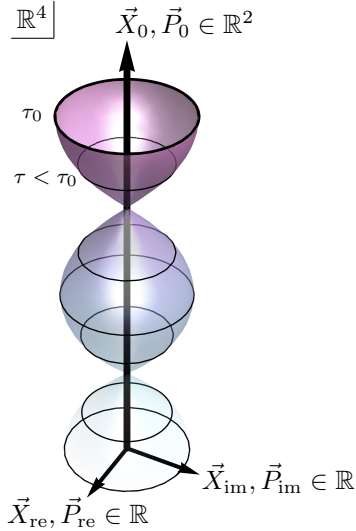
\begin{figure}[!ht] 
	\centering
	\begin{tikzpicture}
	\node[above right] (img) at (0,0) {\includegraphics[width=0.135\textwidth]{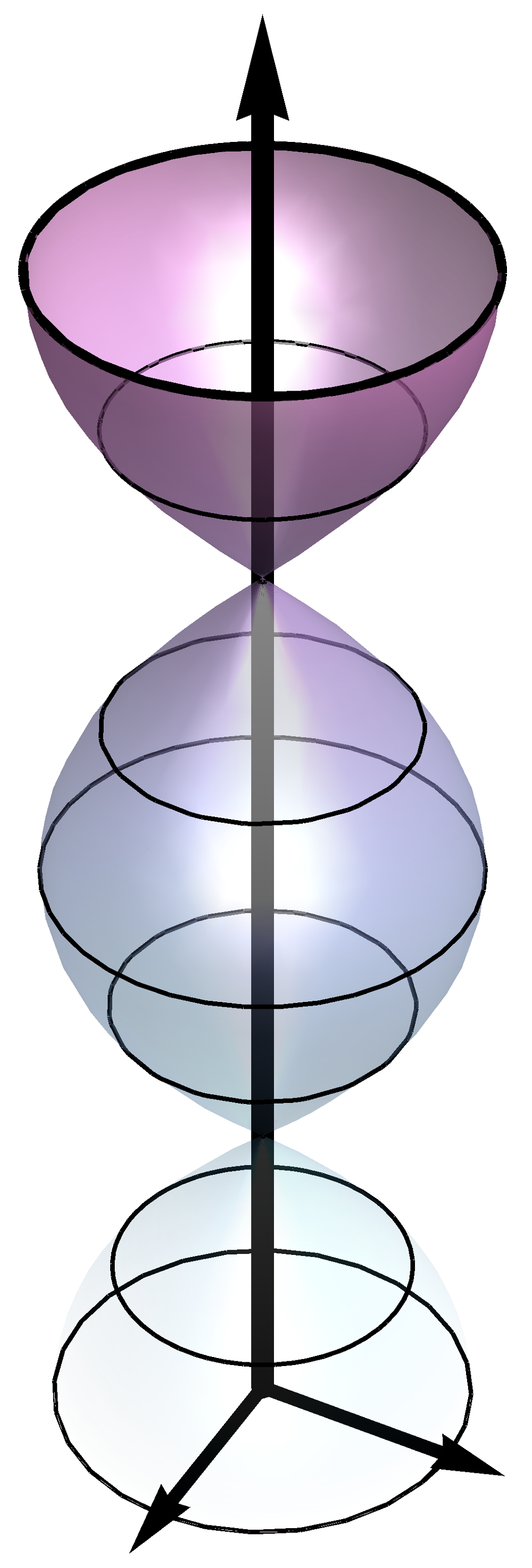}};
	\node at (55pt,185pt) {\footnotesize $\vXz, \vPz \in \Reals^2$};
	\node at (15pt,0pt) {\footnotesize $\vXr, \vPr \in \Reals$};
	\node at (90pt,13pt) {\footnotesize $\vXi, \vPi \in \Reals$};
	\node at (-2pt,187pt) {\footnotesize $\Reals^4$};
     \draw (-10pt,179pt) -- (6pt,179pt);
	\draw (6pt,193pt) -- (6pt,179pt);
	\node at (-2pt,150pt) {\scriptsize $\tau_0$};
	\node at (3pt,126pt) {\scriptsize $\tau < \tau_0$};
	\end{tikzpicture}
					\caption{Single-mode solution, basically a pulsating string, depicted in the four spatial dimensions of ${\rm AdS}_5$, with the zero mode plane projected to a line. The string area is traced out up to some world sheet time $\tau_0$, thin lines showing string configurations for $\tau < \tau_0$. We neglected that the zero modes oscillate themselves.}
					\label{fig:PulsatingString}
	\end{figure}

	This type of string solutions are called pulsating strings and are depicted in \figref{fig:PulsatingString}.
	In AdS spaces they were obtained first for ${\rm AdS}_3$ \cite{deVega:1994yz} and then they have been applied in the context of AdS/CFT in \cite{Minahan:2002rc}, where however in both studies zero modes have been disregarded, $X^k_0 = P_{0,k} = 0$. The virtue of the presented single-mode solution therefore lies in the allowance for zero modes and their exact solution as massive particles in $\AdSxS$.

	Finally, note that by \eqref{eq:SM-IOMS-pqRescale} we have ${\cal M}^2 \propto \sqrt{\l}\,m$. Taking then the rescaling $\v x_0 \mapsto \v x_0/\sqrt{{\cal M}}$, $\v p_0 \mapsto \sqrt{{\cal M}}\,\v p_0$, the leading order zero modes Hamiltonian \eqref{eq:SM-IOMS-EnSquared} becomes a harmonic oscillator
	\be\label{E2leading}
		E^2 = {\cal M}^2 + {\cal M}(\v{p}_0^{\,2}+\v{x}_0^{\,2})+\cdots~.
	\ee
	Of course, this is again nothing but the peculiar scaling of the zero modes observed before in \cite{Passerini:2010xc} and discussed in \secref{sec:PS-Scale}. In particular, the zero modes scale differently in $\l\,m^2$ than the non-zero modes and are not free particles, not even in leading order.


\newpage

\section{Energy Spectrum of Single-Mode Strings} \label{subsec:SM-QNZM}
	Let us now quantize the non-zero mode sector, construct the Casimir operator ${\cal M}^2$, heuristically taking into account corrections due to fermions, and calculate its spectrum perturbatively up to next-to leading order. The energy spectrum of the pulsating string is then extracted similarly to the ${\rm AdS}_5$ particle.

	The energy spectrum of the ${\rm AdS}_5$ scalar particle with Casimir number ${\cal M}^2$ is of harmonic oscillator type (see e.g.~\cite{Dorn:2005vh, Dorn:2005ja, Dorn:2010wt, Jorjadze:2012jk}),
	\be\label{eq:SM-QNZM-MinEnergy}
		E_{(N,{{\cal M}^2})}=2+\sqrt{\mathcal{M}^2+4} + N\,, \qquad N\in \mathbb{N}~.
	\ee
	With help of the canonical transformation established in the last section the energy spectrum of the pulsating string then follows from the spectrum of the operator  ${\cal M}^2$ defined by \eqref{eq:SM-IOMS-MassSquared} depending only on the string-modes and the ${\cal M}_{\rm S}^2$ of the center-of-mass motion on the ${\rm S}^{5}$.
	Expanding \eqref{eq:SM-QNZM-MinEnergy} in powers of $1/{\cal M}$, one gets for the ground state $E_{(0,{{\cal M}^2})}\approx \mathcal{M}+ 2\,$, which is perfectly consistent with what follows from the leading order zero mode Hamiltonian \eqref{E2leading}, with the constant $2$ appearing because the ${\rm AdS}_5$ zero mode vectors are four-dimensional. 
	Note then that the eigenvalues of the SO(6) Casimir number are $J(J+4)$ with non-negative integer $J$ and in the bosonic case we replace ${\cal M}_{\rm S}^2$ in \eqref{eq:SM-IOMS-MassSquared} by this value. There is an  immediate puzzle arising if we consider the single-mode string in its groundstate in the ${\rm AdS}_{5}$ sector, i.e., a point particle rotating on the ${\rm S}^{5}$ only, ${\cal M}^2={\cal M}_{\rm S}^2=J(J+4)$, we obtain $E_{(0,J(J+4))}=J +4$.
	However, $E_{(0,\mathcal{M}^2_S)}$ should be equal to $J$ by comparison to the BMN plane-wave superstring in light-cone gauge with vanishing $E-J$ for the groundstate \cite{Berenstein:2002jq}. This state is dual to the protected gauge theory operator $\tr(Z^{J})$. 

	The resolution of this puzzle lies in our ignorance of fermions. For the point particle, the inclusion of the fermionic degrees of freedom changes the dependence of ${\cal M}_{\rm S}^2$ on $J$ to ${\cal M}_{\rm S}^2=J(J-4)$ giving thus the expected result \cite{Metsaev:1999gz}. For string excitations one would expect in particular, that the eight physical fermionic zero modes included in the reduced Hamiltonian \eqref{eq:SM-IOMS-EnSquared} would produce in the leading order the additional constant $-4$, changing the constant $2$ in  the ground state energy to $-2$.
	Based on this and thecomparison to the plane-wave light-cone groundstate, we thus conclude that the large $\cal M$ expansion of the energy spectrum is given by
	\be\label{eq:SM-QNZM-MinEnergy2}
		E_{(N,{{\cal M}^2})}=\mathcal{M}-2+\frac{2}{\mathcal{M}} + N+\cdots\,, \qquad N\in \mathbb{N}~,
	\ee
	where $\cal M$ may acquire additional $1/\lambda^{1/4}$ corrections due to fermions. 

	Of course, this is only a heuristic argument at this point and should be backed up through an analysis of the full superstring. 

	Let us now turn to the spectrum ${\cal M}^2$ and quantize the string-mode sector.  For this one has to choose either Dirichlet or Neumann boundary conditions for  wave functions $\psi(q)$ which are defined on the half-line $q>0$ at $q=0$. This ambiguity is related to the non-selfadjointness of the momentum operator $p$ on the half-line $q\geq 0$.

	To avoid this ambiguity, we use the quantization scheme proposed in \secref{sec:SM-StringDyn} for the static gauge approach. In our case, this assumes a canonical quantization of the modes $(\v P_0,\, \v X_0)$, $(\v P_\pm,\, \v X_\pm)$ and restriction of the Hilbert space by the quantum version of the constraints \eqref{eq:SM-HRed-ComplConstr}.
	For the analysis of these constraints it is convenient to introduce two 8-dimensional vectors constructed by the pairs ($\v X_{\rm re},$ $\v X_{\rm im}$) and ($\v P_{\rm re},$ $\v P_{\rm im}$), respectively. It has to be noted that, on the constraint surface \eqref{eq:SM-HRed-RealConstr}, these two 8-vectors are parallel to each other, which can be easily seen from the representation \eqref{eq:SM-HRed-DefPQ}.
	The Hamiltonian reduction in the non-zero mode sector is then equivalent to the gauging of the 16 dimensional phase space with coordinates {($\v P_{\rm re},$ $\v P_{\rm im}$; $\v X_{\rm re},$ $\v X_{\rm im}$)} by the group SO(8). As a result, the non-zero mode part of the physical wave functions depends only on the SO(8) scalar $q=\sqrt{\v {X}^2_{\rm re}+ \v {X}^2_{\rm im}}$.

	One can generalize this scheme, introducing two ${{N_q}}$-dimensional vectors ($\v p$, $\,\v q$), which replace the pair ($p$, $\, q$) in \eqref{eq:SM-IOMS-MassSquared} as follows
	\be\label{eq:SM-QNZM-MassSquared}
		{\cal M}^2=(1+\v q^{\,2})\left[\v p^{\,2}+(\v p \cdot \v q)^{2}+{\cal M}^2_S+{\l}\,m^2 \v q^{\,2}\right]~.
	\ee
	Gauging of the phase space ($\v p$, $\v q$) by the SO(${{N_q}})$ group leads to the half-plane $(p,\, q)$ where $q$ is the length of the vector $\v q$ and $p=(\v p\cdot\v q)/q$. The Casimir \eqref{eq:SM-QNZM-MassSquared} is gauge invariant and it obviously reduces to \eqref{eq:SM-IOMS-MassSquared}. Quantum theory on the half-plane $(p,\, q)$ is then identified with the ${N_q}$-dimensional case, restricted to the wave functions $\psi(q)$.

	Note that solving the constraints \eqref{eq:SM-HRed-ComplConstr} partially, one gets similar systems with ${N_q}<8$. For example, taking into account that non-zero modes are orthogonal to zero modes, the pairs ($\v X_{\rm re},$ $\v X_{\rm im}$) and ($\v P_{\rm re},$ $\v P_{\rm im}$) can be treated as 4-vectors and one has to perform the SO(4) gauging.

	Using the rescaling \eqref{eq:SM-IOMS-pqRescale} for ${N_q}$-dimensional vectors $\v p$ and $\v q$, from \eqref{eq:SM-QNZM-MassSquared} we obtain
	\be\label{eq:SM-QNZM-MSqExpand}
		{\cal M}^2=\sqrt\l\,m\left(\v p^{\,2}+\v q^{\,2}\right)+\v p^{\,2}\,\v q^{\,2}+(\v p \cdot \v q)^{2}+(\v q^{\,2})^2+{\cal M}^2_S+{\cal O}(\l^{-1/2})~.
	\ee

	The ${{N_q}}$-dimensional oscillator Hamiltonian
	\be\label{eq:SM-QNZM-HqLead}
		h=\frac{1}{2}(\v p^{\,2}+\v q^{\,2})
	\ee
	has eigenfunctions with zero angular momentum at the even levels $2n$ only. We denote these states by $|n\rangle$. They are obtained by the action of the operators $(\v a^{\,\dag\,\,2})^n$ on the vacuum state. Hence, these wave functions are even in $q$ and they correspond to the Neumann boundary condition at $q=0$.
	In \appref{app:SONq} we present an algebraic construction of these states and calculate mean values of scalar operators. Since the oscillator Hamiltonian \eqref{eq:SM-QNZM-HqLead} describes ${\cal M}^2$ in the leading order, we can use these calculations to find the spectrum of  $\mathcal{M}^2$ perturbatively.

	The construction of the operator ${\cal M}^2$ by its classical expression \eqref{eq:SM-QNZM-MSqExpand} contains operator ordering ambiguities.
	It has to be noted that for a given Hermitian ordering the mean values of a scalar operator essentially depends on the dimension ${{N_q}}$ (see \eqref{eq:SONq_MeanVals1}-\eqref{eq:SONq_MeanVals4} in the \appref{app:SONq}).

	Hence, to fix the operator ${\cal M}^2$, we first rewrite \eqref{eq:SM-QNZM-MSqExpand} in terms of creation and annihilation variables, keeping its scalar structure. For this purpose we introduce complex scalar functions
	\be\label{eq:SM-QNZM-AAdag}
		A=\frac{1}{2}(\v p^{\,2}-\v q^{\,2})-i\v p\cdot\v q~, \qquad
		A^\dag=\frac{1}{2}(\v p^{\,2}-\v q^{\,2})+i\v p\cdot\v q~,
	\ee
	which correspond to the operators $\v a^{\,2}$ and $\v a^{\,\dag\,\,2}$, respectively (see \eqref{eq:SONq_ONqScalars}).

	The Casimir \eqref{eq:SM-QNZM-MSqExpand} in the new variables becomes
	\be\label{eq:SM-QNZM-MSq-hAAdag}
		{\cal M}^2=2\sqrt\l\,m\,h+2\,h^2+\f{1}{2}\,A^\dag\,A-\f{1}{4}(A^{\dag\,2} +A^2) -h(A^\dag+A)+{\cal M}^2_S + \ord{\l^{-1/2}}~.
	\ee
	Choosing a normal ordering with ${\colon\!}h{\colon\!}=h-{{N_q}}/2$ (see \appref{app:SONq}), $h^2$ is substituted by
	\be
		h^2\mapsto {{\colon\!}h^2{\colon\!}=\;}({\colon\!}h{\colon\!})^2-{\colon\!}h{\colon\!}~.
	\ee
	With the help of 
	\be\label{eq:SM-QNZM-MeanValues}
		\langle n|{\colon\!}h{\colon\!}|n\rangle =2\,n~, \qquad \langle n|A^\dag\,A|n\rangle=4\,n^2+2({{N_q}}-2) n~,
	\ee
	the mean values of the operator ${\cal M}^2$ become
	\be\label{eq:SM-QNZM-MSqSpectrum1}
		\langle n|{\cal M}^2|n\rangle = 4\,{\sqrt{\l}}\,m\, n +10\,n^2
			+({{N_q}}-6)n+{\cal M}^2_S + \ord{\l^{-1/2}}~.
	\ee
	Let us now argue for a way to mimick the effect of supersymmetry whose implementation via the quantum symmetry algebra should hopefully fix (some) of the ordering ambiguities encountered in our bosonic string analysis.  Recall that the bosons give rise to the value $N_{q}=8$ in the above equation \eqref{eq:SM-QNZM-MSqSpectrum1} related to the 8 dimensional phase space of the four transverse ${\rm AdS}_{5}$ dimensions of our static gauge. 
	In the spirit of effective ``negative dimensions'' for fermions we  argue that taking them into account will lead to an effective value of ${{N_q}}=0$ in the final expressions for the spectrum. This approach is certainly true for the super-harmonic oscillator present at the leading perturbative order and we will implement it at the orders as well.

	Taking  ${{N_q}}=0$ in \eqref{eq:SM-QNZM-MSqSpectrum1} we obtain
	\be\label{eq:SM-QNZM-MSqSpectrum2}
		\langle n|{\cal M}^2|n\rangle = 4\,{\sqrt{\l}}\,m\,n 
			+ 10\,n^2-6n+{\cal M}^2_S~.
	\ee
	Equation \eqref{eq:SM-QNZM-MinEnergy2} together with ${\cal M}^2_S = J(J+4)$ then leads to  the energy spectrum
	\be\label{eq:SM-QNZM-EnSpectrum}
		E_{N,m,n,J}=2\,\l^{1/4}\sqrt{m\,n} + N-2 + \frac{10\,n^2-6\,n+4+J(J+4)}{4\,\l^{1/4}\sqrt{m \,n}} + \ord{\l^{-3/4}}~,
	\ee
	with $N$ being the excitation of the ${\rm AdS}_{5}$ particle degrees of freedom of \eqref{eq:SM-QNZM-MinEnergy2}. In fact we expect that the inclusion of fermions would change ${\cal M}^2_S$ to $J^2$ as it follows from the flat space spectrum and is supported by numerical studies of the TBA equations.
	For the lowest stringy excitation of $N=0$, $m=1$, $n=1$ and $J=0$ conjecturally dual to the Konishi operator this reduces to
	\be\label{eq:SM-QNZM-En11}
		E_{0,1,1,0}=2\,\l^{1/4}-2+\f{2}{\l^{1/4}}+ \ord{\l^{-3/4}}~,
	\ee
	in accordance with \eqref{eq:In-KoAnDim}.

	It is now important to ask to what extent we can trust the level truncation scheme that we have pursued by restricting the quantum dynamics of the string to a single non-zero mode excitation. Clearly at some order in perturbation theory this minisuperspace approach fails to provide the correct energy of the studied string excitation as the suppressed non-zero modes would contribute in intermediate states in perturbation theory of the full model.
	Minisuperspace quantization for the lowest mode string with $n=m=1$ will yield a better approximation to the energy spectrum than for higher excitations $n,m>1$ as here the fluctuation of lower modes will contribute earlier. In the following section we will quantify this statement in some detail. Indeed the intuitive expectation is confirmed, namely that the result (\ref{eq:SM-QNZM-EnSpectrum}) at order $\mathcal{O}(\lambda^{-1/4})$ can only be trusted for the $m=n=1$ case -- modulo the discussed ordering ambiguity.


\newpage

\section{Decoupling of Other Modes} \label{sec:SM-Decoupling}
	To investigate {to which order and for which states our result \eqref{eq:SM-QNZM-EnSpectrum} can be trusted we have to return again to} the full string action, i.e., consider again all modes, not only the zero modes and $m$th ${\rm AdS}_5$ mode, and then see at which order the modes disregarded in \eqref{eq:SM-HRed-OMAnsatz} could contribute to the energy. Generally, for this one should investigate the full type IIB superstring. However, since a suitable prescription of static gauge for the Green-Schwarz superstring is lacking, we follow the line of thought so far and constrain our investigation to the bosonic subsector, where again we will argue about the effects of supersymmetry heuristically. Hence, we return to \eqref{eq:SM-AdSS-EnSq&Ham}.

	For the spatial ${\rm AdS}_5$ phase space variables as well as the ${\rm S}^5$ non-zero modes we then employ the rescaling suggested in \cite{Passerini:2010xc} and discussed in \secref{sec:PS-Scale}, see also the comment beneath \figref{fig:PulsatingString}. However, we do not rescale the ${\rm S}^5$ zero modes. Hence, the rescaling of the spatial ${\rm AdS}_5$ and ${\rm S}^5$ phase space variables reads
	\begin{align}
		\vec{X}(\tau,\s) &\mapsto \frac{\vec{x}(\tau,\s)}{\l^{1/4}} + \frac{\vec{x}_0 (\tau)}{\l^{1/8}} ~,& \quad
		\vec{P}_X(\tau,\s) &\mapsto \l^{1/4} \vec{p}_x(\tau,\s) + \l^{1/8} \vec{p}_{x,0} (\tau) ~,&
		\label{eq:SM-DoOM-RescX} \\
		\vec{Y}(\tau,\s) &\mapsto \frac{\vec{y}(\tau,\s)}{\l^{-1/4} } + \vec{y}_0 (\tau) ~,& \quad
		\vec{P}_Y(\tau,\s) &\mapsto \l^{1/4} \vec{p}_y(\tau,\s) + \vec{p}_{y,0} (\tau) ~,&
		\label{eq:SM-DoOM-RescY}
	\end{align}
	with $\vx(\tau,\s)$, $\vpx(\tau,\s)$, $\vec{y}(\tau,\s)$ and $\vec{p}_y(\tau,\s)$ comprising the respective ${\rm AdS}_5$ and ${\rm S}^5$ non-zero modes.\footnote{In comparison to \eqref{eq:PS-XPmodeExp} and \eqref{eq:PS-Resc} we omitted marking the non-zero parts by the subscript `${0\hspace{-.51em}/}$'.} We note once more, that by this the
	expansion in powers of phase space variables is {\it
 not} the same as expansion in 't Hooft coupling $\l$. The Hamiltonian density (c.f. \eqref{eq:SM-SSST-EnSquared}) then expands as 
	\begin{align} \label{eq:SM-DoOM-HamDens}
		&\calH 
		=  
			\calH_{X^2,\nz} + \f{\l^{-1/4}}{2} \bigg\{\vpxz^{\,2} + 2\calH_{X^2,\nz}\,\vxz^{\,2} + (\vpx\cdot\vxz)^2 - (\vxpr \cdot\vxz)^2 \\
		& \qqquad\qquad\ \   
		 + \frac{4\,\vy\cdot\vyz}{(1+\vyz^2)}
				\left(\f{(1+\vyz^{\,2})^2}{4} \vpy^{\,2}  - \f{4}{(1+\vyz^{\,2})^2} \vypr^{2} \right)\bigg\}
			\,+\, \ord{\l^{-3/8}}~,\nn\\
		&\calH_{X^2,\nz} =
			\f{1}{2}\left(\vpx^{\,2} + \vxpr^2 + \f{(1+\vyz^{\,2})^2}{4} \vpy^{\,2} + \f{4}{(1+\vyz^{\,2})^2} \vypr^{2}\right)
	\end{align}
	where we dropped terms linear in non-zero modes, e.g., the term $\l^{-1/8}\, \vpx\cdot\vpxz$,  as they obviously do not appear in the Hamiltonian $H = \f{E^2}{2\sqrt{\l}} = \int\!\sf{\dd \s}{2\pi}\calH$.
Nevertheless, \eqref{eq:SM-DoOM-HamDens} looks alarming since at $\ord{\l^{-1/4}}$ in $H$ there will be several operators potentially contributing via second order perturbation theory to $\ord{\l^{-1/4}}$ in the energy $E$. Also, with the single-mode ansatz being a particle on ${\rm S}^5$, the appearance of $\vyz$ at leading order but no $\vpyz$ seems unsatisfactory.

	These problems are overcome by a restricted canonical transformation to a new set of phase space variables, $(\vcpx, \vcx, \vcpxz, \vcxz, \vcpy, \vcy, \vcpyz, \vcyz)$, generated by the following type II generating functional,
	\be \label{eq:SM-DoOM-GenFunc}
		F = \int \frac{\dd \s}{2 \pi}\left( \vcpxz\cdot\vxz + \vcpx\cdot\vx + \k\,(\vcpx\cdot\vxz)(\vxz\cdot\vx)
				+ \vcpyz\cdot\vyz + \f{2}{1+\vyz^{\,2}} \vcpy\cdot\vy \right)~,
	\ee
	with $\k \equiv \f{1}{2\,\l^{1/4}}(1-\f{3}{4\,\l^{1/4}}\vxz^2)$, yielding $\vcxz = \vxz$, $\vcyz = \vyz$ and
	\bea
		&\vcx = \vx + \k (\vx\cdot\vxz) \vxz\ \qquad \Leftrightarrow\qquad\
			\vx = \f{\vcx + \k (\vcxz^{\,2}\,\vcx - (\vcx\cdot\vcxz) \vcxz)}{1+\k\,\vcxz^{\,2}}~,& \nn\\
		{\rm AdS}_5 :\quad&\vpx = \vcpx + \k (\vcpx\cdot\vcxz) \vcxz \qquad\Leftrightarrow\qquad
			\vcpx = \f{\vpx + \k (\vxz^{\,2}\,\vpx - (\vpx\cdot\vxz) \vxz)}{1+\k\,\vxz^{\,2}}~,& \\
		&\vpxz = \vcpxz + \int \frac{\dd \s}{2 \pi} \left(\k\left(\vcpx(\vxz\cdot\vx)+(\vcpx\cdot\vxz)\vx\right)
			+ \f{\p \k}{\p \vxz} (\vcpx\cdot\vxz)(\vxz\cdot\vx)\right)~,& \nn\\[.3em]
		{\rm S}^5:\ \ \quad&\vy = \f{1+\vcyz^{\,2}}{2} \vcy~,\quad~ \vpy = \f{2}{1+\vcyz^{\,2}} \vcpy~,
			\quad~ \vpyz = \vcpyz - \f{2\,\vcyz}{(1+\vyz^{\,2})} \int\!\frac{\dd \s}{2 \pi}(\vcpy\cdot\vcy) ~.&
	\eea
	Notice that the new phase space variables differ from the old ones only by terms vanishing in our single-mode approximation {\eqref{eq:SM-HRed-ComplConstr}, in particular $\vcpx\cdot\vcxz \propto \vpx\cdot\vxz = \vx \cdot \vxz = \vpy = \vy =0$}. Hence, the single-mode solution does not feel the canonical transformation and our result \eqref{eq:SM-QNZM-EnSpectrum} persists.

	As the  canonical transformation \eqref{eq:SM-DoOM-GenFunc} is restricted, the new Hamiltonian density is just the old one expressed in the new phase space variables, and we find
	\begin{align}\label{eq:SM-DoOM-HamDens2}
		&\calH 
			=  
				\calH_{X^2,\nz} + \calH_{X^2,0} + \l^{-1/4} (\vcyz\cdot\vcy) \big(\vcpy^{\,2} - \vcypr^2\big) + \Ord{\l^{-3/8}}~,\\[.3em]
		\label{eq:SM-DoOM-HamDens2nz}
		&\calH_{X^2,\nz} = \frac{1}{2}\left(\vcpx^{\,2} + \vcxpr^2 + \vcpy^{\,2} + \vcypr^{2}\right)~,\\
		\label{eq:SM-DoOM-HamDens2z}
		&\calH_{X^2,0} = \frac{\l^{-1/4}}{2} \left(\vcpxz^{\,2}+ 2\calH_{X^2,\nz}\, \vcxz^{\,2} \right)~,
	\end{align}
	where again we ignored terms linear in non-zero modes.
	Now, in contrast to \eqref{eq:SM-DoOM-HamDens}, the leading term \eqref{eq:SM-DoOM-HamDens2nz} is quadratic in non-zero mode phase space variables only, explaining the notation $\calH_{X^2,\nz}$, and gives nothing but the leading flat space limit {\it without} the zero modes,
	which feel the curvature of the background even at leading order.

	The terms in \eqref{eq:SM-DoOM-HamDens2z} are exactly the diagonal terms considered before determining the ${\rm AdS}_5$ zero modes to be harmonic oscillators in leading order, where for our solution $\vev{\vcpy^{\,2} + \vcypr^2}=0$. In particular, we cast off the off-diagonal terms in \eqref{eq:SM-DoOM-HamDens}.\footnote{This illustrates that the ${\rm AdS}_5$ part of the canonical transformation presented here is the analog of the unitary transformation found in \cite{Passerini:2010xc}.}

	The remaining ${\rm S}^5$ term, the last term in \eqref{eq:SM-DoOM-HamDens2}, has an odd power in non-zero mode phase space variables and {\it will} generally contribute to $\ord{\l^{-1/4}}$ in the energy via second order perturbation theory. 
	However, effectively, this term will only give a contribution to the $\vcyz$ mass term at $\ord{\l^{-1/2}}$ in $H$, which for the ${\rm S}^5$ non-zero modes in the ground state is expected to vanish due to supersymmetry. Hence, from this term and further ${\rm S}^5$ terms at $\ord{\l^{-1/2}}$ in the Hamiltonian effectively only the massless particle on ${\rm S}^5$ should survive.

	Therefore, along the line of thought discussed in \secref{sec:PS-PertTheo}, we split the Hamiltonian into an unperturbed Hamiltonian $H_0$ and a perturbation $\d H$ as 
	\begin{align}
		&H = H_0 + \l^{-1/4}\,\d H = H_0 + \l^{-1/4}\,\d H_{-1/4} + \l^{-3/8}\,\d H_{-3/8}
			+ \ldots ~,\\
		&H_0 = \int \f{\dd \s}{2 \pi} 
			\calH_{X^2,\nz} + \calH_{X^2,0}~,\\
		&\d H_{-1/4} = \int \f{\dd \s}{2 \pi} \left\{(\vcyz\cdot\vcy) \big(\vcpy^{\,2} - \vcypr^2\big)
			\right\}~, \label{eq:SM-DoOM-dH14}\\
		&\d H_{-3/8} = \int \f{\dd \s}{4 \pi} \Big\{\vcpx\cdot\vcxz\left(\vcpx\cdot\vcx + \vcpxz\cdot\vcxz\right)
			+ \vcx\cdot\vcxz \left(\vcpx^{\,2} + 2\,\vcxpr^2 \right) - 2(\vcx\cdot\vcxpr)(\vcxpr\cdot\vcxz)\Big\}~,
			\label{eq:SM-DoOM-dH38}\\
		&\d H_{-1/2} = \int \f{\dd \s}{4 \pi} \Big\{ \vcx^{\,2}\left(\vcpx^{\,2} + \vcxpr^2\right)
			+ \left(\vcpx\cdot \vcx+\vcpxz\cdot \vcxz\right)^2 + \vcpxz^{\,2}\,\vcxz^{\,2} - (\vcx\cdot \vcxpr)^2
			\label{eq:SM-DoOM-dH12}\\
		& \qquad\qquad\qquad\quad\quad + (\vcpx\cdot\vcxz)(\vcpxz\cdot\vcx) + 3 (\vcpx\cdot\vcpxz)(\vcxz\cdot\vcx)
		   \,+\, \text{\{${\rm S}^5$ contributions\}}\Big\}~, \nn
	\end{align}
	where by the previous argument we did not spell out the plethora of ${\rm S}^5$ terms in $\d H_{-1/2}$.
	
	Now the investigation of perturbative correction works similarly to \secref{sec:PS-Scale}. In analogy to \eqref{eq:PS-UnpVirConstrEWeq}, let us denote the eigenvalues of $H$ ($H_0$) of eigenstates $\ket{\Psi}$ ($\ket{\Psi^{(0)}}$) as $\calE_\Psi=\f{E^2}{2\sqrt{\l}}$ ($\calE^{(0)}_\Psi$).

	The terms in the first line of \eqref{eq:SM-DoOM-dH12} then directly correspond to the terms present in \eqref{eq:SM-HRed-OMEnSquared} when plugging in the single mode ansatz \eqref{eq:SM-HRed-OMAnsatz}, where $\vcx\cdot\vcxpr = 0$ followed from the Virasoro constraints. On the other hand, the ${\rm AdS}_5$ operators in the second line are expected to give no contribution to the energy at $\ord{\l^{-1/4}}$, since they are off-diagonal plus potentially a normal ordering constant, which ought to be canceled by supersymmetry.

	What is left is to discuss potential contribution from $\d H_{-3/8}$ \eqref{eq:SM-DoOM-dH38}. Since the operators have an odd power in non-zero modes they will not contribute via first order perturbation theory. In second order perturbation theory \eqref{eq:SM-DoOM-dH38} contributes as
	\be
		\calE^{(2)}_{\Psi} \supset \sum_{\ket{\Phi^{(0)}} \neq \ket{\Psi^{(0)}}}
			\f{\big|\bra{\Psi^{(0)}}\l^{-3/8} \d H_{-3/8} \ket{\Phi^{(0)}}\big|^2}{\calE^{(0)}_{\Psi} -
		\calE^{(0)}_{\Phi}} ~. \label{eq:DoOm_2ndOrdPert}
	\ee
	For states $\ket{\Psi^{(0)}}$ with different level than $\ket{\Psi^{(0)}}$ one has $\calE^{(0)}_{\Psi} - \calE^{(0)}_{\Phi} \propto \l^0$ and the second order contribution will be of order $\ord{\l^{-3/4}}$. However, for states $\ket{\Phi^{(0)}}$ with the same level as $\ket{\Psi^{(0)}}$, which hence only differ in the zero mode state\footnote{Note that states $\ket{\Phi^{(0)}}$ with same level {\it and} number of zero mode excitations as $\ket{\Psi^{(0)}}$ do not couple in \eqref{eq:DoOm_2ndOrdPert}.}, one has $\calE^{(0)}_{\Psi} - \calE^{(0)}_{\Phi} \propto \l^{-1/4}$ such that the second order contribution to $\calE$ is indeed of order $\ord{\l^{-1/2}}$. By this, for general states $\ket{\Psi^{(0)}}$ other modes giving rise to states with the same level do not decouple and will contribute to the energy at $\ord{\l^{-1/4}}$ via \eqref{eq:DoOm_2ndOrdPert}.

	However, acting with $\d H_{-3/8}$ on lowest excited non-zero mode states  $\ket{\Psi^{(0)}}=\alpha^i_{-1} \tilde{\alpha}^j_{-1} \ket{x_{0},p_{0}}$, i.e., on states dual to the Konishi supermultiplet, inevitably changes the level and \eqref{eq:DoOm_2ndOrdPert} does not contribute to the energy at $\ord{\l^{-1/4}}$. Hence, we conclude that our previous results should be trusted for the Konishi state, $n=m=1$, only. {A similar observation has been made} in appendix A.1 of \cite{Passerini:2010xc}.

\section{Comparison with Pure Spinor Calculation} \label{sec:SM-Comp}
	Finally, it seems congenial to compare the case at hand with survey of the pure spinor superstring setup in \chapref{chap:PureSpinor}.

	Recall again that we used coordinates in which the $\text{AdS}_5$ time is cyclic. By this, static gauge immediately yields, that we are calculating the energy squared operator $E^2 = 2\sqrt{\l} H$ and one omits both problems noted in \secref{sec:PS-Energy} at the same time: By use of the Virasoro constraint we are determining the energy of the string $E$, not only the classical energy $E_\cl$ of the point-particle solution, and the energy momentum tensor actually concurs with the Hamiltonian. Equivalently, one can note that we omitted the non-restrictive canonical transformation \eqref{eq:PS-GenFunc}, which in retrospect seems to be incongruous when planing to work in first order formalism. Similarly, we note that imposing static gauge corresponds to cancelling any longitudinal fluctuations.



	Furthermore, in \secref{sec:PS-Scale} the perturbative expension of the Virasoro constraint was investigated and we found an quartic operators with one coordinate $\vX$ taking a zero mode \eqref{eq:PS-VM4corrOrd2}, which could contribute to $\ord{\l^{-1/4}}$ in $E$ via second order perturbation theory. This operator reassembles  the operator \eqref{eq:DoOm_2ndOrdPert} and we established that for the Konishi dual states they in fact do not contribute.

	However, in \secref{sec:PS-Scale} we were also facing cubic operator which for one coordinate taking a zero mode \eqref{eq:PS-VM4corrOrd2} gave rise to second order contributions at the same order. As these depend on two non-zero modes they do not have to change the level  even when acting on dual Konishi states. Hence, in principle we have to expect that second order perturbations of these indeed alter the result \eqref{eq:PS-VMresults2}. In contrast to this, due to the canonical transformation \eqref{eq:SM-DoOM-GenFunc} we end up with \eqref{eq:SM-DoOM-dH38} and we observe that we explicitly got rid of any dangerous operator quadratic in non-zero modes.

	We would like to conclude with an observation. The constitutive idea in \chapref{chap:PureSpinor} was to expand the string around a classical solution using the background field approximation, which in particular leads to the formulas as \eqref{eq:PS-CurrBgPert} and \eqref{eq:PS-CurrBgPert2}. The fact that the spatial ${\rm AdS}_5$ zero modes obtain a mass determined by the non-zero mode excitations now actually prompts a different picture, namely that the non-zero modes distort the space probed by the zero modes. This suggests that the spatial ${\rm AdS}_5$ zero modes ought to be expanded in the background defined by the non-zero modes, which schematically is equivalent to choosing an ${\rm AdS}_5$ coset representative of the form
	\be
		g_\alg{a} = e^{t \S_0} g_\nz\,g_0 = e^{t \S_0} e^{\vX_{{0\!\!\!/}} \cdot \v\S_X} e^{\vX_0 \cdot \v\S_X}~,
	\ee
	where as usual $\vX_{\nz}$ and $\vX_0$ comprise the respective non-zero and zero modes. From \eqref{eq:SM-DoOM-GenFunc} we actually see that something similar happens for the ${\rm S}^5$ modes, as for the non-zero modes the the canonical transformation strips of the zero mode behavior.
	
	We also wonder, whether this point of view is connected to the program of quantizing the orbits of semiclassical string solutions, see \chapref{chap:AdS3}.
	

\chapter{On the {\AdSxSheader} Superstring in Static Gauge} \label{chap:StaticFermions}
	Encouraged by the study \cite{Frolov:2013lva}, see \chapref{chap:SingleMode}, and also with respect to the calculations in \chapref{chap:PureSpinor}, which both have been purely bosonical, it seems pressing to extend the findings to the full supersymmetric case, where one is tempted to investigate the prospects of quantization of the full $\AdSxS$ superstring in static gauge. However, neither quantization of the {\AdSxS} superparticle \cite{Metsaev:1999gz} nor static gauge \cite{Jorjadze:2012iy} for the Green-Schwarz superstring are properly understood, which seem to be prerequisite.

	Nevertheless, one can also take the approach to go ahead and try to apply static gauge to the $\AdSxS$ superstring and then follow along the lines of earlier works. Of course, the hope is then to get an intuition for the arising issues and maybe even solve some of these in analogy to the previous results.

	Therefore, in this chapter we are going to report on ongoing work \cite{privateFHJP2014}, where we followed this logic. In \secref{sec:SF-SString} we are establishing the framework of static gauge for the $\AdSxS$ superstring. Our setup closely follow along the lines of \cite{Frolov:2006cc}, where the superstring was investigated using uniform light-cone gauge. Especially, with a coset representative $g=\L(t) g_\alg{f} g_\alg{b}$, with $g_\alg{f}$ and $g_\alg{b}$ comprising the fermionic and spatial bosonic coordinates, we derive a nice representation of the Lagrangian in terms of the matrix $G_\alg{b} = g_\alg{b} \cK g_\alg{b}^T$, which solely depends on the coset degrees of freedom, as was first observed in \cite{Alday:2005ww}. Furthermore, also the supermatrix of $\psu(2,2|4)$ conserved charges takes an intriguingly simple form.

	In section \secref{sec:SF-SingleMode} we then explore the possibility of a supersymmetric generalization of the bosonic single-mode string \cite{Frolov:2013lva}. We argue for a possible cutting of the modes, which though seems to correspond to a single-mode string on ${\rm S}^5$ rather than ${\rm AdS}_5$.

	Not surprisingly, the investigation in \secref{sec:SF-SString}$\,$  points out that to quantize the $\AdSxS$ superstring perturbatively one first has to understand the scaling and quantization of the zero modes. Moreover, in the case of the bosonic single-mode string we relied heavily on understanding of the massive particle in ${\rm AdS}_5$ \cite{Dorn:2005vh, Dorn:2010wt, Jorjadze:2012jk} and similarly the attempt to find its supersymmetric generalization discussed in \secref{sec:SF-SingleMode} stresses that ultimately understanding of the $\AdSxS$ superparticle will be required.

	Accordingly, in \secref{sec:SF-SParticle} we restrict our interest on the $\AdSxS$ superparticle. Contrary to investigation of the wave function as in \cite{Metsaev:1999gz}, our aim is to quantize canonically. For this we have to find the Poisson structure of the fermions, where we manage to directly diagonalize the kinetic term quadratic in fermions.

	We also follow another route to determine the Poisson structure. For this we start by assuming to have the bosonic and fermionic degrees of freedom in canonical form and then construct the $\psu(2,2|4)$ generators by requiring consistency of the algebra. Comparison with the generators derived from the initial Lagrangian density should then determine a canonical transformation from the initial to the canonical variables.

	The $\psu(2,2|4)$ charges are expanded in fermions $\th$, where the order $\th^0$ are just the bosonic charges stated in \chapref{chap:SingleMode}. We propose an ansatz for the fermionic charges linear in fermions, which is then fixed completely by the symmetry algebra resulting in the coefficients having a nice functional dependence on the $\un(1) \oplus \su(2) \oplus \su(2) \oplus \su(4)$ scalars, the energy squared $E$ as well as Casimir numbers of $\su(2)\oplus\su(2)$ and $\su(4)$, $S^2$ and $M_{\rm S}^2$, only.

\section{The {\AdSxSheader} Superstring in Static Gauge} \label{sec:SF-SString}
	I this section we apply the static gauge to the $\AdSxS$ superstring. Our setup follows the work \cite{Frolov:2006cc}, where a similar setup was conducted using uniform light-cone gauge.

	We start in \ssecref{subsec:SF-SS-CosetParam} by specifying our coset parametrization. In \ssecref{subsec:SF-SS-FirstOrd} transit to the Hamiltonian description, where we also discuss a $\k$-gauge choice which seems suitable for static gauge. In \ssecref{subsec:SF-SS-DensGb} we reexpress the densities in terms of a different bosonic matrix to then comment on the form of the symmetry algebra in \ssecref{subsec:SF-SS-SymOps}.

 \subsection{Coset Parametrization} \label{subsec:SF-SS-CosetParam}
	In contrast to \chapref{chap:SS} and \chapref{chap:PureSpinor}, cf. \eqref{eq:SS-AdSxSgens}, in this chapter for the bosonic $\AdSxS$ generators we use the normalization
	\be \label{eq:SF-SS-Sigmas} 
		\S_0 = \begin{pmatrix} \S & 0 \\ 0 & 0 \end{pmatrix},\qquad
		\S_k = \begin{pmatrix} \g_k & 0 \\ 0 & 0 \end{pmatrix},\qquad
		\S_4+j = \begin{pmatrix} 0 & 0 \\ 0 & i \g_k \end{pmatrix},
	\ee
	with $\S = \g_5$ defined in \eqref{eq:SS-5dgammas2}, hence $\,\str(\S_\m \S_\n)=4\,\d_{\m\n}\,$.

	To generalize the discussion of the static gauge to the Green-Schwarz superstring in $\AdSxS$, we are using the following parametrization  of the coset element
	\be \label{eq:SF-SS-coset}
	g=\Lambda(t)\, g_{\alg{f}}\, g_{\alg{b}}~.
	\ee
	Here, $\L(t)$ and $g_\alg{b}$ comprise the bosonic $\AdSxS$ coordinates,
	\be \label{eq:SF-SS-Lambdagb}
		\L(t) = e^{\sfrac{i}{2} t \S_0} = 
		\begin{pmatrix} \sfrac{i}{2} t \g_5 & 0 \\ 0 & 0 \end{pmatrix},\qquad\qquad
		g_\alg{b} = \begin{pmatrix} g_\alg{a} & 0 \\ 0 & g_\alg{s} \end{pmatrix},
	\ee
	and the blocks $g_\alg{a}$ and $g_\alg{s}$ corresponding to the spatial ${\rm AdS}_5$ and ${\rm S}^5$ directions given by
	\be \label{eq:SF-SS-gags}
		g_{\alg{a}}=\frac{\mI+\sfrac{1}{2}\vz \cdot \vg}{\sqrt{1-\vz^2/4}}~,\qquad\qquad
			g_{\alg{s}} = \frac{(1+Y^6) \mI_4 + i\,\vY\cdot\vg}{\sqrt{2(1+Y^6)}} ~,
	\ee
	and $\vz\cdot\g = \sum^4_{k=1} z^k \g_k$ while $\vY\cdot\g = \sum^5_{j=1} Y^j \g_j$ and $\vY^2 = \sum^5_{j=1} (Y^j)^2 = 1 - (Y^6)^2$.
	
	The matrix $g_\alg{f}$ incorporates the fermions and is given by
	\be \label{eq:SF-SS-gf}
		g_{\alg{f}}=g(\chi)= \chi + \sqrt{1+\chi^2}~,\qquad\qquad \chi = \begin{pmatrix} 0 & \Th \\ - \Th^\dag \S & 0 \end{pmatrix},
	\ee
	where the form of $\chi$ is fixed by the Cartan involution \eqref{eq:SS-RealCond} and {\it a priori} the $4\times4$ block $\Th$ can have arbitrary complex Grassmann-valued entries.

	The coset representative has been chosen such that all fields including the fermions are neutral under the $\Un(1)$ isometry generated by shifts of the ${\rm AdS}_5$ time $t$, which clearly seems favorable when taking static gauge.

	Obviously, $\L(t)^{-1} = \L(-t)$, and the inverse of \eqref{eq:SF-SS-gags} and \eqref{eq:SF-SS-gf} are easily seen to be
	\be \label{eq:SF-SS-gagsgfInv}
		g_{\alg{a}}^{-1} = \frac{\mI - \sfrac{1}{2}\vz \cdot \vg}{\sqrt{1-\vz^2/4}}~,\quad~
			g_{\alg{s}}^{-1} = \frac{(1+Y^6) \mI_4 - i\,\vY\cdot\vg}{\sqrt{2(1+Y^6)}} ~,\quad~
			g_{\alg{f}}^{-1} = -\chi + \sqrt{1+\chi^2}~,
	\ee
	The matrix $g_\alg{a}$ fulfills the important identity
	\be
		\label{eq:SF-SS-gagaS}
		\S\,g_{\alg{a}}^{-1} = g_{\alg{a}}\,\S~,\qquad\qquad \S\,g_{\alg{a}} = g_{\alg{a}}^{-1}\,\S~,
	\ee
	which follows readily from the $\Reals^{1,4}$ Clifford algebra.
	The spatial ${\rm AdS}_5$ coordinates $\vz$ are connected to the ones used in \chapref{chap:SingleMode} as
	\be \label{eq:SF-SS-AdScoordTrafo}
		\vz=\frac{2 \vX}{1+\sqrt{1+\vX^2}}~,\qquad\quad
			g_{\alg{a}}^2 = \sqrt{\vX^2+1}\,\mI_4 + \vX\cdot\vg~.
	\ee

	In the following, in case there is no danger of confusion, we are also using the letters $g_\alg{a}$ and $g_\alg{s}$ to refer to the $8\times8$ matrices $\diag(g_\alg{a},0)$ and $\diag(0,g_\alg{s})$, respectively. Note that this corresponds rather to a Lie-algebra type embedding of the $4 \times 4$ matrices and with this convention we have $g_\alg{b} = g_\alg{a} + g_\alg{b}$.

 \subsection{First Order Formalism and Gauge Fixing} \label{subsec:SF-SS-FirstOrd}
	The Lagrangian density of $\AdSxS$ GS superstring was given in \eqref{eq:SS-AdSxSlag}, with the action being $S=\int d\s \cL$. With view on \chapref{chap:SingleMode} and the cluttering with factors of $2\pi$ in \chapref{chap:PureSpinor} it seems convenient to include a factor of $1/2\pi$ into the measure, $S=\int {d\s\ov 2\pi} \cL$, and work with $\sqrt{\l}$ instead of $\pzg=\sqrt{\l}/2\pi$ \eqref{eq:SS-effST}. By this the Lagrangian density reads
	\be \label{eq:SF-SS-AdSxSlag} 
		\cL = -\frac{\sqrt{\l}}{2}\Big[
		\g^{\a\b} \str\big(A^{(2)}_{\a}A^{(2)}_{\b}\big)+
		\kappa\epsilon^{\a\beta} \str\big(A^{(1)}_{\a}A^{(3)}_{\beta}\big)
		\Big]~, 
	\ee 
	which is the sum of the kinetic and the Wess-Zumino term. Here, $\kappa^2=1$ is required for $\k$-symmetry, we use the convention  $\epsilon^{\tau\sigma}=1$, and $\gamma^{\a\b}= h^{\a\b} \sqrt {-h}$ is the Weyl-invariant combination
	of the world-sheet metric $h_{\a\beta}$ with $\det\gamma=-1$.
	
	To impose the static gauge in the Hamiltonian setting, one has to introduce a Lie-algebra valued auxiliary field $\bp$ and rewrite the superstring Lagrangian in the form
	\begin{eqnarray}\label{eq:SF-SS-Lang}
		\calL =  -\str \bigg( \bp A_0^{(2)} + {\sqrt{\l}\over 2} \kappa \epsilon^{\a\b}A_\a^{(1)} A_\b^{(3)}\, +\,  {\gamma^{01}\over
		\gamma^{00}} \bp  A_1^{(2)}
		- \frac{ \bp^2 + \l (A_1^{(2)})^2}{2 \sqrt{\l} \gamma^{00}} 
		\bigg)\,. 
	\end{eqnarray}

	The last two terms in \eqref{eq:SF-SS-Lang} yield the Virasoro constraints 
	\begin{align}
		\label{eq:SF-SS-C1}
		C_1 &=- \str\left(\bp A_1^{(2)}\right)=0~,\\
		\label{eq:SF-SS-C2} 
		C_2 &= \str\left( \bp^2 + g^2 (A_1^{(2)})^2 \right) =0~.
	\end{align}
	In particular, these ought to be satisfied by a fermionic generalization of the single-mode solution in \chapref{chap:SingleMode}.

	We assume without loss of generality that $\bp$ belongs to the subspace $\pzG^{(2)}$, as the other $\mathbb{Z}_4$-graded components decouple. It therefore admits the decomposition
	\be \label{eq:SF-SS-bpexp} 
		\bp = \bp^{(2)} =- {i\over 2}\bp_0 \Sigma_0 + {1\over 2}\bp_k \Sigma_k + \bp_\mI i\mI_8~ .
	\ee
	with here $k=1,\ldots,9$ and $\S_\m$ the generators of $\pzG^{(2)}$ given in \eqref{eq:SF-SS-Sigmas}.

	Since $A_\a^{(2)}$ belongs to the superalgebra $\su(2,2|4)$, $\str( A_\a^{(2)}) =0$, the quantity $\bp_\mI$ does not contribute to the Lagrangian and will be dropped from $\bp$. Note also that $\str(\bp  A_0^{(2)})=\str(\bp A_0)$ because $\bp\in \pzG^{(2)}$. Plugging \eqref{eq:SF-SS-bpexp} into \eqref{eq:SF-SS-Lang}, the equation of motion (EoM) for $\bp$ reads $\bp = \sqrt{\l} \g^{0 \b} A^{(2)}_{\b}$ and \eqref{eq:SF-SS-AdSxSlag} is easily recovered.

	Let us now make arrangements to impose the static gauge \eqref{eq:BS-StaticGauge}.
	The Virasoro constraints allows one to express $\bp_t = \bp_0$ in terms of other fields
	\be\label{eq:SF-SS-bpt}
		\bp_t^2 = \str\left( \big(\bp^\perp\big)^2 + g^2 (A_1^{(2)})^2 \right),\qquad 
		\bp^\perp =  {1\over 2}\bp_k \Sigma_k = 
			{1\over 2}\bp^\alg{a}_k \Sigma_k + {1\over 2}\bp^\alg{s}_k \Sigma_{4+k}~.
	\ee
	Note that the field $\bp_t$ does not coincide with the momentum $p_t$ canonically conjugate to $t$ but can be expressed in terms of $p_t$. To this end one first computes the current $A$,
	\be\label{eq:SF-SS-A}
		A=-g^{-1}\dd g = -{i\ov2}\dd t\,g_{\alg{b}}^{-1}g_{\alg{f}}^{-1}\S_0 g_{\alg{f}}g_{\alg{b}} -
		g_{\alg{b}}^{-1}g_{\alg{f}}^{-1}\dd g_{\alg{f}}g_{\alg{b}}- g_{\alg{b}}^{-1}\dd g_{\alg{b}}~,
	\ee
	and then one finds
	\be\label{eq:SF-SS-pa2}
		-\str(\bp  A^{(2)}) =
			\str \Big(\bp {i\ov2}g_{\alg{b}}^{-1}g_{\alg{f}}^{-1}\S_0 g_{\alg{f}}g_{\alg{b}} \Big) \dd t +\str \big( g_{\alg{b}}\bp g_{\alg{b}}^{-1}g_{\alg{f}}^{-1}\dd g_{\alg{f}}\big)+\str \big(  \bp g_{\alg{b}}^{-1} \dd g_{\alg{b}}  \big)\,.
	\ee
	It is convenient to introduce
	\begin{align}\label{eq:SF-SS-pt}
		{\bf p}_t &= \str \Big( \bp {i\ov2}g_{\alg{b}}^{-1}g_{\alg{f}}^{-1}\S_0 g_{\alg{f}}g_{\alg{b}}\Big)
		= \str \Big(\frac{\bp_t}{4}\, g_{\alg{a}}^2\S_0 g_{\alg{f}}^{-1}\S_0  g_{\alg{f}} +  {i\ov2}\, g_{\alg{b}} \bp^\perp g_{\alg{b}}^{-1} g_{\alg{f}}^{-1}\S_0 g_{\alg{f}}\Big)\,,
	\end{align}
	where we used \eqref{eq:SF-SS-gagaS}.
	Thus one can express $\bp_t$ in term of ${\bf p}_t$ and $\bp^\perp$. In the absence of fermions $\str \big( \bp^\perp g_{\alg{b}}^{-1}\S_0  g_{\alg{b}}\big)=0$ and ${\bf p}_t$ is just proportional to $\bp_t$.  Then, notice that since 
	$\str \big( \S_0  g_{\alg{b}}^{-1}\dd g_{\alg{b}}\big)=0$, the momenta $p^\perp$, receptively, $p_k$ conjugate to the spatial bosonic coordinates depend only on $\bp^\perp$. We also see that in \eqref{eq:SF-SS-pa2} the term
	\be
		\bp_t\,\str\left(g_{\alg{b}}\S_0 g_{\alg{b}}^{-1} g_{\alg{f}}^{-1} \dd g_{\alg{f}} \right)  
			= \bp_t\,\str\left( g_{\alg{a}}^2\S_0\,g_{\alg{f}}^{-1}\dd g_{\alg{f}} \right)
	\ee
	should give a non-degenerate kinetic term for fermions, reading at quadratic order 
	\be
	\f{1}{2}\bp_t\,\str(g_{\alg{a}}^2\S_0) (d\chi \chi - \chi d\chi)\,.
	\ee

	Next, we have to impose a $\k$-symmetry gauge, which should simplify \eqref{eq:SF-SS-pt} as much as possible and ought to be in compliance with the single-mode ansatz. Up to permutation of rows, the only choice keeping the $\su(4)$-invariance intact seems to be 
	\be\label{eq:SF-SS-chigf}
		\chi_{\rm gf} = \begin{pmatrix} 0 & \Th_{\rm gf} \\ - \Th_{\rm gf}^\dag \S & 0 \end{pmatrix}
		= \begin{pmatrix} 0 & \Th_{\rm gf} \\ - \Th_{\rm gf}^\dag & 0 \end{pmatrix}~,\qquad
		\Th_{\rm gf} = 
		\begin{pmatrix}
		\theta_{11} & \theta_{12} & \theta_{13} & \theta _{14} \\
		\theta_{21} & \theta_{22} & \theta_{23} & \theta _{24} \\
		0 & 0 & 0 & 0 \\
		0 & 0 & 0 & 0 
		\end{pmatrix}\,,
	\ee
	To simplify notation, in what follows we will drop the subscripts and simply denote \eqref{eq:SF-SS-chigf} as $\chi$ and $\Th$, respectively. These are then easily seen to fulfill the identities
	\be
		\S_0 \chi^{2k+1} = \chi^{2k+1} \mI_{s}~, \quad \S_0 \chi^{2k} = \chi^{2k} \mI_{a}~,\quad \chi^{2k+1} \S_0 = \mI_{s} \chi^{2k+1}~, \quad \chi^{2k}\S_0 = \mI_{a} \chi^{2k}~, \nn
	\ee
	with $\mI_\alg{a} = \diag(\mI_4,0)$ and $\mI_\alg{s}=\diag(0,\mI_4)$ defined in \eqref{eq:SS-mIas}, and we find
	\be\label{eq:SF-SS-gfStgf}
		g_{\alg{f}}^{-1}\S_0 g_{\alg{f}} = \S_0 + \Upsilon(\chi^2 +\chi\sqrt{1+\chi^2})~,
	\ee
	with the hypercharge $\Upsilon=\mI_{a}-\mI_{s}$ defined in \eqref{eq:SS-UpsMat}.

	By studying the terms of \eqref{eq:SF-SS-pt} for the different components of $\bp$ one then finds the $\kappa$-gauge fixed form of ${\bf p}_t$ to be
	\be\label{eq:SF-SS-ptgf}
		{\bf p}_t =  \bp_t\,\sqrt{1+\vX^2} \Big(1+{1\ov4}\theta^{\dag }_{ij} \theta_{ij} \Big) +  \str\left( {i\ov8}\bp^\alg{a}_k X^j [\S_j,\S_k] \chi^2 
		- \frac{i\bp_{k}^\alg{s}}{4} g_{\alg{s}}\S_{4+k} g_{\alg{s}}^{-1}\chi^2\right) .
	\ee

	Let us stress that for non-vanishing fermions ${\bf p}_t$ is not the momentum $p_t$ canonically conjugate to $t$, because the WZ term also gives a contribution. To find the contribution we decompose the current \eqref{eq:SF-SS-A} into even and odd parts
	\be
		A = A_{\rm even}+A_{\rm odd}~.
	\ee
	By using \eqref{eq:SF-SS-gfStgf} and splitting $g_{\alg{f}}^{-1} \dd g_{\alg{f}}$ in its bosonic and fermionic parts,
	\begin{align}
		& g_{\alg{f}}^{-1}\dd g_{\alg{f}}= B + F~,\nn\\
		\label{eq:SF-SS-BF}
		& B = \sqrt{1+\chi^2}\dd\sqrt{1+\chi^2} - \chi \dd\chi
			= \frac{1}{2} \left[\sqrt{1+\chi^2}, \dd \sqrt{1+\chi^2}\right] - \frac{1}{2} \left[\chi, \dd \chi\right]~,\\
		& F = \sqrt{1+\chi^2}\dd\chi - \chi \dd\sqrt{1+\chi^2}~.\nn
	\end{align}
	we get
	\be\ba\label{eq:SF-SS-Aevenodd}
	A _{\rm even} &= -g_{\alg{b}}^{-1}\Big( {i\ov 2} \dd t (\S_0+\Upsilon\chi^2)+ B
		+\dd g_{\alg{b}}g_{\alg{b}}^{-1}\Big)g_{\alg{b}}~, \\
	A_{\rm odd} &= -g_{\alg{b}}^{-1}\Big( {i\ov 2} \dd t\,\Upsilon\chi\sqrt{1+\chi^2}
	+ F\Big)g_{\alg{b}}\,.
	\ea\ee
	These formulae can be used to write the WZ term in a more explicit form
	\begin{align} \label{eq:SF-SS-WZ}
		\calL_{\rm WZ}
		&=
		-\frac{\sqrt{\l}\,\k}{4}\epsilon^{\a\beta}\,\str\left((\p_\a t) \Upsilon\chi\sqrt{1+\chi^2}\,G_{\alg b}F_{\beta}^{ st}G_{\alg b}^{-1} - i F_\a\,G_{\alg b}F_{\b}^{st}G_{\alg b}^{-1}\right)
		~, 
	\end{align}
	where we introduce the bosonic matrix field
	\bea \label{eq:SF-SS-Gb}
		&G_{\alg b}=  g_{\alg{b}}{\cal K} g_{\alg{b}}^t=
		\begin{pmatrix}
			G_{\alg{a}}& 0 \\ 
			0 & G_{\alg{s}}
		\end{pmatrix} ,&\\[.2em]
		\label{eq:SF-SS-GaGs}
		&	G_{\alg{a}}=
			\begin{pmatrix}
			0 & -\X^3 & -\X^{1*} & -\X^{2*} \\
			\X^3 & 0 & \X^2 & -\X^1\\
			\X^{1*}& -\X^2 & 0 & -\X^3 \\
			\X^{2*} & \X^1 & \X^3 & 0
			\end{pmatrix}\!,\quad 
		G_{\alg{s}}= i
			\begin{pmatrix}
			0 & -\Y^{3*} & -\Y^{1*} & -\Y^{2*} \\
			\Y^{3*} & 0 & \Y^2 & -\Y^1 \\
			\Y^{1*} & -\Y^2 & 0 & \Y^3 \\
			\Y^{2*} & \Y^1 & -\Y^3 & 0
			\end{pmatrix}\!,&\quad~~\\[.3em]
		&\text{with}\qquad\X^1 = X^1+i X^2~,\quad \X^2 = X^3+i X^4~,
			\quad \X^3 = X^5 = \sqrt{1+\vX^2}~, \qquad\qquad&\\[.2em]
		&\Y^1 = Y^1+i Y^2~,\quad \Y^2 = Y^3+i Y^4~,\quad \Y^3 = Y^5+i Y^6~.&
	\eea
	where $\vX = (X^1,\ldots,X^4)$ were the spatial ${\rm AdS}_5$ coordinates used in \chapref{chap:SingleMode}, see \eqref{eq:SM-AdSS-AdSCoords} and also \eqref{eq:SF-SS-AdScoordTrafo}, while $Y^k$ are the $\Reals^{6}$ embedding coordinates denoted as $\Y^k$ in \eqref{eq:SM-AdSS-SCoords}. In particular, $\X^3 = X^5 = \sqrt{1+\vX^2}$ is the radial component of the temporal plane in $\Reals^{2,4}$ embedding space.
	The matrices $G_{\alg{a}}$ and $G_{\alg{s}}$ belong to $\SU(2,2)$ and $\SU(4)$, respectively,
	\be
		G_{\alg{a}}^{-1} = \S G_{\alg{a}}^\dag \S~,\qquad\qquad
			G_{\alg{s}}^{-1} = G_{\alg{s}}^\dag~.
	\ee 
	As a result all the expressions below are polynomial in $\X^k$ and $\Y^k$.

	Thus, without the Virasoro constraints the Lagrangian takes the form
	\begin{align}\label{eq:SF-SS-Lang2}
		\cL =& p_t \p_\tau t + \str \bigg(\bp^\perp g_{\alg{b}}^{-1} \p_\tau g_{\alg{b}} -{i\ov2}\bp_t \, g_{\alg{a}}^2\,B_\tau +  g_{\alg{b}}\bp^\perp g_{\alg{b}}^{-1}B_\tau\bigg)\\\nonumber
		&+\frac{i \sqrt{\l}\,\k}{2}\, \str\Big(F_\tau\,G_{\alg b} F_{\s}^{st}G_{\alg b}^{-1}\Big)
			+\frac{\sqrt{\l}\,\k}{4} \p_\s t\, \str\left(\Upsilon\chi\sqrt{1+\chi^2}\,G_{\alg b} F_{\tau}^{st}G_{\alg b}^{-1}\right)\,,
	\end{align}
	where the momentum $p_t$ conjugate to $t$ reads
	\be \label{eq:SF-SS-ptcanon}
		p_t={\bf p}_t -\frac{\sqrt{\l}\,\k}{4}\, \str\Big(\Upsilon\chi\sqrt{1+\chi^2}\,G_{\alg b}F_{\s}^{\rm st}G_{\alg b}^{-1}\Big)\,.
	\ee

	If we now fix the static gauge \eqref{eq:BS-StaticGauge},
	\be
		p_t=-E~,\qquad\qquad t=-\frac{p_t}{\sqrt{\l}} \tau = \frac{E}{\sqrt{\l}} \tau~,
	\ee
	the Lagrangian density takes the form
	\be\label{eq:SF-SS-Lstat}
		\cL_{\rm stat} =\str\bigg(\bp^\perp g_{\alg{b}}^{-1} \p_\tau g_{\alg{b}} -\frac{i \bp_t}{2} \, g_{\alg{a}}^2\,B_\tau 
			+  g_{\alg{b}}\bp^\perp  g_{\alg{b}}^{-1}B_\tau
			+ \frac{i \sqrt{\l}\k}{2}F_\tau\,G_{\alg b}F_{\s}^{st}G_{\alg b}^{-1}\bigg) - \frac{p_t^2}{2\sqrt{\l}}\,,
	\ee
	and therefore the static gauge Hamiltonian indeed equals $E^2/2\sqrt{\l}\,$. The fermion dependent WZ terms always contain a time derivative of a fermion fields and therefore modify  
	the Poisson brackets, but they do not contribute to the Hamiltonian. This is an advantage of the Lagrangian density in static gauge compared to the one in light-cone gauge. It will be necessary to redefine the fields to bring the kinetic terms to the canonical form.

	A rather unpleasant fact is that $p_t$ has the form
	\be
		p_t = b\, \bp_t + f~,
	\ee
	such that when substituting $\bp_t$ the Hamiltonian itself depends on $E=-p_t$. By this $E^2$ computed from $E^2=2gH$ is at the same time a function of itself, $E^2 = f(E)$ and the actual value of $E$ is given as a solution of this equation. This is similar to the case encountered in light-cone gauge, where one has to solve $E-J=f(E+J)$. It is furthermore not surprising, as even in the bosonic case we found that the zero mode mass was essentially the energy at leading order.

 \subsection{Densities in Terms of \texorpdfstring{$G_{\alg{b}}$}{Gb}} \label{subsec:SF-SS-DensGb}
	The fact that the WZ term is naturally expressed in terms of the matrix $G_\alg{b}$ instead of $g_\alg{b}$ suggest to express the whole Lagrangian density in terms of $G_{\alg{b}}$ and the matrix $P_\alg{b}$ of momenta canonically conjugate to $G_\alg{b}$. The importance of this matrix was acknowledged in \cite{Alday:2005ww}, see also \cite{Arutyunov:2009ga}, where it was noted that $G_{\alg{b}}$ solely depends on the bosonic coset degrees of freedom. 

	Since $G_{\alg{b}}$ is skew-symmetric, so is $P_{\alg{b}}$ and in our case it is connected with the matrix of spatial momenta $\bp^\perp$ as
	\be
		\bp^\perp = 2 \,g_{\alg{b}}^{-1}G_{\alg{b}}P_{\alg{b}}g_{\alg{b}}~,
			\qquad \str \left(\bp^\perp g_{\alg{b}}^{-1}\dd g_{\alg{b}}\right)  
			= \str \left(P_{\alg{b}}\,\dd G_{\alg{b}}\right)\,.
	\ee
	By use the of identities
	\be \label{eq:SF-SS-gSqGK}
		g_{\alg{a}}^{2} = - G_{\alg{a}} \Kk~,\qquad\qquad g_{\alg{s}}^{2} = - G_{\alg{s}} \Kk~,
	\ee
	one finds for the Lagrangian density in static gauge 
	\eqref{eq:SF-SS-Lstat}
	\be\label{eq:SF-SS-Lstat2}
		\cL_{\rm stat} 
			= \str \left(\!
			P_{\alg{b}}\,\p_\tau G_\alg{b} + {i \bp_t \ov2} \,G_{\alg{a}} \Kk B_\tau + 2G_{\alg{b}}P_{\alg{b}}B_\tau
			+\frac{i \sqrt{\l}\,\k}{2} F_\tau\,G_{\alg b}F_{\s}^{\rm st}G_{\alg b}^{-1}\right) - \frac{p_t^2}{2\sqrt{\l}}~.
	\ee
	The matrix ${\bf p}_t$ can be written in the form
	\begin{align} \label{pt2}
		{\bf p}_t&= -\bp_t\,{1\ov4}\str \big(G_{\alg{a}} \Kk (1+\chi^2)\big) +  {i}\,\str \big(G_{\alg{b}}P_{\alg{b}}(\S_0+\Upsilon\chi^2)\big) \\
		&=\bp_t\,X^5\big(1 + \sfrac{1}{4}\th^\dag_{ij} \th_{ij} \big)
			+ {i}\,\str \big(G_{\alg{b}} P_{\alg{b}}\Upsilon\chi^2\big) \,,\nn
	\end{align}
	where $X^5 = \sqrt{1+\vX^2}$, and we used the ${\rm AdS}_5$ constraint
	\be
	\str (\bp^\perp \mI_\alg{a}) = 2\,\tr( G_{\alg{a}}P_{\alg{a}} )= 0~.
	\ee
	The momentum conjugate to $t$ is then given by
	\be \label{eq:SF-SS-ptf}
		p_t=\bp_t\,X^5 \big(1+\sfrac{1}{4}\th^{\dag}_{ij} \th_{ij} \big) 
			+ {i}\,\str \big(G_{\alg{b}}P_{\alg{b}}\Upsilon\chi^2\big) 
			-\frac{\sqrt{\l}\,\k}{4}\, \str\big(\Upsilon\chi\sqrt{1+\chi^2}\, G_{\alg b}F_{\s}^{\rm st}G_{\alg b}^{-1}\big)\,. 
	\ee

	Also the Virasoro constraints can be expressed in terms of $G_{\alg{b}}$ and $P_{\alg{b}}$. For this, note that in static gauge we have for the current
	\be
		A^{(2)}_\s=
		-{1\ov2}\, g_\alg{b}^{-1}\big(B_\s+G_\alg{b} B_\s^t G_\alg{b}^{-1} + \p_\s G_\alg{b} G_\alg{b}^{-1}\big) g_\alg{b}~ ,
	\ee
	thus \eqref{eq:SF-SS-C1} and \eqref{eq:SF-SS-C2} become
	\begin{align} \label{eq:SF-SS-C1b}
		C_1 =& 
			\str\Big(P_{\alg{b}}
		\p_\s G_{\alg{b}}+{i\ov2}\bp_t\,G_{\alg{a}} \Kk B_\s+2G_{\alg{b}}P_{\alg{b}}B_\s \Big) = 0~,\\
		\label{eq:SF-SS-C2b} 
		C_2 
		=&-\bp_t^2 + \str\left( 4\big(G_{\alg{b}}P_{\alg{b}}\big)^2 
			+ {\l\ov 4}\big(B_\s+G_{\alg{b}}B_\s^tG_{\alg{b}}^{-1}+
		\p_\s G_{\alg{b}}G_{\alg{b}}^{-1}\big)^2 \right)
		=0 ~.
	\end{align}
	Let us finally present the explicit form of $P_{\alg{a}}$ and $P_{\alg{s}}$,
	\bea \label{eq:SF-SS-PaPaExpl}
		&P_\alg{a}=
		\begin{pmatrix}
		0 & \cP_3 & \cP_1^\alg{a} & \cP_2 \\
		-\cP_3^\alg{a} & 0 & -\cP_2^{\alg{a}*} & \cP_1^{\alg{a}*} \\
		-\cP_1^\alg{a} & \cP_2^{\alg{a}*} & 0 & \cP_3^\alg{a} \\
		-\cP_2^\alg{a} & -\cP_1^{\alg{a}*} & -\cP_3^\alg{a} & 0
		\end{pmatrix}
		\!, \quad
		P_\alg{s}= \dfrac{i}{4}
		\begin{pmatrix}
		0 & \cP_3^\alg{s} & \cP_1^\alg{s} & \cP_2^\alg{s} \\
		- \cP_3^\alg{s} & 0 & - \cP_2^{\alg{s}*} & \cP_1^{\alg{s}*} \\
		- \cP_1^\alg{s}& \cP_2^{\alg{s}*} & 0 & - \cP_3^{\alg{s}*} \\
		- \cP_2^\alg{s} & - \cP_1^{\alg{s}*} & \cP_3^{\alg{s}*} & 0
		\end{pmatrix}\!,&\quad~~\\[.4em]
		&\text{with}\qquad\qquad \cP^\alg{a}_1 = P^\alg{a}_1+i P^\alg{a}_2~,\quad \cP^\alg{a}_2 = P^\alg{a}_3+i P^\alg{a}_4~,
			\quad \cP^\alg{a}_3 = P^\alg{a}_5 ~, \qquad\qquad\qquad&\\[.2em]
		&\cP^\alg{s}_1 = P^\alg{s}_1+i P^\alg{s}_2~,\quad \cP^\alg{s}_2 = P^\alg{s}_3+i P^\alg{s}_4~,\quad \cP^\alg{s}_3 = P^\alg{s}_5+i P^\alg{s}_6~.&
	\eea
	where, in accordance with the statement beneath \eqref{eq:SF-SS-GaGs}, the components $P^\alg{a}_i$ are nothing but the momenta used in \chapref{chap:SingleMode}, $\vP=\vP_X = (P^\alg{a}_1,\ldots,P^\alg{a}_4)$, while $P^\alg{s}_k$ are the momentum variables conjugate to the $\Reals^6$ embedding coordinates for ${\rm S}^5$.
	One can check that 
	\be
		\str \left(P_{\alg{a}}\,\dd G_{\alg{a}}\right) = \sum_{i=1}^5 P_i^{\alg{a}}\,\dd X^i~,\qquad\quad
		\str \left(P_{\alg{s}}\,\dd G_{\alg{s}}\right) = \sum_{i=1}^6 P_i^{\alg{s}}\,\dd Y^i\,.
	\ee
	and the coordinates and momenta should fulfill the ${\rm AdS}_5$ and ${\rm S}^5$ constraints
	\bea\label{PXconstr}
		&(X^5)^2=1+\vX^2~,\quad X^5 \dd X^5=\vX\cdot \dd \vX~,\quad 
			\str\left(P_{\alg{a}}G_{\alg{a}}\right) = \sum_{i=1}^5 P_i^{\alg{a}}X^i=0~,&\\[.3em]
		\label{PYconstr}
		&\vY^2+(Y^6)^2=1~,\quad \vY\cdot \dd\vY+Y^6 \dd Y^6=0~,\quad 
		\str\left( P_{\alg{s}}G_{\alg{s}}\right) = \sum_{i=1}^6 P_i^{\alg{s}}\,Y^i=0~.&~~~~
	\eea

\subsection{Symmetry Algebra Operators} \label{subsec:SF-SS-SymOps}
	From \eqref{eq:SS-PSUcharges} the
supermatrix $Q$ of the $\psu(2,2|4)$ conserved charges is given by 
 \be\label{eq:SF-SS-Charges}  
		Q =\int_{-\pi}^{\pi} \dd \s\, \L\,g_\alg{f}\,g_\alg{b}
			\left(\bp +\frac{i\sqrt{\l}\,\k}{2} g_\alg{b}\Kk F_\s^{st} \Kk^{-1} g_\alg{b}^{-1} \right)
			g_\alg{b}^{-1} g_\alg{f}^{-1}\Lambda^{-1}~.
	\ee 
	This expression is very simple and  it has an important property of being explicitly  independent of the world-sheet metric.  It is easy to rewrite $Q$ in terms of $G_{\alg{b}}$ and $P_{\alg{b}}$,
	\be\label{eq:SF-SS-Charges2}  
	Q = \int_{-\pi}^{\pi} \frac{\dd\s}{2\pi}\,\L\,g_{\alg{f}} \left( 
		-\frac{i\,\bp_t}{2}\,G_{\alg{a}}\Kk \S_0
		+ 2 G_{\alg{b}}P_{\alg{b}}
		+\frac{i\sqrt{\l}\,\k}{2} G_{\alg{b}} F_\s^{st} G_{\alg{b}}^{-1} \right)
	 g_{\alg{f}}^{-1}\L^{-1}\,.
	\ee 
	From this it is eminent that the matrix $Q$ schematically has the form
	\bea\label{eq:SF-SS-Qcharges} 
		Q =\int_{-\pi}^{\pi}\frac{\dd\s}{2\pi}\, \L\,
	U\, \L^{-1}~,
	\eea
	where  $U$ does not depend on $t$. 

	As discussed beneath \eqref{eq:SS-PSUcharges}, from $Q$ one obtains the different $\psu(2,2|4)$ charges by multiply $Q$ with the corresponding $8\times 8$ matrix ${\cal M}$ and then taking the
	supertrace 
	\be \label{eq:SF-SS-QM}
		\sQ_{\cal M} = \str\, (Q{\cal M}) ~. 
	\ee 
	It is clear that the diagonal and off-diagonal $4\times 4$ blocks of ${\cal M}$ single out bosonic and fermionic charges of $\psu(2,2|4)$, respectively. In particular, one can check that $p_t$ can be obtained from $Q$ according to \eqref{eq:SS-GSenergy}, recall \eqref{eq:SS-AdSxSgens} and \eqref{eq:SF-SS-Sigmas},
	\be 
		p_t ={i\ov 2}\str\, (\sQ\,\S_0)\,.\label{ptQ}
	\ee 
	In the Hamiltonian setting the conservation laws  have the
	following form
	\be
		\frac{\dd \sQ_{\cal M}}{\dd \tau}=\frac{\p \sQ_{\cal M}}{\p \tau}+\{
		H,\sQ_{\cal M}\}=0~ .
	\ee
	Therefore, generators which do not have explicit dependence on
	$t\propto\tau$ Poisson-commute with the Hamiltonian. 
	They corresponds to matrices ${\cal M}$ which commute with $\S_0$ and the respective $\sQ_{\cal M}$ form the $\un(1)\oplus\su(2)\oplus\su(2)\oplus\su(4)$ subalgebra of $\psu(2,2|4)$. Note furthermore that also the $\k$-gauge \eqref{eq:SF-SS-chigf} was chosen such that it is invariant under this subalgebra.
	
	In particular, by this in later classical expressions one naturally encounters the Casimir numbers, $\Ms^2$ for $\su(4)=\so(6)$, the total spin squared $S^2$ for $\su(2)\oplus\su(2)=\so(4)$ and the energy squared $E^2$ for $\un(1)$.

\section{Single-Mode Superstrings} \label{sec:SF-SingleMode}
	In this section we are investigating the possibility to find a supersymmetric generalization of the bosonic single-mode solution \chapref{chap:SingleMode}. For this, we want to follow the same logic, which is, we first want to cut the modes to then impose the Virasoro constraints. Especially, the generalization of the single mode ansatz \eqref{eq:SM-HRed-OMAnsatz} should get rid of most of the non-zero modes while at the same time keeping the zero modes completely intact. The fact that we had unconstrained zero modes was crucial to recover their non-vanishing contributions to the energy as well as the $\SO(1,4)\times\SO(5)$ invariance. 

	Recall the $\k$-gauge fixing \eqref{eq:SF-SS-chigf} and let us introduce the notation
	\be
		\Th_{(\rm gf)} = \begin{pmatrix} 
			\th_1 & \th_2 & \th_3 & \th_4 \\
			\eta_1 & \eta_2 & \eta_3 & \eta_4 \\
			0 & 0 & 0 & 0 \\
			0 & 0 & 0 & 0
	\end{pmatrix} = \begin{pmatrix} \,\vth\:\, \\ \,\vet\:\, \\ \,\vec{0}\:\, \\ \,\vec{0}\:\, \end{pmatrix} ,
	\ee
	where $\vth$ and $\vet$ transform under $SU(4)$ while $(\th_i, \eta_i)$ transform under $SU(2)$.
	By this
	\begin{align}
		&\chi^2 = - \begin{pmatrix} \Th \Th^\dag & 0 \\ 0 & \Th^\dag \Th \end{pmatrix}, \qquad 
			\sqrt{\mI_8 + \chi^2} = \mI_8 + \sum^\infty_{n=1} c_n \begin{pmatrix} 
				(\Th \Th^\dag)^n & 0 \\
				0 & (\Th^\dag \Th)^n \end{pmatrix}, \\
	\label{eq:SF-SM-ThSq}
	&\text{with}\qquad\Th\,\Th^\dag = \begin{pmatrix} 
		\begin{matrix} 
			\vth \cdot \vth^{\,\dag} & \vth \cdot \vet^{\,\dag} \\
			\vet \cdot \vth^{\,\dag} & \vet \cdot \vet^{\,\dag}
		\end{matrix} & 0 \phantom{\,_{\,_{,}}} \\
		0 & 0 \phantom{\,_{\,_{,}}}
	\end{pmatrix} ,
	\qquad
	(\Th^\dag \Th)_{ij} = (\th^\dag_{i} \th_{j} + \eta^\dag_{i} \eta_{j})~,
\end{align}
	and $c_n = (-1)^n(2n)!/(1-2n)(n!)^2(4^n)$. Note that $\Th\,\Th^\dag$ and $\Th^\dag \Th$ are expressed in terms of inner ($\va^\dag \cdot \vb = \sum^4_{i=1} a^*_i b_i$) and outer $\Complex^4$-products, respectively.

	In analogy to the bosonic ansatz \eqref{eq:SM-HRed-OMAnsatz}, we now cut all but the zero and the $m$th non-zero mode of $\vth$ and $\vet$. But as we are particularly interested in the lowest excited states, dual to the Konishi supermultiplet, and as in \secref{sec:SM-Decoupling} we furthermore saw that only for these states the bosonic setting seems to be consistent, we immediately constrain ourselves to mode number $m=1$, i.e., we take the ansatz
	\be
		\vth(\s) = \vth_0 + \vth_+\,e^{i\,\s} + \vth_-\,e^{-i\,\s}~, \qquad\quad 
		\vet(\s) = \vet_0 + \vet_+\,e^{i\,\s} + \vet_-\,e^{-i\,\s}~.
	\ee
	On top of these we want to find residual constraints such that the Virasoro constraints \eqref{eq:SF-SS-C1} and \eqref{eq:SF-SS-C2} are automatically solved, where in particular we should make sure that all non-zero modes of the Virasoro constraints vanish. 

	Similarly to the observation that the bosonic scalar products ought to be independent of $\s$ one could try to impose $\p_\s (\chi^2) = 0$, that is
	\be
		\p_\s (\Th \Th^\dag) = 0\qquad\quad\text{and}\qquad\quad
			\p_\s (\Th^\dag \Th) = 0~. 
	\ee
	This requirement is however too strong. For example taking the ansatz $\vth_+ = \psi \va$ and  $\vet_+ = \psi^\dag \va$, with $\va$ a constant vector, would set the second mode of $\Th^\dag \Th$ to zero,
	\be
		(\Th^\dag \Th)_{++,ij} = (\th^\dag_{+,i} \th_{+,j} + \eta^\dag_{+,i} \eta_{+,j}) = (\psi^\dag \psi + \psi \psi^\dag) a^\dag_i a_j = 0~.
	\ee
	However, $(\Th \Th^\dag)_{++}=\psi^\dag \psi\,|\va|^2 \diag(1,-1,0,0)$ and also the first mode of $\Th^\dag \Th$ reads
	\be
		(\Th^\dag \Th)_{+,ij} = (\psi^\dag a^*_i \th_{0,j} + \th^\dag_{0,i} \psi a_j + \psi a^*_i \eta_{0,j} + \eta^\dag_{0,i} \psi^\dag a_j) ~,
	\ee
	which seems virtually impossible to cancel without constraining the zero modes.

	Nevertheless, one should try to find simple constraints cutting the number of modes of $\sqrt{\mI_8+\chi^2}$ dramatically. While by the previous we have $\p_\s (\Th^\dag \Th) \neq 0$, note that $(\Th^\dag \Th)^n = \Th^\dag (\Th \Th^\dag)^{n-1} \Th$ and imposing  $\p_\s (\Th \Th^\dag) = 0$ restricts $\sqrt{\mI_4 - \Th^\dag \Th}$ to at most second mode.

	In order to set $\p_\s (\Th \Th^\dag) = 0$ we can choose the ansatz
	\be \label{eq:SF-SM-SMfermAns}
		\vth_\pm = \th_\pm \va_\pm~,\qquad\qquad 
			\vet_\pm = \eta_\pm \vb_\pm ~,
	\ee
	with $\va_\pm, \vb_\pm \in \Complex^4$ constant 
	and $\th_\pm,\,\eta_\pm$ real Grassmann numbers. Omitting the obviously vanishing $2\times2$ blocks in $\Th \Th^\dag$, cf. \eqref{eq:SF-SM-ThSq}, the second and first mode of $\Th \Th^\dag$ then read
	\begin{align}
		\label{eq:SF-SM-TTDplpl}
		&(\Th \Th^\dag)_{++} = \begin{pmatrix} 
				0 & \th_+ \eta_+\,\va_+ \cdot \vb^{\,\dag}_+ \\
				- \th_+ \eta_+\,\vb_+ \cdot \va^{\,\dag}_+ & 0
		\end{pmatrix},\\
		\label{eq:SF-SM-TTDpl}
		&(\Th \Th^\dag)_{+} = \begin{pmatrix} 
				\th_+ (\va_+ \cdot \vth_0^{\,\dag} - \vth_0 \cdot \va_+^{\,\dag}) &
				\th_+\,\va_+ \cdot \vet^{\,\dag}_0 - \eta_+\,\vth_0 \cdot \vb^{\,\dag}_+ \\
				\eta_+\,\vb_+ \cdot \vth^{\,\dag}_0 - \th_+\,\vet_0 \cdot \va^{\,\dag}_+ &
				\eta_+(\vb_+ \cdot \vet^{\,\dag}_0 - \vet_0 \cdot \vb^{\,\dag}_+)
		\end{pmatrix},
	\end{align}
	and similar for $(\Th \Th^\dag)_{--}$ and $(\Th \Th^\dag)_{-}\,$, which are set to zero by (no sum over $\pm$)
	\be \label{eq:SF-SM-abConstr}
		\va_\pm \cdot \vb^{\,\dag}_\pm = 0~,\qquad \va_\pm \cdot\vth^{\,\dag}_0 = \va_\pm\cdot\vet^{\,\dag}_0 = 0~, \qquad \vb_\pm \cdot\vth^{\,\dag}_0 = \vb_\pm\cdot\vet^{\,\dag}_0 = 0~.
	\ee
	To set \eqref{eq:SF-SM-TTDpl} to zero one actually only needs $(\va_+ \cdot \vth_0^{\,\dag} - \vth_0 \cdot \va_+^{\,\dag}) = (\vb_+ \cdot \vet^{\,\dag}_0 - \vet_0 \cdot \vb^{\,\dag}_+) = 0$, but \eqref{eq:SF-SM-abConstr} furthermore ensures that
	\be
		(\Th' \Th^\dag)_{+} = \begin{pmatrix} 
				\th_+\,\va_+ \cdot \vth_0^{\,\dag} &
				\th_+\,\va_+ \cdot \vet^{\,\dag}_0\\
				\eta_+\,\vb_+ \cdot \vth^{\,\dag}_0 &
				\eta_+\,\vb_+ \cdot \vet^{\,\dag}_0
		\end{pmatrix} = 0
	\ee
	and hence also $(\Th' \Th^\dag)^\dag_{+} = (\Th (\Th')^\dag)_{+} = 0$, which implies that
	\be \label{eq:SF-SM-TTDprZero}
		\p_\s \left((\p^k_\s \Th) (\p^l_\s \Th)^\dag\right) = 0
	\ee
	for any $k,l\geq0$. Note that by this even $B_\a B_\b$ has at most second mode and furthermore by cyclicity of the supertrace $\str(B_\a B_\b)$ has only a zero mode. Hence, only the other terms in the Virasoro constraints \eqref{eq:SF-SS-C1b} and \eqref{eq:SF-SS-C2b}, which are linear in $B_\s$, might have non-vanishing zero zero modes and we expect these to yield further constraints.

	Employing \eqref{eq:SF-SM-abConstr} the zero mode $(\Th \Th^\dag)_{0}$
	\be
		\begin{pmatrix} 
			\vth_0\cdot\vth_0^{\,\dag} + \th_+\th_- (\va_+\cdot\va_-^{\,\dag} - \va_-\cdot \va_+^{\,\dag}) &
			\!\vth_0\cdot\vet_0^{\,\dag} + \th_+\eta_-\,\va_+\cdot\vb_-^{\,\dag} - \th_-\eta_+\,\va_-\cdot\vb_+^{\,\dag} \\
			\vet_0\cdot\vth_0^{\,\dag} + \eta_+\th_-\,\vb_+\cdot\va_-^{\,\dag} + \eta_-\th_+\,\vb_-\cdot\va_+^{\,\dag} &
			\vet_0\cdot\vet_0^{\,\dag} + \eta_+\eta_- (\vb_+\cdot\vb_-^{\,\dag} - \vb_-\cdot \vb_+^{\,\dag})
		\end{pmatrix},
	\ee
	Now there might be different interesting choices for the relations between $(\va_+,\vb_+)$ and $(\va_-,\vb_-)$. It seems tempting to define the complex Grassmann numbers
	\be
		\th = \th_+ + i \th_-~,\qquad\qquad \eta = \eta_+ + i \eta_-~,
	\ee
	such that $\th_+ \th_- = \sfrac{i}{2} \th \th^\dag$ and $\eta_+ \eta_- = \sfrac{i}{2} \eta \eta^\dag$. Taking now
	\be
		\va_- = i \va_+~,\qquad \vb_- = i \vb_+~,\qquad
			\abs{\va_\pm}^2 = \abs{\vb_\pm}^ = 1~,
	\ee
	hence $\va_\pm\cdot\vb_\pm^{\,\dag}=0$, the zero mode becomes
	\be
		(\Th\,\Th^\dag)_{0} = 
		\begin{pmatrix} 
			\vth_0\cdot\vth_0^{\,\dag} + \th \th^\dag &
			\vth_0\cdot\vet_0^{\,\dag} \\
			\vet_0\cdot\vth_0^{\,\dag} &
			\vet_0\cdot\vet_0^{\,\dag} + \eta \eta^\dag
		\end{pmatrix},
	\ee
	such that on the diagonal one effectively has the norm of the Grassmann valued $\Complex^5$ vectors $(\vth_0, \th)$ and $(\vet_0, \eta)$, while on the off-diagonal the scalar products still only incorporate the $\Complex^4$ zero mode vectors $\vth_0$ and $\vet_0$.

	Another interesting choice might be
	\be
		\va_- = \frac{i}{\sqrt{2}}(\va_+ + \vb_+)~,\qquad \vb_- = \frac{i}{2}(\vb_+ - \va_+) ~,\qquad
			\abs{\va_\pm}^2 = \abs{\vb_\pm} = \sqrt{2}~,
	\ee
	for which the zero mode becomes
	\be
		(\Th\,\Th^\dag)_{0} = 
		\begin{pmatrix} 
			\vth_0\cdot\vth_0^{\,\dag} + \th \th^\dag &
			\vth_0\cdot\vet_0^{\,\dag} + i (\th_+ \eta_- + \th_- \eta_+)\\
			\vet_0\cdot\vth_0^{\,\dag} - i (\eta_+ \th_- + \eta_- \th_+)&
			\vet_0\cdot\vet_0^{\,\dag} + \eta \eta^\dag
		\end{pmatrix},
	\ee
	i.e., we rescued the imaginary part $\th \eta^\dag$ in the off-diagonal entries. It seems though impossible to have the full $\th \eta^\dag$ and by this the $\Complex^5$ scalar product $(\vth_0,\th)\cdot(\vet,\eta)^\dag$ in the off-diagonal entries.

	There might be yet another problem with the ansatz \eqref{eq:SF-SM-SMfermAns}. The requirement \eqref{eq:SF-SM-abConstr} seems to fix the vectors $\va_\pm$ and $\vb_\pm$ completely, up to rotations and reflections in the $\Complex^2$-plane orthogonal to $\mathrm{span}(\vth_0,\vet_0)$ which keep $\va_\pm\cdot\vb_\pm^{\,\dag} = 0$. Hence, there is no freedom anymore to cancel the remaining modes in the Virasoro constraints \eqref{eq:SF-SS-C1b} and \eqref{eq:SF-SS-C1b} resulting from $\p_\s (\Th^\dag \Th) \neq 0$. 
	
	Therefore, another attractive ansatz might be
	\be
		\vth_\pm = \vet_\pm = \psi_\pm \vc_\pm~,
	\ee
	where $\psi_\pm$ are real Grassmann numbers and $\vc \in \Complex^4$ constant. By this the second modes, $(\Th \Th^\dag)_{++}$ and $(\Th \Th^\dag)_{--}$, automatically vanish and to cancel the first modes $(\Th \Th^\dag)_\pm$ one needs to impose
	\be \label{eq:SF-SM-cConstr}
		\vc_\pm \cdot \vth_0 = \vc_\pm \cdot \vet_0 = 0~.
	\ee
	which does not determine the vector $\vc_+$ completely. Hence one has has more freedom to fulfill additional constraints.
	Defining $\psi = \psi_+ + i \psi_-$ and taking $\vc_- = i \vc_+$ and $\abs{\vc_\pm }^2 =1$ the zero mode becomes
	\be
		(\Th\,\Th^\dag)_{0} = 
		\begin{pmatrix} 
			\vth_0\cdot\vth_0^{\,\dag} + \psi \psi^\dag &
			\vth_0\cdot\vet_0^{\,\dag} + \psi \psi^\dag \\
			\vet_0\cdot\vth_0^{\,\dag} + \psi \psi^\dag&
			\vet_0\cdot\vet_0^{\,\dag} + \psi \psi^\dag
		\end{pmatrix},
	\ee
	which is expressed in scalar products of the $\Complex^5$-vectors $(\vth_0, \psi)$ and $(\vet_0, \psi)$.

	On the other hand, taking $\vc_+ \cdot \vc_-^{\,\dag}=0$ one even finds that the zero mode $(\Th\,\Th^\dag)_{0}$ and hence $\Th\,\Th^\dag$ is completely independent of the non-zero mode fermions $\psi_\pm$. However, for this case the vectors $\vc_\pm$ are again completely fixed.

	For the bosonic case we found that the single-mode solution with a non-zero mode on ${\rm AdS}_5$ effectively leads to a description as an ${\rm AdS}_6 \times {\rm S^5}$ particle. The ans{\"a}tze discussed however rather lead to extension of the zero mode vectors $\vth_0$ and $\vet_0$ transforming under $\SU(4)$, which suggests that the new fermionic degree might correspond to an extension of ${\rm S^5}$ to some higher dimensional version, probably ${\rm S^6}$ or ${\rm S^7}$. One should therefore also consider the bosonic single-mode ansatz with the non-zero mode on ${\rm S}^5$ instead of ${\rm AdS}_5$.

\section{The {\AdSxSheader} Superparticle in Static Gauge} \label{sec:SF-SParticle}
	In the last section we investigated possible ans{\"a}tze for a fermionic mode cutting in order to find a supersymmetric generalization of the bosonic single-mode string, \chapref{chap:SingleMode}. Even though the work is far from complete, it is already apparent that analogously to the bosonic case thorough understanding of the respective superparticle will be of aid.

	Also, in \secref{sec:SF-SString} we presented the general setup for the $\AdSxS$ superstring in static gauge. But to quantize canonically one first has to reduce the symplectic form to the canonical one, where in particular the fermions have to be transformed. The symplectic form can readily be read off from the kinetic term, where even only the terms quadratic in fermions are highly non-trivial. To simplify the problem one could first try to get a better understanding constraining oneself to the zero modes.

	Yet another reason why to concentrate on the zero modes of the $\AdSxS$ superstring is that from the investigation of the bosonic subsector we observed the peculiar scaling of the bosonic zero modes. Therefore, we expect a non-standard scaling for the fermionic zero modes as well, which in analogy to the bosonic case might be connected to the massive $\AdSxS$ superparticle.

	For all these reasons in this section we are constraining our interest to the zero modes of the $\AdSxS$ superstring, which is, we investigate the $\AdSxS$ superparticle. Earlier accounts on this topic are \cite{Metsaev:1999kb, Metsaev:2002vr, Horigane:2009qb}\footnote{Furthermore, one should compare with the initial works on supergravity \cite{Kim:1985ez, Gunaydin:1984fk}.}.

	In contrast to these works, in which mostly the wave functions where investigated, our aim will be to quantize canonically, i.e., to rather work in terms of operators. Again, the main task will be to find a suitable transformation bringing the fermions into canonical from.

	We start in \ssecref{subsec:SF-SP-su22su4} by specifying conventions for the bosonic subalgebra $\su(2,2)\oplus\su(4)$. In \ssecref{subsec:SF-SP-LagHam} we exhibit the Lagrangian an Hamiltonian densities to then, in \ssecref{subsec:SF-SP-FermPoisStruc}, derive the Poisson structure of the fermions at quadratic order. In \ssecref{subsec:SF-SP-PoisStrucPSU} we discuss a different approach, which aims to fix the Poisson structure via the superconformal algebra $\psu(2,2|4)$.

 \subsection{The \texorpdfstring{$\su(2,2)$ and $\su(4)$}{su22su4} Algebras} \label{subsec:SF-SP-su22su4}
	We noted at the end of \secref{sec:SF-SString} that choosing static gauge leaves the subalgebra $\un(1)\oplus\su(2)\oplus\su(2) \oplus\su(4)$ intact. Because of this it shows convenient to assign indices as
	\be
		\a,\b,\g,\ldots = 1,2~,\qquad \da,\db,\dg,\ldots = 1,2~,\qquad
		A,B,C,\ldots = 1,2,3,4~,
	\ee
	such that undotted and dotted Greek letters are indices of fundamentals of the $\su(2)$'s and capital Roman letters correspond to the fundamentals of $\su(4)$. Hence, the $\k$-gauge fixed fermions have the index structure
	\be
		\Th_{(\rm gf)} = \begin{pmatrix} \th_\a{}^A \\ 0_{2\times4}\end{pmatrix}~,\qquad
		\Th_{(\rm gf)}^\dag = \Big( (\th^\dag)_A{}^\a ~\, 0_{4\times2} \Big) = \Big( \vt_A{}^\a ~\, 0_{4\times2} \Big)~.
	\ee
	Let us now discuss the form of the bosonic algebras $\su(2,2)$ and $\su(4)$, which were already stated \eqref{eq:SM-AdSS-SO24Rots} and \eqref{eq:SM-AdSS-SO6Rots} and which will prove useful in the following.

	For our purposes it seems convenient to perform a canonical transforming to a new set of phase space variables
	\be
		\vX=\frac{\vx}{\sqrt{1-\vx^2}}~,\qquad 
			\vP = \sqrt{1-\vx^2}\big(\vp - \vx\,(\vp\cdot\vx)\big)~,
	\ee
	such that at $\tau=0$ the $\so(2,4)$ charges \eqref{eq:SM-AdSS-SO24Rots} take the simplified form
	\begin{align}\label{eq:SF-SP-AdSS_SO24Rots}
		&{J}_{0'\,0} = E ~,&
		 &{J}_{a\,b}=
		p_{a}\,x^b - p_{b}\,x^a~,&\\
	\label{eq:SF-SP-AdSS_SO24Boosts2}
		&{J}_{a\,0'} =E\,x^a~,&
			&{J}_{a 0} =
		-p_a+x^a\,d~,&
	\end{align}
	with $a,b=1,2,3,4$ running over the spatial ${\rm AdS}_5$ directions,
	\be\label{eq:SF-AdSS_EnSq&Ham}
		E^2={2\sqrt\l\, H} = \frac{1}{x^2} \left( (1-x^2) d^2 +S^2 \right) +\frac{\Ms^2}{1-x^2}
		~,\qquad\quad d =(\vp\cdot \vx)~,
	\ee
	and the $\SO(4) \subset \SO(2,4)$ and $\SO(6)$ Casimir numbers $S^2$ and $\Ms^2$,
	\be
		S^2 = J_{a b} J_{a b}~,\qquad\quad 
		\Ms^2 = \sum^6_{i=1} (P^\alg{s}_i)^2 = (1+\v{Y}^2/4)^2\,\v{P}_Y^{\,2}~.
	\ee
	Now we can bring these into an $\su(2,2)$ form by introducing
	\begin{align}
		&\s_{a, \a \da} = (\s_a)_{\a \da}= (- i \vec \s , \mI_2)_{\a\da}~,&\qquad
			&\bar \s_a^{\da \a} = (\bar \s_a)^{\da \a} = (i \vec \s , \mI_2)^{\da \a}~,&\\[.2em]
		\label{eq:SF-SP-xxbDef}
		&x_{\a \da} = (x)_{\a \da} = \frac{x^a}{\sqrt{2}}  \s_{a, \a \da}~,&\qquad 
			&x^{\da \a} = (\bar x)^{\da \a} = \frac{x^a}{\sqrt{2}}  \bar \s^{\da \a}_a~,&\\
		\label{eq:SF-SP-ppbDef}
		&p_{\a \da} = (p)_{\a \da} = \frac{p_a}{\sqrt{2}} \s_{a, \a \da}~,&\qquad 
			&p^{\da \a} = (\bar p)^{\da \a} = \frac{p_a}{\sqrt{2}}  \bar \s^{\da \a}_a~,&
	\end{align}
	which fulfill
	\begin{align} \label{eq:SF-SP-xpdIds}
			&x_{\a \dg} x^{\dg \b} = \frac{\vx^{\,2}}{2} \d_\a{}^\b~,\qquad
			x^{\da \g} x_{\g \db} = \frac{\vx^{\,2}}{2} \d^\da{}_\db~,\qquad
			x_{\a \db} x^{\db \a} = tr(x \bar x) = \vec x^2~,\nn\\
			&~p_{\a \dg} p^{\dg \b} = \frac{\vp^{\,2}}{2} \d_\a{}^\b~,\qquad 
			p^{\da \g} p_{\g \db} = \frac{\vec p^2}{2} \d^\da{}_\db~,\qquad
			p_{\a \db} p^{\db \a} = tr(p \bar p) = \vec p^{\,2}~,\qquad\\
			&\qquad\qquad d = \vec x \cdot \vec p = \tr(x \bar{p}) = x_{\a \db} p^{\db \a} = \tr(p \bar{x}) = p_{\a \db} x^{\db \a}~.\nn
	\end{align}
	The $\su(2)$ indices are raised and lowered by multiplication with the two-dimensional Levi-Civita symbol $\e^{\a\b}$ and $\e^{\da\db}$ as well as their inverse $\e_{\a\b}$ and $\e_{\da\db}$ from the left,
	\be \label{eq:SF-SP-Majo}
		\s_{a,\da \b} = (\s^T_a)_{\da \b} = \eps_{\da \dg} \eps_{\b \d} \bar\s_a^{\dg \d}~,\qquad\qquad
		\bar\s_a^{\a \db} = (\bar\s^T_a)^{\a \db} = \eps^{\a \g} \eps^{\db \dot\d} \s_{a,\g \dot\d}~,
	\ee
	where we use the convention $\eps^{12}=\eps^{\dot1 \dot2} = -\eps_{1 2} = -\eps_{\dot1 \dot2}=1$.

	One can also notice that these follow from the $\SO(4)$ gamma matrices, cf. \eqref{eq:SS-5dgammas},
	\be
		\g_a = \begin{pmatrix} 0 & \s_a\\ \bar\s_a & 0 \end{pmatrix}~,\quad 
			\g_a^\dag = \g_a~,\quad
			\{\g_a , \g_b\} = \begin{pmatrix} \s_{(a} \bar\s_{b)} & 0 \\ 0 & \bar \s_{(a} \s_{b)} \end{pmatrix} = 2 \d_{a b} \mI_4~,
	\ee
	with charge conjugation matrix
	\be 
		 C = \g_4 \g_2 = \begin{pmatrix} \eps & 0 \\ 0 & \eps^{-1} \end{pmatrix} = - C^{-1}~,
	\ee
	where \eqref{eq:SF-SP-xpdIds} and \eqref{eq:SF-SP-Majo} are just the $\SO(4)$ Clifford algebra and the Majorana condition. 

	In terms of \eqref{eq:SF-SP-xxbDef} and \eqref{eq:SF-SP-ppbDef} the $\so(4)=\su(2)\oplus\su(2)$ rotations take the form\footnote{Note that $L_\a{}^\b$ and $L^\da{}_\db$ are the $\so(4) \subset \so(2,4)$ rotations while $L_{IJ}$ in \eqref{eq:SM-AdSS-SO6Rots} where $\so(6)$ rotations.}
	\begin{align} \label{eq:SF-SP-Ldef}
		&L_\a{}^\b = \frac{1}{2i} (p_{\a \dg} x^{\dg \b} - x_{\a \dg} p^{\dg \b}) 
			= p_a x^b \s_{a b,\a}{}^\b 
			=\frac{1}{2} J_{a b} \s_{a b,\a}{}^\b ~,\\
		\label{eq:SF-SP-Lbdef}
		&\tilde L^\da{}_\db = \frac{1}{2i} (p_{\g \db} x^{\da \g} - x_{\g \db} p^{\da \g}) 
			= x^a p_b \bar\s_{a b}{}^\da{}_\db
			= \frac{-1}{2} J_{a b} \bar\s_{a b}{}^\da{}_\db ~,
	\end{align}
	where
	\be
		\s_{a b,\a}{}^\b = (\s_{a b})_\a{}^\b = \frac{1}{2 i} (\s_{[a}\bar\s_{b]})_\a{}^\b 
			~,\qquad
			\bar\s_{a b}{}^\da{}_\db = (\bar\s_{a b})^\da{}_\db = \frac{1}{2 i} (\bar\s_{[a}\s_{b])})^\da{}_\db 
			~.
	\ee
	Froim \eqref{eq:SF-SP-AdSS_SO24Rots} and \eqref{eq:SF-SP-AdSS_SO24Boosts2} the remaining generators are defined in $\su(2,2)$ scheme as
	\begin{align} 
		&E^2 = J_{0'0}^2 = \vec{p}^2 - d^2 + \Ms^2~,\quad
			J_{0'}^{\da \b} =  E\,x^{\da \b}~,\quad
			J_0^{\da \b} = - p^{\da \b} + d\,x^{\da \b}~,\nn\\
		\label{eq:SF-SP-EPKdef}
		& P_{\a \db} = \frac{1}{\sqrt{2}} (- J_0 + i J_{0'})_{\a \db} 
			= \frac{1}{\sqrt{2}} (p_{\a \db} - (d - i E) x_{\a \db})~,\\
		& K^{\da \b} = \frac{1}{\sqrt{2}} (- \bar J_0 - i \bar J_{0'})^{\da \b}\
			= \frac{1}{\sqrt{2}} (p^{\da \b} - (d + i E) x^{\da \b})~,\nn
	\end{align}
	where it can be checked that \eqref{eq:SF-SP-Ldef}, \eqref{eq:SF-SP-Lbdef} and \eqref{eq:SF-SP-EPKdef} fulfill the $\su(2,2)=\so(2,4)$ algebra relations collected in \secref{appsec:OC-su22}. Especially, one should notice that the charges $P_{\a \db}$ and $K^{\da \b}$ defined in \eqref{eq:SF-SP-EPKdef} are not the usual momentum and conformal charges of $\SO(1,3)$, as defined in \eqref{eq:In-ISO2d}, but their $\SO(4)$ analog. This is of course due to the fact that we singled out the energy $E$, i.e., rotations in the temporal $0'0$-plane of $\Reals^{2,4}$, instead of the dilatation $D$, corresponding to boosts in the $0'4$-plane.

	Next, we consider the $\su(4)$ R-symmetry algebra. Let $G_\alg{s}=(y_{AB})$ and $P_\alg{s} = -(p^{AB})$,
	\be\label{eq:SF-SP-ypAB}
		G_{\alg{s}}=
		\begin{pmatrix}
		0 & y_{12} & y_{13}  & y_{14}  \\
		y_{21}  & 0 & y_{23} & y_{24}  \\
		y_{31}  & y_{32}  & 0 & y_{34}  \\
		y_{41} & y_{42} & y_{43}  & 0
		\end{pmatrix},\qquad -P_{\alg{s}}=
		\begin{pmatrix}
		0 & p^{12} & p^{13}  & p^{14}  \\
		p^{21}  & 0 & p^{23} & p^{24}  \\
		p^{31}  & p^{32}  & 0 & p^{34}  \\
		p^{41} & p^{42} & p^{43}  & 0 \\
		\end{pmatrix},
	\ee
	cf. \eqref{eq:SF-SS-GaGs} and \eqref{eq:SF-SS-PaPaExpl}, which satisfy the equations
	\begin{align} \label{eq:SF-SP-ypConstr}
		&y_{AB}=-y_{BA}\,,\quad y^{AB}=-{1\ov2}\e^{ABCD}y_{CD}\,,\quad y_{AB}^*=-y^{AB}\,,\quad y_{AC}y^{CB}=\d_A^B\,,\\
		&p^{AB}=-p^{BA}\,,\quad p_{AB}=-{1\ov2}\e_{ABCD}p^{CD}\,,\quad p_{AB}^*=-p^{AB}\,, \quad p_{AC}p^{CB}={\Ms^2 \ov 16}\d_A^B\,,\\
		\label{eq:SF-SP-M2S5}
		&\Ms^2 =  4p_{AB}p^{BA}=\sum_{k=1}^6 P^{\alg{s}}_kP^{\alg{s}}_k\,,\qquad   y_{AC}p^{CB} + p_{AC}y^{CB} = \frac{1}{2}Y^i P_i^{\alg{s}} \d_A^B=0\,,
	\end{align}
	where $\e^{1234}=\e_{1234}=1$. The $\su(4)$ algebra generators are now defined as
	\be \label{eq:SF-SP-Rsu4Def}
		R_A{}^B = {1\ov i}(p_{AC}y^{CB}-p^{BC}y_{CA}) = \frac{2}{i}p_{AC}y^{CB} = -\frac{2}{i} y_{AC}p^{CB} ~,
	\ee
	where we used the last equation in \eqref{eq:SF-SP-M2S5}. These fulfill the $\su(4)$ algebra Poisson bracket relations
	\be
		{1\ov i}\{ R_A{}^B, R_C{}^D\} = \d_C{}^B R_A{}^D - \d_A{}^D R_C{}^B~.
	\ee
	
	For the following it will be of utmost importance that the square of the $\su(2)$  and $\su(4)$ generators, $L_\a{}^\b$, $\tilde{L}^\da{}_\db$ and $R_A{}^B$, is proportional to the respective identity matrix,
	\be \label{eq:SF-SP-LsqRsq}
		L_\a{}^\g L_\g{}^\b = \frac{S^2}{4} \d_\a{}^\b~,\qquad
		\tilde{L}^\da{}_\dg \tilde{L}^\dg{}_\db = \frac{S^2}{4} \d^\da{}_\db~,\qquad	
		R_A{}^C R_C{}^B = \frac{\Ms^2}{4} \d_A{}^B~.
	\ee

	A comprehensive list of potentially useful relations has been collected in \secref{appsec:OC-UF}.

 \subsection{Lagrangian and Hamiltonian Density} \label{subsec:SF-SP-LagHam}
	The Lagrangian density of the $\AdSxS$ superparticle is obtained from \eqref{eq:SF-SS-Lstat2} by neglecting the $\s$-dependent terms,
	\be\label{eq:SF-SS-LstatPart}
		\cL_{\rm pp} 
			= \str \Big(\!
			P_{\alg{b}}\,\p_\tau G_\alg{b} + {i \bp_t \ov2} \,G_{\alg{a}} \Kk B_\tau + 2G_{\alg{b}}P_{\alg{b}}B_\tau
			\Big) - \frac{p_t^2}{2\sqrt{\l}}~,
	\ee
	where by \eqref{eq:SF-SS-ptf} $\bp_t$ can be expressed in $p_t = -E$, 
	\be\label{eq:SF-SP-pitf0}
		\bp_t= \frac{1}{X^5\big(1+{1\ov4}\vartheta_A{}^\a \theta_\a{}^A \big)}\Big(-E-  {i}\,\str \big(G_{\alg{b}}P_{\alg{b}}\Upsilon\chi^2\big) \Big)\,. 
	\ee

	The second Virasoro constraint \eqref{eq:SF-SS-C2b} becomes
	\be \label{C2b0} 
		C_2 =-\bp_t^2 + 4\,\str\big(G_{\alg{b}}P_{\alg{b}}\big)^2 = 0~,
	\ee 
	such that the Hamiltonian density takes the form
	\begin{align}
		\cH_{\rm pp} =& \frac{1}{2\sqrt{\l}}(X^5)^2\big(1+\sfrac{1}{4}\vartheta_A{}^\a \theta_\a{}^A \big)^2\, 4\,\str\big(G_{\alg{b}}P_{\alg{b}}\big)^2\\
			&+ \frac{i\,E}{\sqrt{\l}}\,\str \big(G_{\alg{b}}P_{\alg{b}}\Upsilon\chi^2\big)
			+\frac{1}{2\sqrt{\l}}\,\Big(\str \big(G_{\alg{b}}P_{\alg{b}}\Upsilon\chi^2\big)\Big)^2 ~.\nn
	\end{align}
	For this it might be useful to notice that
	\be
		4 (X^5)^2\,\str\big(G_{\alg{b}}P_{\alg{b}}\big)^2=(\vec P^a)^2+(\vec P^a)^2(\vec X^a)^2 -(\vec P^a\cdot\vec X^a)^2 + (X^5)^2 \sum^6_{i=1} (P^s_i)^2~.
	\ee

	\subsection{Fermionic Poisson Structure at Quadratic Order} \label{subsec:SF-SP-FermPoisStruc}
	In order to quantize the $\AdSxS$ superparticle canonically we first have to find the Poisson structure of the bosonic and fermionic degrees of freedom. For this we can read off the symplectic form from the kinetic terms in the Lagrangian density, which then have to be brought to canonical form.

	From we see \eqref{eq:SF-SS-LstatPart} that the kinetic term for the fermions is highly nontrivial, which is why we first constrain our interest to the terms quadratic in fermions. For this notice that
	\be
		2 B_\tau = [\dot \chi, \chi] + \ord{\th^4} 
			= - \begin{pmatrix} \dot \Th \Th^\dag - \Th \dot\Th^\dag & 0 \\ 0 & \dot \Th^\dag \Th - \Th^\dag \dot\Th \end{pmatrix} + \ord{\th^4}~.
	\ee
	Hence, the kinetic term quadratic in fermions becomes
	\begin{align}
		\calL_{\rm kin} &= \str \left(\frac{i}{2} \pi_t G_\alg{a} \Kk B_\tau + 2 G_\alg{b} P_\alg{b} B_\tau \right) 
			\approx \frac{i}{4} E\, \str(\mI_\alg{a} [\dot\chi,\chi]) + \str(G_\alg{b} P_\alg{b} [\dot\chi,\chi])\nn\\
			&= - \tr\left((\frac{i}{4} E\,\mI_4 + G_\alg{a} P_\alg{a}) (\dot \Th \Th^\dag - \Th \dot\Th^\dag)\right) 
				+ \tr\left(G_\alg{s} P_\alg{s} (\dot \Th^\dag \Th - \Th^\dag \dot\Th)\right)\\
		&= - \left( \frac{i}{4} E\,\d_\a{}^\b + \frac{1}{2 i} L_\a{}^\b \right)(\dot\th\,\vt - \th\,\dot\vt)_\b{}^\a - \frac{i}{2} R_A{}^B (\dot\vt\,\th  - \vt\,\dot\th)_B{}^A~,\nn
	\end{align}
	where e.g. $(\dot\th\,\vt - \th\,\dot\vt)_\b{}^\a = (\dot\th_\b{}^A \vartheta_A{}^\a - \th_\b{}^A \dot\vartheta_A{}^\a)$.
	The coupling matrices $L_\a{}^\b$ and $R_A{}^B$ are just the matrices of $\su(2)$ and $\su(4)$ generators defined in \eqref{eq:SF-SP-Ldef} and \eqref{eq:SF-SP-Rsu4Def}. By \eqref{eq:SF-SP-LsqRsq} these can be diagonalized by finding matrices $\cV_\a{}^{\ul{\g}}\,$ and $\cW_A{}^{\ul{B}}$, such that
	\begin{align} \label{eq:SF-SP-thphi}
		&L_\a{}^\b = \frac{-S}{2} \cV_\a{}^{\ul{\g}} (\s_3)_{\ul{\g}}{}^{\ul{\d}} \cV_{\ul{\d}}{}^\b 
			= \frac{-S}{2}(\cV \s_3 \cV^\dag)_\a{}^\b~,\nn\\
		&R_A{}^B = \frac{-\Ms}{2} \cW_A{}^{\ul{C}} (\Sigma)_{\ul{C}}{}^{\ul{D}} \cW_{\ul{D}}{}^B
			= \frac{-\Ms}{2} (\cW \S \cW^\dag)_A{}^B ~,\\[.2em]
		&\th_\a{}^A = \cV_\a{}^{\ul{\g}} \phi_{\ul\g}{}^{\ul C} \cW_{\ul C}{}^A = (\cV \phi \cW^\dag)_\a{}^A~,\quad	
			\vartheta_A{}^\a = (\cW \varphi \cV^\dag)_A{}^\a~,\quad 
			\varphi_{\ul A}{}^{\ul\a} = (\phi_{\ul\a}{}^{\ul A})^\ast~, \nn
	\end{align}
	Here, underlined indices should remind us that the new fermions $\phi_{\ul{\a}}{}^{\ul{A}}$ do not transform under $\su(2)$ or $\su(4)$ anymore and one should stress again that as $L_\a{}^\b$ and $R_A{}^B$ the matrices  $\cV_\a{}^{\ul{\g}}\,$ and $\cW_A{}^{\ul{B}}$ only depend on the bosonic phase space variables.

	Dropping terms involving $\dot \cV$ or $\dot \cW$ we then have
	\begin{align} \label{eq:SF-SP-LkinDiag}
		\calL_{\rm kin} &\approx
			- \frac{i}{4}  \left( E\,\d_{\ul\a}{}^{\ul \b} + S (\s_3)_{\ul \a}{}^{\ul\b} \right)(\dot\phi\,\varphi - \phi\,\dot\varphi)_{\ul\b}{}^{\ul{\a}} + \frac{i}{4} \Ms \Sigma_{\ul A}{}^{\ul B} (\dot\varphi\,\phi - \varphi\,\dot\phi)_{\ul B} {}^{\ul A}~,
	\end{align}
	and hence the new fermions $\phi_\ua{}^\uA$ are canonical up to rescaling,
	\be \label{eq:SF-SP-PBphiPre}
		\frac{1}{i}\{\phi_\ua{}^\uA , \varphi_\uB{}^\ub \} = \frac{2\,\d_\ua{}^\ub \d_\uB{}^\uA}{E + (-1)^\ua S + (-1)^{\eps_\uA} M_{\alg{s}}} = \frac{2i\,\d_\ua{}^\ub \d_\uB{}^\uA}{\Delta(\ua,\uA)} ~,
	\ee
	with $\eps_{1,2}=1$, $\eps_{3,4}=0$. 

	An explicit form of the matrices $\cV_\a{}^{\ul{\g}}$ and $\cW_A{}^{\ul{C}}$ is given in \secref{appsec:OC-cVWform}, where especially $\cW_A{}^{\ul{C}}$ turns out to be somewhat cumbersome. At quadratic order in fermions one can however avoid working with the explicit expressions. For this note that
	\begin{align} \label{eq:SF-SP-omegas}
		&\frac{1}{\Delta(\ua,\uA)} = \om_\d + (-1)^\ua\,\om_\s + (-1)^{\eps_\uA}\,\om_\S + (-1)^{\ua+\eps_\uA}\,\om_{\s \S}~,\\[.2em]
		\text{with}\quad&\om = \prod_{\pm, \tpm}(E \pm S \tpm \Ms) ~,\quad
			\om_\d =\frac{2\,E}{\om} (E^2-S^2-\Ms^2)~,\quad
			\om_{\s\Sigma} = \frac{4}{\om} \Ms E S ~, \nn\\
		&\om_\s = \frac{-2\,S}{\om} (E^2-S^2+\Ms^2)~,\quad
			\om_\S = \frac{-2\,\Ms}{\om} \Ms(E^2+S^2-\Ms^2)~, \nn
	\end{align}
	and by $(-1)^\ua \d_\a{}^\b = - (\s_3)_\ua{}^\ub$ and $(-1)^{\eps_\uA} \d_\uA{}^\uB = - \Sigma_\uA{}^\uB$
	\be \label{PBphi}
		\frac{1}{i}\{\phi_\ua{}^\uA , \varphi_\uB{}^\ub \} = \om_\d\,\d_\ua{}^\ub \d_\uA{}^\uB - \om_\s\,(\s_3)_\ua{}^\ub \d_\uA{}^\uB 
			- \om_\S\,\d_\ua{}^\ub \Sigma_\uA{}^\uB + \om_{\s\S}\,(\s_3)_\ua{}^\ub \Sigma_\uA{}^\uB~.
	\ee
	With \eqref{eq:SF-SP-thphi} and neglecting once more terms of order $\ord{\phi^2}$ the Poisson bracket of $\th_\a{}^A$ and $\vartheta_B{}^\b$ becomes
	\begin{align} \label{eq:SF-SP-PBth}
		&\frac{1}{i} \{\th_\a{}^A , \vartheta_B{}^\b \} = 
			\cV_\a{}^\ug \cW_B{}^\uD\, \frac{1}{i}\{\phi_\ug{}^\uC , \varphi_\uD{}^\ud \}\, \cV_\ud{}^\b \cW_\uC{}^A \\
		=~&  \om_\d\,\d_a{}^b \d_A{}^B + \frac{2}{S} \om_\s\,L_\a{}^\b \d_A{}^B 
			+ \frac{2}{\Ms} \om_\S\,\d_a{}^b R_A{}^B + \frac{4}{S \Ms}\om_{\s\S}\,L_a{}^b R_A{}^B~. \nn
	\end{align}

	Furthermore, since $\th_\a{}^A$ only depend on $\phi_\ug{}^\uC$, not $\varphi_\uC{}^\ug$, one still has
	\be \label{eq:SF-SP-PBthvt}
		\{\th_\a{}^A, \th_\b{}^B\} = \{\vartheta_A{}^\a, \vartheta_B{}^\b\} = 0 ~.
	\ee

	We would like to add that at any stage one can recover what would have been the result for canonical Poisson bracket relations,
	\be \label{eq:SF-SP-PBthCan}
		\frac{1}{i}\{\th_\a{}^A,\vt_B{}^\b\}=\d_\a{}^\b \d_B{}^A~,
	\ee
	by setting $\om_\s=\om_\S=\om_{\s\S}=0$ and $\om_\d=1$ instead of \eqref{eq:SF-SP-omegas}.

	Finally, let us comment on the canonical structure of the bosons. As we are working in first order formalism, at first the kinetic terms for the bosonic phase space variables takes a manifestly canonical form. However, transformation of the fermions introduces new kinetic terms, such that the bosons become non-canonical. In particular, the first new kinetic terms for the bosons are the terms involving $\dot \cV$ or $\dot \cW$, which have been dropped in \eqref{eq:SF-SP-LkinDiag}. From this we see that the new boson kinetic terms are at least quadratic in fermions. Especially, if one is only interested in the lowest order in fermions the bosonic variables are effectively still canonical.

 \subsection{Poisson Structure from the Superconformal Algebra} \label{subsec:SF-SP-PoisStrucPSU}
	In the previous subsection we derived the Poisson structure of the fermions $\th_\a{}^A$ at quadratic order. It seems though that going to higher orders is virtually impossible as one is facing both higher order terms in the kinetic Lagrangian $\cL_{\rm kin} $ as well as higher order corrections due to the matrices $\cV_\a{}^\ug$ and $\cW_A{}^\uC$ defined in \eqref{eq:SF-SP-thphi}. In particular, most probably one has to work with an an explicit form of $\cV_\a{}^\ug$ and $\cW_A{}^\uC$, which seem to be rather tedious, see \secref{appsec:OC-cVWform}. Furthermore, as discussed at the end of \ssecref{subsec:SF-SP-FermPoisStruc}, we have not even considered the corrections to the Poisson structure of the bosons due to $\dot\cV_\a{}^\ug$ and $\dot\cW_A{}^\uC$, which where dropped in \eqref{eq:SF-SP-LkinDiag}.

	There is though another route to find the canonical fermions. 
	For this, instead of analyzing the original action one {\it assumes} from the very beginning to work with a symplectic form which is canonical. Hence, the corresponding fermionic and bosonic degrees of freedom, schematically the nine spatial bosonic coordinates and momenta, $q_j$ and $p_j$, and the $16$ real fermions denoted as $\z_k$, $k=1,\ldots,16$, therefore satisfy canonical Poisson bracket relations. In terms of these one constructs the $\psu(2,2|4)$ generators, which then can be compared to the $\psu(2,2|4)$ generators obtained from the initial variables via \eqref{eq:SF-SS-Charges2} and \eqref{eq:SF-SS-QM}. By this one should be able to read off a canonical transformation from the initial non-canonical to the canonical fields.

	We would like to stress that this approach has the additional advantage that one is manifestly working in terms of the $\psu(2,2|4)$ symmetry algebra, which when quantizing the theory will determine ordering ambiguities.

	The $\psu(2,2|4)$ algebra can be expanded in fermions, where bosonic and fermionic generators comprise only even, respectively, only odd powers in fermions. Schematically the expansion takes the form
	\begin{align} \label{eq:SF-SP-JQexp}
		&J_{MN}=J_{MN}^{(0)} + J_{MN}^{jk}(p,q)\z_j\z_k + J_{MN}^{k_1k_2k_3k_4}(p,q)\z_{k_1}\z_{k_2}\z_{k_3}\z_{k_4} +\cdots ~,\\[.2em]
		&Q_\a=Q_\a^{k}(p,q)\z_k + Q_\a^{k_1k_2k_3}(p,q)\z_{k_1}\z_{k_2}\z_{k_3}+
		~,
	\end{align}
	which terminates at $\ord{\z^{16}}$.
	The terms $J_{MN}^{(0)}$ of the bosonic generators independent of fermions are just the ones discussed in \eqref{subsec:SF-SP-su22su4}, where in particular we noted at the end of \ssecref{subsec:SF-SP-FermPoisStruc} that the initial bosonic phase space variables differ from the intrinsically canonical ones by corrections at least quadratic in fermions. 

	The first step is now to look at the terms linear in fermions, the terms $Q_\a^{k}(p,q)\z_k$ in \eqref{eq:SF-SP-JQexp}. For these we will propose an ansatz and then check that taking the Poisson brackets of the fermions gives again the purely bosonic generators $J_{MN}^{(0)}$,
	schematically
	\be \label{eq:SF-SP-JasQQpoisson}
		J^{(0)}_{M N}(p,q) = C^{\a,\b}_{M N}\,Q_\a^{k}(p,q)\,Q_\b^{l}(p,q)\frac{1}{i}\{\z_k,\z_l\}~,
	\ee
	for some appropriate constant matrix $C^{\a,\b}_{M\,N}$.\footnote{Later on, when utilizing the $\un(1)\oplus\su(2)\oplus\su(2)\oplus\su(4)$ symmetry and working with the ansatz \eqref{eq:SF-SP-QStAns} the constant matrix $C^{\a,\b}_{M N}$ is implicitely given by the requirements \eqref{eq:SF-SP-QSshould}.}

	Actually, in the following we are going construct the fermionic charges linear in fermions for both the initial, non-canonical as well as the intrinsically canonical variables. This is only possible as we are interested in the canonical structure at lowest order in fermions, where the bosonical phase space variables stay the same while for the fermions one can transition between initial and canonical variables by setting the $\om$-parameters in \eqref{eq:SF-SP-PBth} either to the values in \eqref{eq:SF-SP-omegas} or to $\om_\s=\om_\S=\om_{\s\S}=0$ and $\om_\d=1$. In particular, both the initial as well as the canonical variables are both denoted by $x_{\a\da}$, $p_{\a\da}$, $y_{AB}$ and $p^{AB}$ for the bosons and $\th_\a{}^A$ and $\vt_A{}^\a$ for the fermions.

	The $\psu(2,2|4)$ generators form the supermatrix \eqref{eq:SF-SS-Charges2} and due to the conventions for the bosonic $\su(2,2)\oplus\su(4)$ generators, this suggests the index structure
	\be \label{eq:SF-SP-psuMat}		
			\begin{pmatrix}L_\a{}^\b + \frac{E}{2} \d_\a{}^\b & P_{\a \db} & Q_{\a}{}^B \\
				K^{\da \b} & - \tilde L^\da{}_{\db} - \frac{E}{2} \d^\da{}_{\db} & i \tilde S^{\da B} \\
				S_{A}{}^{\b} & i \tilde Q_{A \db} & R_A{}^B \end{pmatrix},
	\ee
	where analogously to $P_{\a \dot\b}$ and $K^{\da \b}$ the fermionic charges $Q_{\a}{}^D$, $\tilde Q_{A \db}$, $S_{A}{}^{\b}$, and $\tilde S^{\da B}$ are not the usual $\SO(1,3)$ super- and conformal supercharges but their $\SO(4)$ analog.

	In the following we instead prefer to work with the index structure
	\be \label{eq:SF-SP-psuMat2}
		 \begin{pmatrix}L_\a{}^\b + \frac{E}{2} \d_\a{}^\b & -P_\a{}^\db & Q_{\a}{}^B \\
				K_\da{}^\b & \tilde L_\da{}^{\db} - \frac{E}{2} \d_\da{}^\db & i \tilde S_\da{}^B \\
				S_A{}^\b & -i \tilde Q_A{}^\db & R_A{}^B \end{pmatrix}~,
	\ee
	which is connected to \eqref{eq:SF-SP-psuMat} by a similarity transform with $\,\diag(\d_\a{}^\g , \eps_{\da \dg} , \d_A{}^C )\,$.

	Therefore, we take the following ansatz for fermionic charges linear in fermions
	\begin{align} \label{eq:SF-SP-QStAns}
		&Q_\a{}^A\!=\!\Big(\,V_\a{}^\g \delta_C{}^A + \frac{2}{\Ms}\,W_\a{}^\g R_C{}^A\Big)\th_\g{}^C \,,~~~
		\tilde{S}_{\dot{\a}}{}^A \!=\! \Big(\,\tilde V_\da{}^\g \delta_C{}^A + \frac{2}{\Ms}\,\tilde W_\da{}^\g R_C{}^A\Big)\th_\g{}^C\,,\\
		&S_A{}^\a \!=\! \Big(\,{V_\a{}^\g}^\dag \delta_A{}^C + \frac{2}{\Ms}{W_\a{}^\g}^\dag R_A{}^C\Big)\vt_C{}^\g \,,~~
		\tilde{Q}_A{}^\da \!=\! \Big(\,{\tilde V_\da{}^\g}^\dag \delta_A{}^C + \frac{2}{\Ms} {\tilde W_\da{}^\g}^\dag R_A{}^C\Big)\vt_C{}^\g \,,\nn
	\end{align}
	with $S_A{}^\a = (Q_\a{}^A)^\dag$ and $\tilde Q_A{}^\da = (\tilde S_\da{}^A)^\dag$, 
	\begin{align} \label{eq:SF-SP-defVW}
		&V_{\a}{}^{\g} = c_V\delta_{\a}{}^{\g} + \frac{2}{S}\,b_V L_\a{}^\g~,& 
		&\tV_{\da}{}^{\g} 
			= \sqrt{2\ov x^2} \eps_{\da \db} \bar{x}^{\db \r} \Big(\tc_V\delta_{\r}{}^{\g} + \frac{2}{S}\,\tb_V L_\r{}^\g~\Big)
			~,\\ 
		&W_{\a}{}^{\g} = c_W\delta_{\a}{}^{\g} + \frac{2}{S}\,b_W L_\a{}^\g~,&
		&\tW_{\da}{}^{\g} 
			= \sqrt{2\ov x^2} \eps_{\da \db} \bar{x}^{\db \r} \Big(\tc_W\delta_{\r}{}^{\g} + \frac{2}{S}\,\tb_W L_\r{}^\g~\Big)
			~,&
	\end{align}
	and the four $c$'s and four $b$'s being complex constants. Here, we already included prefactors $\sqrt{2/x^2}$, $2/S$, and $2/\Ms$ to normalize the $\bar{x}^{\db \r}$, $L_\da{}^\g$, and $R_C{}^A$, respectively, cf. \eqref{eq:SF-SP-LsqRsq}. Also, one can use that
	\be
		x^{\da \b} L_\b{}^\g 
			= \frac{1}{2i}\,\Big(\left(d\,\d^\da{}_\dr - p^{\da \b} x_{\b\dr}\right)x^{\dr\g} - \frac{x^2}{2}\,\d^\da{}_\dr p^{\dr \g} \Big) = \frac{i}{2} \left(x^2 p^{\da \g} - d\,x^{\da \g} \right) \nn~.
	\ee
	By 
	$(\d_\a{}^\g)^\dag =\d_\g{}^\a$, $(L_\a{}^\g)^\dag = L_\g{}^\a$, and $(x_\da{}^\g)^\dag = - x_\g{}^\da$, the hermitian conjugate read
	\begin{align}
		&(V_\a{}^\g)^\dag = c_V^* \delta_{\g}{}^{\a} + \frac{2}{S}\,b^*_V L_\g{}^\a~,\qquad
		(\tV_\da{}^\g)^\dag = - \sqrt{2\ov x^2} \Big(\tc_V^* \delta_{\g}{}^{\r} + \frac{2}{S}\,\tb_V^* L_\g{}^\r\Big) x_\r{}^\da
		~,
	\end{align}
	analogously for $W_\a{}^\g$ and $\tW_\da{}^\dg$, and typical products are
	\begin{align}
	  V_\a{}^\g (W_\b{}^\g)^\dag &= (c_V c^*_W + b_V b^*_W) \d_\a{}^\b + \frac{2}{S} (c_V b^*_W + b_V c^*_W) L_\a{}^\b~,\nn\\
		\tV_\da{}^\g (\tW_\db{}^\g)^\dag &= (\tc_V \tc^*_W + \tb_V \tb^*_W) \d_\da{}^\db - \frac{2}{S} (\tc_V \tb^*_W + \tb_V \tc^*_W) \tilde{L}_\da{}^\db~,\nn\\
		\tV_\da{}^\g (W_\b{}^\g)^\dag &= \sqrt{\frac{2}{x^2}} x_\da{}^{\r} \Big((\tc_V c^*_W + \tb_V b^*_W) \d_\r{}^\b + \frac{2}{S} (\tc_V b^*_W + \tb_V c^*_W) L_\r{}^\b \Big) \\
		&= \sqrt{\frac{2}{x^2}} \Big((\tc_V c^*_W + \tb_V b^*_W) x_\da{}^{\b}  + \frac{i}{S} (\tc_V b^*_W + \tb_V c^*_W) (x^2 p_\da{}^{\b} - d\, x_\da{}^{\b}) \Big)\nn\\
		V_\a{}^\g (\tW_\db{}^\g)^\dag 
		&= \sqrt{\frac{2}{x^2}} \Big(-(c_V \tc^*_W + b_V \tb^*_W) x_\a{}^{\db}  + \frac{i}{S} (c_V \tb^*_W + b_V \tc^*_W) (x^2 p_\a{}^{\db} - d\, x_\a{}^{\db}) \Big)\nn
	\end{align}
	where one should note the relative sign in the second line, which stems from
	\be
		x_\da{}^\g x_\g{}^\db = \frac{-x^2}{2}\,\d_\da{}^\db~,\qquad
		x_\da{}^\g L_\g{}^\r x_\r{}^\db = \frac{x^2}{2} \eps_{\da\dg} \eps^{\db \dot{\r}} L^\dg{}_{\dot{\r}}  = \frac{x^2}{2} L_\da{}^\db~.
	\ee
	Also, it is useful to note that by $\tilde{L}_\da{}^\dg x_\dg{}^\b = x_\da{}^\g L_\g{}^\b$ for the ansatz we have \eqref{eq:SF-SP-QStAns}
	\be\ba
		\qquad\quad&L_\a{}^\g Q_\g{}^A = \frac{S}{2}\,Q_\a{}^A|_{c \leftrightarrow b}~,& \qquad
		&Q_\a{}^C R_C{}^A = \frac{\Ms}{2}\,Q_\a{}^A|_{V \leftrightarrow W}~,&\\
		&\tilde{L}_\da{}^\dg \tilde{S}_\dg{}^A = \frac{S}{2}\,\tilde{S}_\da{}^A|_{\tc \leftrightarrow \tb}~,& \quad
		&\tilde{S}_\da{}^C R_C{}^A = \frac{\Ms}{2}\,\tilde{S}_\a{}^A|_{\tilde V \leftrightarrow \tilde W}&
	\ea\ee

	With this we can investigate the $\psu(2,2|4)$ algebra Poisson bracket relations, where we only keep the purely bosonic terms.

	By \eqref{eq:SF-SP-PBthvt}, $\{\th_\a{}^A, \th_\b{}^B\} = 0$, the ansatz \eqref{eq:SF-SP-QStAns} immediately implies that
	\be \label{eq:SF-SP-QQcom}
		\frac{1}{i}\{Q_\a{}^A, Q_\b{}^B\} =  \frac{1}{i}\{\tilde{S}_\da{}^A, \tilde{S}_\db{}^B\} = 
			\frac{1}{i}\{Q_\a{}^A, \tilde{S}_\db{}^B\} =0~.
	\ee
	as it should be.

	One can actually generalize the ansatz \eqref{eq:SF-SP-QStAns} by adding terms such that the fermionic charges $Q_\a{}^B$ and $ \tilde{S}_\da{}^B$ depend on both $\th_\g{}^D$ and $\vt_C{}^{\d}$. By this the amount of complex constants is doubled, viz., one has eight additional constants. It is then interesting to note that for this generalized ansatz the requirement of the commutators \eqref{eq:SF-SP-QQcom} to vanish yields exactly eight complex constraints, which we implicitly solved by taking the ansatz \eqref{eq:SF-SP-QStAns}. 

	The remaining Poisson brackets have to read\footnote{The signs presented here were fixed by requiring consistency.}
	\begin{align} \label{eq:SF-SP-QSshould}
		\frac{1}{i}\{Q_\a{}^A, S_B{}^\b\} &= 
			L_\a{}^\b \d_B{}^A  + \d_\a{}^\b R_A{}^B + \frac{E}{2}\,\d_\a{}^\b \d_B{}^A~,\nn \\
		\frac{1}{i}\{\tilde{S}_\da{}^A, \tilde{Q}_B{}^\db\} &= 
			- \tilde{L}_\da{}^\db \d_B{}^A - \d_\da{}^\db R_B{}^A + \frac{E}{2} \d_\da{}^\db \d_B{}^A ~,\\
		\{Q_\a{}^A, \, \tilde{Q}_B{}^{\db}\} &= \d_B{}^A P_\a{}^{\db} =\d_B{}^A \frac{1}{\sqrt{2}}\big( p_{\a}{}^{\db}-d\, x_{\a}{}^{\db}+iE\, x_{\a}{}^{\db}\big)~.\nn
	\end{align}
	Requiring \eqref{eq:SF-SP-QSshould} to hold gives a system of equations collected in \secref{appsec:OC-SupPoisBra}. As the system shows to be linear in quadratic combinations of the $c$'s and $b$'s, in principle it can be solved directly in terms of these. However, let us introduce
	\be
		2\,\om_{j,k} = \om_\d + (-1)^j \om_\s + (-1)^k \om_\S + (-1)^{j+k} \om_{\s\S}
			= \frac{1}{E + (-1)^j S + (-1)^k \Ms}~,
	\ee
	and the complex constants
	\begin{align} \label{eq:SF-SP-aDef}
		&a_{j,k} = 2\sqrt{\om_{j,k}} \Big(c_V + (-1)^j b_V + (-1)^k c_W + (-1)^{j+k} b_W\Big) ~,\\
		\label{eq:SF-SP-taDef}
		&\ta_{j,k} = 2\sqrt{\om_{j,k}} \Big(\tc_V + (-1)^j \tb_V + (-1)^k \tc_W + (-1)^{j+k} \tb_W\Big)~,
	\end{align}
	with $j,k = 0,1\,$. In terms of these the set of equation \eqref{eq:OC-CoeffQS} to \eqref{eq:OC-CoeffQQt} reduces dramatically to the concise form
	\begin{align} \label{eq:SF-SP-ataEqs}
		&|a_{j,k}|^2 = {\rm sgn}(\om_{j,k}) \big(E + (-1)^j S + (-1)^k \Ms\big)~,\\ 
		&|\ta_{j,k}|^2 = {\rm sgn}(\om_{j,k}) \big(E + (-1)^j S - (-1)^k \Ms\big)~,&\\
		&a_{j,k} \ta^*_{j,k} = \frac{-{\rm sgn}(\om_{j,k})}{\sqrt{x^2}} \left(d (1-x^2) + i (E\,x^2 + (-1)^j S) \right)~.&
	\end{align}
	It is now crucial that these equations are consistent with the requirement
	\be
		|a_{j,k}|^2 \,|\ta_{j,k}|^2 
			= |a_{j,k}\, \ta^*_{j,k}|^2~,
	\ee
	where one can use the expression for the energy squared \eqref{eq:SF-AdSS_EnSq&Ham}.

	Therefore, \eqref{eq:SF-SP-ataEqs} fixes the absolute values of $a_{j,k}$ and $\ta_{j,k}$ as well as their relative phases whereas four overall phases $\xi_{j,k}$ are undetermined, i.e., \eqref{eq:SF-SP-ataEqs} is invariant under
	\be
		\{a_{j,k}, \ta_{j,k}\} \rightarrow \{ e^{i \xi_{j,k}} a_{j,k}, e^{i \xi_{j,k}} \ta_{j,k}\}~. 
	\ee
	
	From this we conclude that the ansatz \eqref{eq:SF-SP-QStAns} for the fermionic charges at linear order in fermions is consistent and completely fixed, up to four overall phases. Note in particular that the final solution takes a very symmetric form and solely depends on the $\un(1)\oplus\su(2)\oplus\su(2)\oplus\su(4)$ Casimir numbers, the energy $E$, the $\su(2)\oplus\su(2)\cong\so(4)$ spin $S$ and the $\su(4)\cong\so(6)$ Casimir number $\Ms$.

	Note, that we did not specify the $\om$-parameters in \eqref{eq:SF-SP-PBth} and the ansatz holds for general values. In particular it holds for both cases, assuming the fermions to have canonical Poisson structure \eqref{eq:SF-SP-PBthCan}, corresponding to $\om_\s=\om_\S=\om_{\s\S}=0$ and $\om_\d=1$, as well as for the Poisson structure found from the kinetic Lagrangian quadratic in the initial, non-canonical fermions, for which the $\om$'s are given in \eqref{eq:SF-SP-omegas}.

	As a final check one should verify that the ansatz \eqref{eq:SF-SP-QStAns} for the non-canonical variables indeed coincides with the fermionic charges linear in fermions obtained from \eqref{eq:SF-SS-Charges2} and \eqref{eq:SF-SS-QM}. This has not be done to date but we are optimistic that this is the case. In particular at lowest order we do not have to derive the canonical map between initial non-canonical and canonical fermions by comparison of the supercharges as we already derived it explicitly in the previous subsection.

	The proposed program seems to work for the simplest case and one can contemplate on the next step. Using again the schematic notation of \eqref{eq:SF-SP-JQexp}, the next order of the algebraic relations comes from linear order in fermions and read
	\begin{align}
		Q^{j}_\a = C^{M N,\b}_{\a}\left(J^{j k}_{MN} Q^{l}_\b \{\z_k,\z_l\} + \{J^{(0)}_{MN}, Q^j_\b\}\right)~,
	\end{align}
	for an appropriate constant matrix $C^{M N,\b}_{\a}$.
	
	Furthermore, from the bosonic subalgebra one has the additional relation
	\be
			J^{jk}_{MN} = C^{R S,T U}_{M N}\left(J^{j m}_{R S} J^{k n}_{T U} \{\z_m,\z_n\} + \{J^{(0)}_{R S}, J^{jk}_{T U}\} + \{J^{j k}_{R S}, J^{(0)}_{T U}\}\right) ~.
	\ee
	As now we know both $J^{(0)}_{MN}$ and $Q^j_\b$ the above equations should suffice to determine $J^{kl}_{MN}$, the coefficients of the terms quadratic in fermions in the bosonic generators.

	After these have been fixed the next relevant algebraic relation is
	\begin{align} 
		&J^{jk}_{MN} = C^{\a,\b}_{M N}\left(Q^{j k m}_\a Q^n_\b \{\z_m,\z_n\} + \{Q^j_\a, Q^k_\b\}\right)~,
	\end{align}
	with the same $C^{\a,\b}_{M N}$ as in \eqref{eq:SF-SP-JasQQpoisson}. This in turn should determine $Q^{j k l}_\a$, the coefficients of the terms cubic in fermions, and so on and so forth.

	The final answer for the fermionic charges linear in fermions turned out to take a very simple and symmetric form and one could hope for something similar to happen for the coefficients of higher order in fermions. Also, to get a better feeling for the underlying structures it might be beneficial to study instead a simpler toy model, for example the superparticle in ${\rm AdS}_2 \times {\rm S}^2$ or ${\rm AdS}_3 \times {\rm S}^3$, which are both connected to the supergroup $\PSU(1,1|2)$. In particular this would have the major advantage that the number of fermions is reduced, such that the algebraic recursion stops significantly earlier.

	Finally, it has been argued \cite{Sahakian:2004gy} that for proper choice of gauge the action will be at most quartic in complex fermions and we are curious whether this is observed in our case. Furthermore, it might be interesting to compare with the related accounts on the $\AdSxS$ isometries \cite{Claus:1998yw, Wulff:2014kja}.

\chapter{The Particle and the Spinning String in {\AdSthreexSthreeheader}} \label{chap:AdS3}
	\vspace{-.5cm}
	In order to quantize the $\AdSxS$ superparticle, in the previous section, \ssecref{subsec:SF-SP-PoisStrucPSU}, we proposed to bring the phase space variables into canonical form by utilizing the $\psu(2,2|4)$ symmetry algebra. Even though this scheme proved to give consistent results at lowest order in fermions, the computational effort is expected to increases drastically for higher powers in fermions. Therefore, we conclude that it might be favorable to first understand some lower dimensional analogue, with obvious contenders being the ${\rm AdS}_2 \times {\rm S}^2$ and ${\rm AdS}_3 \times {\rm S}^3$ superparticles.

	\vspace{-.05cm}
	In this chapter we explore the prospects of yet another quantization scheme, namely, construction of unitary irreducible representations via orbits of the isometry group \cite{Alekseev:1988ce}. For this we first focus on the bosonic particle in ${\rm AdS}_3 \times {\rm S}^3$. After settling the notation in \secref{sec:AdS3-Notation}, in \secref{sec:AdS3-Particle} we consider the ${\rm AdS}_3 \times {\rm S}^3$ particle and analyze its pre-symplectic one-form. We devise Hamiltonian treatment of the isometry group orbits of a fixed solution, giving a one-parameter family of orbits naturally parametrized by creation-annihilation variables. These then imply a Holstein-Primakoff realization of the isometry group generators \cite{Holstein:1940zp, Dzhordzhadze:1994np}. By this, with relative ease we acquire exact quantization of the particle, where our result shows consistency with earlier works, e.g., using static gauge \cite{Dorn:2010wt}.

	\vspace{-.05cm}
	Certainly, one can contemplate on application of this technique to the full $\AdSxS$ superstring. But as elaborated in \secref{sec:SM-Decoupling}, to obtain the first quantum corrections to the Konishi anomalous dimension it might though suffice to supersymmetrize semiclassical solutions and quantize them in a minisuperspace fashion as used in \chapref{chap:SingleMode}.

	\vspace{-.05cm}
	Following this logic we apply the quantization scheme to the multi-spin string solution \cite{Frolov:2003qc, Frolov:2003tu, Arutyunov:2003za}, providing a family of solutions depending on two parameters plus winding numbers. Again, this provides an oscillator type realization of the symmetry generators yielding exact quantization of the system, where in comparison to the particle one has more freedom in the Casimir numbers. Exact formulas for the minimal energy turn out to be rather involved. We therefore conclude by taking the large coupling limit, $\l\gg1$, and identify cases corresponding long and short string solutions.

	\vspace{-.05cm}
	The content of this chapter follows closely the article 
	\cite{Heinze:2014cga}.

\section{Notation and Conventions} \label{sec:AdS3-Notation}

	Let us denote  coordinates of $\Reals^{2,2}$ and $\Reals^4$ by $(X^{0'},X^0,X^1,X^2)$ and $(Y^1,Y^2,Y^3,Y^4),$ respectively. The ${\rm AdS}_3$ and ${\rm S}^3$ spaces are defined by the embedding conditions
	\be\ba\label{eq:AdS3-AdS-S conditions}
		&X\cdot X = (X^1)^2+(X^2)^2-(X^{0'})^2-(X^0)^2=-1~,  \\
		&Y\cdot Y = (Y^1)^2+(Y^2)^2+(Y^3)^2+(Y^4)^2=1~.
	\ea\ee
	Parameterizing  $2\times 2$  matrices by the embedding coordinates in the following form
	\be\label{eq:AdS3-g=Y}
		g = \begin{pmatrix}
			\,X^{0'}+iX^0 &X^1-iX^2\, \\ \,X^1+iX^2& X^{0'}-iX^0\,
		\end{pmatrix}, \qquad\quad 
		\tilde g = \begin{pmatrix}
			Y^4+iY^3 &Y^2+iY^1\, \\-Y^2+iY^1&Y^4-iY^3\,
		\end{pmatrix},
	\ee
	one finds that the conditions \eqref{eq:AdS3-AdS-S conditions} are equivalent to
	$g\in {\rm SU}(1,1)$ and $\tilde g\in {\rm SU}(2)$.
	In this way, ${\rm AdS}_3$ is identified with ${\rm SU}(1,1)$ and ${\rm S}^3$ with ${\rm SU}(2),$ respectively.

	We use the following basis of the $\mathfrak{su}(1,1)$ algebra
	\begin{equation}\label{eq:AdS3-su(1,1) basis}
		\fkT_{0}=i{ \boldsymbol\s}_{3}~, \quad \fkT_{1}={\boldsymbol\s}_{1}~, \quad
		\fkT_{2}={\boldsymbol\s}_{2}~,
	\end{equation}
	with $\{\boldsymbol{\s}_1, \boldsymbol{\s}_2, \boldsymbol{\s}_3\}$ the Pauli matrices. The generators $\fkT_a$ satisfy the relations
	\begin{equation}\label{eq:AdS3-tt=}
		\fkT_a \,\fkT_b =\eta_{a b}\,\mI-\epsilon_{a b}\,^{c}\,
		\fkT_{c}~, \qquad\qquad\text{for}\quad a,b,c=0,1,2~.
	\end{equation}
	Here $\mI$ is the unit matrix, $\eta_{a b}=\text{diag}(-1,1,1)$ and
	$\epsilon_{a b c}$ is the Levi-Civita tensor, with
	$\epsilon_{012}=1$. The inner product defined by
	$\la\, \fkT_a\,\fkT_b\,\ra = \frac{1}{2}\,
	\mbox{tr}(\fkT_a\,\fkT_b)
	=\eta_{a b}$
	provides the isometry between $\mathfrak{su}(1,1)$ and 3d Minkowski space,
	since for ${\mathfrak{u}}=u^a\,\fkT_a$,
	one gets $\la\, {\mathfrak{u}}\,{\mathfrak{u}}\,\ra = u^a\,u_a$.
	Then, ${\mathfrak{u}}$ can be timelike, spacelike or lightlike as the corresponding 3d vector $(u^{_0},u^{_1},u^{_2}).$

	A standard basis in $\su(2)$ is given by  ${\tilde\fkT}_j = i\boldsymbol{\sigma}_j\,$, and one has
	\begin{equation}\label{eq:AdS3-ss=}
		{\tilde\fkT}_i\,{\tilde\fkT}_j=-\delta_{i j}\,{\bf I}-\epsilon_{i j k}\,{\tilde\fkT}_k~,\qquad\qquad\text{for}\quad i,j,k=1,2,3~.
	\end{equation}
	Hence,  $\su(2)$, with inner product $\la{\tilde\fkT}_i\,{\tilde\fkT}_j\ra
	= -\frac{1}{2}\,\mbox{tr}({\tilde\fkT}_i\,{\tilde\fkT}_j)=\delta_{ij}$, is isometric to $\Reals^3$, i.e.
	$\la\, \tilde{\mathfrak{u}}\,\tilde{\mathfrak{u}}\,\ra=\tilde u_j\,\tilde u_j$, where $\tilde u_j=\la\, {\tilde\fkT}_j\,\tilde{\mathfrak u}\,\ra$.

	The matrices $g$ and $\tilde g$ in \eqref{eq:AdS3-g=Y} and their inverse group elements can be written as
	\be\ba\label{eq:AdS3-decomposition}
		&g=X^{0'}\,\mI +X^a \,\fkT_a ~,&
		\qquad\quad&\tilde g=Y^4\,\mI +Y^j\,{\tilde\fkT}_j~,&\\
		&g^{_{-1}}=X^{0'}\,\mI -X^a \,\fkT_a ~,& &\tilde g^{_{-1}}=Y^4\,\mI -Y^j\,{\tilde\fkT}_j~,&
	\ea\ee
	and from \eqref{eq:AdS3-tt=} and \eqref{eq:AdS3-ss=} one obtains the following relations between the length elements
	\be\label{eq:AdS3-dg=dY,dh=dX}
		\la\, g^{_{-1}}\text{d} g\,\,g^{_{-1}}\text{d}g\,\ra=\text{d}X\cdot \text{d}X~,~~~~~~
		\la\,\tilde g^{_{-1}}\text{d}\tilde g\,\,\tilde g^{_{-1}}\text{d} \tilde g\,\ra=\text{d} Y\cdot \text{d} Y~.
		\ee
	The isometry transformations are therefore given by the left-right multiplications
	\be\label{eq:AdS3-isometry tr}
		g\mapsto g_{l}\,g\,g_{r}~,~\qquad \qquad \tilde g\mapsto \tilde g_{l}\,\tilde g\,\tilde g_{r}~.
	\ee

\section{The Particle in SU(1,1)\texorpdfstring{$\times$}{x}SU(2)} \label{sec:AdS3-Particle}
	The dynamics of a particle in ${\rm SU}(1,1)\times {\rm SU}(2)$ is described by the action
	\be\label{eq:AdS3-particle action 1}
		S=\int \text{d}\tau\,\left(\frac{1}{2\xi}\Big(
			\la\, g^{_{-1}}\dot{g}\,g^{_{-1}}\dot{g}\,\ra +
			\la\, \tilde g^{_{-1}}\dot{\tilde g}\,\tilde g^{_{-1}}\dot{\tilde g}\,\ra\Big)
		-\frac{\xi \m_{0}^2}{2}\right)~,
	\ee
	where $\xi$ plays the role of the world-line einbein
	and $\m_{0}$ is the particle mass. In the first order formalism, this action is equivalent to
	\be\label{eq:AdS3-particle action 2}
		S=\int \text{d}\tau\,\left(\la R\,g^{_{-1}}\dot{g}\ra +
		\la \tilde R\,\tilde g^{_{-1}}\dot{\tilde g}\ra- \frac{\xi}{2}
		\left(\la RR\ra+\la \tilde R \tilde R\ra+\m_{0}^2 \right)\right)~,
	\ee
	where $R$ and $\tilde R$ are Lie algebra valued phase space variables, $\xi$ becomes a Lagrange multiplier and its variation defines the mass-shell condition
	\be\label{eq:AdS3-mass-shell}
		\la\, R\,R\, \ra +\la\, \tilde R\, \tilde R\, \ra +\m_{0}^2 =0~
	\ee
	for $R$ being timelike. The Hamilton equations obtained from \eqref{eq:AdS3-particle action 2},
	\be\label{eq:AdS3-Hamilton eq}
		g^{_{-1}}\dot g=\xi R~,   \quad \tilde g^{_{-1}}\dot{\tilde g}=\xi \tilde R~;  \qquad
		\dot R=0~,        \quad    \dot{\tilde R}=0~,
	\ee
	provide the conservation of $R$ and $\tilde R$, as well as of their counterparts
	\be\label{eq:AdS3-L,L_s}
		L=g\,R\,g^{_{-1}}~, \quad  \qquad \tilde L=\tilde g\,\tilde R\,\tilde g^{_{-1}}~.
	\ee
	In the following we will refer to the dynamical integrals $L$ and $\tilde L$ as 'left' and to $R$ and $\tilde R$ as 'right', as they are invariant under right-, respectively, left-multiplication. In particular, they are the Noether charges related to the invariance of the action \eqref{eq:AdS3-particle action 1} with respect to the isometry transformations \eqref{eq:AdS3-isometry tr}.

	The first order action  \eqref{eq:AdS3-particle action 2} defines the pre-symplectic form of the system
	\be\label{eq:AdS3-particle pre-symplectic form}
		\Theta=\la R g^{_{-1}} \mathrm{d}g\ra +
		\la \tilde R \tilde g^{_{-1}} \mathrm{d}\tilde g\ra~,
	\ee
	which leads to the following Poisson brackets 
	\be\ba\label{eq:AdS3-L-R PB}
		\qquad&\{L_a,\,L_b\}=2\e_{a b}\,^c \,L_c ~,&\quad 
			&\{R_a  ,R_b \}=-2\epsilon_{a b}\,^c \,R_{c}~,&\quad
			&\{L_a ,\,R_b\}=0~,&\\[.2em]
		&\{\tilde L_i,\,\tilde L_j\}=2\e_{ijk}\,\tilde L_k~,&
			&\{\tilde R_i,\,\tilde R_j\}=-2\e_{ijk}\,\tilde R_k~,&
			&\{\tilde L_i,\,\tilde R_j\}=0~,&
	\ea\ee
	where $L_a $, $\tilde L_j $, $R_a $, $\tilde R_j$ are the components of the charges in the bases \eqref{eq:AdS3-tt=} and \eqref{eq:AdS3-ss=}
	\be \label{eq:AdS3-components}
		L_a =\la\,\fkT_a \,L\,\ra~,\quad \tilde L_j=\la\,{\tilde\fkT}_j\,\tilde L\,\ra~,\quad
		R_a =\la\,\fkT_a \,R\,\ra~, \quad \tilde R_j=\la\,{\tilde\fkT}_j\,\tilde R\,\ra~.
	\ee

	Since $R=R^a \fkT_a $ and $\tilde R=\tilde R_j\tilde\fkT_j$,
	the mass-shell condition \eqref{eq:AdS3-mass-shell} can be written as
	\be\label{eq:AdS3-constraint}
		R_a  R^a +\tilde R_j \tilde R_j+\m_{0}^2=0~,
	\ee
	and it obviously has vanishing Poisson brackets with components \eqref{eq:AdS3-components}.
	Hence, the components are gauge invariant and, therefore,
	the Poisson brackets algebra \eqref{eq:AdS3-L-R PB} will be preserved
	after a gauge fixing.

	Let us choose the gauge $\xi=1$ and consider a solution of \eqref{eq:AdS3-Hamilton eq} in the ${\rm SU}(1,1)$ part,
	\be\label{eq:AdS3-SU11 particle solution}
	g=e^{\m\tau\fkT_{0}}~, \qquad\qquad\qquad R=\m \fkT_{0}~,
	\ee
	corresponding to a ${\rm AdS}_3$ particle of mass $\m$ in the rest frame.
	The isometry transformations of \eqref{eq:AdS3-SU11 particle solution} provide a class of solutions
	parameterized by $\mu$ and the group variables,
	\be\label{eq:AdS3-isometry map of solutions}
		g= g_{l}\,e^{\m\tau\fkT_{0}}\,g_{r}~, \qquad\qquad R=g_{r}^{_{-1}}\m \fkT_{0}\,g_{r}~.
	\ee
	To find the Poisson bracket structure on the space of
	parameters, we calculate the ${\rm SU}(1,1)$ part of the pre-symplectic form \eqref{eq:AdS3-particle pre-symplectic form}.
	For fixed $\tau$ this calculation yields
	\be\label{eq:AdS3-calculation of AdS one-form}
	\th=\la R g^{_{-1}}\text{d} g\ra=\m\la\, {\fkT_{0}}\, g_{l}^{_{-1}}\,\text{d} g_{l}\,\ra+
	\m\la\, {\fkT_{0}}\, \text{d} g_{r}\,g_{r}^{_{-1}}\,\ra-\tau\m\text{d}\m~,
	\ee
	and we can neglect the exact form $-\tau\m\text{d}\m$. Taking the standard Gauss decomposition,
	\be\label{eq:AdS3-parametrization of g}
		g_{l}=e^{\a_{l}{\fkT_{0}}}\,e^{\g_{l}{\fkT_{1}}}\,e^{\b_{l}{\fkT_{0}}}~,\qquad\qquad
			g_{r}=e^{\b_{r}{\fkT_{0}}}\,e^{\g_{r}{\fkT_{1}}}\,e^{\a_{r}{\fkT_{0}}}~,
	\ee
	the `left' term of the one-form \eqref{eq:AdS3-calculation of AdS one-form} becomes
	\be\label{eq:AdS3-L-one-form}
		\m\la\, \fkT_{0}\, g_{l}^{_{-1}}\,\text{d} g_{l}\,\ra=
		\m\Big(\la\, e^{\g_{l}\fkT_{1}}\,\fkT_{0}\,
		e^{-\g_{l}\fkT_{1}}\, \fkT_{0}\,\ra\,\text{d}\a_{l}
		- \text{d}\b_{l}
		\Big)~.
	\ee
	with the coefficient of $\text{d} \g_{l}$ being $\la\,\fkT_{0}\,\fkT_{1}\,\ra=0$. Similarly, the 'right' term in \eqref{eq:AdS3-calculation of AdS one-form} reads
	\be\label{eq:AdS3-R-one-form}
		\m\la\, \fkT_{0}\,\text{d} g_{r}\, g_{r}^{_{-1}}\,\ra=
		\m\Big(\la\, e^{\g_{r}\fkT_{1}}\,\fkT_{0}\,
		e^{-\g_{r}\fkT_{1}}\, \fkT_{0}\,\ra\,\text{d}\a_{r}
		- \text{d}\b_{r}
		\Big)~.
	\ee
	Taking into then account $\la\,\fkT_a\,\fkT_b\,\ra = \eta_{a b}$ and the adjoint transformation properties
	\be\label{eq:AdS3-tr of b}
		e^{\g\fkT_{1}}\,\fkT_{0}\,e^{-\g\fkT_{1}}=\cosh(2\g)\,\fkT_{0}+\sinh(2\g)\,\fkT_{2}~,
	\quad~
		e^{\a\fkT_{0}}\,\fkT_{2}\,e^{-\a\fkT_{0}}=
		\cos(2\a)\,\fkT_{2}+\sin(2\a)\,\fkT_{1}~,
	\ee
	one can reduce \eqref{eq:AdS3-calculation of AdS one-form} to a canonical one-form
	\be\label{eq:AdS3-canonical AdS one-form}
	\th=\m\text{d}\varphi+H_{l}\text{d}\phi_{l}+H_{r}\text{d}\phi_{r}~,
	\ee
	where we defined
	\bea\label{eq:AdS3-angle variables}
		&\varphi=-(\a_{l}+\b_{l}+\a_{r}+\b_{r})~, \qquad \phi_{l}=\frac{\pi}{2}-2\a_{l}~, \qquad \phi_{r}={\pi}-2\a_{r}~,&\\[.2em]
	\label{eq:AdS3-H_l,r}
		&H_{l}=\frac{\m}{2}\,\big(\cosh (2\g_{l}) -1_{_{\,\!}}\big)~,\qquad\quad
			H_{r}=\frac{\m}{2}\,\big(\cosh (2\g_{r}) -1_{_{\,\!}}\big)~.&
	\eea
	Note that we subtracted 1 from $\cosh(2\g)$ to have $H \geq 0$.

	The conserved Noether charges constructed from \eqref{eq:AdS3-isometry map of solutions} 
	are given by
	\be\label{eq:AdS3-particle L and R}
		L=\mu\, e^{\a_{l}{\fkT_{0}}}\,e^{\g_{l}{\fkT_{1}}}\,{\fkT_{0}}\,
		e^{-\g_{l}{\fkT_{1}}}\,e^{-\a_{l}{\fkT_{0}}}~, \qquad
		R=\mu\, e^{-\a_{r}{\fkT_{0}}}\,e^{-\g_{r}{\fkT_{1}}}\,{\fkT_{0}}\,
		e^{\g_{r}{\fkT_{1}}}\,e^{\a_{r}{\fkT_{0}}}~,
	\ee
	which due to \eqref{eq:AdS3-tr of b} take the form
	\be\ba\label{eq:AdS3-particle L and R=}
		&L= \m\,\Big(\!\cosh(2\g_{l})\,\fkT_{0}+\sinh(2\g_{l}) \big(\cos(2\a_{l})\,\fkT_{2} + \sin(2\a_{l})\,\fkT_{1}\big)_{_{\!}}\Big)~,\\
		&R= \m\,\Big(\!\cosh(2\g_{r})\,\fkT_{0}-\sinh(2\g_{l}) \big(\cos(2\a_{r})\,\fkT_{2} - \sin(2\a_{r})\,\fkT_{1}\big)_{_{\!}}\Big)~.
	\ea\ee
	Hence, the angle variables $\pm\phi_{l}$ and $\pm\phi_{r}$ 
	correspond to the phases of $L_\pm = \frac{1}{2}(L_1 \pm i L_2)$ and $R_\pm = \frac{1}{2}(R_2 \pm i R_1)$, respectively, and we find
	\be\ba\label{eq:AdS3-AdS dynamical integrals}
		&L^0=\m + 2H_{l}~,& \qquad     &R^0=\m + 2H_{r}~,&\\
		&L_\pm=\sqrt{\m H_{l}+H_{l}^2}\,e^{\pm i\phi_{l}}~,&\qquad &R_\pm=\sqrt{\m H_{r}+H_{r}^2}\,e^{\pm i\phi_{r}}~,&
	\ea\ee

	We can follow the same steps for ${\rm SU}(2)$, where we consider the isometry group orbit of the solution $\tilde g=e^{\tilde\m\tau\tilde\fkT_{3}}$. Comparing the adjoint transformation,
	\be\label{eq:AdS3-tr of b tilde}
		e^{\tilde\g\tilde\fkT_{1}}\,\tilde\fkT_{3}\,e^{-\tilde\g\tilde\fkT_{1}}
			= \cos(2\tilde\g)\,\tilde\fkT_{3} + \sin(2\tilde\g)\,\tilde\fkT_{2}~,
		\qquad
		e^{\tilde\a\tilde\fkT_{3}}\ \tilde\fkT_{2}\,e^{-\tilde\a_{l}\tilde\fkT_{0}}
		= \cos(2\tilde\a)\,\tilde\fkT_{2}+\sin(2\tilde\a)\,\tilde\fkT_{1}~,
	\ee
	with \eqref{eq:AdS3-tr of b} we see that the reduced one-form is obtained from the one-form for ${\rm SU}(1,1)$ by substituting untilded with tilded parameters along with the replacements
	\be\label{eq:AdS3-replacement rule}
		\cosh(2\g) \mapsto \cos(2\tilde\g) ~, \qquad 
			\sinh(2\g) \mapsto \sin2\tilde\g ~, \qquad 
			\fkT_{0}\mapsto \tilde\fkT_{3}~.
	\ee
	By this we get
	\be\label{eq:AdS3-su(2) one-form }
		\tilde\theta
			= \tilde\m(\text{d}\tilde\b_{l}+\text{d}\tilde\b_{r})
			+ \tilde\m \cos(2\tilde\g_{l})\,\text{d}\tilde\a_{l}
			+ \tilde\m \cos(2\tilde\g_{r})\,\text{d}\tilde\a_{r}~,
	\ee
	which takes the canonical form 
	\be\label{eq:AdS3-canonical S one-form}
		\tilde\th = \la \tilde R \tilde g^{_{-1}} \mathrm{d}\tilde g\ra=
		\tilde\m\text{d}\tilde\varphi+\tilde H_{l}\text{d}\tilde\phi_{l}+\tilde H_{r}\text{d}\tilde\phi_{r}~,
	\ee
	with the definitions
	\bea\label{eq:AdS3-S canonical coordinates}
		&\tilde\varphi 
			= \tilde\a_{l}+\tilde\b_{l}+\tilde\a_{r}+\tilde\b_{r}~, \qquad
		\tilde\phi_{l}=\frac{\pi}{2}-2\tilde\a_{l}~,
		\qquad \tilde\phi_{r}={\pi}-2\tilde\a_{r}~,&\\[.2em]
		\label{eq:AdS3-tilde H_l,r}
		&\tilde H_{l}=\dfrac{\tilde{\m}}{2}\,[1-\cos (2\tilde\g_{l})]~,\qquad\quad  \tilde H_{r}=\dfrac{\tilde{\m}}{2}\,[1-\cos (2\g_{r})]~.&
	\eea
	Similarly to \eqref{eq:AdS3-AdS dynamical integrals}, the ${\rm SU}(2)$ Noether charges $\tilde L_3$,
	$\tilde L_\pm=\frac{1}{2}(\tilde L_1\pm i\tilde L_2)$ and
	$\tilde R_3$, $\tilde R_\pm=\frac{1}{2}(\tilde R_2\pm i\tilde R_1)$ are parametrized by the canonical coordinates \eqref{eq:AdS3-S canonical coordinates} as
	\be\ba\label{eq:AdS3-S dynamical integrals}
	&\tilde L_3=\tilde\m-2\tilde H_{l}~,& \qquad &\tilde R _3=\tilde\m-2\tilde H_{r}~,&\\
	&\tilde L_\pm=\sqrt{\tilde\m \tilde H_l -\tilde H_{l}^2}\,\,e^{\pm i\tilde\phi_{l}}~,&\qquad &\tilde R_\pm=
	\sqrt{\tilde\m \tilde H_{r}-\tilde H_{r}^2}\,\,e^{\pm i\tilde\phi_{r}}~.&
	\ea\ee

	From the canonical variables $H \geq 0$ and $\phi \in {\rm S}^1$ one naturally defines creation-annihilation variables as
	\be\label{eq:AdS3-a,a^+}
	a^\dag=\sqrt{H}\,\,e^{i\phi}~, ~\qquad\qquad  a=\sqrt{H}\,\,e^{-i\phi}~.~~
	\ee

	The form of the functions \eqref{eq:AdS3-AdS dynamical integrals} and \eqref{eq:AdS3-S dynamical integrals} then dictates the realization of the isometry group generators in terms of creation-annihilation operators, which is known as the Holstein-Primakoff transformation \cite{Holstein:1940zp, Dzhordzhadze:1994np}. Thus, we have
	\begin{align}\nn
		\qquad\qquad&L^0=\m + 2a_{l}^\dag a_{l}~,& 
			&R^0=\m + 2a_{r}^\dag a_{r}~,&\\
		\label{eq:AdS3-AdS operators}
		&L_+=a_{l}^\dag\sqrt{\m +a_{l}^\dag a_{l}}~,&
			&R_+=a_{r}^\dag\sqrt{\m+a_{r}^\dag a_{r}}~,&\\
		\nn
		&L_-=\sqrt{\m +a_{l}^\dag a_{l}}\,\,a_{l}~,&
			&R_-=\sqrt{\m+a_{r}^\dag a_{r}}\,\,a_{r}~,&\\[.4 em]
		\nn
		&\tilde L_3=\tilde\m-2\tilde a^\dag_{l}\,\tilde a_{l}~,&
			&\tilde R _3 = \tilde\m-2\tilde a^\dag_{r}\,\tilde a_{r}~,&
			\\
		\label{eq:AdS3-S operators}
		&\tilde L_+=\tilde a_{l}^\dag\sqrt{\tilde\m -\tilde a^\dag_{l}\,\tilde a_{l}}~,&
			&\tilde R_+=\tilde a^\dag_{r}\sqrt{\tilde \m-
		\tilde a^\dag_{r}\,\tilde a_{r}}~,&\\
		\nn
		&\tilde L_-=\sqrt{\tilde\m -\tilde a^\dag_{l}\,\tilde a_{l}}\,\,\tilde a_{l}~,&
			&\tilde R_-=\sqrt{\tilde\m-\tilde a^\dag_{r}\,\tilde a_{r}}\,\,\tilde a_{r}~.&
	\end{align}
	These yield a representation of $\su_l(1,1)\oplus \su_r(1,1)\oplus \su_l(2)\oplus \su_r(2)$ with basis vectors
	\be
		\ket{\m,\tilde\m\,;\,n_l, n_r, \tilde{n}_l, \tilde{n}_r}
			= \ket{\m, n_{l}} \ket{\m, n_{r}} \ket{\tilde\m,\tilde n_{l}} \ket{\tilde \m, \tilde n_{r}}~,
	\ee
	where $n_{l}$, $n_{r}$, $\tilde\m$, $\tilde n_{l}$, $\tilde n_{r}$ are nonnegative integers and
	$\tilde n_{l}\leq \tilde\m$, $\tilde n_{r}\leq\tilde \m$.
	The representation is characterized by the Casimir numbers
	\be\ba\label{eq:AdS3-Casimir}
		&C_{\rm AdS} = -L_a L^a = -R_a R^a = \m(\m-2)~,& \\
		&\quad~ C_{\rm S}=\tilde L_j \tilde L_j=\tilde R_j \tilde R_j=\tilde\m(\tilde\m+2)~.&
	\ea\ee
	This numbers are related through the mass-shell condition \eqref{eq:AdS3-constraint}
	\be\label{eq:AdS3-mass-shell q}
		C_{\rm AdS}=C_{\rm S}+\m_{0}^2~,
	\ee
	and we find
	\be\label{eq:AdS3-minimal energy}
		\m=1+\sqrt{\m_{0}^2+\tilde\m(\tilde\m+2)+1}~.
	\ee

	Since translations along the ${\rm AdS}_3$ time direction correspond to rotations in the $(X^0, X^{0'})$ plane the energy operator is given by
	\be\label{eq:AdS3-energy}
		E=\frac{1}{2}\left(L^0+R^0\right)~.
	\ee
	and from \eqref{eq:AdS3-AdS operators} we obtain the energy spectrum
	\be\label{eq:AdS3-energy spectrum}
		E=\m+n_{l}+n_{r}~.
	\ee
	Here, $\m$ is defined by \eqref{eq:AdS3-minimal energy} and corresponds to the lowest energy level for a given orbital momentum $\tilde\m$ on ${\rm S}^3$. Equations \eqref{eq:AdS3-minimal energy} and \eqref{eq:AdS3-energy spectrum} reproduces the result obtained in the covariant quantization or in the static gauge approach \cite{Dorn:2010wt}.

\section{The Spinning String in  SU(1,1)\texorpdfstring{$\times$}{x}SU(2)} \label{sec:AdS3-SpinString}
	In this section we use a similar scheme to calculate the energy spectrum of ${\rm AdS}_3 \times {\rm S}^3$ string solutions.

	The Polyakov action for the ${\rm SU}(1,1)\times {\rm SU}(2)$ string is given by
	\be\label{eq:AdS3-Polyakov action}
		S=-\frac{\sqrt\l}{4\pi}\int  \text{d}\tau\,\text{d}\s\,\,
		\sqrt{-h}\,h^{\a\b}\,\Big(\la\,g^{_{-1}} \p_\a g\, g^{_{-1}} \p_\b g\,\ra +
		\la\,\tilde g^{_{-1}} \p_\a \tilde g\, \tilde g^{_{-1}} \p_\b \tilde g\,\ra\Big)~,
	\ee
	where $\l$ is a dimensionless coupling constant. In analogy to the case of the particle, for the closed string this action is equivalent to
	\begin{align} \label{eq:AdS3-Action}
		S=&\int \text{d}\tau \int_0^{2\pi} \frac{\text{d}\s}{2\pi}\,\bigg(\la \calR\,g^{_{-1}}\dot{g}\ra 
			+ \la \tilde \calR\, \tilde g^{_{-1}}\dot{\tilde g}\ra-\xi_2\Big(\la \calR\,g^{_{-1}}{g}'\ra 
			+ \la \tilde \calR\, \tilde g^{_{-1}}{\tilde g}'\ra\Big) \\[.2em] \nn
		&\qquad\qquad\quad
			-\frac{\xi_1}{2}\Big(\frac{\la\, \calR\,\calR\,\ra
			+ \la \tilde \calR\,\tilde \calR\ra}{\sqrt\l}
			+\sqrt\l\la(g^{_{-1}}{g'})^2\ra
			+ \sqrt\l\la(\tilde g^{_{-1}}{\tilde g'})^2\ra\Big)
	\bigg)~.
	\end{align}
	The Lagrange multipliers $\xi_1$ and $\xi_2$  are related to the worldsheet metric by
	\be\label{eq:AdS3-xi-1,2}
		\xi_1=-\frac{1}{\sqrt{-h}\,h^{\tau\tau}}~, \qquad\qquad \xi_2=-\frac{h^{\tau\s}}{h^{\tau\tau}},\qquad
	\ee
	see also \eqref{eq:SF-SS-Lang}, and their variations provide the Virasoro constraints
	\be\ba\label{eq:AdS3-Virasoro}
		&\la\, \calR\,\calR\,\ra+\la \tilde \calR\,\tilde \calR\ra
			+ \l\la(g^{_{-1}}{g'})^2\ra
			+ \l\la({\tilde g}^{_{-1}}{{\tilde g}'})^2\ra=0~, \\[.2em]
		&\qquad\qquad\quad \la \calR\,g^{_{-1}}{g}'\ra
			+ \la \tilde \calR {\tilde g}^{_{-1}}\,{\tilde g}'\ra=0~.
	\ea\ee
	Taking conformal gauge for the worldsheet metric, by \eqref{eq:AdS3-xi-1,2} we have $\xi_1=1$ and $\xi_2=0$. With this the equations of motion obtained from \eqref{eq:AdS3-Action} become
	\be\ba\label{eq:AdS3-EOM}
		\quad\qquad\qquad&\sqrt\l\,\,g^{_{-1}}\dot{g}=\calR~,&\qquad\qquad
			&\dot \calR=\sqrt\l(g^{_{-1}}{g'})'~,&\qquad\qquad\\[.2em]
		&\sqrt\l\,\,\tilde g^{_{-1}}\dot{\tilde g}=\tilde \calR~,&
			&\dot{\tilde \calR}= \sqrt\l({\tilde g}^{_{-1}}{\tilde g'})'~,&
	\ea\ee
	and they are equivalent to
	\be\label{eq:AdS3-Lagrange eq.}
	\p_\tau\left(g^{_{-1}}\dot{g}\right)=\p_\s\left(g^{_{-1}}{g}'\right)~,
	\qquad\quad \p_\tau\left({\tilde g}^{_{-1}}\dot{\tilde g}\right)=\p_\s\left({\tilde g}^{_{-1}}{\tilde g}'\right)~.
	\ee

	We now consider the following solution of these equations \cite{Jorjadze:2012rj}
	\begin{align}\label{eq:AdS3-AdS string solution}
		&g = \begin{pmatrix}
								\cosh\vartheta \,e^{i(e\tau+m\s)}~ & \sinh\vartheta \,e^{i(p\tau+n\s)}\\
								\sinh\vartheta \,e^{-i(p\tau+n\s)}& \cosh\vartheta \,e^{-i(e\tau+m\s)}
							\end{pmatrix}~,\\
		\label{eq:AdS3-S string solution}
		&\tilde g=\begin{pmatrix}
              ~\cos\tilde\vartheta \,e^{i(\tilde e\tau+\tilde m\s)} & i\sin\tilde\vartheta \,e^{i(\tilde p\tau+\tilde n\s)}\\
              i\sin\tilde\vartheta \,e^{-i(\tilde p\tau+\tilde n\s)}& \cos\tilde\vartheta \,e^{-i(\tilde e\tau+\tilde m\s)}
            \end{pmatrix}~,
	\end{align}
	with the parameters fulfilling
	\be\label{eq:AdS3-conditions for e,p}
		p^2-e^2=n^2-m^2~, \qquad\qquad \tilde p^2-\tilde e^2=\tilde n^2-\tilde m^2~,
	\ee
	which turns out to be the multi-spin string solution \cite{Frolov:2003qc, Frolov:2003tu, Arutyunov:2003za}.

	To check that this ansatz satisfies the equations of motion \eqref{eq:AdS3-Lagrange eq.}, one can calculate left-invariant currents and finds
	\begin{align}
		\label{eq:AdS3-AdS_R} 
		&\ba
		&g^{_{-1}}\dot{g}=\frac{i}{2}
			\begin{pmatrix}
				(e-p) + (e+p)\cosh2\vartheta  
				& (e+p)\,e^{-i\,\om_-} \sinh 2\vartheta \\
				-(e+p)\,e^{i\,\om_-} \sinh 2\vartheta 
				& -(e-p) - (e+p)\cosh2\vartheta
			\end{pmatrix}, \\[.2em]  
		&g^{_{-1}}{g}'=\frac{i}{2}
			\begin{pmatrix}
				(m-n) + (m+n)\cosh2\vartheta 
				& (m+n)\,e^{-i\,\om_-} \sinh 2\vartheta \\
				-(m+n)\,e^{i\,\om_-} \sinh 2\vartheta
				& -(m-n)-(m+n)\cosh2\vartheta
			\end{pmatrix},
		\ea \\[.6em]
		\label{eq:AdS3-S_R}
		&\ba
		&\tilde g^{_{-1}}\dot{\tilde g}=\frac{i}{2}
			\begin{pmatrix}
				(\tilde e-\tilde p) + (\tilde e+\tilde p)\cos2\tilde\vartheta
					&i(\tilde e+\tilde p)\sin 2\tilde\vartheta \,e^{-i\,\tilde{\om}_-}\\
				-i(\tilde e+\tilde p)\sin 2\tilde\vartheta\,e^{i\,\tilde{\om}_-}
					&-(\tilde e-\tilde p)-(\tilde p+\tilde e)\cos2\tilde\vartheta
			\end{pmatrix}, \\[.2em] 
		&\tilde g^{_{-1}}{\tilde g}'=\frac{i}{2}
			\begin{pmatrix}
				(\tilde m-\tilde n) + (\tilde m+\tilde n)\cos2\tilde\vartheta&
					i(\tilde m+\tilde n)\,e^{-i\,\tilde{\om}_-}\sin 2\tilde\vartheta\\
				-i(\tilde m+\tilde n)\,e^{ i\,\tilde{\om}_-}\sin 2\tilde\vartheta&
					-(\tilde m-\tilde n) - (\tilde m+\tilde n)\cos2\tilde\vartheta
			\end{pmatrix},
		\ea
	\end{align}
	with the abbreviations $\om_\pm =(e \pm p)\tau + (m \pm n)\s$ and  $\tilde{\om}_\pm =(\tilde e\pm\tilde p)\tau+(\tilde m\pm\tilde n)\s$.

	Since the diagonal components of these matrices are constants, one has to check \eqref{eq:AdS3-Lagrange eq.} for the the off-diagonal entries only, giving the conditions \eqref{eq:AdS3-conditions for e,p}.

	Similar calculations for the right-invariant currents 
	yield
	\be\label{eq:AdS3-dot L}
		\ba
		&\dot{g}\,g^{_{-1}}=\frac{i}{2}
			\begin{pmatrix}
				(e+p) + (e-p)\cosh2\vartheta
				&-(e-p)\sinh 2\vartheta\,e^{i\,\om_+}\\
				(e-p)\sinh 2\vartheta\,e^{-i\,\om_+}
				&-(e+p) -(e-p)\cosh2\vartheta
            \end{pmatrix},\quad~~\\[.2em]
		&\dot{\tilde g}\,\tilde g^{_{-1}}=\frac{i}{2}
			\begin{pmatrix}
				(\tilde e+\tilde p) + (\tilde e-\tilde p)\cos2\tilde\vartheta
				&-i(\tilde e-\tilde p)\,e^{i\,\tilde{\om}_+} \sin 2\tilde\vartheta\\
				i(\tilde e-\tilde p)\,e^{-i\,\tilde{\om}_+}\sin 2\tilde\vartheta
				&-(\tilde e+\tilde p) - (\tilde e-\tilde p)\cos2\tilde\vartheta
			\end{pmatrix}.
	\ea\ee

	The induced metric tensor components obtained from \eqref{eq:AdS3-AdS_R}-\eqref{eq:AdS3-S_R} read
	\begin{align}\nn
		\,&\la (g^{_{-1}}\dot{g})^2\ra = p^2\sinh^2\vartheta-e^2\cosh^2\vartheta~,&
		&\!\!\la (\tilde g^{_{-1}}\dot{\tilde g})^2\ra = \tilde p^2\sin^2\tilde\vartheta + \tilde e^2\cos^2\tilde\vartheta~,& \\
		\label{eq:AdS3-Tr dot-pime}
		&\la (g^{_{-1}}{g}')^2\ra = n^2\sinh^2\vartheta-m^2\cosh^2\vartheta~,& 
		&\!\!\la (\tilde g^{_{-1}}{\tilde g}')^2\ra = \tilde n^2\sin^2\tilde\vartheta+\tilde m^2\cos^2\tilde\vartheta~,& \\ 
		\nn
		&\la g^{_{-1}}\dot{g} g^{_{-1}}{g}')\ra=n p\sinh^2\vartheta-m e\cosh^2\vartheta~,&
		&\!\!\la\tilde g^{_{-1}}\dot{\tilde g} \tilde g^{_{-1}}{\tilde g}'\ra=\tilde n\tilde p\sin^2\tilde\vartheta+\tilde
		m\tilde e\cos^2\tilde\vartheta~.&
	\end{align}
	As the matrices $\calR$ and $\tilde \calR$ are defined by the Hamilton equations \eqref{eq:AdS3-EOM}, the Virasoro constraints \eqref{eq:AdS3-Virasoro} are expressed through these and one gets additional conditions
	\be\ba\label{eq:AdS3-Virasoro 1}
		&(e^2+m^2)\cosh^2\vartheta-(p^2+n^2)\sinh^2\vartheta
			=(\tilde e^2 + \tilde m^2)\cos^2\tilde\vartheta
			+(\tilde p^2 + \tilde n^2)\sin^2\tilde\vartheta~,\\
		&\qquad\qquad\qquad 
			m e \cosh^2\vartheta -np\sinh^2\vartheta
			= \tilde m \tilde e\cos^2\tilde\vartheta
			+ \tilde n\tilde p\sin^2\tilde\vartheta~.
	\ea\ee

	Since we consider a closed string in ${\rm SU}(1,1)\times {\rm SU}(2)$, the parameters $m$, $n$, $\tilde m$, $\tilde n$ have to be integer.
	However, if we unwrap the time coordinate, the polar angle in the $(X^0, X^{0'})$ plane, it has to be periodic in $\s$ itself. This is obviously achieved for $m=0$ only, which is assumed below.

	Thus, our solutions are parameterized by three winding numbers and six continuous variables, which satisfy the four conditions in \eqref{eq:AdS3-conditions for e,p}-\eqref{eq:AdS3-Virasoro 1}.
	Hence, for given winding numbers, we have a two parameter family of solutions.\footnote{Note that the particle solutions in ${\rm SU}(1,1)\times {\rm SU}(2)$ were parameterized by one variable $\tilde\m$.}

	Similarly to the particle dynamics, we consider the isometry group orbits of the solutions, with the aim to find their Hamiltonian description and quantization.
	For this purpose we analyze the pre-symplectic form defined by \eqref{eq:AdS3-Action},
	\be\label{eq:AdS3-string presymplectic}
		\Theta=\int_0^{2\pi} \frac{\text{d}\s}{2\pi}\,\Big(\la \calR\,g^{_{-1}}\text{d}{g}\ra +
		\la \tilde \calR\, \tilde g^{_{-1}}\text{d}{\tilde g}\ra\Big)~.
	\ee
	To calculate this one-form on the space of orbits one has to make the  replacements
	\be\label{eq:AdS3-rplacements}
		\calR\, \mapsto \, \sqrt\l\,g_{r}^{_{-1}}\,g^{_{-1}}\dot g \,g_{r}~, \qquad g \, \mapsto \, g_{l}\,g\,g_{r}~,
		\qquad g^{_{-1}}\, \mapsto \, g_{r}^{_{-1}}g^{_{-1}}g_{l}^{_{-1}}~,
	\ee
	and similarly for the ${\rm SU}(2)$ term, and then identify $g$ with \eqref{eq:AdS3-AdS string solution} and $\tilde g$ with \eqref{eq:AdS3-S string solution}, respectively. For the ${\rm SU}(1,1)$ part this yields
	\be\label{eq:AdS3-AdS strinng one-form 1}
		\theta=\la\, L\, g_{l}^{_{-1}}\,\text{d} g_{l}\,\ra+
		\la\, R\, \text{d} g_{r}\,g_{r}^{_{-1}}\,\ra+
		\sqrt\l\int_0^{2\pi} \frac{\text{d}\s}{2\pi}\,\la \,g^{_{-1}}\dot g \,g^{_{-1}}\text{d}g\,\ra~,
	\ee
	where $L$ and $R$ are the Noether charges related to the isometries \eqref{eq:AdS3-isometry tr} as
	\be\label{eq:AdS3-AdS string charges}
		L=\sqrt\l\int_{0}^{2\pi}\frac{\text{d}\s}{2\pi}\,\,\dot{g}\,g^{_{-1}}~, \qquad\quad
			R=\sqrt\l\int_{0}^{2\pi}\frac{\text{d}\s}{2\pi}\,\,g^{_{-1}}\dot{g}~,
	\ee
	and the differential of $g$ in the last term of \eqref{eq:AdS3-AdS strinng one-form 1} is taken with respect to the parameters of the solution \eqref{eq:AdS3-AdS string solution}.

	Note now that
	\be\ba
		&\la\, g^{_{-1}}\dot{g} \,\,g^{_{-1}}\partial_\vartheta{g})\,\ra=0~,&
		\qquad
		&\la\, g^{_{-1}}\dot{g} \,\left(p\,g^{_{-1}}\partial_e{g}+e\,g^{_{-1}}\partial_p{g})\right)\,\ra=-ep\tau~,&
		\\
		\label{eq:AdS3-one-form calculations}
		&\la\,\tilde g^{_{-1}}\dot{\tilde g} \,\,\tilde g^{_{-1}}\partial_{\tilde\vartheta}{\tilde g})\,\ra=0~, &
		\qquad
		&\la\, \tilde g^{_{-1}}\dot{\tilde g} \,\left(\tilde p\,\tilde g^{_{-1}}\partial_{\tilde e}{\tilde g}+
		\tilde e\,\tilde g^{_{-1}}\partial_{\tilde p}{\tilde g})\right)\,\ra=\tilde e\tilde p\tau~.
	\ea\ee
	Here $g$ and $\tilde g$ are given again by \eqref{eq:AdS3-AdS string solution}-\eqref{eq:AdS3-S string solution} and the
	calculation is straightforward.
	Taking into account then the constraint \eqref{eq:AdS3-conditions for e,p} between the parameters $e$ and $p$,
	the last term in \eqref{eq:AdS3-AdS strinng one-form 1} becomes an exact form -$\tau e\text{d}e$, which shows that it can be neglected.
	The ${\rm SU}(2)$ part is computed in a similar way and alltogether we find the one-form
	\be\label{eq:AdS3-final one-form}
		\Theta=\la\, L\, g_{l}^{_{-1}}\,\text{d} g_{l}\,\ra
			+ \la\, R\, \text{d} g_{r}\,g_{r}^{_{-1}}\,\ra
			+ \la\, \tilde L\,\tilde g_{l}^{_{-1}}\,\text{d}\tilde g_{l}\,\ra
			+ \la\, \tilde R\, \text{d}\tilde g_{r}\,\tilde g_{r}^{_{-1}}\,\ra~,
	\ee
	where $\tilde L$ and $\tilde R$ are the Noether charges similar to \eqref{eq:AdS3-AdS string charges},
	\be\label{eq:AdS3-S string charges}
		\tilde L=\sqrt\l\int_{0}^{2\pi}\frac{\text{d}\s}{2\pi}\,\,\dot{\tilde g}\,\tilde g^{_{-1}}~, \qquad\quad
			\tilde R=\sqrt\l\int_{0}^{2\pi}\frac{\text{d}\s}{2\pi}\,\,\tilde g^{_{-1}}\dot{\tilde g}~.
	\ee
	These charges are easily calculable by the currents given above. However, their matrix form depends on the winding numbers and one has to distinguish between the cases $n\neq 0$ and $n=0$ for ${\rm SU}(1,1)$ as well as $\tilde m^2\neq\tilde n^2$ and $\tilde m^2=\tilde n^2$ for ${\rm SU}(2)$.

	Let us consider the case $n\neq 0$ and $\tilde m^2\neq\tilde n^2$.
	The integration of the off-diagonal terms of the currents \eqref{eq:AdS3-AdS_R}-\eqref{eq:AdS3-dot L} vanish and we obtain
	\be\label{eq:AdS3-string charges 1}
		L=\m_{l}\,\fkT_{0}~, \qquad \tilde L=\tilde\m_{l}\,\tilde\fkT_{3}~, \qquad\qquad
			R=\m_{r}\,\fkT_{0}~, \qquad \tilde R=\tilde\m_{r}\,\tilde\fkT_{3}~,
	\ee
	where
	\begin{align}\label{eq:AdS3-AdS left-right masses}
	\qquad&\m_{l}=\sqrt\l\,(e\cosh^2\vartheta-p\sinh^2\vartheta)~,&\qquad
		&\m_{r}=\sqrt\l\,(e\cosh^2\vartheta+p\sinh^2\vartheta)~,&\\
	\label{eq:AdS3-S left-right masses}
	&\tilde\m_{l}=\sqrt\l\,(\tilde e\cos^2\tilde\vartheta+\tilde p\sin^2\tilde\vartheta)~,&
		&\tilde\m_{r}=\sqrt\l\,(\tilde e\cos^2\tilde\vartheta-\tilde p\sin^2\tilde\vartheta)~.&
	\end{align}
	We can assume that the numbers $\m_{l}$, $\m_{r}$, $\tilde\m_{l}$, $\tilde\m_{r}$ are non-negative.

	Similarly to the particle case, the one-form \eqref{eq:AdS3-final one-form} then becomes
	\be\label{eq:AdS3-AdS strinng one-form}
		\Theta=\m_{l}\la\, {\fkT_{0}}\, g_{l}^{_{-1}}\,\text{d} g_{l}\,\ra+
		\m_{r}\la\, {\fkT_{0}}\, \text{d} g_{r}\,g_{r}^{_{-1}}\,\ra+
		\tilde\m_{l}\la\, {\fkT_{3}}\,\tilde g_{l}^{_{-1}}\,\text{d}\tilde g_{l}\,\ra+
		\tilde\m_{r}\la\, {\fkT_{3}}\, \text{d}\tilde g_{r}\,\tilde g_{r}^{_{-1}}\,\ra~,
	\ee
	and the same parametrization as in \eqref{eq:AdS3-parametrization of g} leads to the canonical one-form
	\be
		\Theta =
			\m_{l}\text{d}\varphi_{l}
			+ H_{l}\text{d}\phi_{l} + \m_{r}\text{d}\varphi_{l}
			+ H_{r}\text{d}\phi_{r} + \tilde\m_{l}\text{d}\tilde\varphi_{l}
			+ \tilde H_{l}\text{d}\tilde\phi_{l} 
			+ \tilde\m_{r}\text{d}\tilde\varphi_{r}
			+ \tilde H_{r}\text{d}\tilde\phi_{r}~.
	\ee
	The components of the symmetry generators have the same form as in \eqref{eq:AdS3-AdS dynamical integrals} and \eqref{eq:AdS3-S dynamical integrals}
	\begin{align}
		&\ba\label{eq:AdS3-AdS string dynamical integrals}
			&L^0=\m_{l} + 2H_{l}~,& \qquad     &R^0=\m_{r} + 2H_{r}~,&\\
			&L_\pm=\sqrt{\m_{l} H_{l}+H_{l}^2}\,e^{\pm i\phi_{l}}~,
			\,&\qquad &R_\pm=
			\sqrt{\m_{r} H_{r}+H_{r}^2}\,e^{\pm i\phi_{r}}~,&
		\ea\\[.2em]
		&\ba\label{eq:AdS3-S string dynamical integrals}
			&\tilde L_3=\tilde\m_{l}-2\tilde H_{l}~,& \qquad &\tilde R _3=\tilde\m_{r}-2\tilde H_{r}~,&\\
			&\tilde L_\pm=\sqrt{\tilde\m_{l} \tilde H_l -\tilde H_{l}^2}\,\,e^{\pm i\tilde\phi_{l}}~,&\qquad &\tilde R_\pm=
			\sqrt{\tilde\m_{r} \tilde H_{r}-\tilde H_{r}^2}\,\,e^{\pm i\tilde\phi_{r}}~,&
		\ea
	\end{align}
	where now the Casimier numbers $\m_{l}$ and $\m_{r}$ as well as $\tilde\m_{l}$ and $\tilde\m_{r}$ are independent.

	Here, as in \eqref{eq:AdS3-AdS operators}-\eqref{eq:AdS3-S operators}, the Holstein-Primakoff transformation provides a realization of the isometry group generators with non-negative integers $\tilde\m_{l}$, $\tilde\m_{r}$.
	The energy given by \eqref{eq:AdS3-energy} is now obtained from \eqref{eq:AdS3-AdS string dynamical integrals} and  \eqref{eq:AdS3-AdS left-right masses} and its spectrum becomes
	\be\label{eq:AdS3-string spectrum}
		E=E_0+m_{l}+m_{r}~,
	\ee
	where $m_{l}$ and $m_{r}$ are non-negative integers and $E_0=\sqrt\l\,e\cosh^2\vartheta$ corresponds to the minimal energy for given $\tilde\m_{l}$, $\tilde\m_{r}$. To find the dependence of this term on $\tilde\m_{l}$, $\tilde\m_{r}$, and the winding numbers one has to use \eqref{eq:AdS3-S left-right masses} and the constraints
	\eqref{eq:AdS3-conditions for e,p}-\eqref{eq:AdS3-Virasoro 1}, which gives
	\be\label{eq:AdS3-e,p tilde}
		\tilde e=\f{1}{\sqrt\l}\,\f{\tilde\m_{l}+\tilde\m_{r}}{1+\cos2\tilde\vartheta}~, \qquad\qquad
		\tilde p=\f{1}{\sqrt\l}\,\f{\tilde\m_{l}-\tilde\m_{r}}{1-\cos2\tilde\vartheta}~.
	\ee
	Inserting these in \eqref{eq:AdS3-conditions for e,p}, one gets a fourth order equation for $\cos2\tilde\vartheta\,$,
	\be\label{eq:AdS3-eq for cos th}
		(\tilde\m_{l}+\tilde\m_{r})^2(1-\cos2\tilde\vartheta)^2 - (\tilde\m_{l} -\tilde\m_{r})^2(1+\cos2\tilde\vartheta)^2 =
		\l(\tilde m^2-\tilde n^2)(1-\cos^22\tilde\vartheta)^2.
	\ee
	The solutions of this equation and \eqref{eq:AdS3-e,p tilde} define the right hand sides of \eqref{eq:AdS3-Virasoro 1}
	as a function of the coupling constant and four integers $(\tilde\m_{l},\tilde\m_{r},\tilde m,\tilde n)$.
	Solving \eqref{eq:AdS3-Virasoro 1} for $e^2$ and $\sinh\vartheta$ one obtains third order equations, which ultimately give a rather complicated answer for $E_0$.
	Therefore, let us analyze this system at strong coupling, $\l \gg 1\,$.

	First we consider the case when both $\tilde m$ and $\tilde n$
	are non-zero and assume  $0<\tilde m^2<\tilde n^2$.
	Using \eqref{eq:AdS3-conditions for e,p} and \eqref{eq:AdS3-S left-right masses}, the system \eqref{eq:AdS3-Virasoro 1} can be written as
	\be\ba\label{eq:AdS3-Vir-1-1}
		&e^2-2n^2\sinh^2\vartheta=\tilde e^2+2\tilde n^2\sin^2\tilde\vartheta+\tilde m^2\cos2\tilde\vartheta~,\\
		&|n|\sqrt{e^2+n^2}\,\sinh^2\vartheta=|\tilde m(\tilde\m_l+\tilde\m_r)+\tilde n(\tilde\m_l-\m_r)|\l^{-{1}/{2}}~.
	\ea\ee

	At large $\l$, from \eqref{eq:AdS3-eq for cos th} and \eqref{eq:AdS3-e,p tilde} we find
	\be\label{eq:AdS3-large lambda}
		\cos2\tilde\vartheta=1-\f{|\tilde\m_{l}-\tilde\m_{r}|}{\sqrt{\tilde n^2-\tilde m^2}}\,\l^{-{1}/{2}}
		+ \mathcal{O}(\l^{-1}) ~,
		\qquad \tilde e=\f{\tilde\m_{l}+\tilde\m_{r}}{2}\,\l^{-{1}/{2}}+ \mathcal{O}(\l^{-1})~,
	\ee
	and \eqref{eq:AdS3-Vir-1-1} yields $\sinh^2\vartheta=\mathcal{O}( \l^{-{1}/{2}})$, $e=|\tilde m|+\mathcal{O}(\l^{-{1}/{2}})$ and  $E_0=|\tilde m|\,\sqrt{\l}+\mathcal{O}(\l^0)$.

	The case $0<\tilde n^2<\tilde m^2$ is analyzed similarly. Its large $\lambda$ behavior
	is govern by
	\be\label{eq:AdS3-large lambda '}
		\cos2\tilde\vartheta=-1+\f{\tilde\m_{l}+\tilde\m_{r}}{\sqrt{ m^2-n^2}}\,\l^{-\f{1}{2}}+ \mathcal{O}(\l^{-1}) ~,
		\qquad \tilde p=\f{\tilde\m_{l}-\tilde\m_{r}}{2}\,\l^{-\f{1}{2}}+ \mathcal{O}(\l^{-1})~,
	\ee
	which again follows from \eqref{eq:AdS3-eq for cos th} and \eqref{eq:AdS3-e,p tilde}.
	Writing now the first equation of \eqref{eq:AdS3-Virasoro 1} as
	\be\label{eq:AdS3-Vir-1-2}
		e^2-2n^2\sinh^2\vartheta=\tilde p^2+2\tilde m^2\cos^2\tilde\vartheta-\tilde n^2\cos2\tilde\vartheta~,
	\ee
	we find $\sinh^2\vartheta=\mathcal{O}( \l^{-{1}/{2}})$,
	$e=|\tilde n|+\mathcal{O}(\l^{-{1}/{2}})$ and  $E_0=|\tilde n|\,{\l}^{{1}/{2}}+\mathcal{O}(\l^0)$.

	The analysis of the case $|\tilde m|=|\tilde n|$ is the most simple and it leads to the same answer.
	Thus, if $\tilde m\neq 0$ and $\tilde n\neq 0$, the leading order behavior of $E_0$ is given by
	\be\label{eq:AdS3-general E}
		E_0=\text{min}(|\tilde m|, |\tilde n|)\,\l^{{1}/{2}}+\mathcal{O}(\l^0)~.
	\ee

	Note that for $\tilde m=0=\tilde n$, from \eqref{eq:AdS3-Virasoro 1} one has $n=0$. The solution then becomes $\s$ independent and it describes the massless particle in ${\rm AdS}_3\times {\rm S}^3$.

	Now it remains to analyze the two cases
 $\tilde m=0$, $\tilde n\neq 0$ and $\tilde m\neq 0$, $\tilde n=0$.
	In the first case the system \eqref{eq:AdS3-Virasoro 1} reduces to
	\be\label{eq:AdS3-eq for e}
		e^2-2n^2\sinh^2\vartheta=\tilde e^2+2\tilde n^2\sin^2\tilde\vartheta~, \quad
		\sqrt{n^2e^2+n^4}\sinh^2\vartheta=\sqrt{\tilde n^2\tilde e^2+\tilde n^4}\sin^2\tilde\vartheta~.
	\ee
	Here, one has to use the same large $\l$ behavior as in \eqref{eq:AdS3-large lambda}
	\be\label{eq:AdS3-large lambda1}
		\cos2\tilde\vartheta=1-\f{|\tilde\m_{l}-\tilde\m_{r}|}{|\tilde n|}\,\l^{-{1}/{2}}+ \mathcal{O}(\l^{-1}) ~,
		\qquad \tilde e=\f{\tilde\m_{l}+\tilde\m_{r}}{2}\,\l^{-{1}/{2}}+ \mathcal{O}(\l^{-1})~.
	\ee
	From \eqref{eq:AdS3-eq for e} we then find
	\bea\label{eq:AdS3-e=1}
		&\sinh^2\vartheta=\f{|\tilde n(\tilde\m_{l}-\tilde\m_{r})|}{2n^2}\,\l^{-{1}/{2}}+ \mathcal{O}(\l^{-1})~, \qquad
		e^2={2|\tilde n(\tilde\m_{l}-\tilde\m_{r})|}\,\l^{-{1}/{2}} + \mathcal{O}(\l^{-1})~,&\quad~\\
		\label{eq:AdS-min E pre1}
		&E_0=\sqrt{2|\tilde n(\tilde\m_{l}-\tilde\m_{r})|}\,\,\l^{{1}/{4}} + \mathcal{O}(\l^{-{1}/{4}})~.&
	\eea
	Note that for $\tilde\m_{l}=\tilde\m_{r}$, the exact solution of the system takes the following simple form
	\be\label{eq:AdS3-m=m}
		\cos2\tilde\vartheta=1~, \qquad \tilde e={\tilde\m_{l}}\,\l^{{1}/{2}}=e~, \qquad \sinh^2\vartheta=0~,\qquad
		E_0=\tilde\m_{l}~,
	\ee
	and it corresponds to a particle solution in \eqref{eq:AdS3-AdS string solution}-\eqref{eq:AdS3-S string solution}.

	In the second case, $\tilde n=0$, the system \eqref{eq:AdS3-Virasoro 1} can be written in the form
	\be\label{eq:AdS3-eq 2 for e}
		e^2-2n^2\sinh^2\vartheta=\tilde p^2+2\tilde m^2\cos^2\tilde\vartheta~, \quad
		\sqrt{n^2e^2+n^4}\sinh^2\vartheta=\sqrt{\tilde m^2\tilde p^2+\tilde m^4}\cos^2\tilde\vartheta~.
	\ee
	The solutions of \eqref{eq:AdS3-eq for cos th} and \eqref{eq:AdS3-e,p tilde} at large $\l$ now are
	\be\label{eq:AdS3-large lambda 2}
		\cos2\tilde\vartheta=-1+\f{|\tilde\m_{l}+\tilde\m_{r}|}{|\tilde m|}\,\l^{-{1}/{2}}+ \mathcal{O}(\l^{-1}) ~,
		\qquad \tilde p=\f{\tilde\m_{l}-\tilde\m_{r}}{2}\,\l^{-{1}/{2}}+ \mathcal{O}(\l^{-1})~,
	\ee
	and \eqref{eq:AdS3-eq 2 for e} leads to
	\bea\label{eq:AdS3-e=2}
		&\sinh^2\vartheta=\f{|\tilde m(\tilde\m_{l}+\tilde\m_{r})|}{2n^2}\,\l^{-{1}/{2}}+ \mathcal{O}(\l^{-1})~, \qquad
			e^2={2|\tilde m(\tilde\m_{l}+\tilde\m_{r})|}\l^{-{1}/{2}} + \mathcal{O}(\l^{-1})~,&\quad~~\\
		\label{eq:AdS-min E pre2}
		&E_0=\sqrt{2|\tilde m(\tilde\m_{l}+\tilde\m_{r})|}\,\l^{{1}/{4}} +\mathcal{O}(\l^{-{1}/{4}})~.&
	\eea
	The case $\m_{l}=\m_{r}$ is again special with the following simple solution in the $SU(2)$ part
	\be\label{eq:AdS3-m=m 2}
		\tilde p=0,  \qquad \tilde e=|\tilde m|~, \qquad \cos^2\tilde\vartheta = \f{\tilde\m_{l}}{|\tilde m|}\,\l^{-{1}/{2}}~,
	\ee
	and the corresponding minimal energy is
	\be\label{eq:AdS3-min E}
		E_0=2\sqrt{|\tilde m \tilde\m_{l}|}\,\l^{{1}/{4}}\,+\mathcal{O}(\l^{-{1}/{4}})~.
	\ee

	In summary, the quantization scheme utilizing the isometry orbits appears to be applicable also for the semiclassical multi-spin string solutions. With considerably little effort, we quantized the system exactly and found an oscillator type energy spectrum \eqref{eq:AdS3-string spectrum}. Taking then the strong coupling limit, $\l \gg 1$, we found that for the case $\{\tilde m, \tilde n\} \neq 0$ the minimal energy \eqref{eq:AdS3-general E} scales as $\sqrt{\l}$, suggesting that the corresponding strings are long, while for $\tilde m = 0$ or $\tilde n=0$ we obtain $E_0 \propto \l^{1/4}$, see \eqref{eq:AdS-min E pre1} and \eqref{eq:AdS-min E pre2}, which is the typical scaling of short strings. Indeed, it is easy to see that only for the latter cases, which could be viewed as a limits of the spinning string \cite{Frolov:2002av}, the string becomes small and for $\cos2\tilde\vartheta = 1$, respectively, $\cos2\tilde\vartheta = -1$ contracts to a point.

	Especially, note that for $|\tilde n (\tilde \m_l - \tilde \m_r)| = 2$ in \eqref{eq:AdS-min E pre1} and $|\tilde m (\tilde \m_l + \tilde \m_r)| = 2$ in \eqref{eq:AdS-min E pre2} we obtain the Energy of strings dual to states in the Konishi supermultiplet \eqref{eq:In-KoAnDim} at leading order. Hence, contemplating on a supersymmetrization of the system, these are the case of interest.

\chapter{Conclusion and Outlook} \label{chap:ConclOutl}

	For more than fifteen years the AdS/CFT correspondence \cite{Maldacena:1997re, Witten:1998qj, Gubser:1998bc} has led to tremendous progress in our perception of both gauge and sting theory, initiating new directions of research in various areas of modern theoretical physics. For the best studied duality pair, the duality between {\NfourSYM} and the type IIB superstring in $\AdSxS$, the quantum integrability has been observed \cite{Minahan:2002ve, Beisert:2003tq} in the planar limit. 
	The conjecture that the quantum integrability holds generally meets all tests and allows to devise powerful techniques, with a recent highlight being the asserted solution of the spectral problem by the quantum spectral curve \cite{Gromov:2013pga, Gromov:2014caa}.

	At the same time, quantization of type IIB string theory in $\AdSxS$ from first principles seems to be under control only in the supergravity sector and in certain semiclassical long string limits. This is somewhat counterintuitive, as for the extension of the string becoming small compared to the radius $R$ of $\AdSxS$, the short string limit, one naively expects to be able to simply quantize the string perturbatively around flat space.
	But even though this idea was proposed more than a decade ago \cite{Gubser:2002tv} there has been little progress on quantization of short strings. This confronts us with the conundrum that apparently, despite all the advance of integrability based methods, we have hardly gained any intuition on quantization of the $\AdSxS$ superstring and of $\mathrm{AdS}$ string theories in general. In turn, it is even the more worrisome that results based on the surmised quantum integrability are not testable in the short string regime.

	In this PhD thesis we have investigated different paths to achieve a consistent perturbative quantization of the $\AdSxS$ superstring in the large 't Hooft coupling limit, giving us a better understanding of the respective energy spectrum and quantum symmetries. The motives for this are actually twofold. Our interest was not only to go beyond the long-string paradigm and devise quantities, which can be tested against the integrability based predictions. But generally, one also hopes to attain a deeper insight into perturbative quantization of strings in general curved space-times. 

	As first observed in \cite{Passerini:2010xc} and as endorsed by this thesis, a key role in this quest seems to be played by the zero modes. Contrary to the `stringy' non-zero modes, these are subject to the curvature of the space-time even at leading order, which requires their scaling behavior in 't Hooft coupling $\l$ to be different from the one of the non-zero modes. More precisely, the non-zero mode excitations appear do distort the curved space time seen by the zero modes.

	Especially, we focused on perturbative quantization of what could be called the prototypical short strings, the lowest excitations of the superstring dual to states in the Konishi supermultiplet. At strong coupling, $R^4\propto\l\gg0$, the Konishi anomalous dimension, respectively, the string energy was known at leading order \eqref{eq:In-GKPresult} since the original work \cite{Gubser:2002tv} and the integrability based predictions were stated in \eqref{eq:In-KoAnDim}.

	After giving a minimal summary of the relevant concepts and terminology of the AdS/CFT correspondence in \chapref{chap:Intro} and after reviewing bosonic and superstring theory in \chapref{chap:BosString} and \chapref{chap:SS}, we started our survey on quantization of the $\AdSxS$ superstring in \chapref{chap:PureSpinor} by reviewing the asserted calculation of the Konishi anomalous dimension in the setting of the pure spinor superstring \cite{Vallilo:2011fj, Mazzucato:2011jt}. 
	Despite the fact that we restrict our analysis to the bosonic subsector we found several inconsistencies. Namely, the particular scaling of the spatial ${\rm AdS}_5$ zero modes is not taking into account, by which ordering ambiguities are not under control and not all possible perturbative contributions were considered. Furthermore, neither scaling of other zero modes is determined, which appears to be especially proplematic for the longitudinal fluctuations along the temporal ${\rm AdS}_5$ direction. 
	Finally, the parameter $E_\cl$ determined by the Virasoro constraint and asserted to match \eqref{eq:In-KoAnDim} to $\ord{\l^{1/4}}$ is actually not the energy of the string and we observed a diviation even at $\ord{\l^0}$.

	Luckily, not all our efforts were in vain. The calculations in \chapref{chap:PureSpinor} emphasized that in principle perturbative quantization of the $\AdSxS$ string ought to be feasible using different coordinates. 
	Also, the fundamental ideas, to determine the string energy via first order perturbative corrections in the Virasoro constraint, still appeals. 
	All these ideas were already formulated in \cite{Passerini:2010xc}. But in contrast to this work we realized that the choice of light-cone gauge, where usually the world-sheet Hamiltonian yields quantities of the form $\sqrt{E \pm J}$, seems unfavorable for short strings, where the energy diverges as $E \propto \l^{1/4}$ while all other charges are finite.

	For this reasons, in \chapref{chap:SingleMode} we aimed for canonical quantization of the bosonic $\AdSxS$ string employing static gauge \cite{Jorjadze:2012iy}, which has the advantage that the world-sheet Hamiltonian immediately determines the spectrum of $E^2$ instead. By the guiding principle to explicitly keep the zero modes we constructed what we called the single-mode string, a semiclassical string solution generalizing the pulsating string \cite{deVega:1994yz, Minahan:2002rc}, which showed to be classical integrable and invariant under the isometries $\SO(2,4)\times\SO(6)$ even at the quantum level. 
	Arguing about SUSY corrections to the ordering ambiguities heuristically, we indeed recovered the Konishi anomalous dimension \eqref{eq:In-KoAnDim} to $\ord{\l^{1/4}}$. We then gave quantitative arguments, why this approach did capture the energy only for string states dual to the Konishi multiplet and not for higher excited states, were our result for the spectrum would have disagreed with integrability based predictions \cite{Frolov:2012zv}.

	This work suggests various possible continuations. A nice exercise might be to see whether the same scheme can be reutilized for other known semi-classical string solutions. It appears rather straight forward to have the non-vanishing non-zero mode of the single-mode string on ${\rm S}^5$ instead of ${\rm AdS}_5$.
	Furthermore, preliminary studies suggest that the same construction is possible for the spinning \cite{Frolov:2002av}, respectively, multi-spin string solution \cite{Frolov:2003qc, Frolov:2003tu, Arutyunov:2003za}, see also \chapref{chap:AdS3}, while adding non-vanishing zero modes to the folded-spinning string \cite{Gubser:2002tv} seems to cause problems.

	Furthermore, in \secref{sec:SM-Decoupling} we have identified the bosonic operators, which contribute to $\ord{\l^{1/4}}$ in the energy of higher excited states via second order perturbation theory. Hence, for the next-to-lowest excited states one could calculate the corresponding contributions and, again heuristically arguing about SUSY corrections, see whether one finds agreement with the prediction in \cite{Frolov:2012zv}.

	The fundamental question has to be of course, whether our heuristic reasoning on the effects of SUSY reflects the actual behavior of the superstring. We have started to address this question in \chapref{chap:StaticFermions}, where we generalized the setup of static gauge from the bosonic to the $\AdSxS$ superstring, leading to nice formulas for the Lagrangian density and the $\psu(2,2|4)$ algebra operators. 
	Next, we contemplated on the form of a corresponding single-mode ansatz including fermionic degrees of freedom, where the found mode cutting seems to correspond to a supersymmetric extension of the single-mode string with non-vanishing non-zero mode in ${\rm S}^5$ rather than in ${AdS}_5$.

	However, as had to be expected, canonical quantization is prevented as the fermions are not in canonical form. To establish access to the problem in \secref{sec:SF-SParticle} we restricted ourselves to the $\AdSxS$ superparticle. Here, additional reasons to do so were that analogously to the bosonic zero modes we also expect the fermionic zero modes to scale differently from the non-zero modes and thorough understanding of the bosonic and fermionic zero modes seems to be a prerequisite to devise a supersymmetric generalization of the single-mode string. 
	By utilizing the residual $\un(1)\oplus\su(2)\oplus\su(2) \oplus\su(4)$ algebra left after assuming static gauge we managed to diagonalize the kinetic Lagrangian quadratic in fermions, viz., to bring the symplectic form at leading order to canonical form.

	But as direct diagonalization of higher order terms in the kinetic Lagrangian seems extremely tedious we proposed and explored another way to derive a transformation to canonical fermions, which is, to construct the $\psu(2,2|4)$ algebra from intrinsically canonical phase space variables and then read off the transformation by comparison with the charges obtained for the initial non-canonical variables. 
	Knowing the bosonic charges at vanishing order in fermions, we derived the fermionic charges at linear order in fermions. The found expressions appeared to be consistent with the previous results and took an intriguingly simple form. We ended the chapter by sketching how to successively continue the scheme with increasing power in fermions.

	The simplicity of the found fermionic charges certainly encourages to continue the proposed scheme, it however also insinuates that the final answer for the $\psu(2,2|4)$ charges to all orders in canonical fermions might take a very simple and symmetric form. To understand the mechanism at work better and to reduce the computational effort it therefore seems beneficial to first look at a simpler toy model, e.g., the ${\rm AdS}_2 \times {\rm S}^2$ or  the ${\rm AdS}_3 \times {\rm S}^3$ superparticle.

	In \chapref{chap:AdS3} we presented the work \cite{Heinze:2014cga}, which should certainly be viewed in this context. There we studied yet another quantization scheme, viz., quantization by use of the symmetry group orbits \cite{Alekseev:1988ce}. Constraining ourselves once more to the bosonic case, this method has been established first for the  ${\rm AdS}_3 \times {\rm S}^3$ particle, resulting in a one-parameter family of orbits, and was then applied to the multi-spin string solution \cite{Frolov:2003qc, Frolov:2003tu, Arutyunov:2003za}, where the orbits depended on two parameters plus winding numbers.
	In both cases it showed convenient to work in terms of the pre-symplectic one-form and we found a description in terms of creation-annihilation variables, naturally resulting in a Holstein-Primakoff realization of the isometry group generators \cite{Holstein:1940zp, Dzhordzhadze:1994np} and yielding oscillator-type energy spectra. We concluded the chapter by identifying multi-spin string solutions corresponding to long and short strings.

	It has been promising to observe that this quantization scheme allows for an exact quantization of both the particle as well as classical string solutions but it is furthermore astounding with what ease this has been achieved. Obviously, the next step would be to generalize the scheme to the ${\rm AdS}_3 \times {\rm S}^3$ superparticle, i.e., to orbits of the superisometries. This will then not only serve as a benchmark for the $\AdSxS$ superparticle but hopefully also for a generalization of the method to supersymmetrized semiclassical string solutions, which we hope are capable of determining the Konishi anomalous dimension to $\ord{\l^{-1/4}}$.

	Another topic of current relevance are the superstring theories in ${\rm AdS}_2\times{\rm S}^2 \times {\rm T}^6$ and ${\rm AdS}_3\times{\rm S}^3 \times {\rm M}_4$, see for example \cite{Hoare:2014kma}, respectively,  \cite{Borsato:2014exa, Sfondrini:2014via}. In this context, it is natural to explore, whether our findings also lead to interesting applications for these cases. It might show useful to compare with the most recent works on semiclassical strings, see for example \cite{Ahn:2014tua, David:2014qta, Hernandez:2014eta} and references therein.

	As commented on in \chapref{chap:StaticFermions}, there is yet another question to be answered. In the discussion above the complexity of the $\AdSxS$ superstring has been reduced by focusing instead on quantization of semiclassical strings or even the respective particles in curved space-time. But with the aim to understand static gauge quantization of the $\AdSxS$ superstring another intermediate step is to first understand the static gauge quantization for the superstring in flat-space. 
	Preliminary studies \cite{privateGeorge2014} show that for the RNS-superstring this works analogously to the bosonic string in flat space \cite{Jorjadze:2012iy}. Of course, the interesting case is the Green-Schwarz superstring and we would like to continue our investigation in this direction.

\chapter*{Acknowledgements}
\addcontentsline{toc}{chapter}{Acknowledgements}
	First of all, I would like to thank my supervisor Jan Plefka for giving me the possibility to write this thesis in his group, for always being open for discussions and for fruitful collaboration. Throughout my time as a PhD student he has led the quantum field and string theory group at Humboldt-University Berlin most competently resulting in a pleasant working atmosphere, which I enjoyed very much.

	I would like to thank Stefan Fredenhagen, who has organizing the International Max Planck Research School during most of my PhD and who has been a competent and most pleasant person of contact at the Albert Einstein Institute in Potsdam-Golm. Especially, I am indebted to both Jan Plefka and Stefan Fredenhagen for supporting me in my personal situation as a young parent.

	I am grateful to the administrative staff at Humboldt-University and the Albert Einstein Institute, especially Sylvia Richter and Christine Gottschalkson, for being always a proficient support concerning any bureaucratic matters.

	Furthermore, I am indebted to the staff at Nordita in Stockholm, and in particular to Paolo Di Vecchia, Larus Thorlacius, 	and Konstantin Zarembo, who gave me the opportunity to perform parts of my PhD as a visitor at Nordita. 

	I would like to thank Sergey Frolov, Luka Megrelidze, and George Jorjadze for fruitful collaborations. Especially, I would like to thank Sergey Frolov and George Jorjadze for enjoyable discussions and for sharing their expertise with me, from which I profited immensely.

	I am grateful to Brenno Vallilo for constructive discussion of the analysis in \chapref{chap:PureSpinor}.

	For interesting and mostly pleasant discussions I am thankful to Marco Stefano Bianchi, Harald Dorn, Alessandra Cagnazzo, Pawel Caputa, Amit Dekel, Livia Ferro, Valentina Forini, Blaise Gout{\'e}raux, Sven Bjarke Gudnason, Ben Hoare, Chrysostomos Kalousios, Johan K{\"a}ll{\'e}n, Alexander Krikun, Thomas Klose, Florian Loebbert, Daniele Marmiroli, Tristan McLoughlin, Carlo Meneghelli, Vladimir Mitev, Matin Mojaza, Christoph Sieg and Gang Yang. 

	I am grateful to my fellow PhD students Louise Anderson, Mohammad Assadsolimani, Lorenzo Bianchi, Benedikt Biedermann, Xinyi Chen-Lin, Jan Fokken, Rouven Frassek, Ilmar Gahramanov, James Gordon, Nils Kanning, Dennis Müller, Hagen Münkler, Daniel Medina Rincon, Paul Radtke, Andreas Rodigast, Ralf Sattler, Theodor Schuster, Edoardo Vescovi, Matthias Wilhelm and particularly my roommates Konstantin Wiegandt, Sebastian Wuttke and Jonas Pollok for enjoyable conversations and for always providing a nice working environment.

	Especially, I would like to thank Jonas Pollok for reading the manuscript of this thesis and suggesting improvements, for helping to finalize it, for so many other favors, but most importantly for his companionship. I sincerely wish you and Anna all the best for the upcoming months.

	I am grateful to Forough Ebadian for helping me and my wife Ghazaleh so much with our son Noa during the last months and years, which enabled me to spend more time on scientific work.

	I would like to thank my friends Volker Barg, Petter Cronlund, Manuel Freis, Lukas Gilz, Yoshi Herz, David Hollertz, Lars Jahnke, John Kaneko, Alexander Knauf, Hugo Neto, Denis N{\"o}sges, Christopher Rausch, Ellika Steenberg, and Marius Wagner for supporting me emotionally and for occasionally distracting me from my thesis.

	Mein besonderer Dank gilt meiner Familie, meinem Vater Wolfgang, der sich immer um das Voranschreiten meiner Arbeit sorgte, meinem Bruder Georg, der obwohl auch er sich r{\"a}umlich weiter von mir entfernt hat mir immer noch mit am n{\"a}chsten steht, und meiner Mutter Agnes, die mir stets ein Vorbild war die mich trotz all meiner Entscheidungen immer bedingungslos unterst{\"u}tzt hat. Ich vermisse euch sehr.

	Finally, I want to thank my wife Ghazaleh, without whom this thesis would not have been possible. Thank you for baring most responsibilities during the last months and for taking care not only of our son Noa but also of me. I love you.

\appendix
\chapter{Notation} \label{app:Notation}
	Apart from one obvious exceptions, throughout the text we use natural units
	\be
		c = \hbar = G = 1~.
	\ee
	and Einstein sum convention, i.e., repeated upper and lower indices are summed over.
	
	For the metric $G_{\m\n}(x)$ we adopt the 'mostly plus' convention, which is that spatial (temporal) directions have positive (negative) eigenvalues. Hence, the metric of flat space with $d_t$ temporal and $d_s=d-d_t$ spacial dimensions\footnote{We mostly use the character $d$ for the number of total dimensions as $D$ will denote the covariant derivative as well as the dilatation operator. In \secref{sec:SF-SParticle}, also $d$ will denote a dilatation like operator.} $\Reals^{d_t, d_s}$ reads
	\be
		\eta_{M N} =\diag(\underbrace{-1,-1,\ldots,-1}_{d_t\ \text{times}},\underbrace{1,1,\ldots,1}_{d_s\ \text{times}})
	\ee
	where usually the indices run as $M,N=-(d_t-1),\ldots,0,1,\ldots,d_s$.

	Cases of particular interest are flat Minkowski space $\Reals^{1,d-1}$, where we rather tend to use greek indices $\m,\n,\ldots\ $, the embedding space $\Reals^{d+1} = \Reals^{0,d+1}$ of the $d$-sphere ${\rm S}^d$ and the embedding space $\Reals^{2,d-1}$ of $d$ dimensional Anti-de Sitter space ${\rm AdS}_{d}$. For the latter we will mostly denote the temporal index $-1$ as $0'$, see for example \eqref{eq:In-AdSdef}.

	The relations between the 't Hooft coupling $\l$, the Yang-Mills coupling $\gYM$, the sting coupling $g_s$, the string tension $T_0$, the Regge slope $\a'$, and the the effective dimensional string tension $\pzg$ are found in \eqref{eq:In-tHooft} and \eqref{eq:SS-effST}.

\chapter{Single-Mode Strings -- \texorpdfstring{$\text{SO}({N_q})$}{SONq} Gauging } \label{app:SONq}

Here we consider a ${N_q}$-dimensional harmonic oscillator and calculate mean values of SO$({N_q})$ scalar operators
for the states with vanishing angular momentum.

Let us introduce the  annihilation and creation operators
\be\label{eq:SONq_aadag}
a_\a=\frac{p_\a-iq_\a}{\sqrt 2}~, \qquad a^\dag_\a=\frac{p_\a+iq_\a}{\sqrt 2}~, \quad (\a=1,\dots ,{N_q})~,
\ee
which satisfy the standard commutation relations
$[a_\a, a_\b]=[a_\a^\dag, a_\b^\dag ]=0,$ $\,[a_\a, a_\b^\dag ]=\d_{\a\b}.$

Three operators constructed by quadratic scalar combinations in $\v a$ and $\v a^{\,\dag}$,
\be\label{eq:SONq_ONqScalars}
h=\frac{1}{2}(\v a^{\,\dag}\cdot\v a + \v a\cdot\v a^{\,\dag})~, \qquad A= \v a^{\,2}~, \qquad A^\dag = \v a^{\,\dag\,2}~,
\ee
form the ${\mathfrak su}(1,1)$ algebra
\be\label{eq:SONq_so21}
[h, A]=-2A~, \qquad [h, A^\dag]=2A^\dag~,\qquad [A,\,A^\dag]=4h~.
\ee
The corresponding Casimir relates to the total angular momentum operator ${\cal J}^2=\frac{1}{2}\,{\cal J}_{\a\b}{\cal J}_{\a\b}$ as
\be\label{eq:SONq_ScalarOpRels}
h^2-\frac{1}{2}(A\,A^\dag+A^\dag\,A)={{\cal J}^2}+\f{{N_q}^2}{4}-{N_q}~,
\ee
where the SO$({N_q})$ rotation generators ${\cal J}_{\a\b}=i(a^\dag _\a a_\b-a^\dag _\b a_\a)$ annihilate the states
\be\label{eq:SONq_ScalarStates}
|n\rangle \propto \, (A^\dag)^n |0\rangle~,
\ee
$|0\rangle$ being the standard vacuum.

Hence, $h |n\rangle=(2n+{N_q}/2) |n\rangle$, and
 by  \eqref{eq:SONq_ScalarOpRels} one finds
\be\label{eq:SONq_ExpVal-AAdag}
\langle n |(A\,A^\dag+A^\dag\,A)|n\rangle=8\,n^2+4\,{N_q}\,n+2{N_q}~.
\ee

With the help of \eqref{eq:SONq_so21}, from \eqref{eq:SONq_ScalarOpRels} we also obtain
\be\label{eq:SONq_ExciteAnnihilate}
A|n\rangle=\sqrt{2n(2n+{N_q}-2)}\,\,|n-1\rangle ~,  \quad A^\dag|n\rangle=
\sqrt{(2n+{N_q})(2n+2)}\,\,|n+1\rangle~,
\ee
which define the normalization of the states \eqref{eq:SONq_ScalarStates}.

By \eqref{eq:SONq_aadag}-\eqref{eq:SONq_ONqScalars} one has $\v p^{\,2}=h+\f{1}{2}(A+A^\dag)$, $\,\v q^{\,2}=h-\f{1}{2}(A+A^\dag)$, $D\equiv\frac{1}{2}(\v p\cdot\v q+\v q\cdot\v p)=\f{i}{2}(A-A^\dag),$ and using \eqref{eq:SONq_ScalarOpRels} once more,
one obtains the following mean values
\begin{align}
	\label{eq:SONq_MeanVals1}
	&\langle n |\v p^{\,2} |n\rangle=\langle n |\v q^{\,2} |n\rangle=2\,n+\f{N_q}{2}~,\\ 
	\label{eq:SONq_MeanVals2}
	&\langle n |D^2 |n\rangle = 2\,n^2+{N_q}\,n+\f{N_q}{2}~,~~~~~~\\
	\label{eq:SONq_MeanVals3}
	&\langle n | (\v q^{\,2})^2 |n\rangle
	=6\,n^2+3\,{N_q}\,n+\f{{N_q}^2}{4}+\f{N_q}{2}~,\\
	\label{eq:SONq_MeanVals4}
	&\langle n |\frac{1}{2}\left(\v p^{\,2}\v q^{\,2}+\v q^{\,2}\v p^{\,2}\right) |n\rangle
	=2\,n^2+{N_q}\,n+\f{{N_q}^2}{4}-\f{{N_q}}{2}~.
\end{align}

These equations show the dependence of mean values for different SO$({N_q})$ scalar operators
on the dimension of space ${N_q}$.

\chapter{Static Gauge Superstring -- Additional Material} \label{app:OmCalc}

\section{\texorpdfstring{$\su(2,2)$}{su22} Algera Relations} \label{appsec:OC-su22}
	The generators defined in \eqref{eq:SF-SP-Ldef}, \eqref{eq:SF-SP-Lbdef} and \eqref{eq:SF-SP-EPKdef} fulfill the $\su(2,2)$ algebra Poisson bracket relations
	\begin{align} 
		&\{E^2 , J_{0'}^{\da \b}\} = - E J_0^{\da \b}~,\quad \{E^2 , J_{0}^{\da \b}\} = + E J_{0'}^{\da \b}~,\quad
			\{J_{0',\a \da} , J_{0}^{\db \b}\} =  E\,\d_\a{}^\b \d^\db{}_\da~, \\
		&\{J_{0',\a \da} , J_{0'}^{\db \b}\} = \{J_{0,\a \da} , J_{0}^{\db \b}\} 
			= -i (L_\a{}^\b \d^\db{}_\da + \d_\a{}^\b \tilde L^\db{}_\da)~,\\[.2em]
		&{1\ov i}\{ E, P_{\a\db}\} = + P_{\a\db}~,\qquad{1\ov i}\{ E, K^{\da\b}\} = -K^{\da\b}~,\\
		&{1\ov i}\{ P_{\a \da}, K^{\db \b}\} = -L_{\a}{}^{\b} \d^\db{}_\da - \d_\a{}^\b \tilde L^{\db}{}_{\da} 
			- E\,\delta_\a{}^\b \delta^\da{}_\db~,\\[.2em]
		&{1\ov i}\{ L_\a{}^\b, L_\g{}^\d\} = \d_\g{}^\b L_\a{}^\d - \d_\a{}^\d L_\g{}^\b\,,\qquad
			{1\ov i}\{\tilde L^\da{}_\db, \tilde L^\dg{}_{\dot\d}\} 
				= \d^\da{}_{\dot\d} \tilde L^\dg{}_\db - \d^\dg{}_\db \tilde L^\da{}_{\dot\d}\,, \label{eq:LLbrak}\\
		&{1\ov i}\{ L_\a{}^\b, P_{\g \dg}\} = \d_\g{}^\b P_{\a \dg} - \frac{1}{2} \d_\a{}^\b P_{\g \dg}~,\qquad
			{1\ov i}\{ \tilde L^\da{}_\db, P_{\g \dg}\} = \d^\da{}_\dg P_{\g \db}-\frac{1}{2} \d^\da{}_\db P_{\g \dg}~,\\
		&{1\ov i}\{ L_\a{}^\b, K^{\dg \g}\} = - \d_\a{}^\g K^{\dg \b} + \frac{1}{2} \d_\a{}^\b K^{\dg \g}~,\quad
			{1\ov i}\{ \tilde L^\da{}_\db, K^{\dg \g}\} = \frac{1}{2}\d^\da{}_\db K^{\dg \g} -\d^\da{}_\dg K^{\db \g}~,
	\end{align}
	where the last two lines are invariant under exchange of $P$ and $K$, $P_{\g \dg} \rightarrow K_{\g \dg}$ and $K^{\dg \g} \rightarrow P^{\dg \g}$. In particular we have the vanishing Poisson brackets
	\be
		\{L_\a{}^\b,  \tilde L^\da{}_{\db}\} =
			\{L_\a{}^\b,E\}=\{\tilde L^\da{}_\db,E\} = 0~,\quad
			\{ P_{\a\da},P^{\db\b}\} = \{ K_{\a\da},K^{\db \b}\} = 0~.
	\ee

\newpage

\section{Useful Formulae} \label{appsec:OC-UF}
	The quantities defined in \ssecref{subsec:SF-SP-su22su4} fulfill the following relations
	\bea
	&p^2=p_{\g\dg}\, p^{\g\dg}\,,\quad x^2=x_{\g\dg}\, x^{\g\dg}\,,\quad d=p_{\g\dg}x^{\g\dg}\,,&\\[.3em]
	&p_{\a\dot\g}p^{\b\dot\g}={1\ov2}p^2\d_\a{}^\b\,,\quad x_{\a\dot\g}x^{\b\dot\g}={1\ov2}x^2\d_\a{}^\b\,,&\\[.3em] &p_{\g\dot\a}p^{\g\dot\b}={1\ov2}p^2\d_{\dot\a}{}^{\dot\b}\,,\quad x_{\g\dot\a}x^{\g\dot\b}={1\ov2}x^2\d_{\dot\a}{}^{\dot\b}\,,&\\[.3em]
	&p_{\a\dot\g}x^{\b\dot\g}={d\ov2}\d_{\a}{}^{\b} +i\, L_{\a}{}^{\b}\,,\quad x_{\a\dot\g}p^{\b\dot\g}={d\ov2}\d_{\a}{}^{\b} -i\, L_{\a}{}^{\b}\,,&\\[.3em]
	&p_{\g\dot\a}x^{\g\dot\b}={d\ov2}\d_{\dot\a}{}^{\dot\b} +i\,\tilde L_{\dot\a}{}^{\dot\b} \,,\quad x_{\g\dot\a}p^{\g\dot\b}={d\ov2}\d_{\dot\a}{}^{\dot\b} -i\,\tilde L_{\dot\a}{}^{\dot\b} 
	\,,&\\[.3em]
	&\e_{\g\r}L_\alpha{}^{\r}p^{\g\dot\b}={i\ov2}d\,p_{\a}{}^\db-{i\ov2}p^2x_{\a}{}^\db\,,\quad \e_{\g\r}L_\alpha{}^{\r}x^{\g\dot\b}=-{i\ov2}d\,x_{\a}{}^\db+{i\ov2}x^2p_{\a}{}^\db\,,&\\[.3em]
	&L_\a{}^\g L_\g{}^\b = {S^2\ov4}\,\d_\a{}^\b \,,\quad L_\a{}^\g \e_{\g\r}L_\b{}^\r = -{S^2\ov4}\,\e_{\a\b} \,,\quad S^2=p^2x^2-d^2\,,&\\[.3em]
	&(x^2p_{\g\da}-d\,x_{\g\da})(x^2p^{\g\db}-d\,x^{\g\db})={1\ov2}x^2S^2\delta_\da^\db\,,&\\[.3em]
	&(x^2p_{\g\da}-d\,x_{\g\da})x^{\g\db}=ix^2\tilde L_\da{}^\db\,,
	\quad 
	x_{\g\da}(x^2p^{\g\db}-d\,x^{\g\db})=-ix^2\tilde L_\da{}^\db\,,&\\[.3em]
	&y_{AC}y^{CB}=\d_A^B\,,
	\quad p_{AC}p^{CB}={\Ms^2\ov 16}\d_A^B\,,
	\quad   C_{py}=p_{AB}y^{BA} \approx 0\,,&\\[.3em]
	&p_{AC}y^{CB}+p^{BC}y_{CA}={1\ov2}C_{py} \d_A^B \approx 0\,.&\\[.3em]
	&R_A{}^B = {1\ov i}(p_{AC}y^{CB}-p^{BC}y_{CA})\approx {2\ov i}p_{AC}y^{CB}\approx -{2\ov i}p^{BC}y_{CA}\,,&\\[.3em]
	&R_A{}^C R_C{}^B= {\Ms^2\ov 4}\delta^{B}_{A}\,,&\\[.3em]
	&p^{AC}R_C{}^B = -i{\Ms^2\ov 8}y^{AB}\,,\quad R_A{}^Cp_{CB} = i{\Ms^2\ov 8}y_{AB}\,,&\\[.3em]
	&y^{AC}R_C{}^B = 2i\,p^{AB}\,,\quad R_A{}^Cy_{CB} = - 2i\,p_{AB}\,,&
	\eea
	where $\approx$ marks weak equality.

\newpage
\section{Convenient form for \texorpdfstring{$\cV$}{cV} and \texorpdfstring{$\cW$}{cV}} \label{appsec:OC-cVWform}
	Especially since $\{E,S\}=\{L_\a{}^\b, S\}=\{\tilde{L}_\da{}^\db, S\}=0$ it seems convenient to work with
	\be
		\cV_\a{}^\b = \sqrt{\frac{2}{S(S - 2 L_1{}^1)}} \Big(L_\a{}^\b - \frac{S}{2} (\s_3)_\a{}^\b\Big)~.
	\ee
	Using matrix notation, with $L^2 = \frac{S^2}{4} \mI_2$ and $L^\dag = L$ for this we have
	\begin{align}
		&\cV = \cV^\dag~,\qquad\{\s_3, L\}_+ = 2 L_1{}^1\,\mI_2 = (S_1 + S_2)\,\mI_2~,\nn\\
		&\cV \cV^\dag = \cV^\dag \cV = \frac{2}{S(S - 2 L_1{}^1)} \left(\frac{S^2}{4} (1+1) \mI_2 - \frac{S}{2} 2 L_1{}^1\,\mI_2 \right) = \mI_2~,\\
		&L \cV = \sqrt{\frac{2}{S(S - 2 L_1{}^1)}} \left(\frac{S^2}{4} \mI_2 - \frac{S}{2}\,L \s_3 \right) = \calV \left(\frac{-S}{2} \s_3\right)~,\nn
	\end{align}
	hence $L = \cV (\frac{-S}{2} \s_3) \cV^\dag$, as anticipated.

	For $R$, $R^2=\frac{\Ms^2}{4}\mI_4$, we can attempt to use the same trick
	\be
		\tilde{\cW} =\tilde{\cW}^\dag = R - \frac{\Ms}{2} \S~,\quad\Rightarrow\quad 
			R \tilde{\cW} = \frac{-\Ms}{2} \tilde{\cW} \Sigma~.
	\ee
	but due to $\tilde{\cW}$ being $4\times4$ instead of $2\times2$ as $\cV$ it is not unitary. In particular, let us split up $R$ similar to the $\text{AdS}_5$ isometries as
	\begin{align}
		& R = \begin{pmatrix} L_\alg{s} + \frac{J_3}{2}\,\mI_2 & P_\alg{s} \\ K_\alg{s} & \tilde{L}_\alg{s} -  \frac{J_3}{2}\,\mI_2 \end{pmatrix}~,\quad 
			\{\Sigma, R\}_+ = J_3\,\mI_4 + 2 \begin{pmatrix} L_\alg{s} & 0 \\ 0 & -\tilde{L}_\alg{s}\end{pmatrix} \\
		&L_\alg{s} = \frac{1}{2} \begin{pmatrix} J_1 + J_2 & i(\bar\calP^\alg{s}_1 \bar\calY^2 - \bar\calP^\alg{s}_2 \bar\calY^1) \\ -i(\calP^\alg{s}_1 \calY^2 - \calP^\alg{s}_2 \calY^1) & - J_1 - J_2 \end{pmatrix}
		~,\quad L_\alg{s}^2 = \frac{J^2}{4} \mI_2 ~,\nn\\
		&\tilde{L}_\alg{s} = \frac{1}{2} \begin{pmatrix} J_1 - J_2 & -i(\bar\calP^\alg{s}_1 \calY^2 - \calP^\alg{s}_2 \bar\calY^1) \\ i(\calP^\alg{s}_1 \bar\calY^2 - \bar\calP^\alg{s}_2 \calY^1) & -J_1 + J_2\end{pmatrix}
		~,\quad \tilde{L}_\alg{s}^2 = \frac{J^2}{4} \mI_2 ~,\nn\\
		&J^\alg{s}_{k\,l} = P^\alg{s}_k Y^l - P^\alg{s}_l Y^k~,\quad J_k = J^\alg{s}_{(2k-1),2k}~,
			\quad J = \frac{1}{2} \sum_{k,l=1}^4 (J^\alg{s}_{k\,l})^2 ~,\nn
	\end{align}
	with $J$ being the Casimir of $\text{SO}(4)\subset\text{SO}(6)$, the analogue of $S$. Then we get
	\be
		\tilde{\cW}^\dag \tilde{\cW} = \frac{\Ms}{2}\left(\Ms \mI_4 - \{\Sigma, R\}_+\right)
			= \frac{\Ms}{2}\left((\Ms - J_3) \mI_4 - 2 \begin{pmatrix} L_\alg{s} & 0 \\ 0 & -\tilde{L}_\alg{s} \end{pmatrix}\right) ~,
	\ee
	and we are left with diagonalization of $L_\alg{s}$ and $\tilde{L}_\alg{s}$, so we use the same trick once more.
	\begin{align}
		&\tilde{w} = \tilde{w}^\dag = \begin{pmatrix} \sqrt{\frac{2}{J(J+2 (L_\alg{s})_1{}^1)}} \left(\frac{J}{2} \s_3 + L_\alg{s}\right) & 0 \\ 0 & \sqrt{\frac{2}{J(J-2 (\tilde{L}_\alg{s})_1{}^1)}} \left(\frac{J}{2} \s_3 - \tilde{L}_\alg{s}\right)\end{pmatrix}~,\\
		&\tilde{w}^\dag \tilde{w} = \!\!\begin{pmatrix} \frac{2}{J(J+2 (L_\alg{s})_1{}^1)} \left( \frac{J^2}{2} \mI_2 + \frac{J}{2} \{L_\alg{s}, \s_3\}_+ \right) & 0 \\ 0 & \frac{2}{J(J-2 (\tilde{L}_\alg{s})_1{}^1)} \left(\frac{J^2}{2} \mI_2  - \frac{J}{2} \{\tilde{L}_\alg{s}, \s_3\}_+ \right)\end{pmatrix}\!\! = \mI_4~,\nn\\[.4em]
		&(\tilde{w}^\dag \tilde{\cW}^\dag \tilde{\cW} \tilde{w})_A{}^B
			= \frac{\Ms}{2}\left((\Ms - J_3) \mI_4 - J \begin{pmatrix} \s_3 & 0 \\ 0 & \s_3 \end{pmatrix}\right)_A^{\ \ B} \nn\\
		&\qquad=  \frac{\Ms}{2}\left(\Ms + J_3 + (-1)^A J\right) \d_A{}^B~.\nn
	\end{align}
	By $2 (L_\alg{s})_1{}^1 = J_1 + J_2$ and $2 (\tilde{L}_\alg{s})_1{}^1 = J_1 - J_2$ we therefore have 
	\be \label{eq:cWfinal}
		\cW_A{}^B = \frac{\left(R-\frac{\Ms}{2}\Sigma\right)_A{}^C \Big(J\,(\mI_2\times\s_3) + \{R,\Sigma\}_+ - J_3\, \mI_4\Big)_C{}^B}{\sqrt{\Ms J (\Ms + J_3 + (-1)^A J) (J + \Sigma_A{}^A\,J_1 +J_2)}} \propto \left(\tilde{\cW} \tilde{w}\right)_A{}^B ~,
	\ee
	where we want to stress again that $\tilde{\cW}=\tilde{\cW}^\dag$ and $\tilde{w}=\tilde{w}^\dag$ but $\tilde{\cW} \tilde{w} \propto \cW \neq \cW^\dag \propto \tilde{w} \tilde{\cW}$.
	
	It is unsatisfacory that $\cW$ does not possess higher symmetry, e.g., in \eqref{eq:cWfinal} $J_3$ plays a different role than $J_1$ and $J_2$. We could generalize the previous by noting that
	\be
		R \Big(R A - \frac{\Ms}{2} A \S\Big) = \Big(R A - \frac{\Ms}{2} A \S\Big) \Big(\frac{-\Ms}{2} \S\Big) ~,
	\ee
	i.e., $(R A - \frac{\Ms}{2} A \S)$ is a matrix of eigenvectors for any matrix $A$. Hence, one would like to find an $A$ such that
	\be
		(R A  \frac{\Ms}{2} A \S)^\dag (R A - \frac{\Ms}{2} A \S) 
		= \frac{\Ms^2}{4}(A^\dag A + \S A^\dag A \S) - \frac{\Ms}{2} \{A^\dag R A,\S\}_+ = \mI_4~.
	\ee
	We haven't been able to find a more convenient $A$. Taking $A$ a permutation matrix indeed interchanges the angular momenta $J_i$, as has to be expected.

\newpage
\section{Supercharge Poisson Brackets} \label{appsec:OC-SupPoisBra}
	Introducing abbreviations
	\begin{align} \label{eq:OC-lrangles}
		&\langle c_V c^*_V \rangle = \left(b_V b^*_V + b_W b^*_W + c_V c^*_V + c_W c^*_W\right)~,\\
		&\langle c_V b^*_V \rangle = \left(c_V b^*_V + b_V c^*_V + c_W b^*_W + b_W c^*_W\right) ~, \nn\\
		&\langle c_V c^*_W \rangle = \left(b_W b^*_V + b_V b^*_W + c_W c^*_V + c_V c^*_W\right) ~, \nn\\
		&\langle c_V b^*_W \rangle = \left(c_W b^*_V + c_V b^*_W + b_W c^*_V + b_V c^*_W\right) ~, \nn
	\end{align}
	and similar for $\langle c_V \tc^*_V \rangle$ and  $\langle \tc_V \tc^*_V \rangle$, etc., from the Poisson brackets which ought to read \eqref{eq:SF-SP-QSshould} we find the constraints
	\begin{align}\label{eq:OC-CoeffQS}
		\{Q,S\}:\quad&\om_\d \langle c_V c^*_V\rangle +\om_\s \langle c_V b^*_V\rangle +\om_\S \langle c_V c^*_W\rangle +\om_{\s\S} \langle c_V b^*_W\rangle  = \frac{E}{2} ~,\\
		&\om_\d \langle c_V b^*_V\rangle +\om_\s \langle c_V c^*_V\rangle +\om_\S \langle c_V b^*_W\rangle +\om_{\s\S} \langle c_V c^*_W\rangle  = \frac{S}{2} ~,\nn\\
		&\om_\d \langle c_V c^*_W\rangle +\om_\s \langle c_V b^*_W\rangle +\om_\S \langle c_V c^*_V\rangle +\om_{\s\S} \langle c_V b^*_V\rangle  = \frac{\Ms}{2} ~,\nn\\[.2em]
		&\om_\d \langle c_V b^*_W\rangle +\om_\s \langle c_V c^*_W\rangle +\om_\S \langle c_V b^*_V\rangle +\om_{\s\S} \langle c_V c^*_V\rangle  = 0 ~,\nn\\[.6em]
		\label{eq:OC-CoeffStQt}
		\{\tilde{S},\tilde{Q}\}:\quad&\om_\d \langle \tc_V \tc^*_V\rangle +\om_\s \langle \tc_V \tb^*_V\rangle +\om_\S \langle \tc_V \tc^*_W\rangle +\om_{\s\S} \langle \tc_V \tb^*_W\rangle  = \frac{E}{2} ~,\\
		&\om_\d \langle \tc_V \tb^*_V\rangle +\om_\s \langle \tc_V \tc^*_V\rangle +\om_\S \langle \tc_V \tb^*_W\rangle +\om_{\s\S} \langle \tc_V \tc^*_W\rangle  = \frac{S}{2} ~,\nn\\
		&\om_\d \langle \tc_V \tc^*_W\rangle +\om_\s \langle \tc_V \tb^*_W\rangle +\om_\S \langle \tc_V \tc^*_V\rangle +\om_{\s\S} \langle \tc_V \tb^*_V\rangle  = -\frac{\Ms}{2} ~,\nn\\
		&\om_\d \langle \tc_V \tb^*_W\rangle +\om_\s \langle \tc_V \tc^*_W\rangle +\om_\S \langle \tc_V \tb^*_V\rangle +\om_{\s\S} \langle \tc_V \tc^*_V\rangle  = 0 ~,\nn\\[.6em]
		\label{eq:OC-CoeffQQt}
		\{Q,\tilde{Q}\}:\quad&\om_\d \langle c_V \tb^*_W\rangle +\om_\s \langle c_V \tc^*_W\rangle +\om_\S \langle c_V \tb^*_V\rangle +\om_{\s\S} \langle c_V \tc^*_V\rangle  = 0 ~,\\[.3em]
		&\om_\d \langle c_V \tc^*_W\rangle +\om_\s \langle c_V \tb^*_W\rangle +\om_\S \langle c_V \tc^*_V\rangle +\om_{\s\S} \langle c_V \tb^*_V\rangle  = 0 ~,\nn\\
		&\om_\d \langle c_V \tb^*_V \rangle +\om_\s \langle c_V \tc^*_V\rangle +\om_\S \langle c_V \tb^*_W\rangle +\om_{\s\S} \langle c_V \tc^*_W\rangle  = \frac{S}{2}\sqrt{\frac{2}{x^2}} \frac{-i}{\sqrt{2}}~,\nn\\
		&\om_\d \langle c_V \tc^*_V\rangle +\om_\s \langle c_V \tb^*_V\rangle +\om_\S \langle c_V \tc^*_W\rangle +\om_{\s\S} \langle c_V \tb^*_W\rangle  \nn\\ 
		&\qquad= - \sqrt{\frac{x^2}{2}} \frac{1}{\sqrt{2}}\left(\frac{d}{x^2}(1-x^2) + i E\right) \nn~.
	\end{align}
	The quantities in \eqref{eq:OC-lrangles} are connected to $a_{j,k}$ and $\ta_{j,k}$ defined in \eqref{eq:SF-SP-aDef} and \eqref{eq:SF-SP-taDef} by (no sum over $j$ and $k$)
	\begin{align}
		|a_{j,k}|^2 = 4\,\om_{j,k} \Big(\langle c_V c^*_V\rangle + (-1)^j \langle c_V b^*_V\rangle 
			+ (-1)^k \langle c_V c^*_W\rangle + (-1)^{j+k} \langle c_V b^*_W\rangle\Big) ~,\nn\\
		|\ta_{j,k}|^2 = 4\,\om_{j,k} \Big(\langle \tc_V \tc^*_V\rangle + (-1)^j \langle \tc_V \tb^*_V\rangle
			+ (-1)^k \langle \tc_V \tc^*_W\rangle + (-1)^{j+k} \langle \tc_V \tb^*_W\rangle\Big) ~,\\
		a_{j,k}\,\ta^*_{j,k} = 4\,\om_{j,k} \Big(\langle c_V \tc^*_V\rangle + (-1)^j \langle c_V \tb^*_V\rangle 
			+ (-1)^k \langle c_V \tc^*_W\rangle + (-1)^{j+k} \langle c_V \tb^*_W\rangle\Big) ~.\nn
	\end{align}


\bibliographystyle{../../../Latex/bibSpires/nb}
\bibliography{../../../Latex/bibSpires/bibSpires}

\newpage
\section*{Software}
	\begin{itemize}
		\item This thesis has been typeset in {\LaTeXe} using the {\TeX}/{\LaTeX} editor {\it Kile}\\
		(\url{http://kile.sourceforge.net/}).
		\item For referencing we used the Java based BibTex manager {\it JabRef}\\ 
			(\url{http://jabref.sourceforge.net/}).
		\item Computations were aided by the use of {\it Mathematica 9.0} by {\it Wolfram Research}.
		\item Parts of the calculations used the Mathematica package {\tt grassmann.m} \cite{grassmann.m}.
	\end{itemize}



\end{document}